\documentclass[11pt,a4paper,openany]{thesis}
\usepackage{amsmath,amssymb,amsfonts,aas_macros,graphicx}
\usepackage{natbib}

\usepackage{epsf}
\bibpunct[,]{(}{)}{;}{a}{}{,}

\begin{document}
\pagenumbering{roman}
%
\begin{titlepage}
 \begin{center}
  \vspace*{3.0cm}
  {\Large\bfseries Strong Gravitational Lenses in a Cold Dark Matter
  Universe\\}
  \vspace{2cm}
  {\Large A dissertation submitted to the University of Tokyo\\
  in partial fulfillment of the requirements\\
  for the degree of Doctor of Science in Physics\\}
  \vspace{5.5cm}
  {\Large\scshape Masamune Oguri\\}
  \vspace{1.0cm}
  {\Large Department of Physics, University of Tokyo\\}
  \vspace{2cm}
  {\Large April, 2004}
 \end{center}
\end{titlepage}
\thispagestyle{empty}
\vspace*{9.0cm}
 \begin{center}
\copyright \hspace*{0.1cm} Copyright 2004 by Masamune Oguri\\ \medskip
All rights reserved.
\end{center}
\begin{abstract}
 We present theoretical and observational studies of strong
 gravitational lenses produced by clusters of galaxies. Our purpose is
 to test the Cold Dark Matter (CDM) model at small and highly non-linear
 scales where it has been claimed that the CDM model may confront
 several difficulties. We concentrate our attention on the statistics of
 strong gravitational lenses because the strong lensing is sensitive to
 the mass distributions of central, high-density regions of lensing
 clusters where the cold and collisionless hypotheses on dark matter are
 crucial. We use two complementary statistics, lensed arcs and quasars,
 to probe the mass distributions. 

 First, we construct a triaxial lens model, and develop a new method to
 include triaxiality of dark halos in the lens statistics. We find that
 the effect of triaxiality is significant; it enhances lensing
 probabilities by a factors of a few to ten, assuming the degree of
 triaxiality predicted in the CDM model. Thus it is essential to take
 triaxiality into account in the lens statistics.  In particular, we
 argue that both central concentration and large triaxiality of dark
 halos are required to reproduce the observed number of arcs in
 clusters; thus the result can be interpreted as a strong evidence for
 the cold and collisionless dark matter. 

 One of the most notable advantages of the triaxial modeling over the
 spherical modeling is that the triaxial modeling allows us to predict
 image multiplicities. We find that the CDM halos predict significant
 fraction (more than 20\%) of naked cusp lenses, unlike lensing by
 isothermal galaxies where naked cusp configurations are rare.  In
 addition, we point out the image multiplicities depend strongly on the
 central concentration of dark halos. Therefore we propose image
 multiplicities as a new powerful test of  the CDM model.

 While many lensed arcs are known, no quasar strongly lensed by clusters
 of galaxies has been discovered. We searched for large-separation
 lensed quasars from the data of the Sloan Digital Sky Survey, and
 succeeded in discovering the first large-separation lensed quasar SDSS
 J1004+4112. The system consists of four lensed images of a quasar at
 $z=1.73$. We identify the lensing cluster at $z=0.68$, from the deep
 imaging and spectroscopy of galaxies in the cluster. We calculate the
 expected probabilities and image multiplicities for lensed quasars in
 the SDSS, and find that the discovery of the large-separation quadruple
 lens SDSS J1004+4112 is quite consistent with the theoretical
 predictions based on the CDM model. 
\end{abstract}
\tableofcontents
\begin{acknowledgments}
Needless to say, this thesis would have not been possible without the
help of many people. First I have to apologize for not being able to
list all of them due to limited space and time.

I would like to thank my supervisor, Yasushi Suto, for his continuous 
encouragement and support. He has taught me many things, not only
specific scientific topics but also more general idea of how to conduct
the research; he has taught me how to find a research theme, how to
manage a collaboration well, how to conquer difficulties, and how to
advertise my research. The scientific topics might become out-of-fashion
some day, but such a general idea will continue to be useful throughout
my life. In addition, he always cares me and my future, and helps me to
determine the course to be taken. He also read the draft version of this
thesis carefully, and gave me many useful comments. Thanks.

I was lucky enough to have an opportunity to work with Naohisa Inada on 
the lens search in the SDSS data. This changed my research life
drastically. Before I met him, I was a pure theorist: But I learned from
him how exciting it is to handle real observational data. I never forget
the midnight of May 2, 2003 when we discovered SDSS J1004+4112 in the
SDSS data -- it was one of the most exciting events I have ever
experienced. I wish to thank him for inviting me to such an exciting
research field. 

Also, I would like to thank the other members of the SDSS lens search
team. In particular, I thank Bart Pindor, Joe Hennawi, Michael Gregg, 
Bob Becker, Fransico Castander, Gordon Richards, Daniel Eisenstein, Josh
Frieman, and Dave Johnston for the follow-up observations of lens
candidates. I am also grateful to Ed Turner, Michael Strauss, 
Pat Hall, Don Schneider, Paul Schechter, Tomotsugu Goto, Hans-Walter
Rix, Bob Nichol and Don York for useful comments and suggestions. We
couldn't discover so many lenses including SDSS J1004 without their
supports. I also thank Shin-Ichi Ichikawa for allowing me to use his
observing time at the Subaru 8.2-meter telescope for SDSS J1004; indeed,
it was a great experience to do observations at the summit of Mauna Kea.

The discovery of SDSS J1004 led to the collaboration with Chuck Keeton
on large-separation lensed quasars;  it was quite exciting to me. I was
impressed many times by the quality of his analysis, and by his
extensive knowledge of gravitational lensing. I'm sure that the thesis
work is greatly improved by the collaboration with him. 

The discovery also expanded my research field; it led to the
collaborations with radio/X-ray/optical observational groups. Masato
Tsuboi and Takeshi Kuwabara kindly taught me radio observations at
Nobeyama 45-meter telescope; Kazuhisa Mitsuda and Naomi Ota made a great
effort to write up a Chandra proposal which I am involved with. I thank
Satoshi Miyazaki for the ongoing collaboration on weak lensing analysis
of the cluster.

I enjoyed the collaboration with Jounghun Lee on the triaxial lens
modeling and also on dark halo substructures. She is always cheerful 
and friendly to everyone including shy persons like me, so I did enjoy
 discussing with her. 

I have been benefitted by other collaborations, though they are not
included in this thesis: I thank Ed Turner for many suggestions and
discussions. I really like the work of time delay statistics which made
his inspired vision a reality.  Atsushi Taruya taught me how to conduct
the research through the collaboration. I also thank Yozo Kawano for
interesting discussions on lens modeling, which has expanded my
understanding of strong lensing.
 
I could not have accomplished this thesis without the help,
discussion, and encouragement of many people. Some of these people
include: Y.~P.~Jing, Masahiro Takada, Eiichiro Komatsu, Takashi Hamana,
Tetsu Kitayama, Shin Sasaki, Naoki Yoshida, Takahiko Matsubara, Kaiki
Taro Inoue, Massimo Meneghetti, Matthias Bartelmann, Toshiyuki
Fukushige, and Ryuichi Takahashi.  I am also grateful to members of
observational cosmology group at UTAP, including Issha Kayo, Chiaki
Hikage, Mamoru Shimizu, Kohji Yoshikawa, Atsunori Yonehara, Kazuhiro 
Yahata, and other colleagues, for stimulating discussions. 

I'm blessed with good friends, Keitaro Takahashi, Kei Kotake, Kiyotomo
Ichiki, and Hiroshi Ohno, with whom I have collaborated on several
topics such as decaying cold dark matter model. We not only discussed
many cosmological and astrophysical issues, but also shared the dark
side. Indeed, it has been a fun to randomly discuss wild ideas with 
them. They are always very active, and I am always stimulated by their
activities. I was lucky to have such outstanding colleagues.  

I would like to thank all the members of UTAP for providing a
comfortable research environment to me. In particular, I would like to
thank Katsuhiko Sato for continuous encouragement during my graduate
student life. I have been able to concentrate on my research thanks to
the environmental effect. 

I would appreciate financial supports from JSPS through JSPS Research
Fellowship for Young Scientists.

Finally, I would like to thank my parents for encouraging me to go my
own way. Actually I know I'm not a dutiful son, but I'd appreciate
their support.  

\end{acknowledgments}
\newpage
\vspace*{9.0cm}
\newpage
\pagenumbering{arabic}
\chapter{Introduction}
\label{chap:intro}
\def\mychapheadname{Introduction}
\markboth{CHAPTER \thechapter.
{\MakeUppercase\mychapheadname}}{}

\section{The Dark Side of the Universe} 
\markboth{CHAPTER \thechapter.
{\MakeUppercase\mychapheadname}}{\thesection.
\MakeUppercase{The Dark Side of the Universe}}

One of the major goals of {\it cosmology} is to answer the simple
questions: What is the physical origin of the universe? How has the
universe evolved? What is the final fate of the universe? Surprisingly,
modern cosmology can partly answer these fundamental questions: During
the past decades, cosmologists were able to find a standard cosmological
model, namely a ``concordance'' model. In this model the universe is
homogeneous and isotropic, and the geometry is flat. The universe
consists of ordinary matter, radiation, dark matter, and dark energy. 
The structure and objects have been generated from small adiabatic
Gaussian fluctuations through gravitational instability.

\begin{figure}[tb]
\begin{center}
\includegraphics[width=0.5\hsize]{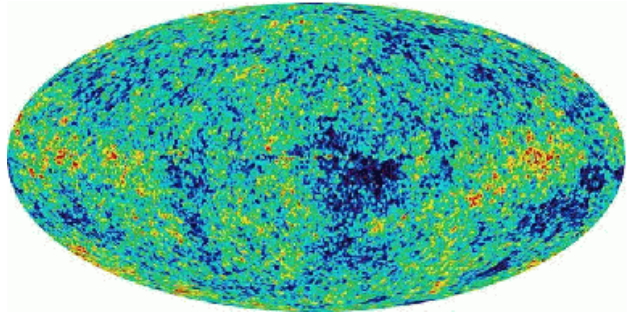}
\caption{A sky map of CMB anisotropies measured by {\it WMAP}
 \citep{bennett03}. 
\label{fig:intro_cmb}}
\end{center}
\end{figure}

This model is quite successful. The most representative observation which
demonstrates the success of the standard model would be Cosmic Microwave
Background (CMB) anisotropies. The patterns of tiny temperature
fluctuations on the $2.7$K background radiation represents the seeds of
the current cosmic structure. Recently, {\it Wilkinson Microwave
Anisotropy Probe} {\it (WMAP)} observed this tiny temperature
fluctuations in detail, and showed that the observed fluctuation
patterns (see Figure \ref{fig:intro_cmb}) show an excellent agreement
with the standard model predictions. In addition, the model is also
consistent with many independent cosmological observations, such as
distant type-Ia supernovae, clusters of galaxies, large-scale structure,
and big bang nucleosynthesis. Almost all observations can be explained
by the model, suggesting that we have reached a correct view of the
universe. 

However, the model we have reached looks unsatisfactory, since in the
model our universe is dominated by dark matter ($\sim 30$\%) and
dark energy ($\sim 70$\%) both of which we haven't understood yet. 
The ordinary matter we now know accounts for only $\sim 4$\% of the
total density of the universe, and the rest, i.e., $\sim 96$\% of the
universe is ``dark''. We know that dark matter should be non-baryonic, 
but it's still unknown what dark matter is. Good candidates for dark
matter particles include supersymmetric particles (e.g., neutralino) and
axion, but they could be totally brand-new particles. We don't know the
nature of dark energy, neither.  The cosmological constant has been thought
to be a candidate of dark energy, it is quite difficult to achieve such
small constant in the early universe. Dark energy may be a slowly rolling 
scalar field, but even if so we don't know what the scalar field is.
Thus, the current standard model is unstable in the sense that we have
to resort to unknown energy components. One of the main goals of
cosmology over the next decades would be, therefore, to find out what
the dark components are. In this thesis, we concentrate our attention on
dark matter, because the nature of dark matter is still very controversial;
indeed, the standard Cold Dark Matter (CDM) model may have confronted
several difficulties on small non-linear scales, such as
over-concentration of dark halos and over-production of substructures in
galaxy-scale dark halos. Since such small, highly non-linear structures
are sensitive to the nature of dark matter, these difficulties are often
regarded as an evidence that our current assumptions on dark matter is
wrong. 

\section{Gravitational Lensing: Revealing the Dark Side} 
\markboth{CHAPTER \thechapter.
{\MakeUppercase\mychapheadname}}{\thesection.
\MakeUppercase{Gravitational Lensing: Revealing the Dark Side}}

Although it is undoubtedly important to reveal the nature of dark
matter, the main problem lies in the fact that dark matter is {\it
dark}, i.e., it cannot be observed by usual methods. However, there is
one way to probe the distribution of dark matter {\it directly} --- 
gravitational lensing. 

General relativity predicts that gravitational fields around massive
objects, such as galaxies and clusters of galaxies, distort space-time
and curve passing light rays. This phenomenon is called gravitational
lensing. Gravitational lensing is sensitive only to the intervening
mass, whether dark or luminous, thus is a powerful probe of the
distribution of dark matter. If the gravitational fields are very
strong, they can bend light rays so much that light can take different
paths to the observer. In this case, we observe multiple images or
highly distorted image of a distant source. Such drastic phenomena are
called strong lensing. On the other hand, even if the gravitational
fields are not so strong, they can be detected through systematic
distortions of background galaxies. This is called weak lensing. Now
both strong and weak lensing are indispensable tools for cosmology.

Historically, it was \citet{einstein36} who first predicted strong
lensing phenomena.  He calculated formation of multiple images due to a
foreground star, but in the paper it was concluded that {\it ``there is
no hope of observing this phenomenon directly''}.
\citet{zwicky37a,zwicky37b} pointed out that the phenomena are more
likely to be observed if we consider a foreground galaxy rather than
star. However, it was still premature for strong lensing to be observed.

At long last, strong gravitational lensing was first discovered by
\citet*{walsh79}. They showed that twin quasars Q0957+561A, B have
almost the same spectra, and concluded that they are likely to be
gravitational lensing. After that, $\sim 80$ gravitationally lensed
quasars have been found so far. The first gravitationally lensed arc was 
also found in a rich cluster A370 \citep{lynds86,soucail87}, and
$\sim 40$ giant arcs have been detected in rich clusters. Therefore,
strong gravitational lensing is now practically useful tool to explore
the dark side of the universe.

For weak lensing signal, i.e., small systematic distortions of galaxies
in response to the foreground mass distributions, to be detected, many
background galaxies are needed to reduce the intrinsic ellipticities.
Such weak lensing signal was first detected by \citet*{tyson90} by make
use of a high surface density of faint blue galaxies \citep{tyson88}.
Now weak lensing is one of the most popular method to study clusters of
galaxies. In addition, the weak lensing signal due to large-scale
structure also has been detected
\citep*{vanwaerbeke00,bacon00,wittman00} and is now regarded as a
powerful tool to study the large-scale structure of the universe.   

In summary, gravitational lensing can be a powerful tool to study the
distribution of dark matter which is not seen with usual methods.

\section{Lensed Arcs and Quasars: Complementary Probes} 
\markboth{CHAPTER \thechapter.
{\MakeUppercase\mychapheadname}}{\thesection.
\MakeUppercase{Lensed Arcs and Quasars: Complementary Probes}}

In this thesis, we study strong gravitational lenses as a test of the CDM
paradigm. We use strong lensing because it is sensitive to mass
distributions at innermost regions of dark halos, and the mass
distributions at the regions are particularly sensitive to the nature of
dark matter. To probe mass distributions of dark halos, we focus on
lensing by clusters of galaxies. In practice, effects of baryonic infall
are expected to be small in clusters, compared with galaxies where inner
structures are significantly affected by baryon cooling.  

\begin{figure}[tb]
\begin{center}
\includegraphics[width=0.48\hsize]{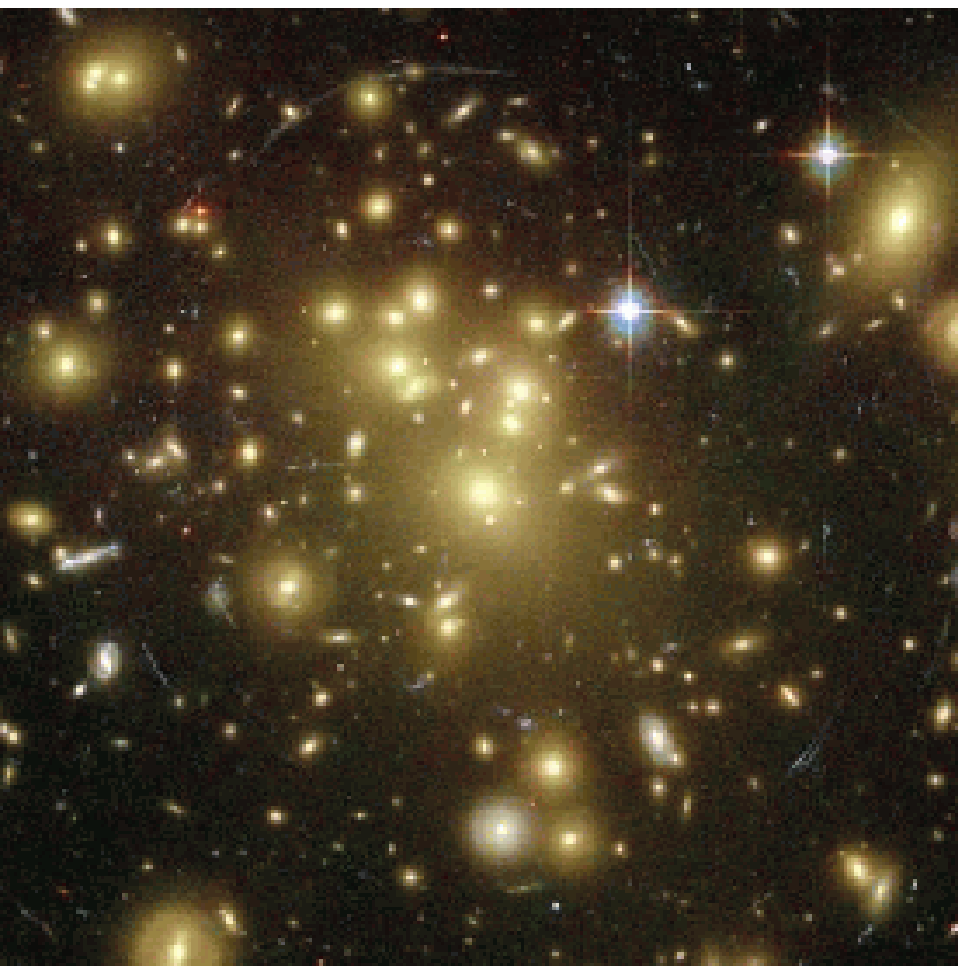}
\includegraphics[width=0.48\hsize]{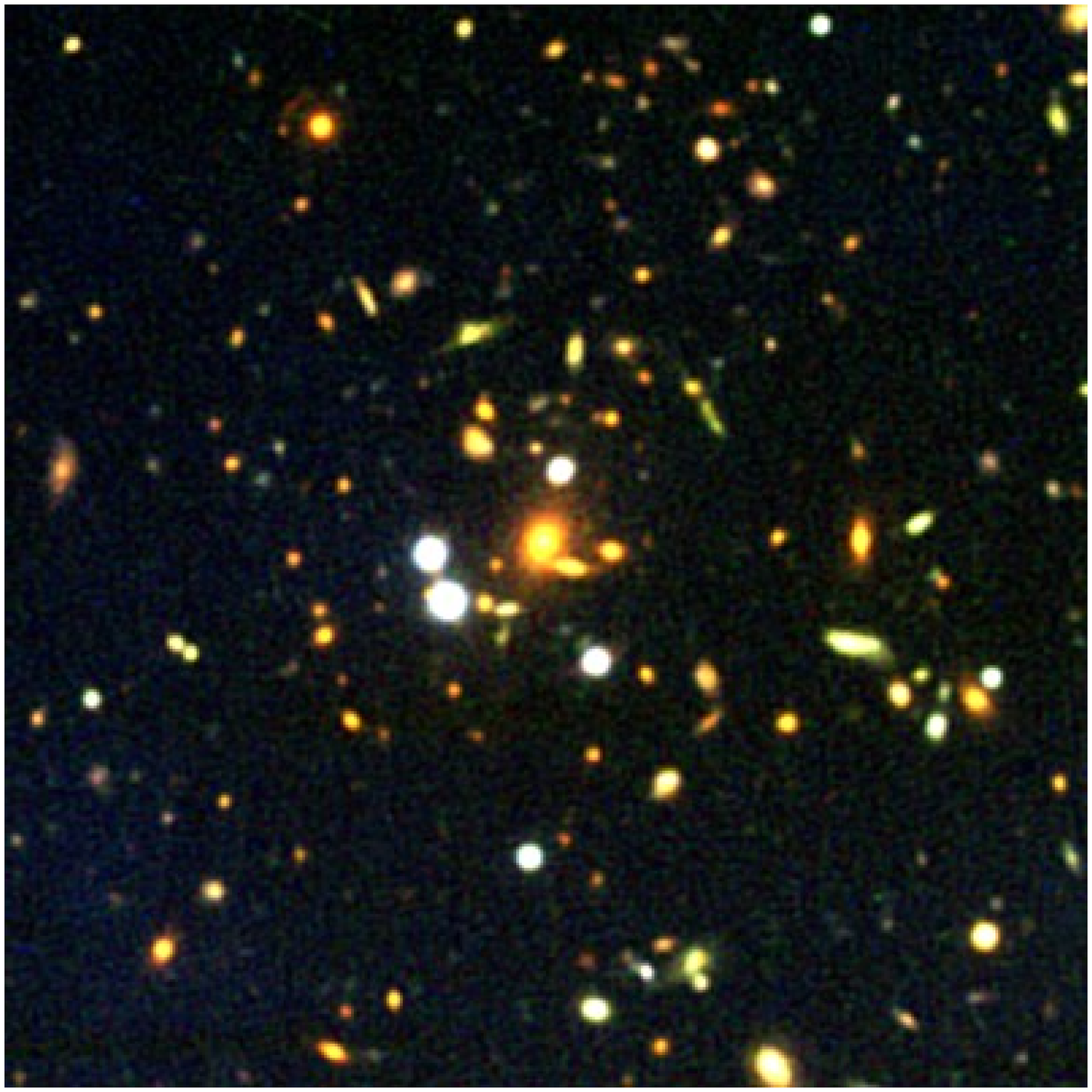}
\caption{Left: Image of galaxy cluster Abell 1689, taken with Advance
 Camera for Surveys (ACS) of {\it Hubble Space Telescope} ({\it HST}).
 Many (curved) lensed galaxies are seen around the cluster. Taken form a
 webpage at http://antwrp.gsfc.nasa.gov/apod/ap030109.html. Right: First
 discovered gravitationally lensed quasar due to a cluster of galaxies.
 Four images around the center of the cluster represent the quadruple
 lensed images of a quasar. See Chapter \ref{chap:sdss} for details. 
\label{fig:intro_gl}}
\end{center}
\end{figure}

There are two types of strong gravitational lensing due to clusters of
galaxies: lensed arcs and quasars (see Figure
\ref{fig:intro_gl}).\footnote{As an another possibility, strong
gravitational lensing of distant supernovae might be observed in the
future. In particular, strong lensing of type-Ia supernovae has several
interesting applications which make use of the standard-candle nature of
type-Ia supernovae \citep*{oguri03a,oguri03b}.} 
The differences of these are summarized in Table \ref{table:intro_comp}.
The most notable difference is the selection of lenses. For instance, in
lensed quasar surveys one first identifies source quasars and then
checks whether they are lensed, while in searching for lensed arcs one
selects massive clusters and then searches for lensed arcs in them.  In
other words, surveys for arcs are biased toward high mass
concentrations, while lensed quasars probe random lines of sight.
Clusters selected by the presence of lensed quasars could, therefore,
differ from those selected as having giant arcs. In addition, there are
many differences between lensed arcs and quasars. For instance, sources
of lensed arc systems are high-$z$ galaxies, which are extended and have
poorly known source population (e.g., luminosity function). While the
complexity of lensed arcs offers detailed constraints on the lens
potential, the simplicity of lensed quasar systems can be an advantage
because there is no confusion from unrelated background objects. Lensed
quasars also make it easier to measure the source redshift and convert
from dimensionless lensing quantities to physical units. The main
disadvantage of lensed quasars is that such lens systems are quite rare
due to the sparseness of quasars. In short, these two types of strong
lensing, i.e., lensed arcs and quasars, have many different
characteristics, and hence the conclusion would be much more robust when
these complementary probes yield the similar results. 

\begin{table}[tb]
 \begin{center}
  \begin{tabular}{ccc}\hline\hline
Lensed Arcs &  & Lensed Quasars\\
\hline
Lens-selected           & Selection             & Source-selected \\
Extended source         & Source size           & Point source \\
Less known              & Source population     & Well known \\
Difficult               & Source-$z$ estimation & Easy \\
Difficult               & Image identification  & Easy \\
Not variable            & Time-variability      & Variable \\
(Mostly) Massive        & Mass of lens cluster  & Small -- Massive \\
$\sim 40$ (giant arcs)  & Observed number       & 1 \\
\hline
\end{tabular}
\caption{Comparison of two types of cluster-scale strong lensing; lensed
  arcs and lensed quasars.} 
\label{table:intro_comp}
 \end{center}
\end{table}

In this thesis, we concentrate our attention on the {\it statistics} of
strongly lensed arcs and quasars using the semi-analytic method that we
developed. Why statistics? Statistics allow us to probe the mean mass
distributions of clusters. Although individual modeling of lensing
clusters can measure their mass distributions precisely, it may suffer
from the special selection function and the scatter around the mean mass
distribution. In addition, individual mass modeling sometimes confronts
difficulties including uncertainties of the center of the mass
distribution and the degeneracy between the ellipticity and central
concentration. Why analytic approach? It is quite demanding for the
numerical simulations to resolve the precise inner structure of lensing
halos while keeping the reasonable number of those objects sufficient
for statistical discussion.  Moreover, an analytic approach has
advantages of the ease of taking the selection function into account and
the ability to clarify the key ingredients which dominate the statistics.

\section{Organization of the Thesis} 
\markboth{CHAPTER \thechapter.
{\MakeUppercase\mychapheadname}}{\thesection.
\MakeUppercase{Organization of the Thesis}}

Part of this thesis is based in the published work
\citep*{oguri03e,oguri04a,oguri04c}. 
This thesis is organized as follows.

In Chapters \ref{chap:conc} and \ref{chap:halo}, we review the current
status of the CDM model. First, in Chapter \ref{chap:conc} we show how
successful the standard model is. We review important cosmological
observations to see how well the cosmology has converged toward a
``concordance'' model. In Chapter \ref{chap:halo}, in contrast, we
see that the CDM model has difficulties to overcome; we summarize
structures of dark halos in the CDM model and their difficulties in
comparing with observations. 

Chapter \ref{chap:tri} is devoted to present the triaxial dark halo
model and its lensing properties which we adopt in the thesis.
Specifically, we review the triaxial model presented by \citet{jing02},
and then we study their lensing properties, such as convergence and
deflection angles.

Our main work is presented in Chapters  \ref{chap:arc}-\ref{chap:sdss}.
First we explore arc statistics. We compute the numbers of arcs and
compare them with observations in Chapter \ref{chap:arc}. 
Next we study lensed quasars. In Chapter \ref{chap:lat}, we
theoretically predict the probabilities and image multiplicities of
large-separation lensed quasars. Then we search for such lensed quasars
from the Sloan Digital Sky Survey (SDSS) data and discover the first
one, SDSS J1004+4112, which is shown in Chapter \ref{chap:sdss}. 
Implications of the discovery are also shown in this Chapter.

Finally, we draw our conclusion in Chapter \ref{chap:sum}.

\chapter{A ``Concordance'' Model of Cosmology}
\label{chap:conc}
\def\mychapheadname{A ``Concordance'' Model of Cosmology}
\markboth{CHAPTER \thechapter.
{\MakeUppercase\mychapheadname}}{}

\section{The Case for a Flat Universe with
Dark Matter and Dark Energy} 
\markboth{CHAPTER \thechapter.
{\MakeUppercase\mychapheadname}}{\thesection.
\MakeUppercase{The Case for a Flat Universe with Dark Matter and Dark Energy}}

Once we accept the cosmological principle (i.e., homogeneous and
isotropic universe) as well as general relativity, the evolution of our
universe is governed by the Friedman equation which is derived by
solving the Einstein equation (see Appendix \ref{chap:cosmo}). One of
the most fundamental parameters in this equation is the curvature of the
universe. The curvature has attracted many attentions because it is
(partly) related to fundamental questions about our universe:  
Is the universe finite or infinite?  Or, is it expanding forever or
not?\footnote{Actually the curvature does not necessarily determine the
finiteness of the universe if we allow more complicated,
multiply-connected universe (imagine a torus which is flat
but compact, for instance). In addition, the curvature does not
necessarily determine the fate of the universe, neither, because of the
existence of dark energy (see \S \ref{sec:conc_de}).}

Other important parameter is the density content of the universe.
The most surprising fact we have learned from cosmology is that the
universe is dominated by two unknown components: Dark matter and dark energy. 
Dark matter is a non-relativistic matter component and is thought to be
non-baryonic. Dark energy is a energy component which accelerates the
expansion of the universe. In this section we review the observational
case for a flat universe with the energy components dominated by 
dark matter and dark energy.

\subsection{Flat Universe}

Whether our universe is flat or not has been debated for a long time.
Some theorists have claimed that the universe is likely to be flat,
because the flat universe is {\it most unlikely}: If we consider the
universe consists of ordinary matter only, then the evolution of the
curvature term becomes
\begin{equation}
 \Omega_K(a)=\Omega_M(a)-1=\frac{\Omega_Ka^{-2}}{H^2(a)/H_0^2}\propto a,
\label{conc_cuv}
\end{equation}
during the matter dominated era ($H^2(a)\propto a^{-3}$).
Therefore, if the universe is not a flat universe, then the curvature
had to be fine-tuned in the very early universe, because we know that 
the curvature term is, if it exists, close to unity $\Omega_K={\mathcal
O}(1)$ from the fact that $\Omega_M={\mathcal O}(1)$.  
But it is also extremely unlikely that $\Omega_K(a)$ is {\it exactly}
zero, because the above discussion suggests that such a spacetime is 
unstable. This problem is known as flatness problem.

One possible solution of this flatness problem is to consider an
accelerating phase in the early universe. The inflationary scenario,
in which the universe experiences acceleration due to the domination 
of the potential energy of a scalar field (called inflaton because
it is still unclear what the scalar field is), gives a natural
explanation of why the universe is (nearly) flat. This is easily
understood as follows. From the fact that the potential energy density
of a scalar field does not change as the universe expands, we derive
$H(a)\sim {\rm const.}$ and $a\propto e^{Ht}$ during inflation (see also
\S \ref{sec:conc_de}). In this case, equation (\ref{conc_cuv}) reduces
to $\Omega_K(a)\propto a^{-2}$, which means that $|\Omega_K(a)|$ rapidly
decreases during inflation. 

\begin{figure}[tb]
\begin{center}
 \includegraphics[width=0.6\hsize]{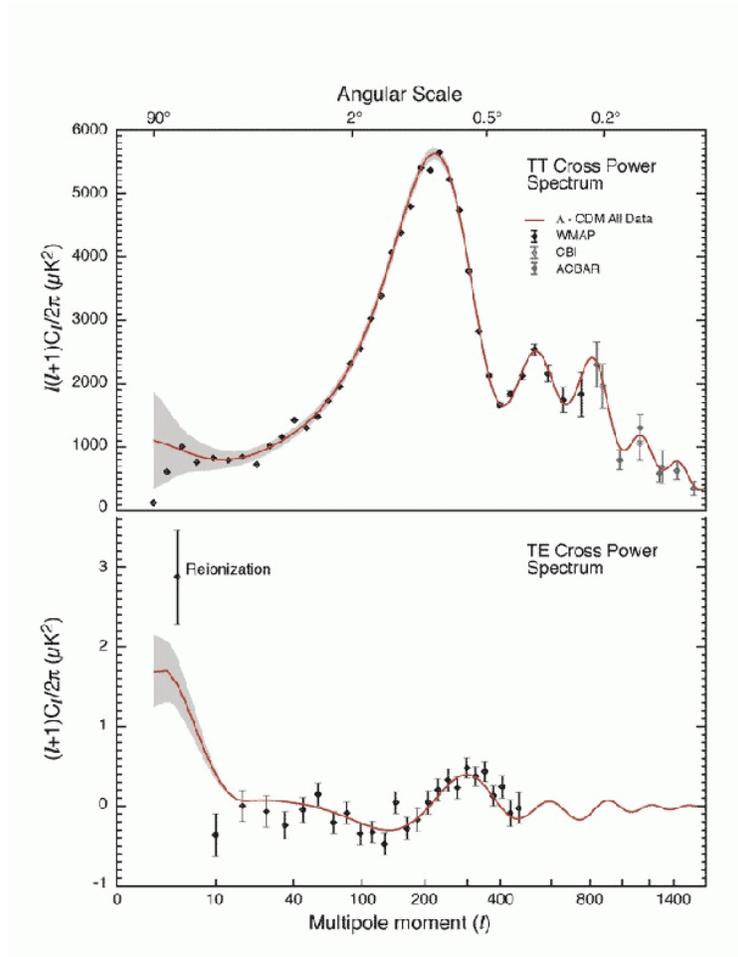}
\caption{Top: {\it WMAP} temperature angular power spectrum is compared with
 best-fit Lambda-dominated CDM model. The small-scale data from ACBAR
 \citep{kuo04} and CBI \citep{pearson03} are also shown. 
 Bottom: Temperature-polarization cross-power spectrum is compared with
 best-fit Lambda-dominated CDM model. Strong correlation seen at very
 large scale is generated by the reionization of the universe. This
 Figure is taken from \citet{bennett03}. 
\label{fig:conc_wmap}}
\end{center}
\end{figure}

While these progresses on theoretical understanding of the flatness of
the universe, observational case for the flat universe had not been so
strong. However, observations of CMB anisotropies changed the situation:
The angular scale of peaks in angular power spectrum of CMB can be an
excellent 
indicator of the curvature of the universe because of the following two
reasons; (1) physical scale of the peak is determined by the sound horizon
scale at the decoupling, which depends on the cosmological parameters
only weakly; (2) on the other hand, the angular diameter 
distance from us to the last scattering surface of CMB strongly depends
on the curvature of the universe. The measurements of the first peak has
been reported by many groups such as BOOMERanG \citep{debernardis00},
MAXIMA \citep{hanany00}, DASI \citep{halverson02}, and {\it WMAP}
\citep{bennett03}. All these groups claimed the detection of the peak at
multipole moment $l\sim 200$, which is roughly corresponding to
$1^\circ$ scale, indicating that the universe is (nearly) flat.  Figure 
\ref{fig:conc_wmap} shows angular power spectrum measured by {\it WMAP}
\citep{bennett03}. A distinct peak around $l=200$ constrains the
curvature of the universe to be $-0.02<\Omega_K<0.08$ (95\% C.L.) if we
include a weak prior $h>0.5$ \citep{spergel03}. 

\subsection{Dark Matter}
\label{sec:conc_dm}
Dark matter is a dust component ($w=0$) which does not interact (or only
weakly interacts) electromagnetically. Dark matter is first proposed by 
\citet{zwicky33}: He estimated the mass of the Coma cluster from
peculiar velocities of galaxies in the cluster and found that it is 400
times larger than that estimated by adding up all of the galaxy masses.
Although the idea had not been taken seriously, it was revived in
1970's and 1980's. For instance, \citet{rubin70} found that the
velocities of the ionized clouds in the Andromeda galaxy do not decrease
with increasing distance from the center and that the extra mass has to
be in the outer part of the galaxy; \citet{rubin85} confirmed that the 
phenomena is commonly seen in spiral galaxies; \citet{ostriker73} pointed
out that the spherical halo component is needed to stabilize the flatten
disk galaxy. From these studies, people began to accept the idea of dark
matter. 

Now, one of the strongest case for dark matter comes from observations of
cluster of galaxies. First of all, clusters of galaxies are X-ray luminous;
thus the mass of the cluster can be estimated under the assumption of
hydrodynamic equilibrium, which turns out to be much larger than mass of
the visible matters (i.e., gas + stars). For instance, \citet{white93}
obtained the fraction of the visible mass in Coma cluster to be
$M_b/M_{\rm tot}\simeq 0.01+0.05h^{-3/2}$. This means that the cluster
of galaxies must be dominated by invisible dark matter. More direct
evidence is offered by gravitational lensing, because it allows one to
measure the mass of clusters directly. \citet{sqires96} estimated an
upper bound for the fraction of the gas mass to be $M_{\rm gas}/M_{\rm
tot}<(0.04\pm0.02)h^{-3/2}$ using weak lensing method. Both X-ray and
lensing data consistently show that visible matters cannot account for
the total masses of clusters of galaxies. Given the total baryon density
of $\Omega_bh^2\sim 0.02$ (see \S \ref{sec:conc_omegab}), these results
imply $\Omega_M\sim 0.3$.

Another evidence for the existence of dark matter is the CMB anisotropy. 
The existence of non-baryonic dark matter can be concluded from the
detailed observations of CMB angular power spectrum as follows; (1)
relative peak heights of even peaks (second peak, fourth peak, ..) to those of
odd peaks (first peak, third peak, ..) tightly constrain the baryon matter
density $\Omega_bh^2$; (2) the amount of boost of angular power spectrum
around first peak is caused by the potential decay during radiation
dominated era (early integrated Sachs-Wolfe effect), and therefore is
sensitive to matter-radiation equality, 
i.e., $\Omega_Mh^2$. The detailed angular power spectrum measured by
{\it WMAP} revealed that $\Omega_Mh^2$ is about six times larger than
$\Omega_bh^2$ \citep{spergel03}. This indicates that the most of the
matter in the universe should be non-baryonic and dark. In addition, the
need for dark matter can be also said from much simpler discussions; the
CMB anisotropies of the order of $10^{-5}$ cannot be achieved from the
baryonic matter only because of the slow linear growth rate $D_+\propto
a$ combined with the decoupling at $z\sim 10^{-3}$. For the enough
amounts of non-linear objects to be observed today, we need an energy
component which was not coupled to baryon-photon fluid before decoupling
and had already grown to much larger than $10^{-5}$ at the last
scattering surface. 

\begin{figure}[tb]
\begin{center}
 \includegraphics[width=0.6\hsize]{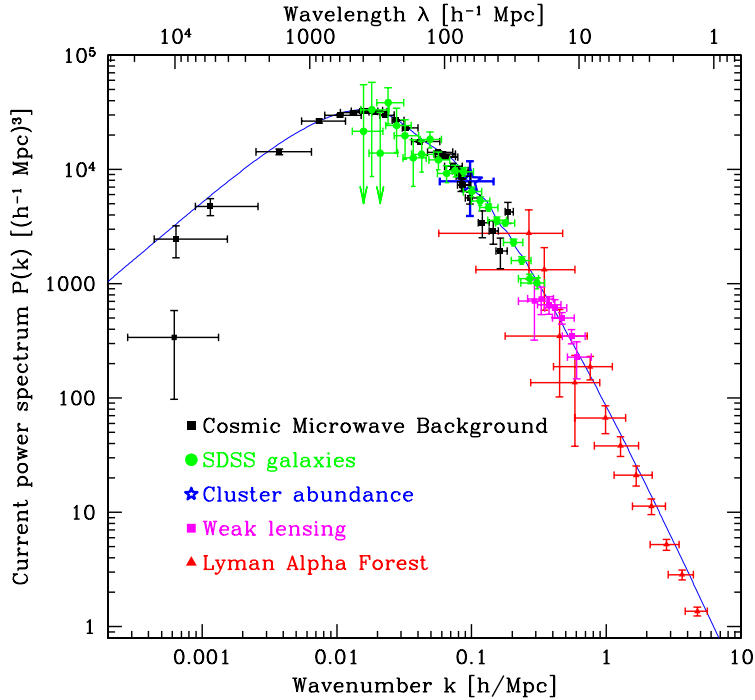}
\caption{Comparison of several comparisons on the power spectrum. The line
 is the CDM prediction with $\Omega_M=0.28$, $h=0.72$, and
 $\Omega_b/\Omega_M=0.16$. This Figure is taken from \citet{tegmark04}.  
\label{fig:conc_pk}}
\end{center}
\end{figure}

Dark matter candidates can be classified according to their
collisionless damping (freestreaming) scales. If we regard massive
neutrinos ($m_\nu\gtrsim 10{\rm eV}$) as dominant component of dark
matter, then they were relativistic until the horizon scale of
$\sim$Mpc; therefore fluctuations below $\sim$Mpc were smoothed out due
to their relativistic motions. Such dark matter is called {\it hot dark
matter (HDM)}. On the other hand, one can consider a possibility of very
massive dark matter so that it became non-relativistic long time 
ago; this time collisionless damping scale will be much smaller than
important scales for structure formation. This is called {\it cold dark
matter (CDM)}. There is also a possibility of {\it warm dark matter
(WDM)} which has collisionless damping scale of $\sim$kpc. The
difference of these dark matter models is well understood by their power
spectra (see Appendix \ref{chap:pk}). Now observations support the cold
dark matter model; Figure \ref{fig:conc_pk} shows the comparison of
observed power spectrum with cold dark matter predictions. They are in
good agreement at $\gtrsim$Mpc scales. Therefore now it is believed that
most of dark matter is non-baryonic and cold. 

\subsection{Dark Energy} 
\label{sec:conc_de}

Dark energy is an unknown energy component which accelerates the
expansion of the universe. In terms of the equation of state, 
accelerating universe is possible if the (effective) equation of state
satisfies $w<-1/3$. Cosmological constant, which is one of candidates
for dark energy, is first proposed by Einstein to make the universe
static. However, soon after the proposal Einstein discarded the idea of
cosmological constant because it turned out that the universe is not
static but expanding {\it (he regretted the idea as ``the biggest
blunder of my life'')}.  

Since then, cosmological constant had not been taken seriously.
However things began to change in 1990's. A reason to invoke the
cosmological constant is the age of the universe; cosmological constant
was needed to reconcile the possible large Hubble constant
\citep[e.g.,][]{aaronson86,tonry91} with the lower limit of the cosmic age
inferred from the evolution of the globular cluster
\citep[e.g.,][]{vandenberg83}. In addition, the best fit of the number
count of faint galaxies was also obtained only with a cosmological
constant \citep{fukugita90b}. Moreover, cosmological constant was
favorable from the theoretical point of view because a flat universe,
which is predicted by inflation model, can be reconciled with the
observed low-matter universe if we assume the large cosmological
constant (see \S \ref{sec:conc_dm}).  

However, the idea of cosmological constant included several
difficulties such as {\it coincidence problem} {\it (``why now?'' problem)};
that is, it is highly unnatural for the density of cosmological constant
to be comparable to that of matter components today, given the different
dependence of densities on the scale factor $a$ ($\propto a^{-3}$ for
matter and $\propto a^0$ for cosmological constant). This means that the
cosmological constant must be fine-tuned in order to be important in the
current universe. One solution is to consider a dynamical scalar field
on the analogy of the inflation model \citep*[often called
quintessense;][]{caldwell98}. The energy and pressure density 
of a scalar field $\Phi$ are 
\begin{eqnarray}
 \rho_\Phi=\frac{1}{2}\dot{\Phi}^2+V(\Phi),\\
 p_\Phi=\frac{1}{2}\dot{\Phi}^2-V(\Phi),
\end{eqnarray}
where $V(\Phi)$ denotes the potential of the scalar field. Hence,
$w_\Phi=p_\Phi/\rho_\Phi\sim -1$ can be achieved if the
potential term dominates.
The advantage of this model is that there does exist a model in which
the solution is attractive, i.e., it is asymptotic solution for a broad
range of initial conditions \citep[e.g.,][]{ratra88}. Observationally
the model may be discriminated to cosmological constant because 
the effective equation of state $w_\Phi$ is not necessarily $w_\Phi\sim
-1$, and also is not necessarily time-independent. Now the energy
components with the negative equation of state are collectively called
{\it dark energy}; it includes cosmological constant, dynamical scalar
field, and topological defects which also have the negative equation of
state. 

\begin{figure}[tb]
\begin{center}
 \includegraphics[width=0.45 \hsize,angle=90]{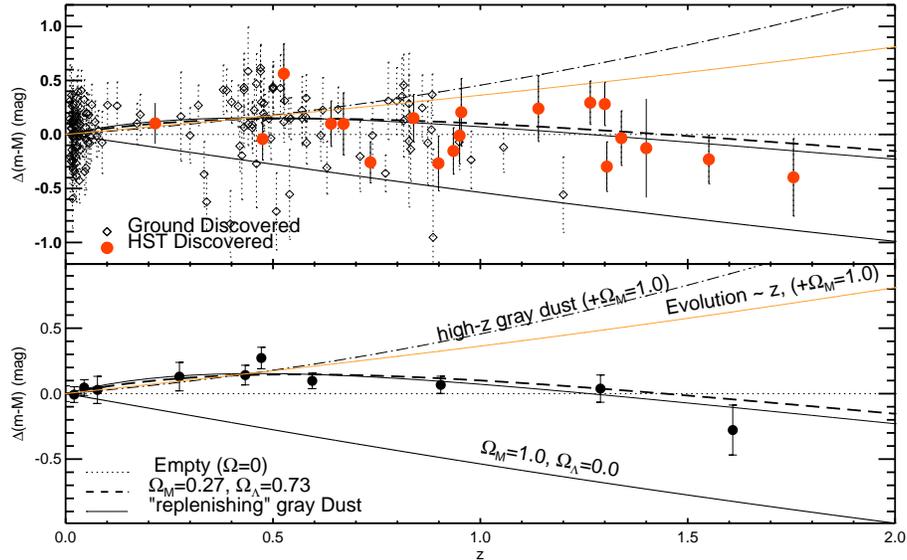}
\caption{The Hubble diagram of type-Ia supernovae. $\Delta(m-M)$ denotes
 the distance modulus relative to an empty universe. The data are in
 good agreement with the cosmological constant dominated model;
 $\Omega_M=0.23$ and $\Omega_\Lambda=0.73$. As seen, the data
 exclude simple gray dust model and evolution model, though
 ``replenishing'' (i.e., physical density of dust is constant) dust
 model can account for the observed behavior. This Figure  is taken from
 \citet{riess04}. 
\label{fig:conc_snia}}
\end{center}
\end{figure}

The most ``direct'' evidence of dark energy is thought to be distant
type-Ia supernovae. Empirically, absolute magnitudes of type-Ia
supernovae have turned out to be almost constant with small
dispersion.\footnote{Actually, there is a tight correlation between
the luminosity decline rate and absolute magnitude \citep{phillips93}.
This empirical relation allows us to reduce the dispersion from
$\sigma\sim 0.6$ mag to $\sigma<0.2$ mag. } Therefore, once the absolute
magnitudes are 
calibrated in a local universe, then type-Ia supernovae can be a
distance indicator at high-$z$ ($z\sim 1$) universe. Actually, two
groups independently showed that the apparent magnitude at $0.1\lesssim
z\lesssim 1$ is fainter than empty universe, and that we need dark
energy component to account for the observed magnitude-redshift relations
\citep{riess98,perlmutter99}. Now the case is much stronger because
it has turned out that at $z\gtrsim 1$ supernovae become brighter, which
indicates that the universe was decelerating in the past
\citep{riess01,riess04}. This different behavior before and after $z\sim
1$ excludes many of alternative explanations of the observed supernovae
magnitude-redshift relations (see Figure \ref{fig:conc_snia}).

Gravitational lensing statistics are known to offer similar cosmological
test. The lensing probability is sensitive to the volume of the
universe, so it can be used to place interesting constraints on the
cosmological constant $\Omega_\Lambda$
\citep*{turner90,fukugita90a,kochanek96,chiba99,chae02,mitchell04}.\footnote{In
contrast, \citet{keeton02}
argued that the lensing rate becomes insensitive to $\Omega_\Lambda$
when the number density of galaxies at high-$z$ is calibrated by observations.} 
For instance, \citet{chae02} derived the constant on the cosmological
constant assuming the flat universe as
$\Omega_\Lambda=0.69^{+0.14}_{-0.27}{}^{+0.10}_{-0.12}$ (68\% C.L.).

Another evidence is again offered by CDM anisotropies. First, as
discussed before, the peak position and peak height of first peak strongly
constrain $\Omega_K$ and $\Omega_Mh^2$, respectively. Even if we combine
these constraints, however, the strong degeneracy between $\Omega_M$
(or $\Omega_\Lambda$) and $h$ still remains. This degeneracy is known as
geometric degeneracy \citep{efstathiou99}. However, if we add one more
constrain, such as the distance ladder, supernova Ia, or power
spectrum, as a prior, then we obtain the finite cosmological constant
at high statistical significance \citep[see, e.g.,][]{spergel03}.
Although the above example is indirect, more direct evidence has been
obtained through positive correlation between CDM anisotropies and
galaxies distribution which is caused by a potential decay in a
dark energy dominated universe \citep*[late integrated Sachs-Wolfe
effect;][]{fosalba03,boughn04,afshordi04,nolta04,scranton04}.  

\begin{figure}[tb]
\begin{center}
 \includegraphics[width=0.55\hsize]{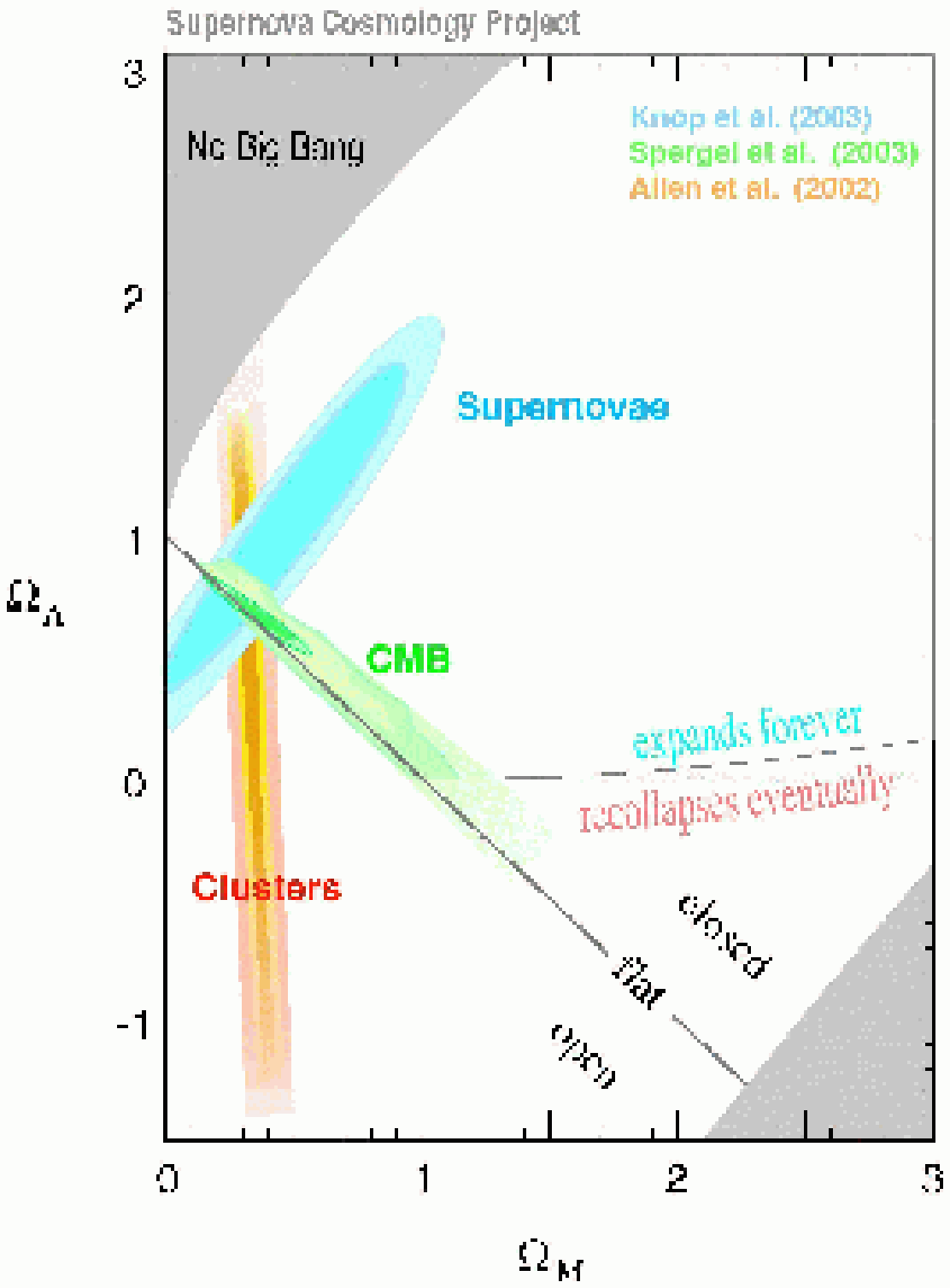}
\caption{Three independent constraints on cosmological parameters
 projected in the $\Omega_M$-$\Omega_\Lambda$ plane. ``CMB'' is
 constraints from CMB anisotropy observed by {\it WMAP} \citep{spergel03}. 
 ``Supernovae'' denote results of type-Ia supernovae observations by
 \citet{knop03}. Constraints denoted by ``Clusters'' come from X-ray
 measurements of distant clusters \citep*{allen02}. It is surprising that
 these three independent observations are well explained by a model with
 $\Omega_M\sim 0.3$ and $\Omega_\Lambda\sim 0.7$. 
 This Figure is taken from http://www-supernova.lbl.gov/ 
\label{fig:conc_conc}}
\end{center}
\end{figure}

\section{A ``Concordance'' Model}
\label{sec:conc_conc}
\markboth{CHAPTER \thechapter.
{\MakeUppercase\mychapheadname}}{\thesection.
\MakeUppercase{A ``Concordance'' Model}}

\begin{table}[tb]
 \begin{center}
  \begin{tabular}{ccl}\hline\hline
Parameter & Fiducial & Meaning \\
\hline
$\Omega_M(=1-\Omega_\Lambda)$ & $0.3$ &
Current matter density of the universe (CDM$+$baryon) \\
$\Omega_b$ & $0.04$ &
Current baryon density of the universe \\
$h$ & $0.7$ &
The Hubble constant in units of $100{\rm km\,s^{-1}\,Mpc^{-1}}$\\
$n_s$ & $1.0$ &
Index of the primordial power spectrum \\
$\sigma_8$ & $0.9$ &
Normalization of the density fluctuation\\
\hline
\end{tabular}
\caption{Model parameters of the concordance model described in
  \S \ref{sec:conc_conc}. The entry ``Fiducial'' denotes the value we adopt
  in this thesis (unless otherwise specified). } 
\label{table:conc_para}
 \end{center}
\end{table}

As discussed in the previous section, now there are lots of evidences
that the universe is flat and dominated by dark matter and dark energy. 
As shown in Figure \ref{fig:conc_conc}, several independent observations
point a model with $\Omega_M\sim 0.3$ and $\Omega_\Lambda\sim 0.7$. This
remarkably successful model is now called a ``concordance'' model.
Although the definition of the concordance model may not be unique, we
define it by the followings:
\begin{itemize}
 \item The evolution of the universe is governed by general relativity. 
       The topology of the universe is simple.
 \item The universe is flat ($E^3$). The matter components are baryon,
       dark matter, and dark energy. We also assume the densities of 
       radiation components of photons and neutrinos as inferred from
       the CMB temperature and that calculated from the standard thermal
       history. We assume the three massless species of
       neutrinos.\footnote{Under these assumptions, the current
       radiation densities can be calculated as 
       $\Omega_\gamma h^2\simeq 2.4\times 10^{-5}$ and 
       $\sum\Omega_\nu h^2\simeq 1.8\times 10^{-5}$. Hence they are not
       free parameters.} 
 \item We assume the adiabatic primordial fluctuations. The fluctuations
       obey the Gaussian statistics. The primordial power spectrum can
       be described by a power law, $P_i(k)\propto k^{n_s}$.
 \item The dark matter is cold, i.e., collisionless damping of the power
       spectrum due to freestreaming is negligible.
 \item The equation of state $w$ of dark energy is assumed to be
       $w=-1$ and time-independent.
\end{itemize}
The assumption of the adiabatic Gaussian primordial fluctuations are
supported by the high-resolution measurements of CMB anisotropies
\citep{spergel03,komatsu03,peiris03}. In particular, large-scale
temperature-polarization anti-correlation (see Figure
\ref{fig:conc_wmap}) offers a strong evidence for adiabatic superhorizon
fluctuations \citep{peiris03}. Theoretically, such fluctuations can be
explained in the simple inflation model, in which the fluctuations are 
generated by quantum fluctuations of the inflaton field. 

We assume the above concordance model throughout the thesis. In this
model, the number of basic (independent) parameters is only five, which
are listed in Table \ref{table:conc_para}. We also show fiducial values
of these parameters which will be adopted in this thesis. In the next
section, we will see that the adopted cosmological parameters do explain
so many observations. To see how successful it is, we show an example; 
1,348 data points of the angular power spectrums (both
temperature-temperature and temperature-polarization) measured by {\it
WMAP} (see Figure \ref{fig:conc_wmap}) are well fitted by the
concordance model (plus one more parameter, optical depth $\tau$) which
contains only 6 parameters. The reduced chi-square is $\chi^2/{\rm
dof}=1.066$ for 1,342 degrees of freedom \citep{spergel03}. 

\section{Constraints on Model Parameters}
\markboth{CHAPTER \thechapter.
{\MakeUppercase\mychapheadname}}{\thesection.
\MakeUppercase{Constraints on Model Parameters}}

\subsection{Current Matter Density $\Omega_M$}
\label{sec:conc_omegam}

\begin{table}[tb]
 \begin{center}
  \begin{tabular}{ccl}\hline\hline
Method & $\Omega_M$: Mean (68\% C.L.) & Ref. \\
\hline
Type-Ia supernovae & $0.29^{+0.05}_{-0.03}$ & 1\\
Lensing statistics & $0.31^{+0.27}_{-0.14}{}_{-0.10}^{+0.12}$ & 2\\
Cluster X-ray gas mass (+HST $h$, BBN) & $0.30^{+0.04}_{-0.03}$ & 3 \\
Galaxy power spectrum (+HST $h$) & $0.36\pm0.07$ & 4 \\
CMB anisotropy & $0.29\pm0.07$ & 5 \\
\hline
\end{tabular}
\caption{Recent $\Omega_M$ determinations.\protect\\
\footnotesize{
\hspace*{5mm}Ref. --- (1) \citealt{riess04}; (2) \citealt{chae02}; 
(3) \citealt{allen02}; (4) \citealt{pope04}; (5) \citealt{spergel03}}} 
\label{table:conc_omegam}
 \end{center}
\end{table}

As discussed before, type-Ia supernovae and lensing statistics offer 
good tests for $\Omega_\Lambda$ (and also $\Omega_M$). Assuming a flat
universe, \citet{riess04} derived constraints on $\Omega_M$ with 10\%
accuracy; $\Omega_M=0.29^{+0.05}_{-0.03}$. \citet{chae02} constrained
$\Omega_M$ from the lensing statistics of flat-spectrum $\sim$10,000
radio sources; $\Omega_M=0.31^{+0.27}_{-0.14}{}_{-0.10}^{+0.12}$
(stat.+syst.). Note that both methods might suffer from systematic
effects, such as intergalactic dusts and evolution of the number density
of galaxies. Nevertheless, the excellent agreement suggests that the
result of $\Omega_M \sim 0.3$ seems quite robust.

Clusters of galaxies also can be used to put tight constraints on
$\Omega_M$. \citet{allen02} derived $\Omega_M$ using the (apparent)
redshift evolution of cluster gas-mass fraction, which is the method
described by \citet{sasaki96}, and found that
$\Omega_M=0.30^{+0.04}_{-0.03}$, once the priors on $h$ and
$\Omega_bh^2$ are taken into account (see \S \ref{sec:conc_omegab} and
\S \ref{sec:conc_h}). The result shows an excellent agreement with
type-Ia supernovae and lensing statistics. 

Another method to determine $\Omega_M$ is the galaxy power spectrum (see 
also Appendix \ref{chap:pk}). Usually the power spectrum has a
characteristic scale corresponding to the horizon size at
matter-radiation equality $\lambda_{\rm eq}$, because of the following
reason. Fluctuations with scales larger  than $\lambda_{\rm eq}$ enter
the horizon when the universe is matter dominant. Such fluctuations grow
as soon as they enter the horizon. Therefore the power spectrum at those
scales does not change the shape. On the other hand, fluctuations with
scales less than $\lambda_{\rm eq}$ enter the horizon when the universe
is still radiation dominant. Fluctuations in that epoch do not grow in
practice because of the rapid expansion of the universe. Smaller
fluctuations suffer from the longer period of the suppression, and
result in the modification of the spectrum as $P(k)\sim P_i(k)k^{-4}$.
Therefore, the measurement of  this characteristic scale will constrain
the matter-radiation equality epoch and thus $\Omega_M h$. If combined
with constraints on $h$, this allows one to determine $\Omega_M$.
\citet{pope04} derived $\Omega_M=0.36\pm0.07$ based on the galaxy power
spectrum measured by the SDSS.

Finally, we mention constraints from CMB anisotropies. Even CMB alone
does constrain $\Omega_M$ if we assume a flat universe, by combining the
precise peak position (determined by $\Omega_Mh^{3.4}$) and early ISW
constraints on $\Omega_Mh^2$. From the WMAP measurements,
\citet{spergel03} constrained $\Omega_M=0.29\pm0.07$ assuming the
concordance model. 
 
These recent determinations are summarized in Table
\ref{table:conc_omegam}. It is surprising that all measurements are
consistent with $\Omega_M=0.3$. The matter density can be also
determined from e.g., cluster abundances and weak lensing, but we use
these to constrain $\sigma_8$ because of tight correlation between
$\Omega_M$ and $\sigma_8$ in these measurements. 

\subsection{Current Baryon Density $\Omega_b$}
\label{sec:conc_omegab}

\begin{table}[tb]
 \begin{center}
  \begin{tabular}{ccl}\hline\hline
Method & $\Omega_bh^2$: Mean (68\% C.L.) & Ref. \\
\hline
Big bang nucleosynthesis (D)  & $0.0214\pm0.0020$ & 1 \\
Big bang nucleosynthesis ($^7$Li) & $0.006-0.016$ (95\%) & 2, 3 \\
CMB anisotropy  & $0.0237^{+0.0013}_{-0.0012}$ & 4 \\
Ly$\alpha$ forests & $0.02\pm0.01$ & 5\\
\hline
\end{tabular}
\caption{Recent $\Omega_bh^2$ determinations.\protect\\
\footnotesize{
\hspace*{5mm}Ref. --- (1) \citealt{kirman03}; (2) \citealt{ryan00}; 
(3) \citealt{coc02}; (4) \citealt{spergel03}; (5) \citealt{scott00}}} 
\label{table:conc_omegab}
 \end{center}
\end{table}

The baryon density $\Omega_b$ is usually constrained in the combination
of $\Omega_bh^2$. Therefore, in this subsection we review current
determinations of $\Omega_bh^2$ rather than $\Omega_b$.

Traditionally $\Omega_bh^2$ has been determined by the Big Bang
Nucleosynthesis (BBN). Since the light element abundances produced in
the early universe depends only on baryon-to-photon ratio $\eta$ and
thus on $\Omega_bh^2$, by observing the primordial light element
abundances, such as $^4$He, D, and $^7$Li, one can determine
$\Omega_bh^2$. Among the light elements, D is perhaps the best element
to constrain $\Omega_bh^2$ because of the small uncertainties. For
instance, \citet{kirman03} combined absorption systems of five quasars,
and derived $\Omega_bh^2=0.0214\pm0.0020$. On the other hand, \citet[][
see also \citealt{ryan00}]{coc02} determined the baryon density using $^7$Li
as $\Omega_bh^2=0.006-0.016$ (95\%). This value seems much lower than
that derived from D.

Another precise measurement comes from CMB anisotropies. As discussed,
relative peak heights of even peaks to those of odd peaks constrain
$\Omega_bh^2$. This method is quite robust because no other parameters
can mimic such behavior. Using up to the third peak, {\it WMAP} data
only determined $\Omega_bh^2=0.0237^{+0.0013}_{-0.0012}$
\citep{spergel03}. 

The ionizing background $J_H$ also allows us to
determine  $\Omega_bh^2$. This is because the Ly$\alpha$ optical depth,
which can be measured from the transmission power spectrum of
Ly$\alpha$ forest, is proportional to $(\Omega_bh^2)^2/J_H$. For
instance, the ionizing background from the proximity effect measured by
\citet{scott00} implies $\Omega_bh^2=0.02\pm0.01$.
 
These recent determinations are summarized in Table
\ref{table:conc_omegab}. It is indeed striking that BBN and CMB are
consistent with each other ($\Omega_bh^2\sim0.02$, which implies
$\Omega_b\sim 0.04$ if we adopt $h\sim 0.7$), since they probe baryon
densities at totally different epochs ($t\sim3$ minutes for BBN and
$t\sim 4\times 10^5$ years for CMB). However, it should be kept in mind
that $^7$Li seems inconsistent with the other results. This discrepancy
might be ascribed to the uncertainties of nucleon reaction rate
\citep{coc04} or the baryon input after BBN \citep*{ichikawa04}.

\subsection{The Hubble Constant $h$}
\label{sec:conc_h}

\begin{table}[tb]
 \begin{center}
  \begin{tabular}{ccl}\hline\hline
Method & $h$: Mean (68\% C.L.) & Ref. \\
\hline
{\it HST} Key Project  & $0.72\pm0.03\pm0.07$ & 1 \\
Cluster SZ + X-ray & $0.60^{+0.04}_{-0.04}{}^{+0.13}_{-0.18}$ & 2 \\
Gravitational lens time delay & $0.48\pm0.02$ & 3\\
CMB anisotropy  & $0.72\pm0.05$ & 4 \\
\hline
\end{tabular}
\caption{Recent Hubble constant determinations.\protect\\
\footnotesize{
\hspace*{5mm}Ref. --- (1) \citealt{freedman01}; (3) \citealt{reese02}; 
(3) \citealt{kochanek02a}; (4) \citealt{spergel03} }} 
\label{table:conc_h}
 \end{center}
\end{table}

The Hubble constant has been regarded as the most important cosmological
parameter because it is directly related with the distances to the
objects (and the size of the universe). The traditional way to measure
the Hubble constant is the distance ladder. Cepheid distances are used
to calibrate the second distance indicators such as type Ia supernovae,
Tully-Fisher relation, fundamental plane, type II supernovae, and
surface brightness fluctuations. The {\it HST} Key Project
\citep{freedman01} is aimed to find many Cepheids to calibrate the
secondary distances. They concluded that the Hubble parameter is
determined with the accuracy of 10\%, $h=0.72\pm0.03\pm0.07$
(stat.+syst.). 

Since the above method can measure only the {\it local} ($z\lesssim
0.1$) value of the $h$, it is important to check it from
independent direct methods. The distant clusters offer one of such methods. 
The flux of Sunyaev-Zel'dovich effect in a cluster is $\propto n_eT_e$,
where $n_e$ and $T_e$ are electron number density and temperature, while
X-ray flux is $\propto n_e^2T_e^{1/2}$. From these different dependences
on $n_e$, we can determine the luminosity distance to the cluster.
Although at high redshift ($z\sim 1$) derived Hubble constant is somewhat
sensitive to assumed cosmological model, by assuming $\Omega_M=0.3$ and
$\Omega_\Lambda=0.7$, \citet{reese02} determined
$h=0.60^{+0.04}_{-0.04}{}^{+0.13}_{-0.18}$ (stat.+syst.) from 18 distant
clusters.  

Another direct method to probe $h$ is gravitational lens time delays,
because time delays are dimensional quantity (i.e., $\Delta t\propto
h^{-1}$) while other observables (image separations, etc.) are
dimensionless. \citet{kochanek02a} applied this method to five quasar
lens systems, and derived the value of $h=0.48\pm0.02$ assuming the
singular isothermal mass distribution of the lensing galaxy. The value
is significantly lower than those derived from the other method.
However, the important point is that there is a strong degeneracy
between mass distributions in the lens galaxies and the derived $h$.
Thus, this low value may be interpreted that the lens galaxy is more
complicated than the simple singular isothermal mass distribution.

CMB observations give the value of $h$ by combining the early ISW and
the peak position, as discussed in \S \ref{sec:conc_omegam}.
\citet{spergel03} constrained $h=0.72\pm0.05$ based on the {\it WMAP}
measurements. 

These recent determinations are summarized in Table
\ref{table:conc_h}. The value of $h$ is also basically converging. 
The values derived from clusters and gravitational lens time delays seem
lower than the others, but those methods actually may be dominated by
systematic effects. Thus, more and more theoretical and observational
studies are needed to reduce these systematic uncertainties (not
statistical uncertainties). 

\subsection{Index of the Primordial Power Spectrum $n_s$}

\begin{table}[tb]
 \begin{center}
  \begin{tabular}{cc}\hline\hline
Data & $n_s$: Mean (68\% C.L.)\\
\hline
{\it WMAP} & $0.99\pm0.04$ \\
{\it WMAP}ext & $0.97\pm0.03$ \\
{\it WMAP}ext+2dF & $0.97\pm0.03$ \\
{\it WMAP}ext+2dF+Ly$\alpha$ & $0.96\pm0.02$\\
\hline
\end{tabular}
\caption{Index determinations presented by \citet{spergel03}. ``ext''
  denotes small-scale CMB data from ACBAR \citep{kuo04} and CBI
  \citep{pearson03}. ``2dF'' is the power spectrum measurements by
  two-degree Field system \citep{percival01}. The measurements of
  Ly$\alpha$ power spectrum \citep{croft02,gnedin02} are denoted by
  ``Ly$\alpha$''. See Figures \ref{fig:conc_wmap} and \ref{fig:conc_pk}
  to understand which data  corresponds to which scales.} 
\label{table:conc_ns}
 \end{center}
\end{table}

To determine $n_s$, it is essential to see superhorizon fluctuations
which have not been affected by any physical processes after the
fluctuation generation. Therefore, here we restrict our attention on
constraints from the CMB anisotropies (plus some other data sets).

The constraints from {\it WMAP} \citep{spergel03} are summarized in
Table \ref{table:conc_ns}. We show how the results are changing as we
add small-scale data sets. First of all, the results indicate that the
index is very close to 1.\footnote{Historically, the spectrum with
$n_s=1$ has been called Harrison-Zel'dovich spectrum
\citep{harrison70,zeldovich72}. It has an interesting feature that
fluctuations for all wavelengths come into horizon with the same
amplitude. This is understood from $k^3P_\Phi(k)\sim k^{n_s-1}$, where
$P_\Phi(k)$ denotes the power spectrum of curvature perturbation
$\Phi$.} This is indeed consistent with what inflation model did
predict. Basically, the small deviation from $n_s=1$ and its
scale-dependence reflects the shape of potential during the inflation.
Therefore, in principle we can reconstruct the potential of inflation
from the precise measurement of the primordial power spectrum.
\citet{peiris03} showed that the current date are not so good as to
constrain inflation models severely, but do have an ability to reject
some of inflation models.  

\subsection{Normalization of the Density Fluctuation $\sigma_8$}

\begin{table}[tb]
 \begin{center}
  \begin{tabular}{cccl}\hline\hline
Method & $\sigma_8$: Mean (68\% C.L.) & $\Omega_M$ dependence  & Ref. \\
\hline
Cluster (local)  & $0.69\pm0.04$ & $(\Omega_M/0.3)^{-0.25}$ & 1\\
& $0.77\pm0.04$ & $(\Omega_M/0.3)^{-0.60}$ ?  & 2\\
& $0.68\pm0.06$ & $(\Omega_M/0.3)^{-0.60}$ & 3\\  
& $0.74\pm0.04$ & $(\Omega_M/0.3)^{-0.60}$ ? & 4\\  
& $0.77\pm0.07$ & $(\Omega_M/0.3)^{-0.44}$ & 5\\
Cluster (high-$z$) & $0.92\pm0.09$ & $(\Omega_M/0.3)^{-0.14}$ & 6\\
& $1.04\pm0.06$ & $(\Omega_M/0.3)^{-0.00}$ ? & 7\\
Weak lensing & $1.09\pm0.12$ & $(\Omega_M/0.3)^{-0.51}$ & 8\\
& $1.02\pm0.16$ & $(\Omega_M/0.3)^{-0.46}$ & 9\\
& $0.67\pm0.10$ & $(\Omega_M/0.3)^{-0.60}$ & 10\\
& $0.78^{+0.27}_{-0.12}$ & $(\Omega_M/0.3)^{-0.37}$ & 11\\
& $0.97\pm0.13$ & $(\Omega_M/0.3)^{-0.68}$ & 12\\
& $0.72\pm0.09$ & $(\Omega_M/0.3)^{-0.49}$ & 13\\
& $0.71^{+0.06}_{-0.08}$ & $(\Omega_M/0.3)^{-0.57}$ & 14\\
& $0.86^{+0.05}_{-0.07}$ & $(\Omega_M/0.3)^{-0.52}$ & 15\\
& $0.94\pm0.17$ & $(\Omega_M/0.3)^{-0.44}$ & 16\\
CMB ({\it WMAP}) &  $0.90\pm0.10$ & $\cdots$ & 17 \\
CMB ({\it WMAP}ext) & $0.80\pm0.10$ & $\cdots$ & 17\\
\hline
\end{tabular}
\caption{Recent $\sigma_8$ determinations.\protect\\
\footnotesize{
\hspace*{5mm}Ref. --- (1) \citealt{allen03}; 
(2) \citealt{pierpaoli03};  (3) \citealt{bahcall03a};
(4) \citealt{schuecker03};  (5) \citealt{seljak02};  
(6) \citealt{bahcall03b};   (7) \citealt{komatsu02}; 
(8) \citealt{massey04};     (9) \citealt{rhodes04};     
(10) \citealt{heymans04};   (11) \citealt{hamana03};    
(12) \citealt{bacon03};     (13) \citealt{brown03};     
(14) \citealt{jarvis03};    (15) \citealt*{hoekstra02}; 
(16) \citealt*{refregier02}; (17) \citealt{spergel03}}}  
\label{table:conc_sigms8}
 \end{center}
\end{table}

The amplitude of current linear fluctuations within $8h^{-1}{\rm Mpc}$
sphere, $\sigma_8$, has been determined from cluster abundances, weak
lensing surveys, and CMB anisotropies. Cluster abundances and weak
lensing surveys are more direct methods in the sense that they probe
current fluctuations at the scale near $8h^{-1}{\rm Mpc}$. On the other
hand, from CMB anisotropies we know large-scale fluctuations at $z\sim
10^3$; therefore we have to extrapolate the result both in the scale and
time. Usually, constraints from cluster abundances and weak lensing
surveys show strong $\sigma_8$-$\Omega_M$ correlation, thus below we
normalize the values to $\Omega_M=0.3$. Results are summarized in Table
\ref{table:conc_sigms8}.

Since the mass function of clusters of galaxies is sensitive to density
fluctuations, cluster abundances are powerful tool to determine
$\sigma_8$. Weak lensing shear power spectrum (or shear 2-point
correlation function) is also directly related with the matter power
spectrum, thus it allows us to constrain $\sigma_8$. As seen in Table
\ref{table:conc_sigms8}, however, different groups presented rather
different values. Actually the differences are much larger than
statistical errors, indicating that systematic errors may dominate. 
The source of the systematic effects is unknown, but it could be
uncertainties (or inaccuracies) of the theoretical power spectrum (see
Appendix \ref{chap:pk}). In cluster surveys, it is sometimes difficult
to convert observable quantities (temperature, luminosity, richness,
etc.) to masses of clusters. One of the biggest shortcomings in weak
lensing surveys is that the constraints are dependent on the redshift
distribution of the source galaxies which is quite hard to know from
observations. Therefore, we conclude that the current constraints on the
value of $\sigma_8$ are very roughly $0.7\lesssim \sigma_8\lesssim 1.0$.

It should be noted that cluster abundances at high-$z$ require
significantly larger $\sigma_8$ ($\sigma_8\sim 1$) than those at local
universe ($\sigma_8\sim 0.7$). While this discrepancy can be resolved by
lowering $\Omega_M$ (see ``$\Omega_M$ dependence'' in Table
\ref{table:conc_sigms8}), it might imply something beyond the
concordance model; \citet{oguri03d} introduced decaying CDM model to resolve
the discrepancy; \citet*{lokas04} considered the dark energy model with
$w\neq-1$ (see \S \ref{sec:conc_beyond}), and found that the data might
be better explained. 

Finally, we see constrains from CMB anisotropies. The result from {\it
WMAP} alone is $\sigma_8=0.90\pm0.10$ , and when we add small-scale CMB
data sets it changes to $\sigma_8=0.80\pm0.10$. In both cases, the
results are roughly consistent with those of cluster abundances and weak
lensing surveys.

\section{Beyond the Concordance Model?}
\label{sec:conc_beyond}
\markboth{CHAPTER \thechapter.
{\MakeUppercase\mychapheadname}}{\thesection.
\MakeUppercase{Beyond the Concordance Model?}}

Although the concordance model has achieved remarkable success, the
possibility that we will need the model beyond the concordance model in
the future still remains. Below we pursuit some of the possibilities.

\subsection*{Dark Energy}
We have already seen in \S \ref{sec:conc_de} that the dynamical model is
more natural than the cosmological ``constant''. To discriminate these, it is
important to determine the dark energy equation of state $w$: If it
turns out that $w=-1$ and $w$ is time-independent, then the dark energy
is likely to be the cosmological constant. If not, the accelerating
universe might be caused by a scalar field. In this case, the degree of
the deviation from $w=-1$ and its time evolution is directly related
with the shape of the potential of the scalar field. Thus it is often
said that the measurements of $w$ offer clues to the nature of dark
energy.  

Because of this importance, until now a number of methods are proposed
to probe the dark energy equation of state
\citep*[e.g.,][]{matsubara03,jain03,blake03,cooray04,takahashi04b}
besides the methods described above. In addition, there are so many
plans to catch up the nature of dark energy, such as
SNAP\footnote{See webpage at http://snap.lbl.gov/} which will determine
$w$ with $\sim$5\% accuracy.

However, the current status seems not so exciting: None of the results
does require the dark energy with $w\neq-1$. For instance,
assuming constant $w$ \citet{spergel03} concluded $w=-0.98\pm0.12$ from
the {\it WMAP}ext+large scale structure data. \citet{riess04} derived
$w=-1.02^{+0.13}_{-0.19}$ using type-Ia supernova data with a prior
$\Omega_M=0.27\pm0.04$. The data are also consistent with the static
(i.e., time-independent) nature of dark energy. Thus the property of
dark energy should be very close to that of the cosmological constant,
even if it is not the cosmological constant.

\subsection*{Running Spectral Index}

\begin{figure}[tb]
\begin{center}
 \includegraphics[width=0.5\hsize]{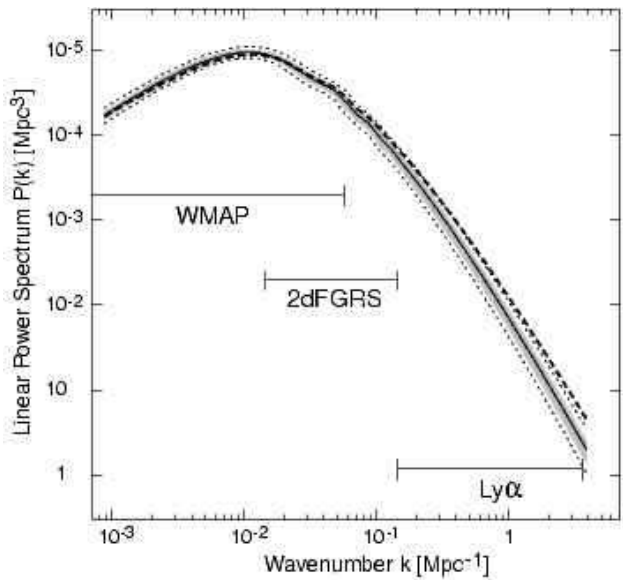}
\caption{The fitted power spectrum by {\it WMAP} results. Shaded regions
 and dotted lines show the $1\sigma$ contours and $2\sigma$ limit for
 the running spectral index model. The dashed line is the best-fit power
 spectrum of the concordance model described in \S \ref{sec:conc_conc}.
This Figure  is taken from \citet{spergel03}. 
\label{fig:conc_wmappk}}
\end{center}
\end{figure}

The somewhat strange behavior seen in Table \ref{table:conc_ns} is
that the value of $n_s$ becomes smaller and smaller as we add more and
more small-scale data. This may indicate that we need a new parameter
beyond the concordance model.

Motivated by this, \citet{spergel03} considered a running spectral index
model in which the primordial power spectrum is described by 
\begin{equation}
 P_i(k)=P_i(k_0)\left(\frac{k}{k_0}\right)^{n_s+(1/2)dn_s/d\ln k\ln(k/k_0)},
\end{equation}
where $dn_s/d\ln k$ is a constant parameter which means the degree of
running (recover power-law $P_i(k)$ when $dn_s/d\ln k=0$). They found
$dn_s/d\ln k=-0.031^{+0.016}_{-0.017}$ from the combined data set ({\it
WMAP}ext+2dF+Ly$\alpha$). Thus the data favor (but not require) the
running spectral index. See Figure \ref{fig:conc_wmappk} for the
difference between the best-fits of the concordance model and running
spectral index model. Theoretically, such running of the power spectrum
can easily be achieved by taking account of the second-order slow-roll
parameters \citep[see][]{peiris03}. 

One of the important consequence of the running spectral index model is
significantly lower amplitudes of fluctuations on small scales (see
Figure \ref{fig:conc_wmappk}). This might be a good news for possible
problems of CDM model on small scales (Chapter \ref{chap:halo}).
However, things are more complicated; a simple extrapolation of the
running spectral index to smaller scale cannot explain the early
reionization found by the temperature-polarization cross-power spectrum
at low-$l$ \citep{yoshida03}. Therefore, it is still unclear whether we
really need the running spectral index or not.

\subsection*{Non-Gaussian Density Fluctuations}

The Gaussianity of initial density fluctuations is quite important
because of the following reasons: (1) It is related with how the
fluctuations were generated. Thus it sheds light on the very early
universe (perhaps inflationary phase). (2) Practically it is important in
studying the structure formation in the universe, because it gives
the initial condition of density fluctuations. Indeed small input of
non-Gaussianity largely changes the evolution and formation of
non-linear objects.

The most simple, direct way to test the Gaussianity is to explore the
CMB map. \citet{komatsu03} explicitly showed the CMB map obtained by
{\it WMAP} is consistent with the Gaussian fluctuations. Specifically,
they quantified the degree of Gaussianity by adding a quadratic term to
the curvature perturbation $\Phi$:
\begin{equation}
 \Phi(\vec{x})=\Phi_{\rm lin}(\vec{x})+f_{\rm NL}\left[\Phi_{\rm lin}^2(\vec{x})-\langle\Phi_{\rm lin}^2(\vec{x})\rangle\right],
\label{conc_curv}
\end{equation}
where $\Phi_{\rm lin}$ is the Gaussian linear perturbation, and $f_{\rm NL}$
describes the amplitude of non-Gaussianity. This functional form is
motivated by inflation models, though simple inflation models predict
quite small non-Gaussianity, $f_{\rm NL}=\mathcal{O}(1)$. Their results
are $f_{\rm NL}=38\pm48$ from the angular bispectrum, and $f_{\rm
NL}=22\pm81$ from the Minkowski functionals. Both results are consistent
with $f_{\rm NL}=0$, i.e., Gaussian fluctuations. Since $\Phi\sim(\Delta
T/T)\sim10^{-5}$, these results mean that the second (non-Gaussian) term
in equation (\ref{conc_curv}) should be at least  $10^{-3}$ smaller than
the first (Gaussian) term.  

Although CMB anisotropies are basically consistent with Gaussian
fluctuations, small non-Gaussianity might be seen in {\it WMAP} data;
\citet{park04} found non-Gaussian signatures using genus statistics; 
\citet{vielva04} also detected non-Gaussian signals at $\sim 4^\circ$
scales from simple one-point statistics (skewness and kurtosis). In both
case, the non-Gaussian signals are significant only on the southern
hemisphere. In addition, the non-Gaussianity may explain the evolution
of cluster abundances better \citep*{mathis04}. However, it is still
controversial whether we have really detected non-Gaussianity. 
In either case, it is very important to test Gaussianity further, using
higher-resolution CMB data or large-scale structure
\citep*[e.g.,][]{scoccimarro04}. 

\chapter{Structures of Dark Matter Halos in a Cold Dark Matter Universe:
Concord or Conflict?}
\label{chap:halo}
\def\mychapheadname{Structures of Dark Matter Halos in a CDM Universe}
\markboth{CHAPTER \thechapter.
{\MakeUppercase\mychapheadname}}{}

\section{Has CDM Confronted Difficulties?} 
\markboth{CHAPTER \thechapter.
{\MakeUppercase\mychapheadname}}{\thesection.
\MakeUppercase{Has CDM Confronted Difficulties?}}

\subsection{Testing the CDM Paradigm on Small Non-Linear Scales}

As extensively discussed in Chapter \ref{chap:conc}, the CDM model has 
been quite successful in explaining the large-scale structure of the
universe. However, this just means that the CDM model has passed the
tests at large scale ($\gtrsim$Mpc) where the linear theory can be
applied. It is therefore of great importance to check whether the CDM
model is still successful on much smaller scales. The tests on small
scales allow us (1) to study the nature of dark matter because structures
on very small scales (i.e., dense and highly non-linear regions) are
sensitive to possible interactions such as collisions and annihilations,
and (2) to probe the power spectrum on small scales and to test the
hypothesis that the dark matter is cold. Note that the standard CDM
model is assumed to be collisionless, since good candidates of the
non-baryonic cold dark matter, such as WIMPs (see Appendix
\ref{chap:dm}), have only small cross sections of interactions.

\begin{figure}[tb]
\begin{center}
 \includegraphics[width=0.5\hsize]{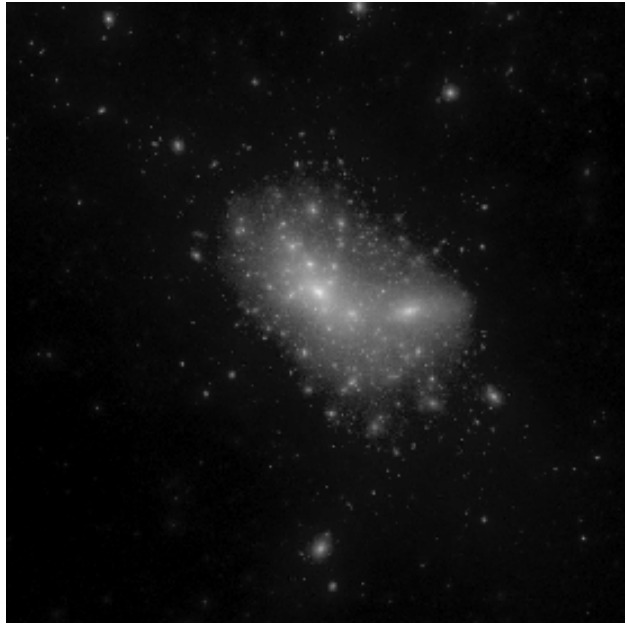}
\caption{The structure of a dark halo at $z=0$ in $N$-body simulation
 based on the CDM model, using 30 million particles. This Figure is
 taken from \citet*{fukushige04}. 
\label{fig:halo_fkm}}
\end{center}
\end{figure}

Dark matter halos, highly nonlinear self-gravitating systems of dark
matter and plausible sites hosting a variety of astronomical objects
such as galaxies and clusters, are desirable tool to test the CDM on
small scale, mainly because structures of dark halos are predicted from
theory by using $N$-body simulations (see Figure \ref{fig:halo_fkm}).
Changing the nature of dark matter drastically alters structures of dark
halos, such as central concentrations, shapes, and abundances of
substructures. Therefore by comparing structures of dark halos in
observations with theoretical predictions, one can in principle check the
validity of the collisionless CDM hypothesis. It is also known that the
central concentrations of dark halos are affected by the amplitude of
power spectrum on corresponding scales. Hence increasing the velocity of
dark matter in the early universe (i.e., relaxing the assumption of
``cold'') has a significant effect on structures of dark halos.

It was only a decade ago that we began to be able to compute structures
of dark halos in detail, because in $N$-body simulations we need
numerous number of particles to achieve sufficient mass and force
resolutions. However, even if we come to know structures of dark halos
in theory, it's no picnic to observe structures of dark halos. It is
difficult because dark matter is dark, i.e., cannot be observed by
normal methods. Thus, sometimes we have to resort to indirect methods
based on several assumptions, which often turn out to be wrong or
inaccurate.  

\subsection{The Crisis?}
\label{sec:halo_crisis}

The CDM model predicts centrally concentrated mass distribution of dark
halos. \citet*{navarro95,navarro96,navarro97} found in their $N$-body
simulations that the spherically averaged density profiles of dark halos
are well fitted by the following form: 
\begin{equation}
 \rho(r)=\frac{\rho_{\rm crit}(z)\delta_{\rm c}(z)}
{\left(r/r_{\rm s}\right)\left(1+r/r_{\rm s}\right)^2},
\label{halo_nfw}
\end{equation}
where $\delta_{\rm c}(z)$ and $r_{\rm s}$ are characteristic density
contrast and the scale radius, respectively, and both depend on the mass
of dark halos. However, they claimed that the functional form of
equation (\ref{halo_nfw}) is universal, i.e., it can be applicable to
dark halos of any masses (and it is independent of cosmological
parameters). This density profile is sometimes called the NFW density
profile. The NFW density profile has attracted many attentions, because
it seems quite unnatural for dark halos to have such universal forms,
and also because it is practically useful tool in making theoretical
predictions based on the CDM model. In practice, most higher-resolution
$N$-body simulations indicate steeper inner profiles $\rho\sim r^{-\alpha}$
where $1\lesssim\alpha\lesssim 1.5$
\citep*{fukushige97,fukushige01,fukushige03,moore99b,ghigna00,jing00a,  
klypin01,power03,fukushige04,hayashi04,navarro04}. In any case, all
$N$-body simulations predict the cuspy, centrally concentrated density
profile of dark halos.

Another important prediction of the CDM model is many substructures in
dark halos (see Figure \ref{fig:halo_fkm}). In the CDM model,
substructures contribute 10\%-15\% of the total mass of the host halos
\citep*[e.g.,][]{tormen98}; thus they can affect many
astrophysical/astronomical phenomena in several ways. In addition, it
turned out that the substructure mass function depends on the mass of the
host halo only weakly, once the mass of substructures is normalized by
the mass of their host halo \citep[e.g.,][]{moore99a}. 

\begin{figure}[p]
\begin{center}
 \includegraphics[width=0.7\hsize]{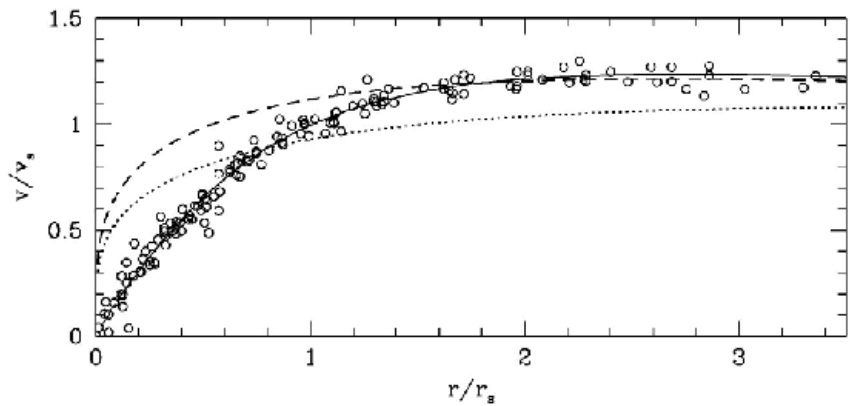}
\caption{The rotation curves of galaxies, scaled to fit the universal
 density profile. The solid line indicates the rotation curve that
 results from the density profile with a constant density core. Dashed
 and dotted curves are those predicted by the $N$-body simulations.This
 Figure is taken from \citet{moore99b}. 
\label{fig:halo_rot}}
\end{center}
\end{figure}
\begin{figure}[p]
\begin{center}
 \includegraphics[width=0.6\hsize]{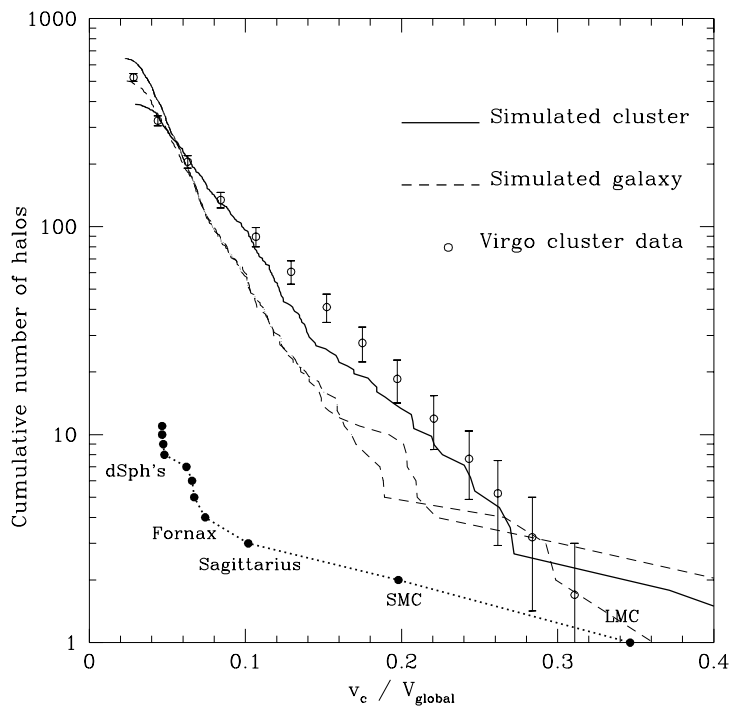}
\caption{Abundances of substructures within the Milky Way and the Virgo
 Cluster are compared with those in $N$-body simulations. The cumulative 
 numbers of substructures are plotted as a function of their circular
 velocity. The dotted line shows the observed distribution within the
 Milky Way, while open circles plot the data for  the Virgo
 Cluster. This Figure is taken from \citet{moore99a}.
\label{fig:halo_subs}}
\end{center}
\end{figure}

Are the properties of dark halos consistent with observations? The
answer may be no; many possible problems have been raised in both
galaxy- and cluster-mass scales.

The central concentrations of dark halos have been tested using the
dwarf/Low Surface Brightness (LSB) galaxy systems, mainly because in
such systems dark matter dominate even near the center and the effects
of baryons are thought to be small. Specifically, the rotation curves at
the inner part of dwarf galaxies have been measured, and have been
compared with CDM predictions. Surprisingly, it has been claimed that the
cuspy profile found in $N$-body simulations cannot explain the slow
rises of the rotation curves in observations. For instance,
\citet{moore99b} explicitly showed that the universal density profile
failed to reproduce observed rotation curves of dwarf/LSB galaxies
(Figure \ref{fig:halo_rot}). They claimed that the density profile with
a nearly flat core is needed to fit the data. 

Another problem in galaxy-mass scale is that substructures are not so
common as the CDM model predicts. \citet{moore99a} compared abundances
of substructures within the Milky Way and the Virgo Cluster with those
in $N$-body simulations (Figure \ref{fig:halo_subs}). They showed that 
the CDM model clearly over-predicts the number of substructures in the
Milky Way, but it is consistent in the Virgo Cluster.

\begin{figure}[tb]
\begin{center}
 \includegraphics[angle=-90,width=0.6\hsize]{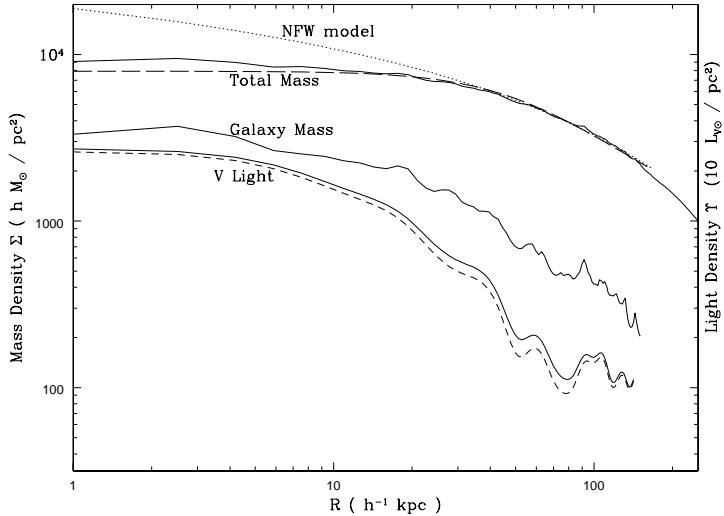}
\caption{Reconstructed mass density and light density  of the lensing
 cluster CL0024+1654 at $z=0.39$. Total ({\it thick}) and galaxy-only
 ({\it  thin}) components of the mass are shown. Total mass profile is
 quite different from the one NFW predicted ({\it dotted}). This Figure
 is taken from \citet{tyson98}. 
\label{fig:halo_0024}}
\end{center}
\end{figure}

In addition, there might be difficulties in the cluster-mass scale as
well as in the galaxy-mass scale. \citet*{tyson98} reconstructed the mass
distribution from the strong/weak gravitational lensing seen in the
cluster CL0024+1654, and concluded that the mass distribution differs
from the NFW density profile. The reconstructed mass distribution also
has a nearly flat core in the mass center, as in the case of dwarf/LSB
galaxies.  

These results suggest that the CDM model might be wrong on small scales.
A number of solutions to this problem has been proposed, including the
modification of the nature of dark matter (see Appendix \ref{chap:dm}). 
For instance, \citet{spergel00} proposed interactions between dark
matter particles. They showed that the elastic cross section of
$\sigma_{XX}/m_X\sim 10^{-24}{\rm cm^2GeV^{-1}}$ is needed to solve the
discrepancy. This, in turn, implies that detailed comparison of
structures of dark halos can constrain the nature of dark matter. 
In the next sections, we will see the current status of the
comparison to check how significant the claimed discrepancy is. 

\section{Testing the CDM Paradigm: Halo Concentration} 
\markboth{CHAPTER \thechapter.
{\MakeUppercase\mychapheadname}}{\thesection.
\MakeUppercase{Testing the CDM Paradigm: Halo Concentration}}

\subsection{Rotation Curves}

\begin{figure}[tb]
\begin{center}
 \includegraphics[width=0.5\hsize]{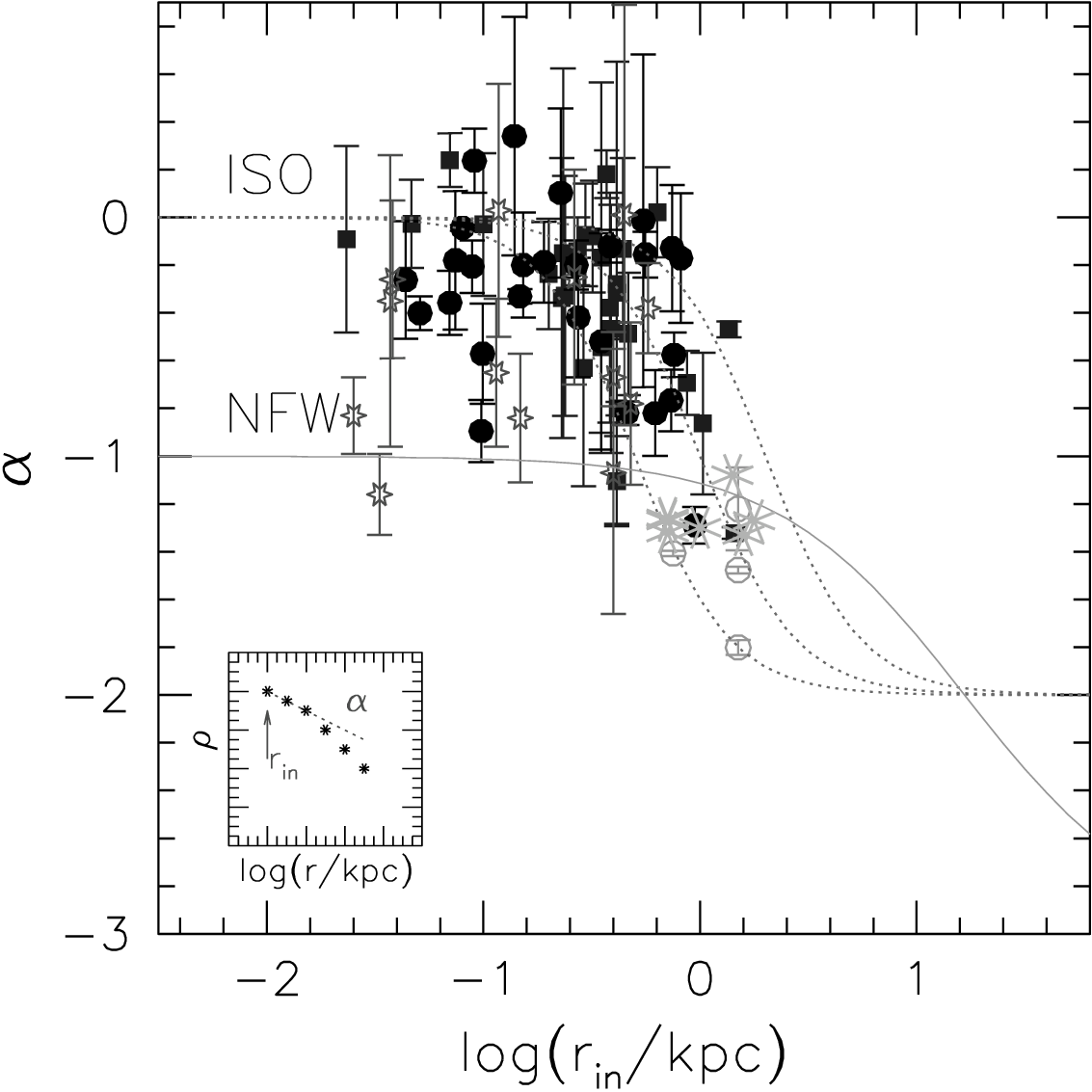}
\caption{Inner mass density versus resolution $r_{\rm in}$ of
 LSB rotation curves. Unlike the text, $\alpha$ is defined by the local
 slope of the density profile, $\alpha=d\ln\rho/d\ln r$. Filled circles
 and squares show the results of \citet{deblok01} and \citet{deblok02},
 respectively. Open starts denote the results of \citet{swaters03}. The
 large asterisks near $\alpha\sim -1$ and $r_{\rm in}\sim 1$kpc are the
 simulations by \citet{hayashi04}. This Figure is taken from \citet{deblok04}.
\label{fig:halo_alpha}}
\end{center}
\end{figure}

Rotation curves in dwarf/LSB galaxies are one of the most popular methods
to test the halo concentration. Many observations followed the claim of
\citet{moore99b}; \citet{deblok01} and \citet{deblok02} found
core-dominated structures of LSB galaxies using H$\alpha$/\ion{H}{1}
rotation curves, and claimed that they are clearly inconsistent with the
CDM model. \citet*{deblok03} showed
that systematic effects, such as non-circular motion and off-center, are
not so significant as to change the conclusions. \citet{simon03} reached
the similar conclusion from high-resolution measurements of the dwarf
spiral galaxy with H$\alpha$/CO. On the other hand,
\citet{vandenbosch01} and \citet{swaters03} also analyzed H$\alpha$
rotation curves and claimed that the current data poorly constrain the
inner density profile, and that it is difficult to discriminate between
cusp and core.  See Figure \ref{fig:halo_alpha} for the summary of
rotation curve measurements. 

Therefore, current observations seem to favor core rather than cusp,
although the arguments against core interpretations still remain. Actually,
\citet{hayashi04} pointed out that the discrepancy might simply reflect
the difference between circular velocity and gas rotation speed. If this
is true, rotation curves cannot be a good test of the halo
concentration. Moreover, it might be possible that a disk bar, which
should be ubiquitous in forming galaxies, produces cores in cuspy
CDM halos \citep{weinberg02}. Thus we need to understand how dwarf/LSB
galaxies are formed, and also to clarify the relation between gas dynamics
and gravitational potential in a realistic situation, before we conclude
that the CDM model is inconsistent with observations.

It also should be noted that the observations probe the central regions
smaller than those current $N$-body simulations are accessible.
Therefore, the discrepancy could be just due to extrapolation of results of
$N$-body simulations beyond the resolution. 
 
\subsection{Clusters of Galaxies}

First we review follow-up studies of CL0024+1654 which mass distribution
was claimed to be inconsistent with the CDM model by \citet{tyson98}.
\citet{broadhurst00} claimed that a cuspy mass distribution also can
reproduce lensed images, but \citet{shapiro00} pointed out in such a cuspy
model the velocity dispersion is too large to be consistent with the
observation. \citet{czoske02} suggested that the flat density core might
be produced by the high-speed collision along the line of sight which
are inferred from the spectroscopy of $\sim 300$ galaxies in the
cluster. However, X-ray data showed that the gas in the cluster seems to
be in equilibrium \citep{ota04}. The lesson to be drawn from these
studies, therefore, is that individual modeling of specific cluster is
difficult and may suffer from the special selection function.

Besides CL0024+1654, there has been many attempts to constrain the halo
concentration with lensing clusters. \citet{smith01} found steep inner
profile ($\alpha\sim 1.3$) in A383. \citet{gavazzi03} analyzed
MS2137$-$2353 and found that cored profile better reproduce the lensed
images. Weak lensing analyses have been also done in several clusters.
Basically they are consistent with the NFW density profile
\citep*{clowe01,clowe02,dahle03}, although the cored profile tends to
fit the data equally. 

\begin{figure}[tb]
\begin{center}
 \includegraphics[width=0.4\hsize]{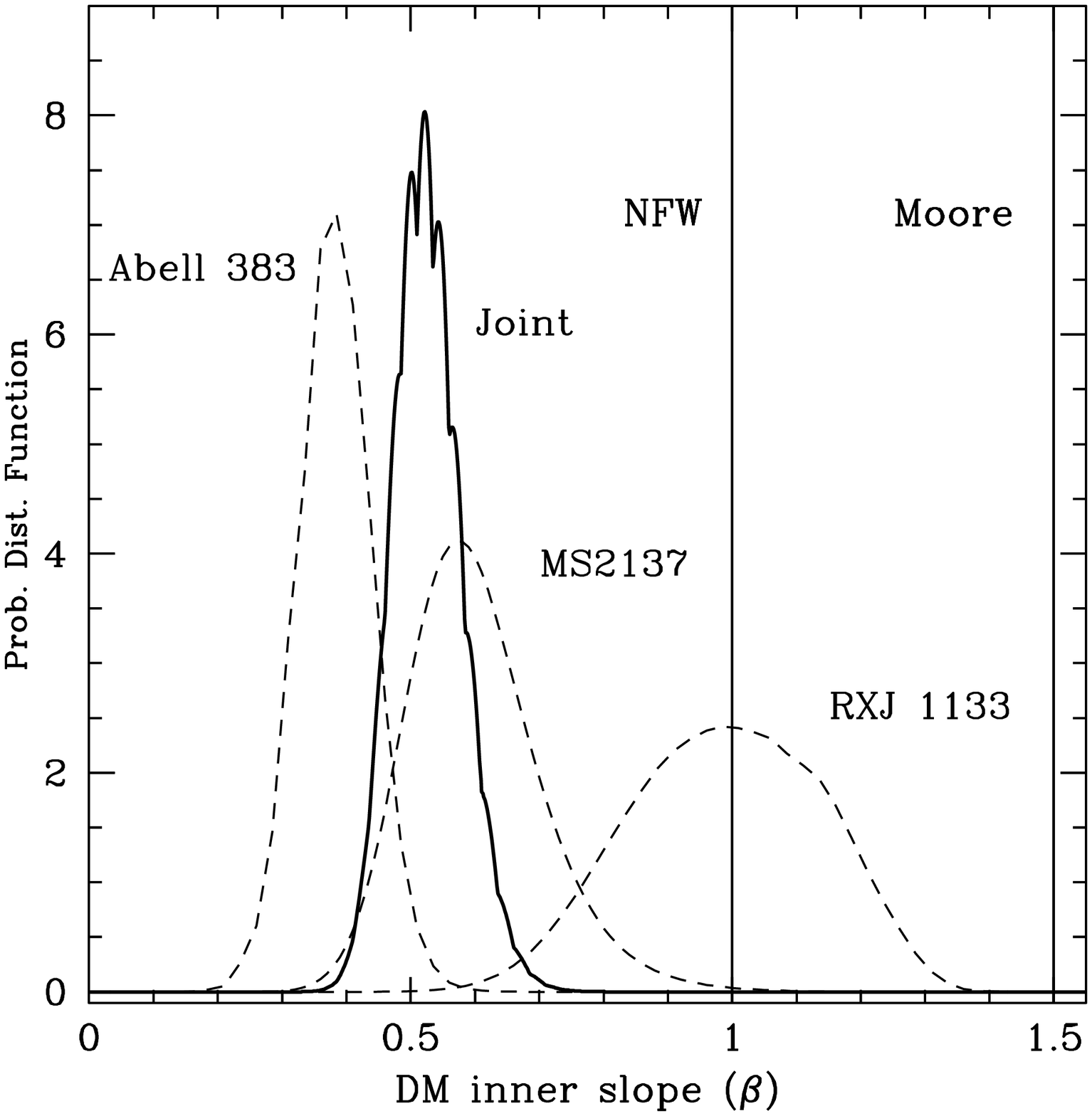}
\hspace{5mm}
 \includegraphics[width=0.4\hsize]{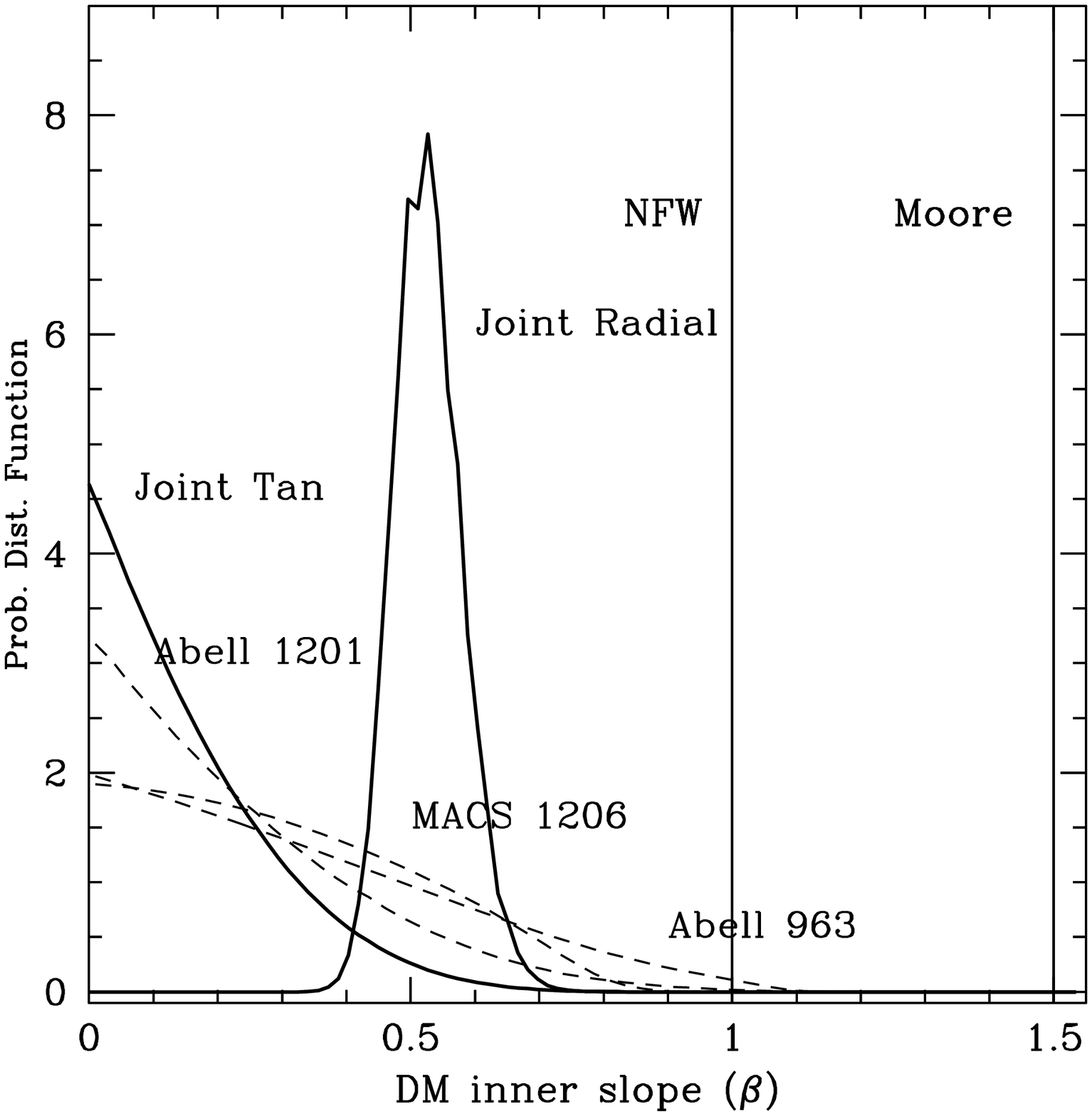}
\caption{PDFs of the inner density profile ($\beta$ should be read
 $\alpha$). Left: PDFs for radial arc clusters. Right: PDFs for tangential
 arc clusters. These Figures are taken from \citet{sand04}. 
\label{fig:halo_sand}}
\end{center}
\end{figure}

Recently, \citet{sand04} studied 6 lensing clusters in detail, and
claimed that they are inconsistent with the CDM model on average
(see Figure \ref{fig:halo_sand}). They showed that
clusters with radial arcs are well constrained to $\alpha\sim0.5$,
though scatter is very large, and that clusters with only tangential
arcs give upper limit on the inner slope, $\alpha\lesssim 0.5$.
Both tangential and radial arc clusters strongly disfavor the NFW
density profile. However, the result strongly relies on several
simplified assumptions, such as the spherical symmetry and the fixed
value of the scale radius; \citet{bartelmann04} and \citet{dalal04b} showed
that steep inner density profiles can be reconciled with the data if we
relax the assumptions. Their arguments clearly demonstrate that the
degeneracy between mass distributions is so significant that it is quite
difficult to draw conclusions from the modeling of lensing clusters
only.

An additional constraint comes from X-ray observations, by assuming
hydrostastic equilibrium. \citet{tamura00} found that ASCA and ROSAT
measurements of the cluster A1060 are consistent with the NFW density
profile. \citet{lewis03} and \citet{buote04} measured X-ray luminosity
and temperature profiles of regular, relaxed clusters A2029 and A2589,
and found that the density profiles show good agreements with the
CDM predictions. Specifically, the inner slope is constrained to
$\alpha=1.19\pm0.04$ for A2029 and to $\alpha=1.35\pm0.21$ for A2589,
respectively. It seems like that X-ray data basically support for the
CDM model, but analyses in more clusters will be important to draw a
robust conclusion.

\section{Testing the CDM Paradigm: Halo Shape} 
\markboth{CHAPTER \thechapter.
{\MakeUppercase\mychapheadname}}{\thesection.
\MakeUppercase{Testing the CDM Paradigm: Halo Shape}}

\begin{figure}[tb]
\begin{center}
 \includegraphics[width=0.45\hsize]{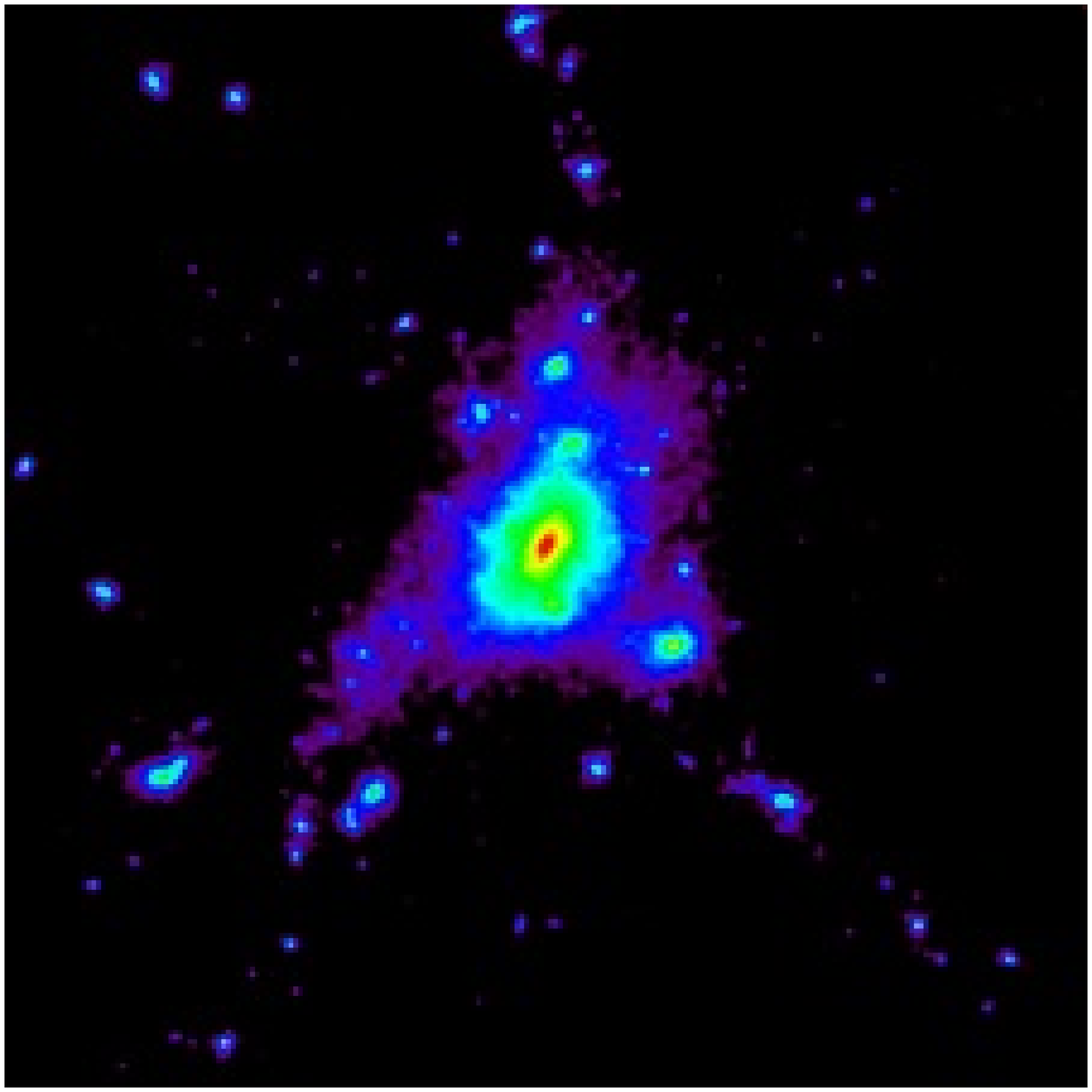}
\hspace{1mm}
 \includegraphics[width=0.45\hsize]{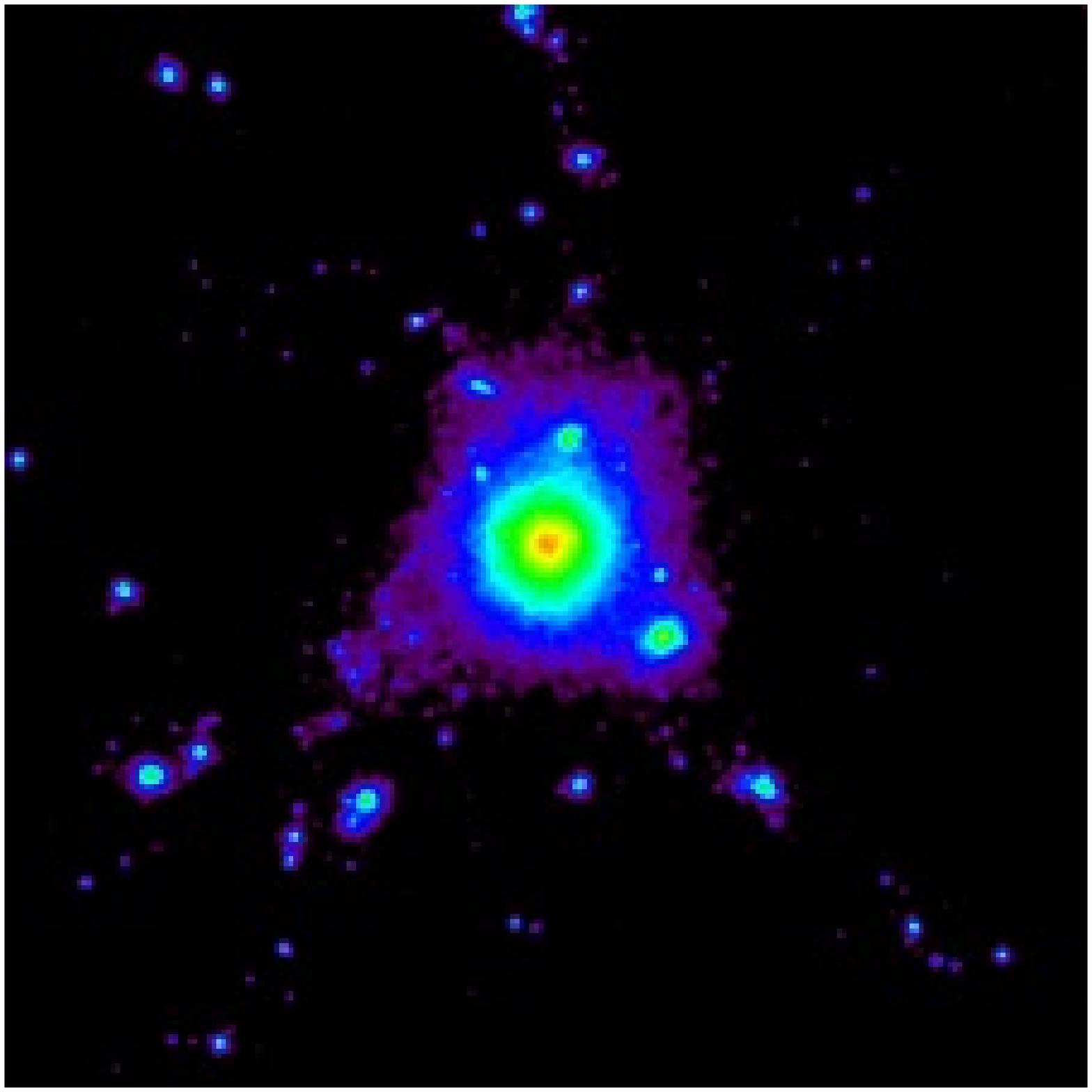}
\caption{Dark matter distributions in simulated clusters. Left: The
 shape of dark halos in the standard CDM model. Right: The
 shape of dark halos in the weakly self-interacting dark matter model
 with the cross section of $\sigma_{XX}=10{\rm cm^2}g^{-1}$. This Figure
 is taken from \citet{yoshida00b}.
\label{fig:halo_ny}}
\end{center}
\end{figure}

Shapes of dark halos also give us insight on the nature of dark matter.
In the CDM model, dark halos are not spherical, but rather triaxial
\citep[e.g.,][]{jing02}. Collisions between dark matter particles always 
make dark halos rounder, thus observations of elongated triaxial
dark halos would support for the CDM model. Indeed,
\citet{yoshida00a,yoshida00b} found in their series of $N$-body
simulations that self-interactions of dark matter do make the core of
dark halos rounder. Figure \ref{fig:halo_ny} clearly demonstrate how
collisions affect the shape of dark halos.

Observations seem inconsistent with this round halo; \citet{buote02}
observed the elliptical galaxies NGC720 with X-ray, and found that the
axis ratio of $\sim 0.4$ gives the best fit to the data;
\citet{miralda02} analyzed the lensing cluster MS2137$-$2353 and
discussed that the projected ellipticity of $\gtrsim 0.2$ is required to
fit the lensed images; \citet*{hoekstra04} derived the average
ellipticity of dark halos from weak lensing as $\langle
e\rangle=0.33^{+0.07}_{-0.09}$. These results suggest that collisions are
not so significant as to modify the shape of dark halos. However, it
should be noted that axis ratios have quite broad probability distribution
in the CDM model \citep{jing02}, and this might be also the case for
self-interacting dark matter model. If so, we need large number of
dark halos to test the shapes so as not to be affected by sample variance.

\section{Testing the CDM Paradigm: Halo Substructures} 
\markboth{CHAPTER \thechapter.
{\MakeUppercase\mychapheadname}}{\thesection.
\MakeUppercase{Testing the CDM Paradigm: Halo Substructures}}

\subsection{Satellite Galaxies}

The over-abundance of substructures in galactic halos, as shown in
Figure \ref{fig:halo_subs}, has raised many discussions. The abundant
substructures are firm prediction of the CDM model, because many
independent numerical simulations have confirmed the fact that the CDM
model predicts roughly 10\%-15\% of mass in a dark halo is bound to
substructures \citep{tormen98,klypin99,okamoto99,ghigna00,springel01,
zentner03,delucia04,kravtsov04}. The fraction of substructures is also
supported by theory \citep[e.g.,][]{oguri04d}. Popular ways to resolve
the conflict include modifying the nature of dark matter (see Appendix
\ref{chap:dm}) and introducing new inflationary models that can produce
density fluctuations with small-scale power cut-off such as  an
inflation model with broken scale-invariance \citep{kamionkowski00} and
a double hybrid inflation \citep{yokoyama00}. 

\begin{figure}[tb]
\begin{center}
 \includegraphics[width=0.6\hsize]{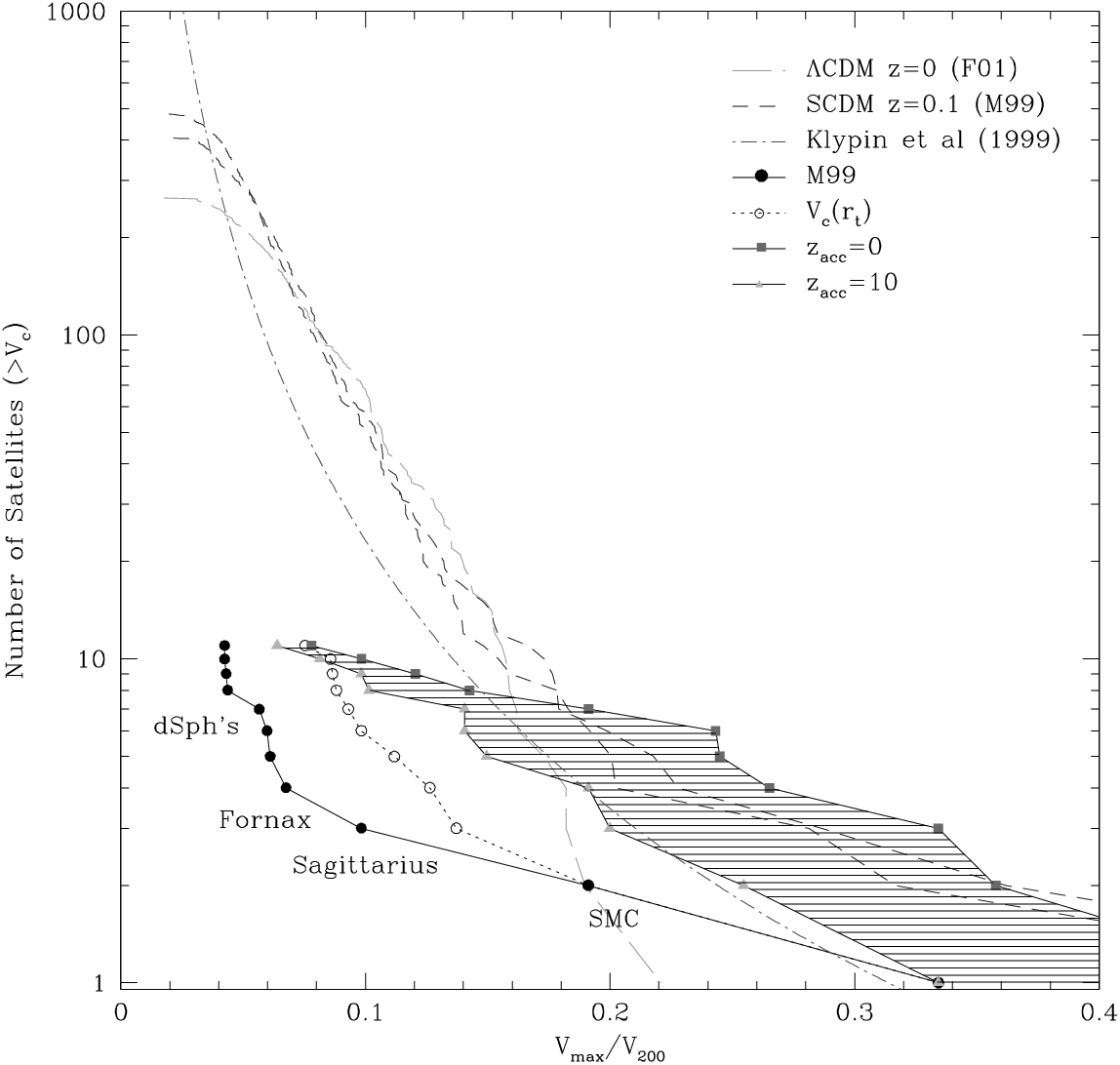}
\caption{Abundance of substructures within the Milky Way are compared
 with those in $N$-body simulations (see also Figure
 \ref{fig:halo_subs}). Numerical simulations of \citet{font01},
 \citet{moore99a}, and \citet{klypin99} are plotted to show the
 robustness of simulation results. Filled circles show the observed
 distribution assuming the isothermal potential as in \citet{moore99a}.
 Open circles plot circular velocities at luminosity cutoff assuming
 the NFW density profile. In the shaded region, observed velocities are
 converted to peak velocities of unstripped NFW halos (the uncertainty
 reflects different accretion time). Now there is no major discrepancy
 for massive substructures, $V_{\rm max}/V_{200}\gtrsim 0.16$.
 This Figure is taken from \citet{hayashi03}. 
\label{fig:halo_hayashi}}
\end{center}
\end{figure}

However, the problem may be resolved also by taking account of
astrophysical processes such as photo-ionizing background
\citep{somerville02} and inefficient star formation in small mass halos
\citep{stoehr02}. These ideas claim that the observed number of
satellite galaxies is small because only very massive substructures
contain stars and most substructures are {\it dark}. These ideas based
on the dark substructures, however, may not be consistent with
observations, either: Recent high-resolution numerical simulations have
found that the massive substructures tend to place in the outer part of
host halos, which is not the case for the satellites of the Milky Way
\citep[e.g.,][]{delucia04}.  

On the other hand, \citet{hayashi03} claimed that the apparent
discrepancy of abundances of massive substructures is caused by the
large difference between tidal radii of substructures in simulations and
radial cutoff observed in surface brightness profiles. This implies that
it may be possible to account for observed abundance of substructures
without invoking the nature of dark matter and/or photo-ionizing
background. Figure \ref{fig:halo_hayashi} illustrates the result. This
result suggests that we must be careful in comparing abundances of
satellite galaxies with $N$-body simulations. Even in their results,
low-mass substructures shows the difference, and this may be caused by
complicated astrophysical processes, such as the efficient feedback and
evaporation of gas.

\subsection{Gravitational Lensing}

As suggested in the previous subsection, one of main difficulties in the
comparison between simulations and observation is that substantial
fraction of substructures may be dark. It is quite hard to test the
existence of such dark substructures observationally. 

\begin{figure}[p]
\begin{center}
 \includegraphics[width=0.7\hsize]{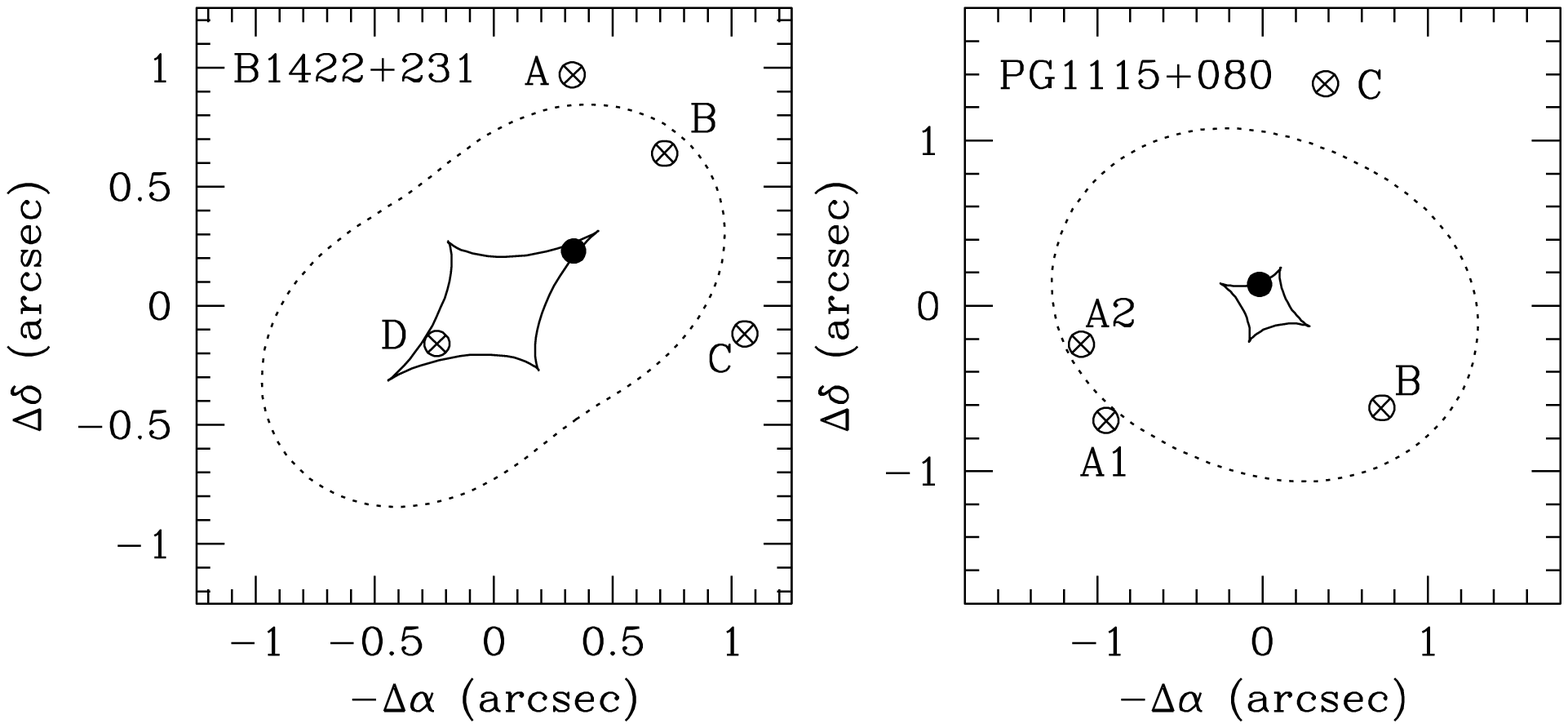}\\ \bigskip
 \hspace*{5mm}\includegraphics[width=0.7\hsize]{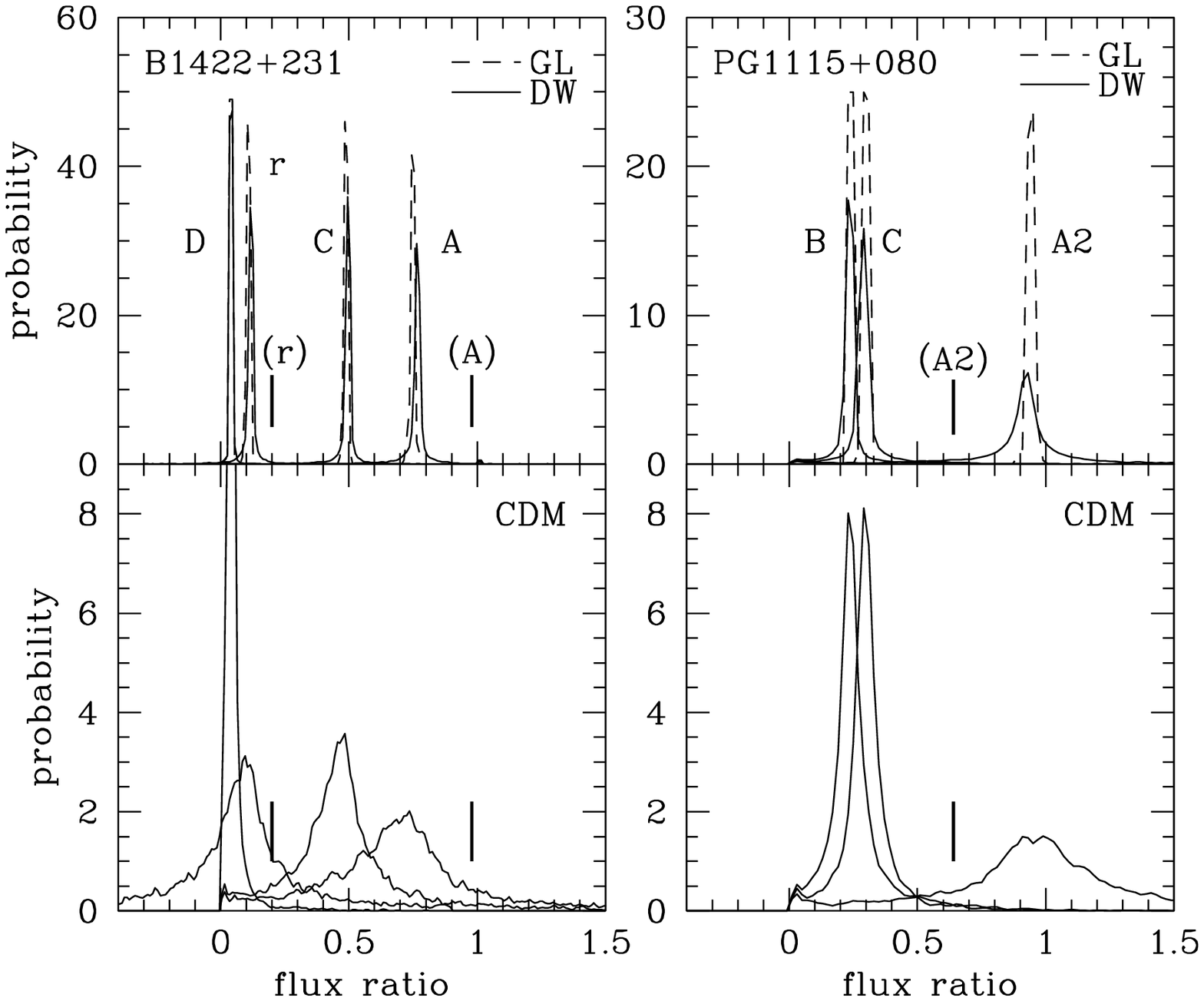}
\caption{Upper: Image configurations of lensed quasar
 systems B1422+231 and PG1115+080. Lower: Probability distributions of
 images in the lensed quasar systems. As models of substructures, globular
 clusters (GL), dwarf satellites (DW), and CDM substructures (CDM) are
 considered. Fluxes are normalized by those of image B (B1422+231) and
 image A1 (PG1115+080). The flux ratio $r$ for B1422+23 is defined by
 $r=(A+B+C)/(|A|+|B|+|C|)$. Observed flux ratios are denoted by short
 vertical bars. These Figures are taken from \citet{chiba02}.
\label{fig:halo_sublens}}
\end{center}
\end{figure}

Gravitational lensing can avoid such problem; it can detect
substructures directly even if they are dark. The existence of
substructures in lensed quasar systems was first suggested by
\citet{mao98}. They claimed that the anomalous flux ratio in the
quadruple lens B1422+231 is due to substructures in the lens galaxy.
Indeed, it has been shown that the large amount of substructures
predicted in the CDM model is needed to account for flux anomalies in
several lens systems
\citep{metcalf02,chiba02,dalal02,bradac02,kochanek04a}. As an example,
Figure \ref{fig:halo_sublens} shows how much the CDM substructures can
change the flux ratios between multiple images. Although the observed
flux ratios are very different from the median flux ratios predicted in
modeling, the probability distributions of flux ratios induced by the
CDM substructures are so broad that the anomalous flux ratios can be
reconciled with the mass modeling. 

Caveats about this method are that the results are sensitive to the
spatial distribution of substructures \citep*{chen03b}, and that there
is a degeneracy with the complexity of the smooth components
\citep*{evans03,keeton03c,kochanek04a,kawano04}. Actually, substructures
we need to fit flux rations might be even larger than the CDM model
predicts, when we take account of the spatial distribution
\citep{evans03,mao04}.  Therefore these flux anomalies might be caused
by stellar components in the lens galaxies \citep{schechter02} or
massive black holes ($\sim 10^5-10^6M_\odot$) in the halos
\citep{mao04}, rather than the CDM substructures. To avoid these
problems, it may be needed to develop more sophisticated methods, such
as spectroscopic lensing \citep[e.g.,][]{metcalf04} and mesolensing,
i.e., additional strong lensing of multiple images by substructures
\citep*{yonehara03}. 

\section{Need for More Studies} 
\markboth{CHAPTER \thechapter.
{\MakeUppercase\mychapheadname}}{\thesection.
\MakeUppercase{Need for More Studies}}

We have reviewed various tests of the CDM paradigm at small non-linear
scales. Although the situation is not so bad as first insisted (\S
\ref{sec:halo_crisis}), it is still inconclusive whether these
observations are well explained by the CDM model or not. In particular, 
a confusion comes from the fact that similar approaches sometimes yield 
different conclusions. This implies that systematic effects, e.g.,
the selection effect, the treatment of astrophysical processes,
degeneracy with other parameters, etc., are very important. The
understanding of astrophysical processes is especially important in
using indirect methods such as rotation curves of galaxies. 

In conclusion, we need more studies; we need more independent tests in
order to come to a firm conclusion on the validity of the CDM model on
small scales, as well as the improvements of each test. Think of the
reason why the concordance model (Chapter \ref{chap:conc}) is now
accepted widely; this is because many independent tests point to the
concordance model! We believe results on small scales also converge to
somewhere as we add more observations, though no one knows where it is. 

\chapter{Gravitational Lensing by Triaxial Dark Halos} 
\label{chap:tri}
\def\mychapheadname{Gravitational Lensing by Triaxial Dark Halos}
\markboth{CHAPTER \thechapter.
{\MakeUppercase\mychapheadname}}{}

\section{Why Do We Need Non-Spherical Lens Models?}
\label{sec:tri_intro}
\markboth{CHAPTER \thechapter.
{\MakeUppercase\mychapheadname}}{\thesection.
\MakeUppercase{Why Do We Need Non-Spherical Lens Model?}}

In this thesis, we newly construct a non-spherical lens model for lens
statistics (see Appendix \ref{chap:lens} for basics of gravitational
lensing). Actually, there has been no work on cluster-scale lens 
statistics that adopts non-spherical lens models (see Introductions of
Chapters \ref{chap:arc} and \ref{chap:lat}), despite CDM halos are not
spherical at all (see Chapter \ref{chap:halo}). The main reasons are (1)
non-spherical modeling makes it much more difficult to compute lensing
cross sections and hence lens statistics, and (2) we didn't have a
reliable model of non-spherical descriptions of lens objects, i.e., dark
halos. As for (2), however, it is now possible to construct such
non-spherical model using high-resolution $N$-body simulations. For
instance, \citet{jing02} fitted dark halos by the triaxial model and derived
the probability distribution functions (PDFs) of the triaxiality from
their cosmological simulations. These modelings enable us to
incorporate the non-sphericity in the lens statistics. To overcome (1),
we will develop semi-analytic methods to compute the number of lenses in
triaxial dark matter halos in Chapters \ref{chap:arc} and
\ref{chap:lat}. This combines the lensing cross section from the Monte
Carlo ray-tracing simulations and the probability distribution function
of the axis ratios evaluated from the cosmological simulations.   

\begin{figure}[t]
\begin{center}
 \includegraphics[width=0.8\hsize]{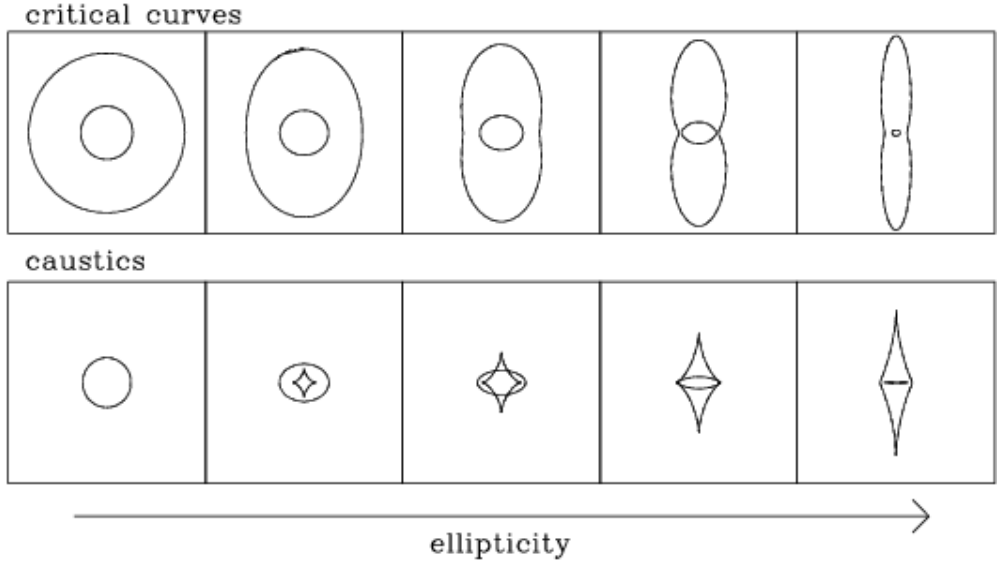}
\caption{Illustration on how critical curves ({\it upper}) and caustics
 ({\it lower}) changes as the ellipticity of the projected mass density
 distribution is increased (from $e=0$ to $e=0.8$). In this plot, we
 adopt the elliptical NFW density profile. The first column shows those
 of the spherical mass distribution.  
\label{fig:tri_crit_e}}
\end{center}
\end{figure}

However, why do we need non-spherical lens models in statistical
studies? One of the most important reasons is that the deviation from
the spherical symmetry affects gravitational lensing drastically. We
illustrate this in Figure \ref{fig:tri_crit_e}. In the spherical mass
distribution, there is only one caustic curve (radial caustic), because
the tangential caustic degenerates at the center of mass distribution.
But, once we introduce non-sphericity in the mass distribution, the
tangential caustic no longer degenerates; it grows as increasing the
ellipticity, and at last it becomes much larger than the radial caustic.
Since the probability for multiple images is proportional to the strong
lensing cross section which is given by the area enclosed by caustics,
it is expected that the non-sphericity has a great impact on lens statistics.

Another important reason that we need to include the non-sphericity is
image multiplicities. Since it is shown that the number of images increases
(or decreases) by 2 when the source crosses a caustic, spherical halos
can produce 3 images at most. This is not the case for non-spherical
halos; the elliptical halos shown in Figure \ref{fig:tri_crit_e} can
produce more than 3 images due to non-degenerate tangential caustics.
In addition, the topology of caustics is sensitive to the degree of
the non-sphericity, as seen in Figure \ref{fig:tri_crit_e}. In the
observational side, many lensed quasar systems with more than 3 images
have been discovered so far. Thus the non-spherical modeling conveys us
qualitatively new information on mass distributions of lens objects.  

In short, the non-spherical lens modeling is an essential ingredient
for lens statistics, rather than a minor upgrade; it can change lens
probabilities drastically, and it offers us new information on the shape
of clusters. 

\section{Description of Triaxial Dark Matter Halos}
\label{sec:tri_halo}
\markboth{CHAPTER \thechapter.
{\MakeUppercase\mychapheadname}}{\thesection.
\MakeUppercase{Description of Triaxial Dark Matter Halos}}

In this section, we briefly summarize the triaxial model of dark matter
halos proposed by \citet[][ referred as JS02 in the rest of this
chapter]{jing02}.  They obtained the detailed triaxial modeling on 
the basis of their high-resolution individual halo simulations as well
as large-scale cosmological simulations. Most importantly, they provided
a series of useful fitting formulae for mass- and redshift-dependence
and the PDFs of the axis ratio and the concentration parameter.  Such
detailed and quantitative modeling enables us to incorporate the
non-sphericity of dark matter halos in a reliable manner.

\subsection{Isodensity Surfaces}

\begin{figure}[tb]
\begin{center}
 \includegraphics[width=0.5\hsize]{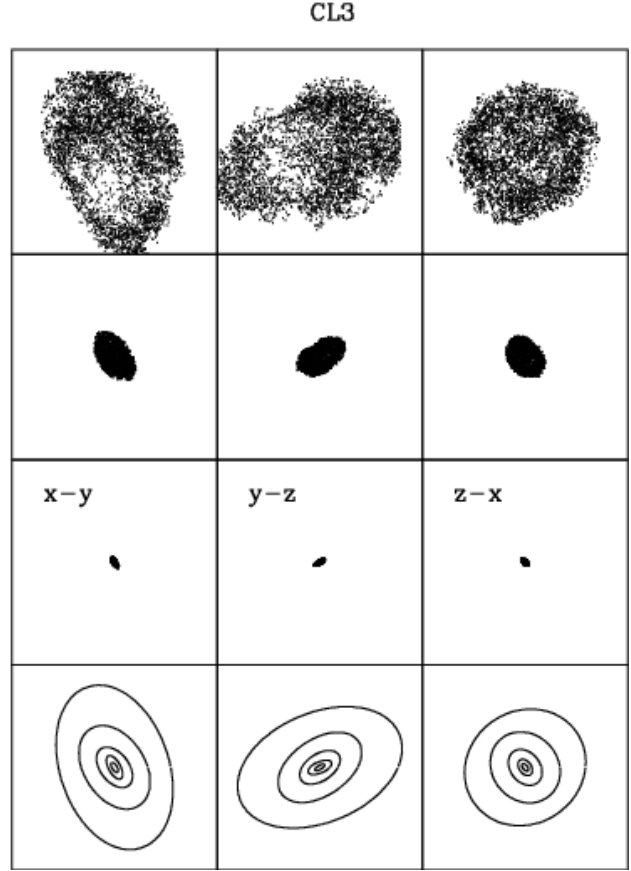}
\caption{Examples of projected particles distributions in a cluster-size
 halo. From top to bottom, particles in the isodensity shells with
 $A=100$, $2500$, $6.25\times 10^4$ are plotted. The bottom panels show
 triaxial fits to five isodensity surfaces defined in equation
 (\ref{tri_iso}). This Figure is taken from JS02.
\label{fig:tri_iso}}
\end{center}
\end{figure}

JS02 adopted the following method to find isodensity surfaces. First they
begin with the computation of a local density at each particle's
position by using the smoothing kernel \citep[e.g.,][]{hernquist89}:
\begin{eqnarray}
  W(r,h_i)=\frac{1}{\pi h_i^3} \left\{
      \begin{array}{ll}
       {\displaystyle 1-\frac{3}{2}\left(\frac{r}{h_i}\right)^2+\frac{3}{4}\left(\frac{r}{h_i}\right)^3} & \mbox{($r\leq h_i$)} \\ 
       {\displaystyle \frac{1}{4}\left(2-\frac{r}{h_i}\right)^3} & \mbox{($h_i<r<2 h_i$)} \\
 {\displaystyle 0} & \mbox{otherwise,}
            \end{array}
   \right. 
\end{eqnarray}
with $h_i$ being the smoothing length for $i$-th particle. They use 32
nearest neighbor particles to compute the local density $\rho_i$. The
smoothing length $h_i$ is set to be one-half the radius of the sphere
that contains those 32 neighbors. Then, from $\rho_i$ they construct
the isodensity surfaces corresponding to the five different thresholds:
\begin{equation}
\rho_s^{(n)}=A^{(n)}\rho_{\rm crit},\;\;\;\;\;A^{(n)}=100\times 5^{n-1}\;\;(n=1\sim 5).
\label{tri_iso}
\end{equation}
Actually they collected all particles with
$0.97\rho_s^{(n)}<\rho_i<1.03\rho_s^{(n)}$ and determined $n$-th
isodensity surface. To obtain the isodensity surfaces of the overall
density profile, they eliminate small distinct regions caused by the
substructures. 

\begin{figure}[p]
\begin{center}
 \includegraphics[width=0.7\hsize]{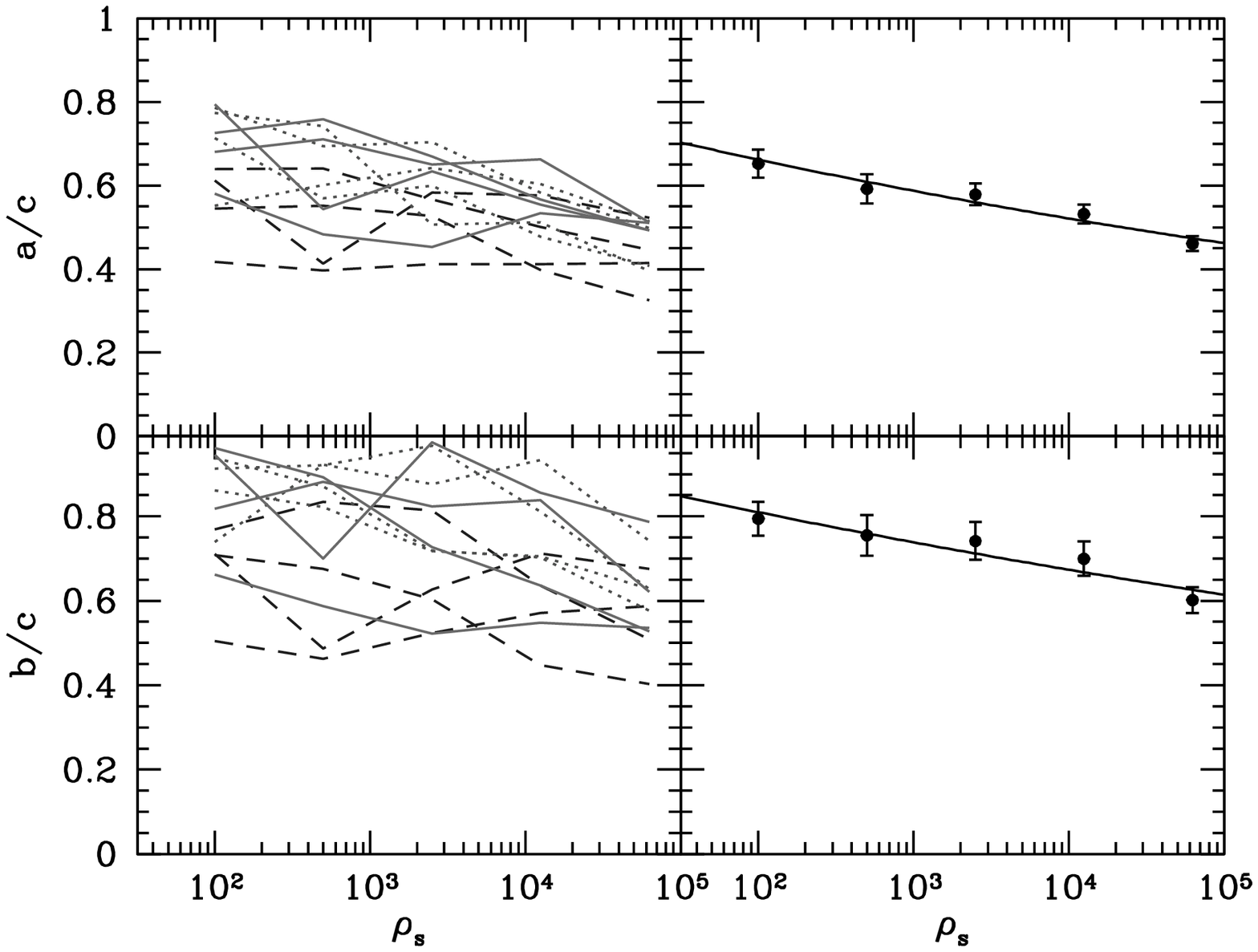}\\ \bigskip
 \includegraphics[width=0.7\hsize]{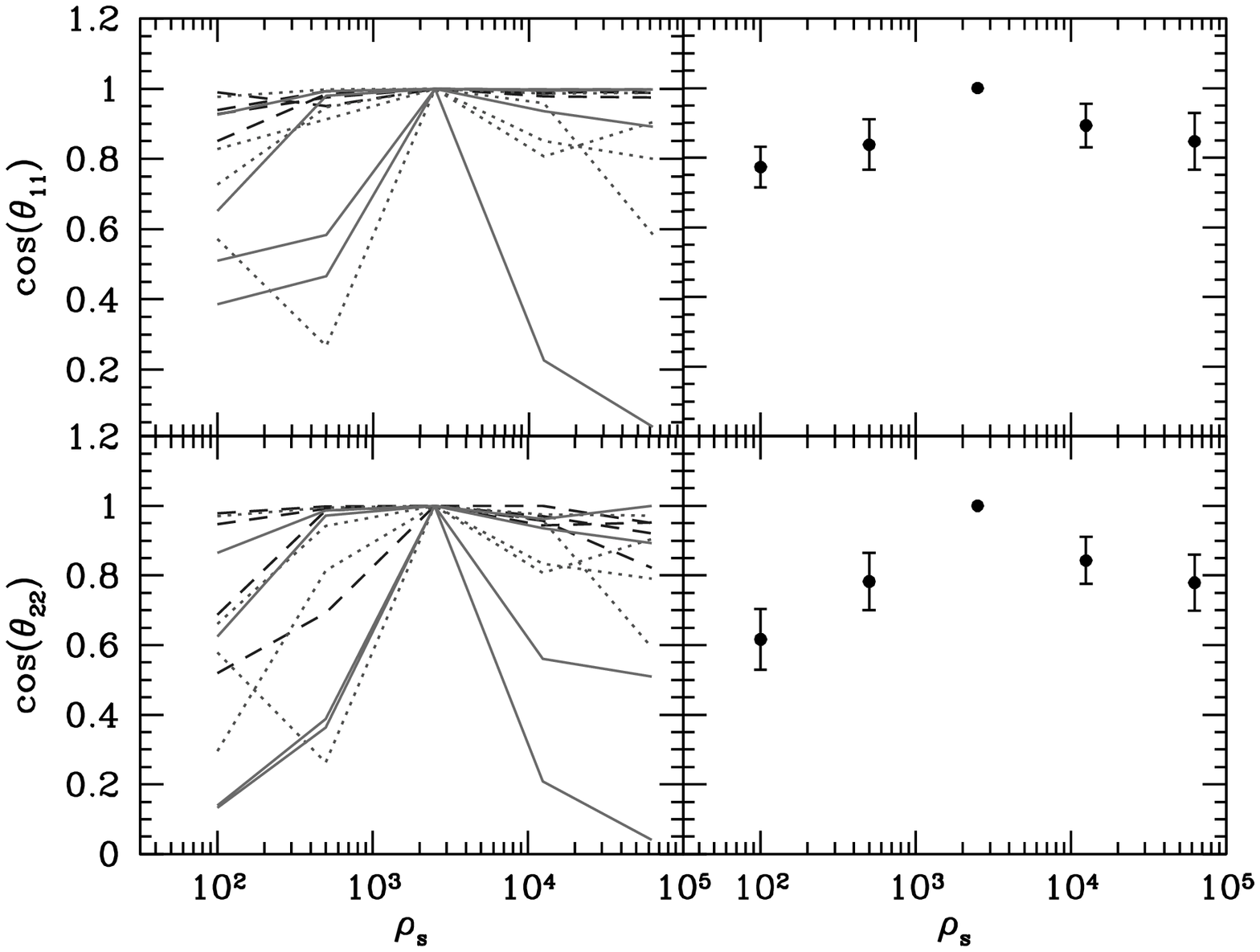}
\caption{Axis ratios ({\it top}) and degree of alignments ({\it bottom})
 on $\rho_s$, from fits to 12 halos. Right panels show the mean and its
 $1\sigma$ error. The angle $\theta_{11}$ is defined by the angle
 between major axis of the isodensity surface and that of the
 $A^{(3)}=2500$ isodensity surface. Similarly, $\theta_{22}$ is defined
 with respect to the middle axes. These Figures are taken from JS02.
\label{fig:tri_rhos}}
\end{center}
\end{figure}

An example is shown in Figure \ref{fig:tri_iso}. Since the isodensity
surfaces are well approximated by triaxial ellipsoids, JS02 fitted the
isodensity surfaces by the following form:
\begin{equation}
R^2(\rho_s)=\frac{X^2}{a^2(\rho_s)}+\frac{Y^2}{b^2(\rho_s)}+\frac{Z^2}{c^2(\rho_s)},
\end{equation}
where the lengths of principal vectors are chosen as $a\leq b\leq c$.
The dependence of axis ratios and the degree of alignments on $\rho_s$
is summarized in Figure \ref{fig:tri_rhos}. Axis ratios weakly depend on
$\rho_s$; the isodensity surfaces become more elongated in the central
region than in the outer region. It is also found that the alignments
are basically good; $\cos\theta_{11}\gtrsim 0.7$ and
$\cos\theta_{22}\gtrsim 0.6$. These agreement makes sense to approximate
a dark halo as the triaxial ellipsoid with the same axis ratios and axis
directions for the entire halo. Thus JS02 measured axis ratios from the
isodensity surface for $A^{(3)}=2500$, and regards them as the axis
ratios of the whole dark halo.

\subsection{Density Profile of Triaxial Dark Matter Halos}
\label{sec:tri_tnfw}

By considering triaxial shells in the way discussed in the previous
subsection, JS02 found that the following density profiles of triaxial dark
halos fit quite well: 
\begin{equation}
 \rho(R)=\frac{\delta_{\rm ce}\rho_{\rm crit}(z)}
{(R/R_0)^\alpha(1+R/R_0)^{3-\alpha}},
\label{tri_gnfw}
\end{equation}
where
\begin{equation}
 R^2\equiv c^2\left(\frac{x^2}{a^2}+\frac{y^2}{b^2}
+\frac{z^2}{c^2}\right)\;\;\;(a\leq b\leq c).
\label{tri_rdef}
\end{equation}
The precise value of the inner slope, $\alpha$, is still controversial,
but almost all the N-body simulations based on the collisionless CDM 
scenario indicate values between $1$ and $1.5$
\citep{navarro96,navarro97,fukushige97,fukushige01,fukushige03,moore99b,
ghigna00,jing00a,klypin01,power03,fukushige04,hayashi04,navarro04}.
Thus in this thesis we consider both $\alpha=1$ and $\alpha=1.5$ so as
to cover a possible range of the CDM predictions. 

JS02 defined the concentration parameter in the triaxial model as
\begin{equation}
 c_e\equiv \frac{R_e}{R_0},
\end{equation}
where $R_e$ is chosen so that the mean density within the ellipsoid of
the major axis radius $R_e$ is $\Delta_e\Omega_M(z)\rho_{\rm crit}(z)$
with\footnote{Note that our definitions 
of $\Delta_{\rm vir}$ and $\Delta_e$ are slightly different
from those of JS02; $\Delta_{\rm vir}({\rm JS02})= 
\Omega_M(z)\Delta_{\rm vir}$,
and $\Delta_e({\rm JS02})= \Omega_M(z)\Delta_e$. 
Of course this does not change the definition of $R_e$.}
\begin{equation}
 \Delta_e=5\Delta_{\rm vir}\left(\frac{c^2}{ab}\right)^{0.75}.
\end{equation}
Here $\Omega_M(z)$ and $\rho_{\rm crit}(z)$ denote the matter density
parameter and the critical density of universe at redshift $z$,
respectively, and $\Delta_{\rm vir}(z)$ denotes the overdensity of
objects virialized at $z$ (see \S \ref{sec:cosmo_mf}).

Then the characteristic density $\delta_{\rm ce}$ in equation
(\ref{tri_gnfw}) is written in terms of the concentration parameter
$c_e$ as 
\begin{equation}
 \delta_{\rm ce}=\frac{\Delta_e \Omega_M(z)}{3}\frac{c_e^3}{m(c_e)},
\end{equation}
where $m(c_e)$ is
\begin{equation}
\label{tri_mce}
 m(c_e)\equiv\frac{c_e^{3-\alpha}}{3-\alpha}
\, {}_2F_1\left(3-\alpha, 3-\alpha; 4-\alpha; -c_e\right),
\end{equation}
with ${}_2F_1\left(a, b; c; x\right)$ being the hypergeometric function.
For $\alpha=1$ and $1.5$, equation (\ref{tri_mce}) simply reduces to
 \begin{eqnarray}
  m(c_e)= \left\{
      \begin{array}{ll}
        \displaystyle{\ln(1+c_e)-\frac{c_e}{1+c_e}} & 
        \mbox{($\alpha=1$)}, \\ 
        \displaystyle{2\ln(\sqrt{c_e}+\sqrt{1+c_e})-2\sqrt{\frac{c_e}{1+c_e}}} &
        \mbox{($\alpha=1.5$)} .
      \end{array}
   \right. 
\end{eqnarray}
Since $R_e$ is empirically related to the (spherical) virial radius 
$r_{\rm vir}$ as $R_e/r_{\rm vir}\simeq 0.45$ (JS02), the scaling radius
in the triaxial model, $R_0$, for a halo of a mass $M_{\rm vir}$ is
given as 
\begin{equation}
 R_0=0.45 \frac{r_{\rm vir}}{c_e}
=\frac{0.45}{c_e} \left(\frac{3M_{\rm vir}}
{4\pi \Delta_{\rm vir}\Omega_M(z)\rho_{\rm crit}(z)}\right)^{1/3}.
\end{equation} 

\begin{figure}[p]
\begin{center}
 \includegraphics[width=0.65\hsize]{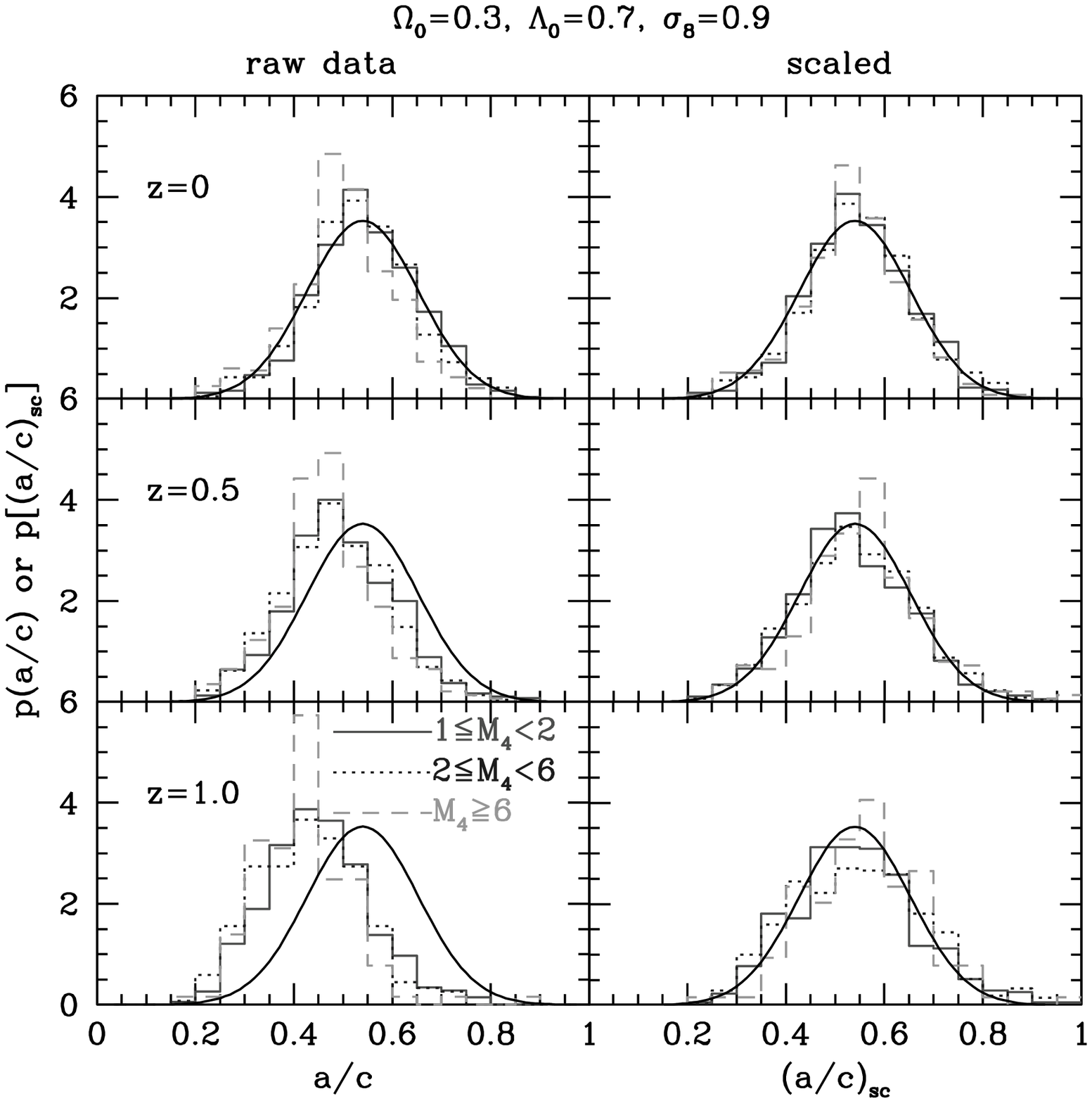}\\ \bigskip
 \includegraphics[width=0.65\hsize]{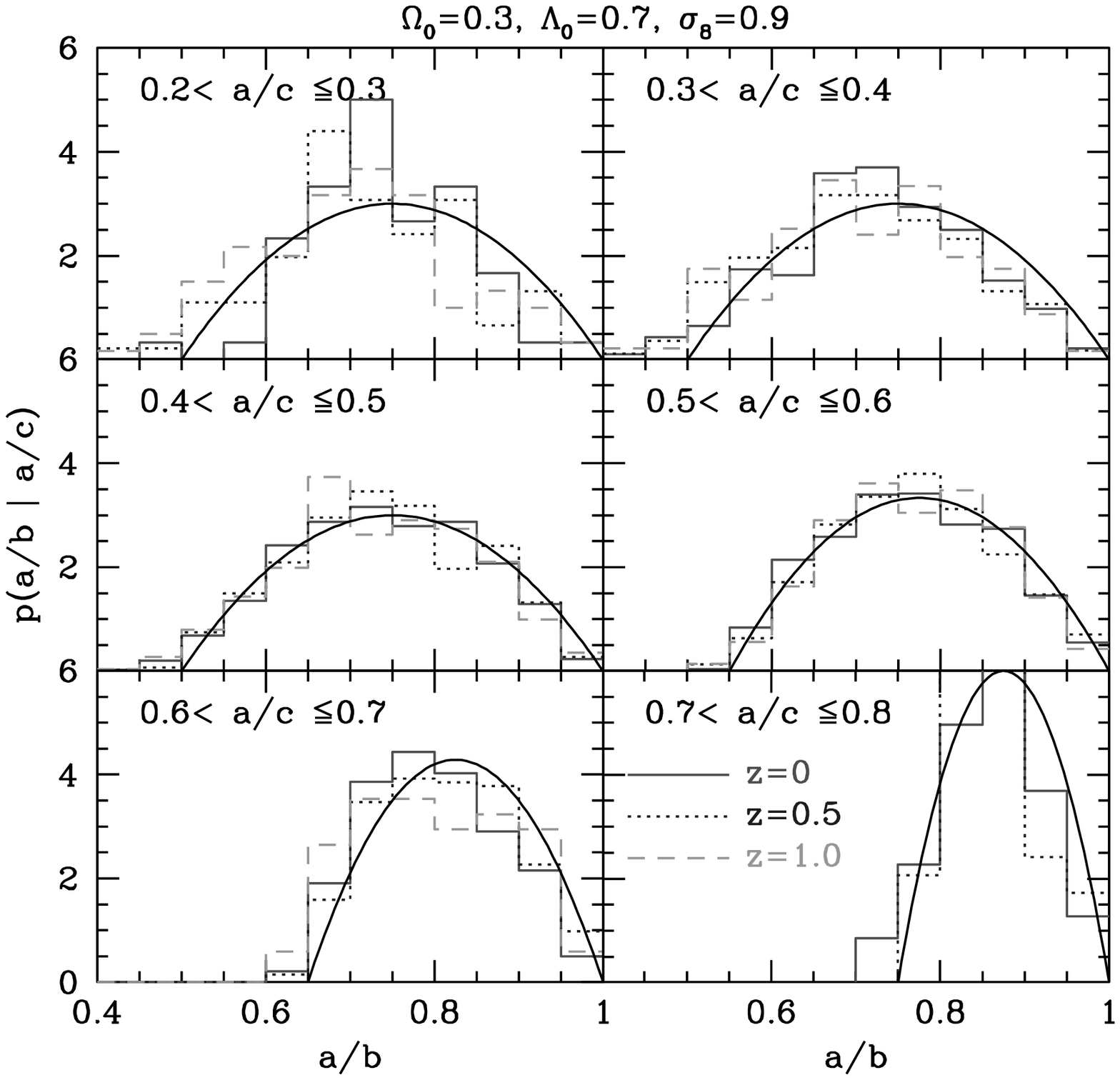}
\caption{Distributions of axis ratios, $a/c$ ({\it top}) and $a/b$ ({\it
 bottom}). Different lines denote different number of particles (i.e.,
 mass). Specifically, the number of particles within virial radius is
 described by $M_4\equiv (N_{\rm halo}/10^4)$. The scaled axis ratio
 $(a/c)_{\rm sc}$ is defined by equation (\ref{tri_r_ac}).
 Lines show fits to data (eqs. [\ref{tri_p_a}] and [\ref{tri_p_ab}]).
 These Figures are taken from JS02.
\label{fig:tri_pdf}}
\end{center}
\end{figure}

\begin{figure}[tb]
\begin{center}
 \includegraphics[width=0.5\hsize]{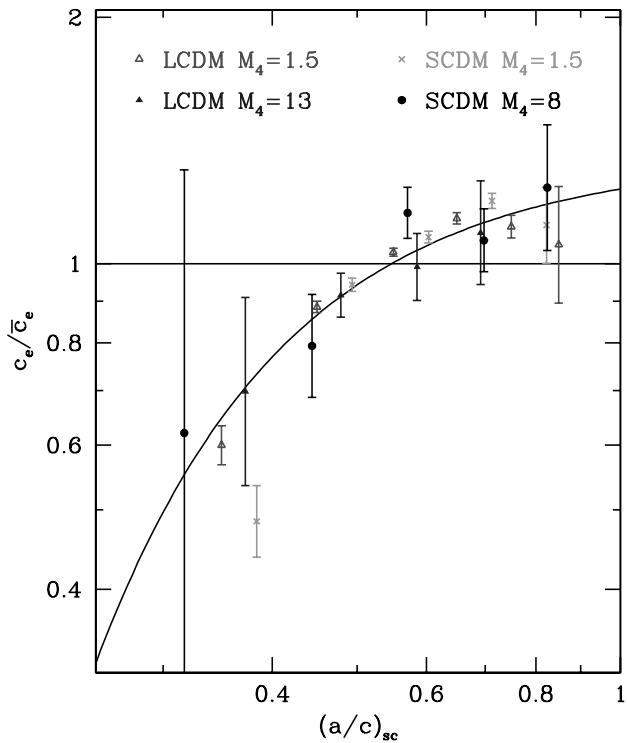}
\caption{Dependence of the concentration parameter $c_e$ on the axis
 ratio. The fitting formula (prefactor in equation
 [\ref{tri_median-ce}]) is also shown by the solid line. This Figure is
 taken from JS02.
\label{fig:tri_c_ac}}
\end{center}
\end{figure}

Since we do not know the properties of the density profile of an
individual lensing halo, our prediction for lensing probabilities is
necessarily statistical in a sense that it should be made after
averaging over appropriate PDFs of the properties of halos. For this
purpose, we need the PDFs of axis ratios. JS02 empirically derived such
PDFs from their cosmological simulations. For the axis ratios, they are
given as 
\begin{equation}
 p(a/c)=\frac{1}{\sqrt{2\pi}\times 0.113}\exp\left[-\frac{\left\{(a/c)_{\rm sc}-0.54\right\}^2}{2(0.113)^2}\right]\frac{(a/c)_{\rm sc}}{a/c},
\label{tri_p_a}
\end{equation}
and 
\begin{equation}
 p(a/b|a/c)=\frac{3}{2(1-\max(a/c,0.5))}\left[1-\left(\frac{2a/b-1-\max(a/c,0.5)}{1-\max(a/c,0.5)}\right)^2\right],
\label{tri_p_ab}
\end{equation}
for $a/b \ge \max(a/c,0.5)$, and $p(a/b|a/c) = 0$ otherwise (JS02). 
Here $M_*$ is the characteristic nonlinear mass so that the rms top-hat
smoothed overdensity at that mass scale is $1.68$. The scaled axis ratio
$(a/c)_{\rm sc}$ is defined by 
\begin{equation}
\left(\frac{a}{c}\right)_{\rm sc}\equiv \left(\frac{a}{c}\right)\left(\frac{M_{\rm vir}}{M_*}\right)^{0.07[\Omega_M(z)]^{0.7}},
\label{tri_r_ac}
\end{equation}
and represents the mass and redshift dependences of axis ratios. In Figure
\ref{fig:tri_pdf}, we show how well these fitting formulae work. 
For the concentration parameter, 
\begin{equation}
 p(c_e)=\frac{1}{\sqrt{2\pi}\times 0.3}
\exp\left[-\frac{(\ln c_e-\ln \bar{c}_e)^2}{2(0.3)^2}\right]\frac{1}{c_e},
\label{tri_p_ce}
\end{equation}
where the fit to the median concentration parameter $\bar{c}_e$ for
$\alpha=1$ is given as \footnote{This expression looks different from
its counterpart (eq. [21]) of JS02 for two reasons. One is due to a typo
in JS02 who omitted the factor $\sqrt{\Delta_{\rm vir}(z_c;{\rm
JS02})/\Delta_{\rm vir}(z;{\rm JS02})}$. Since $\Delta_{\rm vir}({\rm
JS02})= \Omega_M(z)\Delta_{\rm vir}$ according to the notation of this
thesis, this recovers the difference in the latter part. The other is the
fact that we also incorporate the additional axis ratio dependence of
$\bar{c}_e$ which is noted in equation (23) of JS02. See Figure
\ref{fig:tri_c_ac} for the accuracy of the fitting formula of the
additional axis ratio dependence. This explains the prefactor before
$A_e$ in equation (\ref{tri_median-ce}) of this thesis.}: 
\begin{equation}
\label{tri_median-ce}
 \bar{c}_e=1.35\exp\left[-\left\{\frac{0.3}{(a/c)_{\rm sc}}\right\}^2\right]A_e\sqrt{\frac{\Delta_{\rm vir}(z_c)}{\Delta_{\rm vir}(z)}}\left(\frac{1+z_c}{1+z}\right)^{3/2},
\end{equation}
with $z_c$ being the collapse redshift of the halo of mass $M_{\rm vir}$
(JS02). In the case of $\alpha=1$, we simply use the above expression,
and for $\alpha=1.5$, we use the relation
$\bar{c}_e(\alpha=1.5)=0.5\bar{c}_e(\alpha=1)$ (\citealt{keeton01a};
JS02).  JS02 estimated $A_e=1.1$ in the Lambda-dominated CDM model, but 
this value is likely to be dependent on the underlying cosmology to some 
extent. As we stressed in Chapter \ref{chap:conc}, however, we fix 
cosmological parameters to those of the ``concordance'' cosmology.
Therefore in this thesis we mostly fix the value to $A_e=1.1$. 

\section{Lensing Properties of Triaxial Dark Matter Halos}
\label{sec:tri_lens}
\markboth{CHAPTER \thechapter.
{\MakeUppercase\mychapheadname}}{\thesection.
\MakeUppercase{Lensing Properties of Triaxial Dark Matter Halos}}

In this section, we present several expressions for triaxial dark
matter halos which are useful in calculating gravitational lensing
properties. Under the thin lens approximation, gravitational lensing
properties are fully characterized by the matter density projected along 
the line-of-sight (see Appendix \ref{chap:lens}). We have to calculate
the mass density profile projected along the arbitrary line-of-sight
directions, because the line-of-sight, in general, does not coincide
with the principal axis of a triaxial dark matter halo. 
 
\subsection{Coordinate Systems}

We introduce two Cartesian coordinate systems, $\vec{x} = (x,y,z)$ and 
$\vec{x'} = (x',y',z')$, which represent respectively the principal
coordinate system of the triaxial dark halo and the observer's
coordinate system. The origins of both coordinate systems are set at the
center of the halo.  It is assumed that the $z'$-axis runs along the
line-of-sight direction of the observer, and that the $z$-axis lies
along the major principal axis.  In general, the relative orientation
between the two coordinate systems can be specified by the three Euler
angles. However, in our case, it is only the line-of-sight direction
that is fixed while the rotation angle of the $x'$-$y'$ plane relative
to $x$-$y$ plane is arbitrary, and thus we may need only two angles to
specify the relative orientation of the two coordinate systems. Here we
make a choice of $x'$-axis lying in the $x$-$y$ plane. Then the relative
orientation of the two coordinate systems can be expressed in terms of
the line-of-sight direction in the halo principal coordinate system. 

\begin{figure}[t]
\begin{center}
 \hspace*{20mm}
 \includegraphics[width=0.5\hsize]{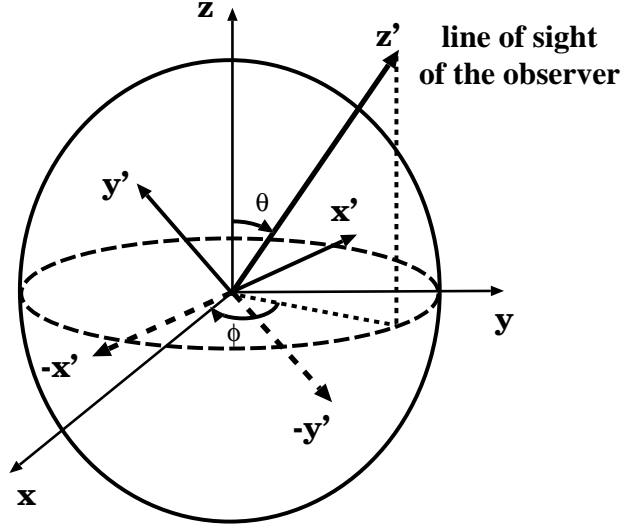}
\caption{ The orientations of the coordinate systems. The Cartesian axes
($x,y,z$) represent the halo principal coordinate system while the axes 
($x',y',z'$) stand for the observers coordinate system with $z'$-axis
 aligned with the line-of-sight direction. The $x'$-axis lies in the
 $x$-$y$ plane. The angle $(\theta,\phi)$ represent the polar angle of
 the line-of-sight direction in the $(x,y,z)$-coordinate system. 
\label{fig:tri_shape}}
\end{center}
\end{figure}

Let $(\theta,\phi)$ be the polar coordinates of the line-of-sight 
direction in the $\vec{x}$-coordinate system. Then the relation 
between the two coordinate systems can be expressed in terms of the 
rotation matrix $A$ \citep{binney85} as 
\begin{equation}
 \vec{x}=A\vec{x'},
\end{equation}
where 
\begin{equation}
A\equiv\left(
\begin{array}{ccc}
 -\sin\phi & -\cos\phi\cos\theta & \cos\phi\sin\theta \\
 \cos\phi  & -\sin\phi\cos\theta & \sin\phi\sin\theta \\
 0         & \sin\theta          & \cos\theta\\
\end{array}
\right) .
\end{equation}
Figure \ref{fig:tri_shape} represents the relative orientation between
the observer's coordinate system and the halo principal coordinate system. 

\subsection{Lensing Properties}

For simplicity, in this section we redefine $\vec{x}/R_0$ and
$\vec{x'}/R_0$ as $\vec{x}$ and $\vec{x'}$, respectively. In this case, 
\begin{equation}
 \rho(R)=\frac{\delta_{\rm ce}\rho_{\rm crit}(z)}
{R^\alpha(1+R)^{3-\alpha}} ,
\end{equation}
with $R$ being defined by equation (\ref{tri_rdef}). In terms of the
observer's coordinates $(x',y',z')$, $R$ is written as
\begin{equation}
\label{tri_Robs}
 R = \sqrt{fz'^{2} + gz' + h},
\end{equation}
where
\begin{eqnarray}
f &=& \sin^{2}\theta\left(\frac{c^2}{a^2}\cos^{2}\phi + 
\frac{c^2}{b^2}\sin^{2}\phi\right)
+ \cos^{2}\theta, \label{tri_f}\\
g &=& \sin\theta\sin2\phi\left(\frac{c^2}{b^2}-\frac{c^2}{a^2}\right)x' + 
\sin 2\theta\left(1-\frac{c^2}{a^2}\cos^{2}\phi-
\frac{c^2}{b^2}\sin^{2}\phi\right)y', \label{tri_g}\\
h &=& \left(\frac{c^2}{a^2}\sin^{2}\phi + \frac{c^2}{b^2}\cos^{2}\phi\right)x'^2 + \sin 2\phi\cos\theta\left(\frac{c^2}{a^2}-\frac{c^2}{b^2}\right)x'y' 
\nonumber \\
&& + \left[\cos^{2}\theta\left(\frac{c^2}{a^2}\cos^{2}\phi + 
\frac{c^2}{b^2}\sin^{2}\phi\right) + 
\sin^{2}\theta\right]y'^{2}\label{tri_h}.
\end{eqnarray}
Defining two new variables $z'_*$ and $\zeta$  
\begin{eqnarray}
z'_*{} &\equiv& \sqrt{f}\left(z' + \frac{g}{2f}\right), \label{tri_newz}\\
\zeta &\equiv& h - \frac{g^{2}}{4f}\label{tri_zeta}, 
\end{eqnarray}
we rewrite equation (\ref{tri_Robs}) as 
\begin{equation}
R = \sqrt{z'_*{}^2 + \zeta^{2}}. 
\end{equation}

Then the convergence $\kappa$ can be expressed as a function of $\zeta$: 
\begin{equation}
 \kappa=\frac{R_0}{\Sigma_{\rm crit}}
\int_{-\infty}^{\infty}\rho(R)dz'
=\frac{R_0}{\Sigma_{\rm crit}}
\int_{-\infty}^{\infty}\frac{1}{\sqrt{f}}\rho
\left(\sqrt{z'_*{}^2+\zeta^2}\right)dz'_* 
\equiv\frac{b_{\rm TNFW}}{2}f_{\rm GNFW}(\zeta),
\label{tri_convergence}
\end{equation}
where 
\begin{equation}
 b_{\rm TNFW} \equiv \frac{1}{\sqrt{f}}
\frac{4\delta_{\rm ce}\rho_{\rm crit}(z)R_0}{\Sigma_{\rm crit}},
\end{equation}
and
\begin{equation}
 f_{\rm GNFW}(r) \equiv 
\int_{0}^{\infty}\frac{1}{\left(\sqrt{r^2+z^2}\right)^\alpha
\left(1+\sqrt{r^2+z^2}\right)^{3-\alpha}}dz.
\label{tri_f_nfw} 
\end{equation}
The critical surface mass density (see Appendix \ref{chap:lens}) is
denoted by $\Sigma_{\rm crit}$.

The meaning of the variable $\zeta$ can be easily understood by 
substituting equations (\ref{tri_f})-(\ref{tri_h}) into equation
(\ref{tri_zeta}): 
\begin{equation}
\zeta^{2} = \frac{1}{f}\left(Ax'^{2} + Bx'y' + Cy'^{2}\right),
\label{tri_zeta2}
\end{equation}
where 
\begin{eqnarray}
A &\equiv& \cos^{2}\theta\left(\frac{c^2}{a^2}\sin^{2}\phi + 
\frac{c^2}{b^2}\cos^{2}\phi\right) + \frac{c^2}{a^2}\frac{c^2}{b^2}\sin^{2}\theta, 
\label{tri_A} \\
B &\equiv& \cos\theta\sin 2\phi\left(\frac{c^2}{a^2}-\frac{c^2}{b^2}\right), 
\label{tri_B} \\
C &\equiv& \frac{c^2}{b^2}\sin^{2}\phi+\frac{c^2}{a^2}\cos^{2}\phi.
\end{eqnarray}
The quadratic form of equation (\ref{tri_zeta2}) implies that the
iso-$\zeta$ curves are ellipses, and that the position angle of ellipses
$\psi$ is 
\begin{equation}
\psi = \frac{1}{2}\arctan\frac{B}{A-C}.
\label{tri_psi}
\end{equation}

\begin{figure}[tb]
\begin{center}
 \includegraphics[width=0.5\hsize]{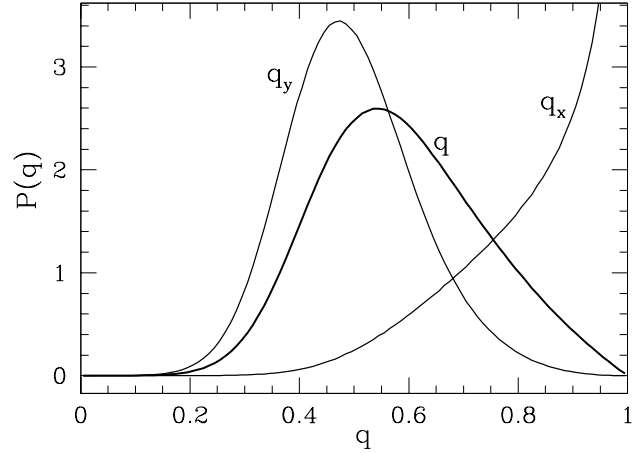}
\caption{PDFs of $q$ (eq. [\ref{tri_q}]), $q_x$ (eq.
 [\ref{tri_qx}]), and $q_y$ (eq. [\ref{tri_qy}]). Here we consider a
 halo with mass $M_{\rm vir}=10^{15}h^{-1}M_{\odot}$ and redshift
 $z=0.3$, but the result only weakly depends on the halo mass and
 redshift. These PDFs are calculated from PDFs of axis ratios $p(a/c)$
 and $p(a/b)$ for which we use equations (\ref{tri_p_a}) and
 (\ref{tri_p_ab}). We assume that the orientations of dark halos are
 random. \label{fig:tri_probq}} 
\end{center}
\end{figure}

By rotating the $x'y'$-plane by the angle $\psi$, we diagonalize 
equation (\ref{tri_zeta2}) such that
\begin{equation}
\zeta^{2} = \frac{x'^{2}}{q_{x}^{2}} + \frac{y'^{2}}{q_{y}^{2}}, 
\end{equation}
where
\begin{eqnarray}
q_{x}^{2} &\equiv& \frac{2f}{A + C - \sqrt{(A - C)^{2} + B^{2}}}, 
\label{tri_qx} \\ 
q_{y}^{2} &\equiv& \frac{2f}{A + C + \sqrt{(A - C)^{2} + B^{2}}}.
\label{tri_qy} 
\end{eqnarray}
Note that $q_{x} \ge q_{y}$ for the given $\psi$. We further define the 
axis ratio $q$ as
\begin{equation}
q \equiv \frac{q_y}{q_x} = 
\left(\frac{A + C - \sqrt{(A - C)^{2} + B^{2}}}
{A + C + \sqrt{(A - C)^{2} + B^{2}}}\right)^{1/2}, 
\label{tri_q}
\end{equation}
which represents the ellipticities of the projected isodensity curves of
the triaxial dark halos. In this case, the convergence $\kappa$ is
expressed as
\begin{equation}
 \kappa=\kappa(\xi),\;\;\;\;\;\;\mbox{where }\xi^2=x'{}^2+\frac{y'{}^2}{q^2}.
\end{equation}
The advantage of this diagonalization is that we can apply the previous
method to calculate lensing properties \citep{schramm90,keeton01d} where
the deflection angle $\vec{\alpha}=(\alpha_{x'},\alpha_{y'})$ is expressed
as a one-dimensional integral of the convergence $\kappa(\xi)$: 
\begin{eqnarray}
 \alpha_{x'}(x',y')  &=& qx'J_0(x',y'),\\
 \alpha_{y'}(x',y')  &=& qy'J_1(x',y'),
\end{eqnarray}
where the integral $J_n(x,y)$ is
\begin{equation}
J_n(x,y) = \int_0^1 \frac{\kappa\left(\xi(v)\right)}{\left[1-(1-q^2)v\right]^{n+1/2}}dv, 
\end{equation}
and $\xi(v)$ is
\begin{equation}
  \xi^2(v)=v\left(x^2+\frac{y^2}{1-(1-q^2) v}\right).
\end{equation}
Figure \ref{fig:tri_probq} plots PDFs of $q$, $q_{x}$, and $q_{y}$. They
were computed numerically using equations
(\ref{tri_p_a})-(\ref{tri_p_ab}) and (\ref{tri_qx})-(\ref{tri_q}) under
the assumption that the triaxial halo orientations (i.e., the angles
$\theta$ and $\phi$) are randomly distributed. In this plot we set
$M_{\rm vir}=10^{15}h^{-1}M_{\odot}$ and $z=0.3$, which are s typical
mass scale and a redshift of lensing clusters. It is clear from Figure
\ref{fig:tri_probq} that the axis ratio of projected isodensity contours
strongly deviates from unity, having maximum around $q \sim 0.6$. This
large degree of ellipticity suggests that the triaxial dark halos in
realistic cosmological models significantly enhances lensing cross
sections compared with the conventional spherical model predictions. 

\begin{figure}[tb]
\begin{center}
 \includegraphics[width=0.7\hsize]{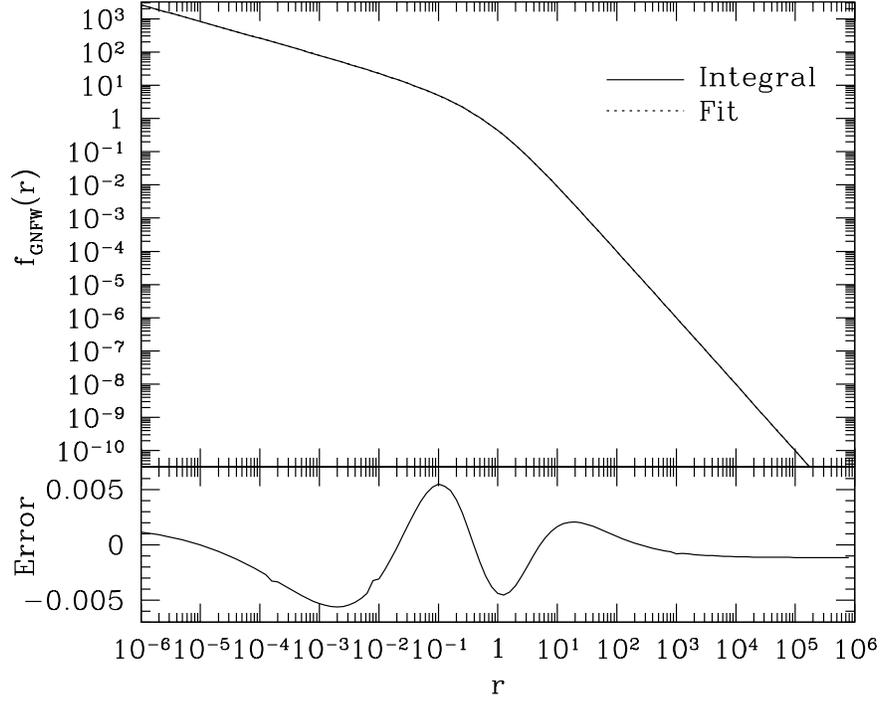}
\caption{Comparison of $f_{\rm GNFW}(r)$ for $\alpha=1.5$ (denoted by
 ``Integral''; see eq. [\ref{tri_f_nfw}]) and its fitting formula
 (denoted by ``Fit''; see eq. [\ref{tri_fit_a15}]). ``Error'' plotted in
 the bottom panel is defined by 
$[f_{\rm GNFW}({\rm Integral})-f_{\rm GNFW}({\rm Fit})]/f_{\rm GNFW}({\rm Integral})$.   
\label{fig:tri_fit}} 
\end{center}
\end{figure}

For $\alpha=1$, $f_{\rm GNFW}(r)$ defined in equation (\ref{tri_f_nfw}) 
is analytically expressed as \citep{bartelmann96}:
\begin{eqnarray}
 f_{\rm GNFW}(r)
=\left\{\begin{array}{ll}
{\displaystyle \frac{1}{1-r^2}
\left[-1+\frac{2}{\sqrt{1-r^2}}{\rm arctanh}\sqrt{\frac{1-r}{1+r}}\right]}
&\mbox{($r<1$),}\\
{\displaystyle \frac{1}{r^2-1}
\left[1-\frac{2}{\sqrt{r^2-1}}\arctan\sqrt{\frac{r-1}{r+1}}\right]}
&\mbox{($r>1$),}
\end{array}\right.
\end{eqnarray}
but it does not has a simple analytical expression for $\alpha=1.5$.
Thus we use the following fitting formula in this case: 
\begin{equation}
 f_{\rm GNFW}(r)
=\frac{2.614}{r^{0.5}\left(1+2.378r^{0.5833}+2.617r^{1.5}\right)}.
\label{tri_fit_a15}
\end{equation}
The error of the above fit is $\lesssim 0.6$\% (see Figure
\ref{fig:tri_fit}). 

We note that the triaxial dark halo model we presented in this section
can be improved in several ways. First, we assumed that the axis
ratios of the triaxial ellipsoids are constant with radius. However,
JS02 showed that the axis ratios decrease slightly toward the halo
centers: $a/c$ decreases by $\sim\!0.2$ as the mean radius decreases
from $\sim\!0.6 r_{\rm vir}$ to $\sim\!0.06 r_{\rm vir}$ (see Figure
\ref{fig:tri_rhos}).  Since strong lensing is most sensitive to the
inner parts of dark halos, it is possible that we have actually
underestimated the effects of triaxiality on the statistics of
large-separation lenses.  On the other hand, including baryons (which
were neglected in the simulations of JS02) would tend to make dark halos
rounder by $\Delta(a/c) \sim 0.1$--$0.2$ \citep[e.g.,][]{kazantzidis04}.  

\section{Effect of Triaxiality on Strong Gravitational Lensing}
\label{sec:tri_eff}
\markboth{CHAPTER \thechapter.
{\MakeUppercase\mychapheadname}}{\thesection.
\MakeUppercase{Effect of Triaxiality on Strong Gravitational Lensing}}

\subsection{Projection Effect}

\begin{figure}[tb]
\begin{center}
 \includegraphics[width=0.8\hsize]{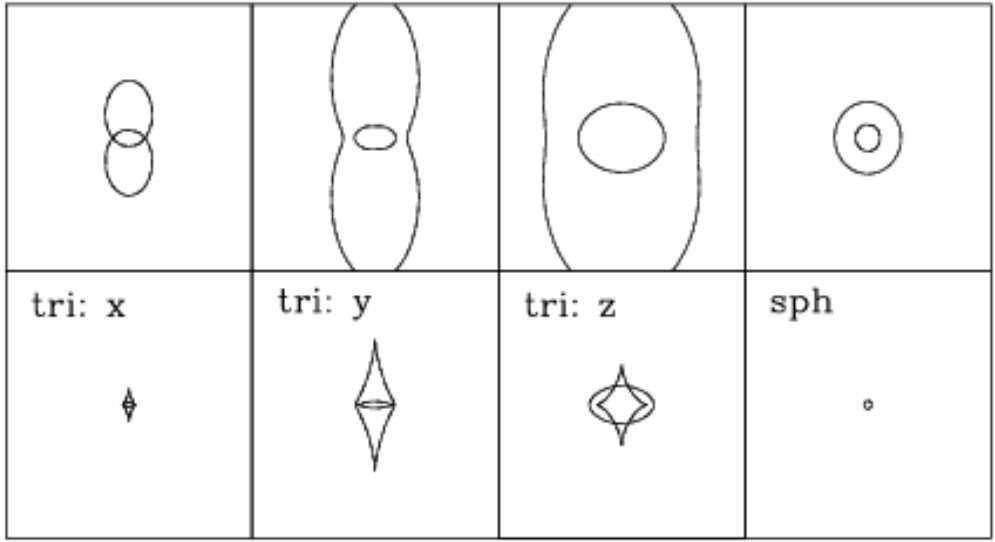}
\caption{Critical curves ({\it upper}) and caustics ({\it lower}) of
 triaxial dark halos projected along three principal vectors (see Figure
 \ref{fig:tri_probq}). We consider a halo with mass $M_{\rm vir}=10^{15}h^{-1}M_\odot$ and redshift $z=0.3$, and the source plane is placed at $z=2.0$. 
 The axis ratios are fixed to $a/c=0.5$ and $b/c=0.7$. For the
 concentration parameters, we use the median value given by equation
 (\ref{tri_median-ce}). The size of all boxes is the same. The inner
 slope is $\alpha=1$. The case for the corresponding spherical dark halo
 is also shown  for reference.   
\label{fig:tri_proj}}
\end{center}
\end{figure}

One of the most important consequences of the triaxial lens model is that 
lensing properties are different for different line-of-sight directions, even
if the lensing halo is the same. Since strong gravitational lensing is
characterized by critical curves and caustics, first we see how they
change as we change the direction of the projection. 

Figure \ref{fig:tri_proj} shows the results. In this Figure, we consider
a halo with mass $M_{\rm vir}=10^{15}h^{-1}M_\odot$ and redshift
$z=0.3$, and the source plane is placed at $z=2.0$. We also fix the axis
ratios to $a/c=0.5$ and $b/c=0.7$, which are typical values for the CDM
halos (see \S \ref{sec:tri_tnfw}). For the concentration parameters, we
use the median value given by equation (\ref{tri_median-ce}). We show
critical curves and caustics of this identical halo projected along
three principal vectors. First of all, both sizes and structures of
critical curves and caustics are quite different for different
projections. The size is the largest when projected along the major
axis ($z$), and the smallest along the minor axis ($x$). This is
reasonable because the deflection angle is proportional to the surface
mass density, i.e., density integrated along the line-of-sight (see
Appendix \ref{chap:lens}), and thus the mass distributions elongated
along line-of-sight are expected to have larger deflection angles.
Moreover, the size seems larger compared with that of the corresponding
(i.e., with the same mass and redshift, and have the median concentration
parameter) spherical dark halo. Thus it is expected that triaxiality
enhances the number of lenses significantly.

\subsection{Lensing Cross Sections}

\begin{figure}[tb]
\begin{center}
 \includegraphics[width=0.7\hsize]{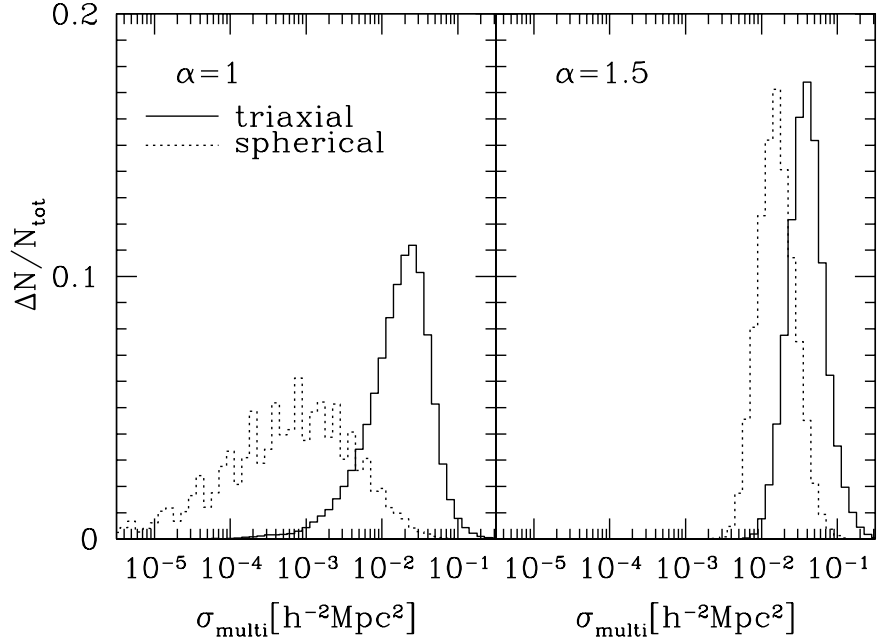}
\caption{Distributions of cross sections for multiple images (defined in
 the lens plane), both for $\alpha=1$ and $1.5$. We consider a halo with
 mass  $M_{\rm vir}=10^{15}h^{-1}M_\odot$ and redshift $z=0.3$, and the
 source plane is placed at $z=2.0$. We took account of PDFs of $a/c$ (eq.
 [\ref{tri_p_a}]), $a/b$ (eq. [\ref{tri_p_ab}]), and $c_e$ (eq.
 [\ref{tri_p_ce}]).  We assume that the orientations of dark halos are
 random. Distributions of cross sections for the corresponding spherical
 dark halo is also shown for reference. 
\label{fig:tri_csdis}}
\end{center}
\end{figure}

Next we see cross sections for multiple images. Again, we consider a
halo with mass $M_{\rm vir}=10^{15}h^{-1}M_\odot$ and redshift $z=0.3$,
and the source plane is placed at $z=2.0$. This time, however, we do not
fix axis ratios but randomly choose them according to their
corresponding PDFs (eqs.[\ref{tri_p_a}] and [\ref{tri_p_ab}]). The
concentration parameters are also distributed according to the PDF (eq.
[\ref{tri_p_ce}]). The orientations of dark halos are assumed to be
random. We also consider the spherical case  for reference; in this case
the distribution of cross sections are caused only by the PDF of
the concentration parameter. 
 
We compute distributions of cross sections both for $\alpha=1$ and
$1.5$, which are shown in Figure \ref{fig:tri_csdis}. Most importantly, 
the triaxiality systematically shifts the distributions toward larger cross
sections. The amount of the enhancement depends on the value of
$\alpha$, but it is more than one order of magnitude for $\alpha=1$.
Therefore, the triaxiality does increase the number (or probability) of
strong gravitational lensing, as expected.

Another effect we can read off from this Figure is the comparable (or
even shallower) width of the distribution of cross sections with the
triaxial dark halo. At first sight, this seems strange because in the
triaxial model we incorporate PDFs of axis ratios and orientations, in
addition to that of concentration parameters that are also taken into
account in the spherical model. In fact, the width is affected by the
correlation between axis ratios and concentration parameters; Figure
\ref{fig:tri_c_ac} implies that the decrease of cross sections due to
smaller concentration parameters tends to be compensated by smaller axis
ratios.  Therefore a correlation shown in Figure \ref{fig:tri_c_ac} is
quite important and should be taken into account in applying the
triaxial model to lens statistics.

\chapter{Arc Statistics in Triaxial Dark Halos: Theoretical Predictions
and Comparison with Observations}
\label{chap:arc}
\def\mychapheadname{Arc Statistics in Triaxial Dark Halos}
\markboth{CHAPTER \thechapter.
{\MakeUppercase\mychapheadname}}{}

\section{Introduction}
\markboth{CHAPTER \thechapter.
{\MakeUppercase\mychapheadname}}{\thesection.
\MakeUppercase{Introduction}}

The discovery of a lensed arc in a rich cluster A370
\citep{lynds86,soucail87} opened a direct window to probe the dark mass
distribution in clusters of galaxies.  Since gravitational lensing
phenomena are solely dictated by intervening mass distributions, they
are not biased by the luminous objects unlike other conventional
observations. Indeed, previous work \citep*{wu93,miralda93a,miralda95, 
miralda02,bartelmann96,hattori97b,molikawa99,williams99,meneghetti01,
molikawa01,oguri01,gavazzi03,sand04,wambsganss04,dalal04a,maccio04}
showed that the number, shapes,  
and positions of lensed arcs are sensitive to the mass distribution of
clusters. For instance, \citet{oguri01} calculated the number of arcs
using the generalization of the universal density profile proposed by
NFW and pointed out that it is extremely sensitive to the inner slope
and the concentration parameter of the density profile; the number of
arcs changes by more than an order of magnitude among different models
that are of cosmological interest. Therefore lensing arc surveys provide
an important probe of density profiles of clusters in a manner
complementary to the statistics of large-separation lensed quasars
(Chapters \ref{chap:lat} and \ref{chap:sdss}). 

While most previous studies of lensed arcs have aimed at constraining
the cosmological parameters
\citep*{wu96,bartelmann98,cooray99,sereno02,golse02,bartelmann03}, we
rather focus on extracting information of the density profiles of dark
matter halos. Thus we fix cosmological parameters to those of our
fiducial concordance model (see Table \ref{table:conc_para}). In fact,
arc statistics depend on the assumed set of cosmological parameters in
two ways; directly through the geometry of the universe and somewhat
indirectly through properties of density profiles which also depend on
the cosmology. For instance, \citet{bartelmann98} found that the numbers
of arcs significantly change among different cosmological models, and
concluded that only open CDM models can reproduce the high frequency of
observed arcs. \citet{oguri01} showed, however, that the result largely
comes from the larger concentration parameter of halo profiles in the
open CDM model than in the Lambda-dominated CDM model. Thus this may be
more related to the small-scale behavior of the CDM model than the
``global'' effect of the cosmological constant. 

\begin{figure}[tb]
\begin{center}
 \includegraphics[width=0.7\hsize]{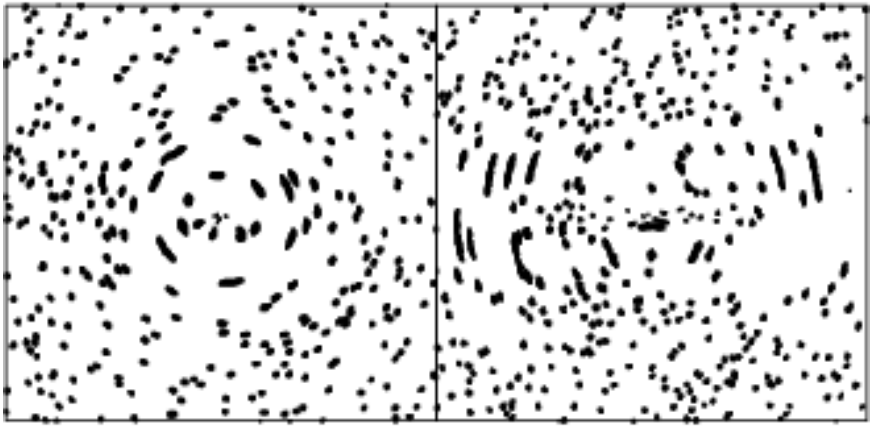}
\caption{Snapshots of simulated arc images with ({\it right}) and without
 ({\it left}) the lens ellipticity, produced from the same spatial
 distribution of source galaxies. This Figure is taken from
 \citet{oguri02b}. 
 \label{fig:arc_demo}}
\end{center}
\end{figure}

The values of cosmological parameters are determined fairly accurately
now, thus our primary interest here is to confront the density profiles
of dark matter halos with the arc statistics, and thereby we would like
even to test the collisionless CDM paradigm.  For this purpose, a
non-spherical description for the lensing halos is the most essential
since cross sections for arcs are quite sensitive to the non-sphericity  
of mass distribution (Chapter \ref{chap:tri}; see also
\citealt*{bartelmann95a}; \citealt*{bartelmann95b};
\citealt*{meneghetti01}; \citealt*{oguri02b}). Figure \ref{fig:arc_demo}
demonstrate how large the effects of non-sphericity on arc statistics are.
Indeed, previous analytic models adopting spherical lens models failed
to reproduce the observed high frequency of arcs
\citep{hattori97b,molikawa99}. 

Recently, new methods to constrain the mass profile of individual
clusters also have been developed
\citep{smith01,sand02,sand04,clowe01,clowe02,gavazzi03}. Although such
methods can measure mass distributions of individual clusters precisely,
it may suffer from the special selection function and the scatter around
the mean mass distribution. For instances, analysis of clusters only with
giant arcs  may result in more elongated clusters than average because
\citet{jing02} showed that triaxial axis ratios have fairly broad 
distributions. Therefore it is of great importance to study statistics
of lensed arcs which allow us to obtain information on the mean profile. 

In this chapter, we develop and study in detail, for the first time,
such an analytical model of the non-spherical lensing objects for the
arc statistics. Specifically we adopt the triaxial description of dark
matter halos proposed by \citet[][ see Chapter \ref{chap:tri}]{jing02}.
They have presented detailed triaxial modeling of halo density profiles,
which enables us to incorporate the asymmetry of dark matter halos
statistically and systematically.  We first compute the lensing cross
sections for arcs on the basis of the Monte Carlo simulations. Then we
make systematic predictions of the number of arcs by averaging the cross
sections over the probability distribution functions (PDFs) of the axis
ratios and the concentration parameters and assuming the random
orientation of the dark halos along the line-of-sight of the observer.
Those theoretical predictions are compared with the number of observed
arcs in a sample of $38$ X-ray selected clusters compiled by
\citet{luppino99}.  We pay particular attention to several selection
functions of clusters and arcs which may systematically affect our
results \citep[e.g.,][]{wambsganss04}. 

\section{Modeling the Number of Arcs in Triaxial Dark Halos}
\label{sec:arc_arcstat}
\markboth{CHAPTER \thechapter.
{\MakeUppercase\mychapheadname}}{\thesection.
\MakeUppercase{Modeling the Number of Arcs in Triaxial Dark Halos}}

\subsection{Cross Sections for Arcs from the 
Monte Carlo Simulation}
\label{sec:arc_mc}

First we compute the cross section for arcs without distinguishing
tangential and radial arcs mainly because of the computational cost.  
In fact, the previous analyses indicate that while the number ratio 
of radial to tangential arcs offers another information on the density
profile, the ratio is rather insensitive to the non-sphericity
\citep{molikawa01,oguri02b}.

Since the analytical computation of the cross sections is not
practically feasible except for spherical models, we resort to the
direct Monte Carlo method
\citep{bartelmann94,miralda93b,molikawa01,oguri02b}. We showed that the 
convergence of triaxial dark matter halos is expressed by equation
(\ref{tri_convergence}). Thus the corresponding lensing deflection angle
$\vec{\alpha}$, and therefore the cross section $\tilde{\sigma}$, are
fully characterized by the two parameters, $b_{\rm TNFW}$ and $q$, as
long as the finite size of source galaxies is safely neglected. Thus we
perform the Monte Carlo simulations on the dimensionless $X$-$Y$ plane,
where $X$ and $Y$ are $X\equiv x'/(R_0q_x)$ and $Y\equiv y'/(R_0q_x)$,
and tabulate the deflection angle and the dimensionless cross section
\begin{eqnarray} 
 \vec{\alpha}&=&\vec{\alpha}(b_{\rm TNFW},q),\\
\tilde{\sigma}&=&\tilde{\sigma}(b_{\rm TNFW},q) ,
\end{eqnarray}
in $50\times19$ bins ($\alpha=1$) or $70\times19$ bins ($\alpha=1.5$) 
for $b_{\rm TNFW}$ and $q$, respectively. The dimensionless cross section is
translated to the dimensional one in the source plane as
\begin{equation}
 \sigma=\tilde{\sigma}(b_{\rm TNFW},q)
\times \left(R_0q_x\frac{D_{\rm OS}}{D_{\rm OL}}\right)^2 .
\end{equation}

\begin{figure}[tb]
\begin{center}
 \includegraphics[width=0.7\hsize]{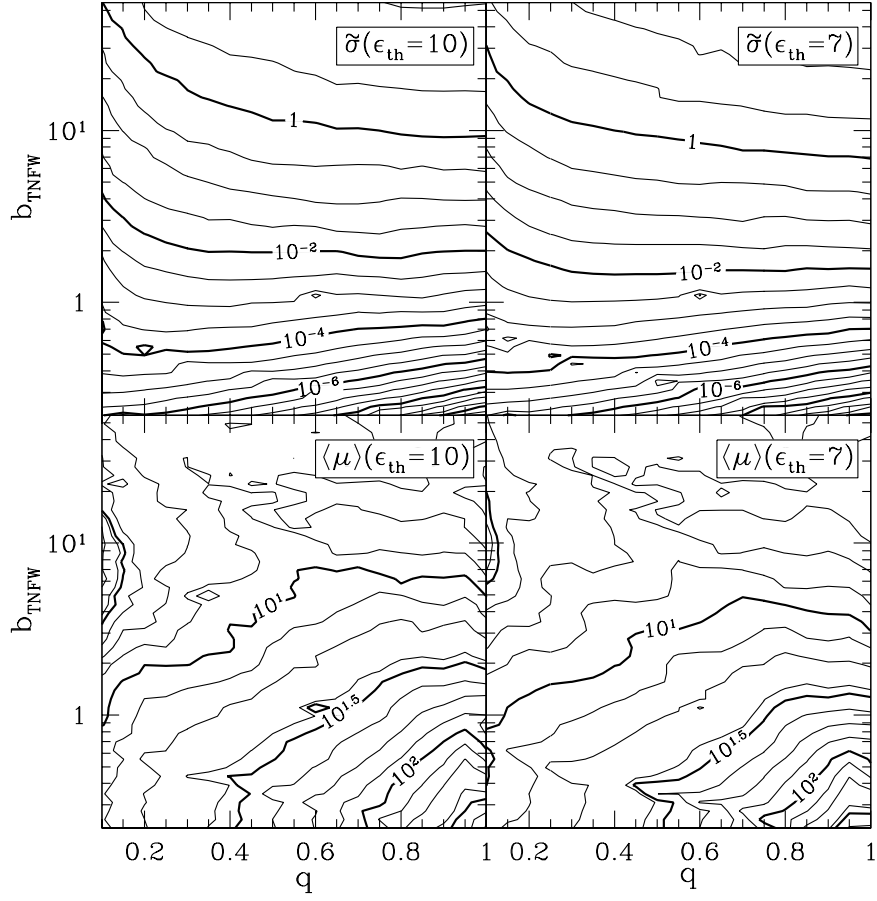}
\caption{Contours of dimensionless cross sections $\tilde{\sigma}$ and
 average magnification factors $\langle \mu\rangle$ in the 
 $q$-$b_{\rm TNFW}$ plane for $\alpha=1$. The threshold axis ratios for
 arcs are set to $\epsilon_{\rm th}=10$ ({\it upper}) and $7$ ({\it
 lower}), respectively. Contours are drawn at $10^{0.5n}$ for
 $\tilde{\sigma}$ and at $10^{0.125n}$ for  $\langle \mu\rangle$, where
 $n$ is integer. When $n$ is in multiples of $4$, contours are drawn by
 thick lines. These cross sections and magnification factors are derived
 from Monte Carlo simulations described in \S \ref{sec:arc_mc}. 
\label{fig:arc_cross_a10}}
\end{center}
\end{figure}

\begin{figure}[tb]
\begin{center}
 \includegraphics[width=0.7\hsize]{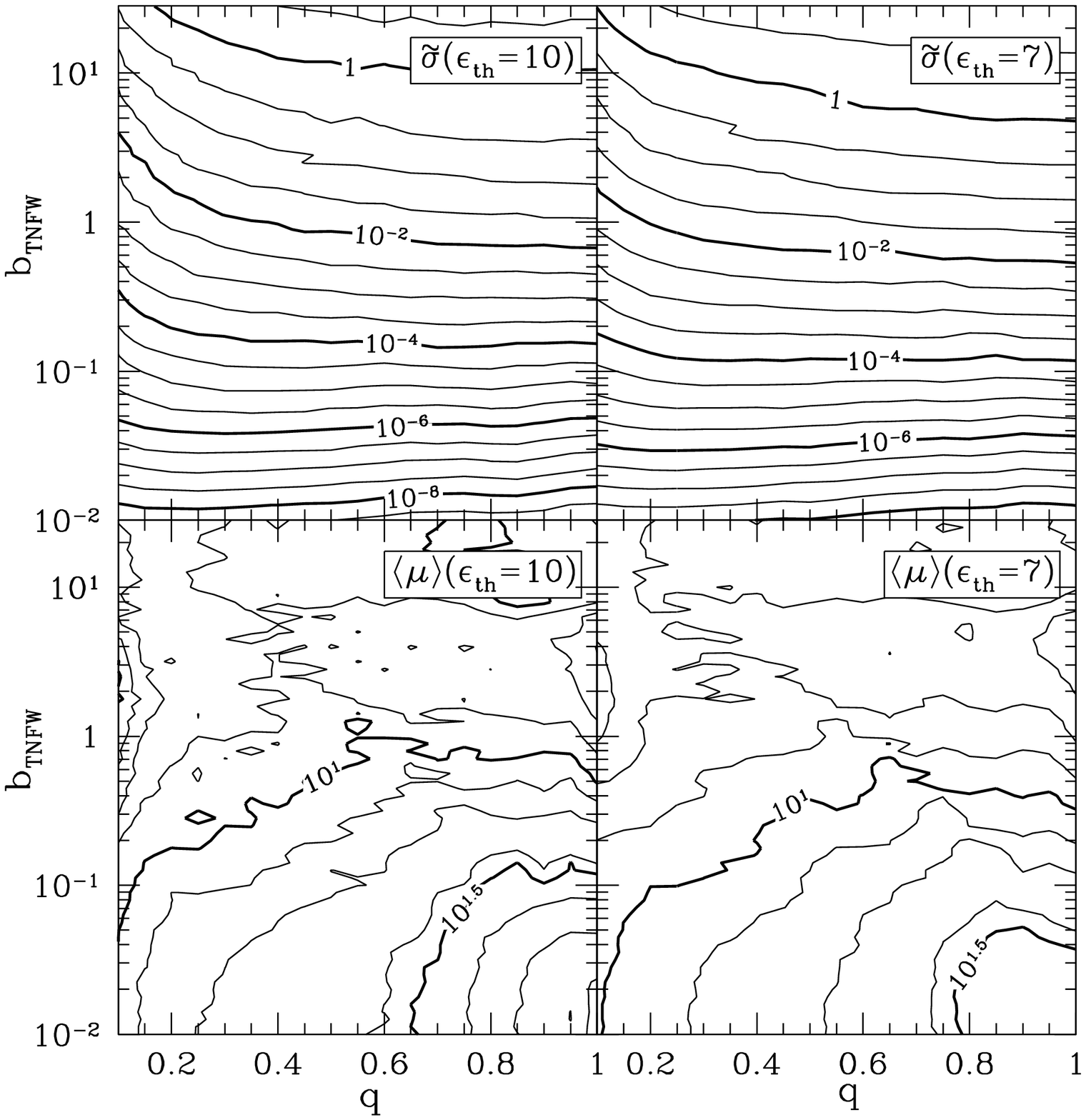}
\caption{Same as Figure \ref{fig:arc_cross_a10}, except for $\alpha=1.5$.
 \label{fig:arc_cross_a15}}
\end{center}
\end{figure}

We follow the simulation method by \citet{oguri02b} which is briefly
summarized below. For more details, please see Appendix \ref{chap:arcsim}.
We use a $2048 \times 2048$ regular grid on the
$X$-$Y$ plane and calculate the deflection angle at each grid point.
The box size is adjusted so as to include all arcs in the box for
each $(b_{\rm TNFW}, q)$. Therefore, the box size almost scales as the
tangential critical line for each $(b_{\rm TNFW}, q)$. After those
deflection angles are obtained at each grid point, we trace 
back the corresponding position in the source plane, and see whether or
not it constitutes a part of lensed images. In order to take account of
the source ellipticity which is also important in arc statistics
\citep{keeton01c}, we assume that it distributes randomly in the range
of [0,0.5], where source ellipticity is defined by $1-b_{\rm s}/a_{\rm
s}$ with $a_{\rm s}$ and $b_{\rm s}$ being semi-major and semi-minor
axes, respectively. We adopt this distribution of intrinsic ellipticities
in order to compare our results with the previous works 
\citep[e.g.,][]{bartelmann98} in which the same distribution was
assumed. Moreover, the distribution is roughly consistent with the
observed distribution \citep*[e.g.,][]{lambas92}. Once we identify a
lensed image, we compute its length $l$ and width $w$ as described in
Appendix \ref{chap:arcsim}. Finally we define a lensed arc if the ratio
of $l$ and $w$ exceeds the threshold value $\epsilon_{\rm th}$ that we
set:  
\begin{equation}
 \frac{l}{w}\geq \epsilon_{\rm th} .
\end{equation}
In practice, we consider $\epsilon_{\rm th}=7$ and $10$ to check the
robustness of the conclusion. We also compute the average magnification
of the arcs $\langle\mu\rangle$ for each set of $(b_{\rm TNFW}, q)$
which is required in estimating the magnification bias
\citep*{turner80,turner84}. The contours of the lensing cross sections
and the average magnification are plotted in Figures
\ref{fig:arc_cross_a10} and \ref{fig:arc_cross_a15} for $\alpha=1$ and
$\alpha=1.5$, respectively.  We confirmed that the cross sections for
$q=1$ cases reproduce the analytic result of spherical lens models for
point source.
 
We should note that our current method does not take account of the
finite size effect of source galaxies, and thus our results are,
strictly speaking, applicable only to a sufficiently small source. Since
the number of tangential arcs, which dominates the total number of arcs,  
is known to be insensitive to the source size
\citep{hattori97b,bartelmann98,molikawa01,oguri01,oguri02b}, this should
not change our conclusion. 

\subsection{Predicting Numbers of Arcs}

The next step is to average the cross section for arcs corresponding to 
a halo of $M_{\rm vir}$ at $z_{\rm L}$ and a galaxy at $z_{\rm S}$ over
the halo properties (its orientations and axis ratios):
\begin{equation}
 \overline{\sigma}(M_{\rm vir}, z_{\rm L}, z_{\rm S})
=\int d(a/c)\int dc_e \int d(a/b) \int d\theta \int d\phi 
~p(a/c)p(c_e)p(a/b|a/c)p(\theta)p(\phi)\sigma.
\label{arc_cs_av}
\end{equation}
PDFs of axis ratios and concentration parameters are summarized in
\S\ref{sec:tri_tnfw}. In what follows, we assume the orientations of
triaxial dark matter halos are completely random: 
\begin{eqnarray}
 p(\theta)&=&\frac{\sin\theta}{2},\label{arc_p_theta}\\
p(\phi)&=&\frac{1}{2\pi}.
\end{eqnarray}
The realistic prediction for the number of arcs also requires to
properly take account of the magnification bias. Thus we use the
average of the cross section times number density of galaxies above the
magnitude limit:
\begin{eqnarray}
 \overline{\sigma n_{\rm g}}(M_{\rm vir}, z_{\rm L}, z_{\rm S})
&=&\int d(a/c)\int dc_e \int d(a/b) \int d\theta \int d\phi \cr
&&\times p(a/c) \, p(c_e) \, p(a/b|a/c) \, p(\theta) \, 
p(\phi) \, \sigma \int_{L_{\rm min}}^\infty dL n_g(L,z_{\rm S}),
\label{arc_csng}
\end{eqnarray}
where $n_g(L,z)$ is the luminosity function of source galaxies for which
we adopt the Schechter form \citep{schechter76}:
\begin{equation}
 n_g(L,z)dL =  \phi^*\left(\frac{L}{L^*}\right)^{\alpha_{\rm s}}
\exp\left(-\frac{L}{L^*}\right)\frac{dL}{L^*}.
\label{arc_schechter}
\end{equation} 
Its integral over $L$ simply reduces to
\begin{equation}
 \int_{L_{\rm min}}^\infty dL \, n_g(L,z_{\rm S})
=\phi^*\Gamma(\alpha_{\rm s}+1,L_{\rm min}/L^*),
\end{equation}
with $\Gamma(a,x)$ being the incomplete gamma function of the second
kind.  The lower limit of the integral, $L_{\rm lim}$, may be computed
from limiting magnitude of observation, $m^*$, and the lensing
magnification factor $\langle\mu\rangle$ (see Figures
\ref{fig:arc_cross_a10} and \ref{fig:arc_cross_a15}): 
\begin{eqnarray}
 \frac{L_{\rm min}}{L^*}
&=& \frac{10^{-0.4(m_{\rm lim}-m^*)}}{\langle\mu\rangle}, \\
 m^* &=& M^*+5\log
\left[\frac{D_{\rm OS}(1+z_{\rm S})^2}{10{\rm pc}}\right]+K(z_{\rm S}) .
\end{eqnarray}
We adopt the K-correction in B-band for spiral galaxies \citep{king85}:
\begin{equation}
K(z)=-0.05+2.35z+2.55z^2-4.89z^3+1.85z^4 .
\label{arc_kcor}
\end{equation}

Finally the number distribution of lensed arcs for a halo of mass
$M_{\rm vir}$ at $z_{\rm L}$ is given by 
\begin{equation}
 \frac{dN_{\rm arc}}{dz_{\rm S}}(z_{\rm S};M_{\rm vir},z_{\rm L})=\overline{\sigma n_{\rm g}}(M_{\rm vir}, z_{\rm L}, z_{\rm S})\frac{cdt}{dz_{\rm S}}(1+z_{\rm S})^3 ,
\label{arc_dnds}
\end{equation}
and the total number of lensed arcs for the halo is
\begin{equation}
 N_{\rm arc}(M_{\rm vir},z_{\rm L})
=\int_{z_{\rm L}}^{z_{\rm S,max}} dz_{\rm S}\,
\frac{dN_{\rm arc}}{dz_{\rm S}}(z_{\rm S};M_{\rm vir},z_{\rm L}).
\label{arc_n_arc}
\end{equation}
While the upper limit of redshifts of source galaxies, $z_{\rm S,max}$,
is in principle arbitrary, it is practically limited by the validity of
the input luminosity function of source galaxies and the applied
K-correction at high redshifts. In the present analysis, we
conservatively set $z_{\rm S,max}=1.25$ because of the K-correction
(eq.[\ref{arc_kcor}]) and the luminosity function (\S \ref{sec:arc_lf}).
Nevertheless we stress here that our methodology can be applied to at
higher redshifts if they are replaced by any reliable models valid there.
 
\subsection{Luminosity Function of Source Galaxies}
\label{sec:arc_lf}

\begin{table}[tb]
 \begin{center}
  \begin{tabular}{lccccccl}\hline\hline
 Name & Model & $z$ Range & $\alpha_{\rm s}$ &
 $M^*_{AB}-5\log h{}^{\rm a}$ & $\phi^*[h^3{\rm Mpc^{-3}]}$ 
 & Ref.\\ \hline
   HDF1 & Lambda${}^{\rm b}$ & $0.00$ -
 $0.50{}^{\rm c}$ & $-1.19$ & $-20.26$ & $2.5 \times 10^{-2}$ & 1\\
 & & $0.50$ - $0.75$ & $-1.19$ & $-19.97$ & $2.9 \times 10^{-2}$ & \\
 & & $0.75$ - $1.25$ & $-1.25$ & $-20.61$ & $1.2 \times 10^{-2}$ & \\
 HDF2 & EdS${}^{\rm d}$ & $0.00$ -
 $0.50{}^{\rm c}$ & $-1.40$ & $-21.20$ & $9.0 \times 10^{-3}$ & 2\\
 & & $0.50$ - $1.00$ & $-1.30$ & $-19.90$ & $4.2 \times 10^{-2}$ & \\
 & & $1.00$ - $1.25$ & $-1.60$ & $-22.10$ & $6.0 \times 10^{-3}$ & \\
 SDF & EdS${}^{\rm d}$ & $0.00$ -
 $1.00{}^{\rm c}$ & $-1.07$ & $-19.78$ & $4.2 \times 10^{-2}$ & 3\\
 & & $1.00$ - $1.25$ & $-0.92$ & $-20.13$ & $4.3 \times 10^{-2}$ & \\
 CFRS & EdS${}^{\rm d}$ & $0.00$ -
 $0.50{}^{\rm c}$ & $-1.03$ & $-19.53$ & $2.7 \times 10^{-2}$ & 4\\
 & & $0.50$ - $0.75$ & $-0.50$ & $-19.32$ & $6.2 \times 10^{-2}$ & \\
 & & $0.75$ - $1.00$ & $-1.28$ & $-19.73$ & $5.4 \times 10^{-2}$ & \\
 & & $1.00$ - $1.25{}^{\rm e}$ & $-2.50$ & $-21.36$ & 
 $9.6 \times 10^{-4}$ & \\ 
\hline
\end{tabular}
\caption{B-band luminosity functions of source galaxies used in
 this paper.\protect\\
\footnotesize{\hspace*{5mm}${}^{\rm a}$ B-band AB magnitude can be converted to conventional
 Johnson-Morgan magnitude via $B_{AB}=B-0.14$ \citep*{fukugita95}.\hspace{10mm}\protect\\
\hspace*{5mm}${}^{\rm b}$ $\Omega_M=0.3$, $\Omega_\Lambda=0.7$.\hspace{70mm}\protect\\
\hspace*{5mm}${}^{\rm c}$ Extrapolated to $z=0$.\hspace{70mm}\protect\\
\hspace*{5mm}${}^{\rm d}$ $\Omega_M=1$, $\Omega_\Lambda=0$.\hspace{70mm}\protect\\
\hspace*{5mm}${}^{\rm e}$ The luminosity function for blue galaxies
  only.\hspace{70mm}\protect\\
\hspace*{5mm}Ref. --- (1) \citealt{poli01}; (2) \citealt{sawicki97}; (3)
 \citealt{kashikawa03}; (4) \citealt{lilly95} }
}
\label{table:arc_lf}
 \end{center}
\end{table}

\begin{figure}[tb]
\begin{center}
 \includegraphics[width=0.7\hsize]{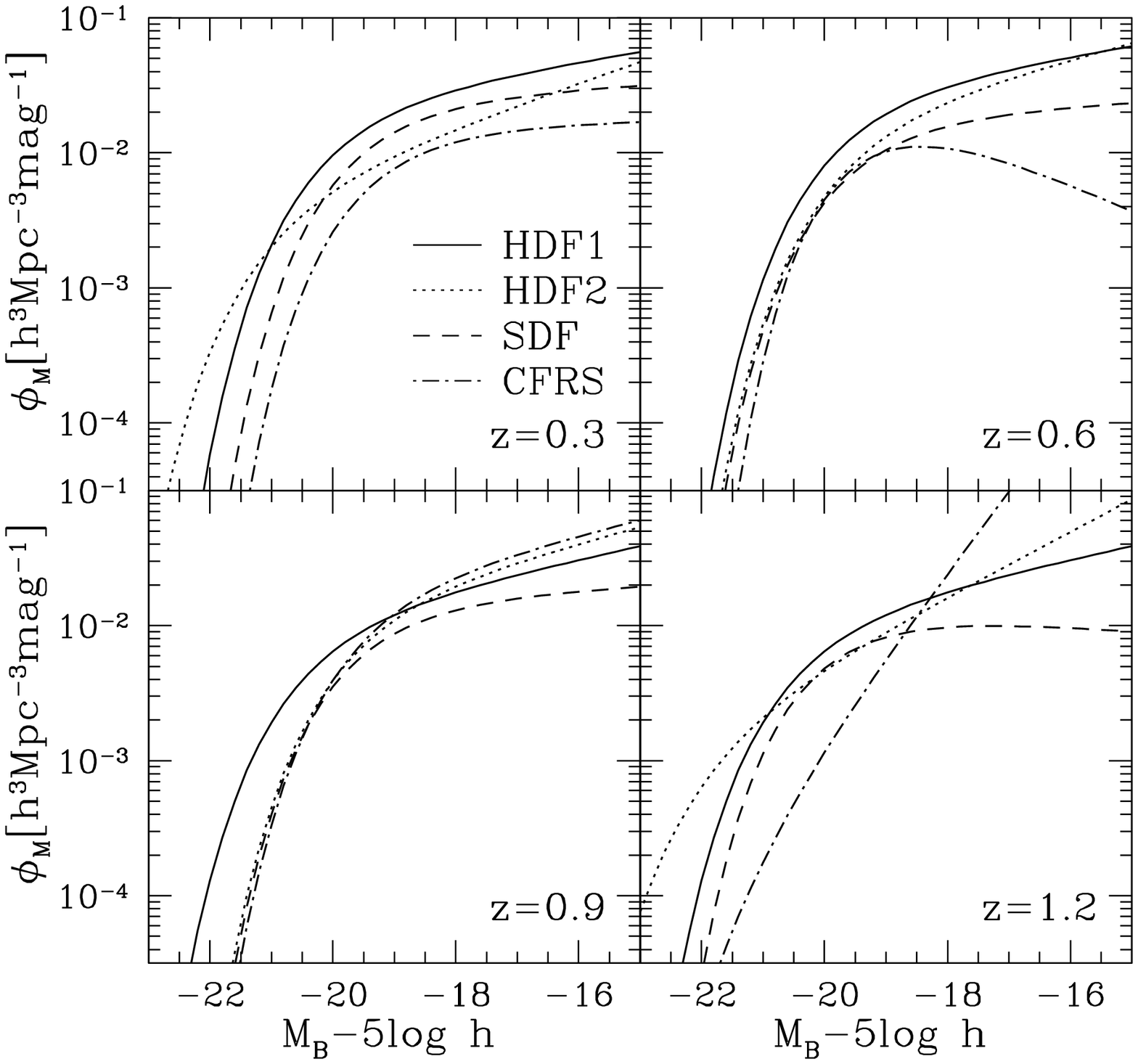}
\caption{Luminosity functions of source galaxies (eq.
 [\ref{arc_schechter_m}]) for $z=0.3$, $0.6$,  $0.9$, and $1.2$. Parameters
 of these luminosity functions are summarized in Table \ref{table:arc_lf}.
 \label{fig:arc_lf}}
\end{center}
\end{figure}

While the predicted number of arcs sensitively depends on the luminosity
function of source galaxies \citep[e.g.,][]{hamana97}, $n_{\rm g}(L.z)$
is still fairly uncertain especially at high $z$. Thus we consider the
following four luminosity functions measured up to $z=1.25$: HDF1 from
the Hubble Deep Field and the New Technology Telescope Deep Field
\citep{poli01}, HDF2 from the Hubble Deep Field \citep*{sawicki97}, SDF
from the Subaru Deep Field \citep{kashikawa03}, and CFRS from the
Canada-France Redshift Survey \citep{lilly95}. They are summarized
in Table \ref{table:arc_lf}. Although the Schechter fits to those luminosity
functions are valid only at $z> (0.2 \sim 0.6)$, we simply extrapolate
the values even down to $z=0$ if necessary. This does not affect our
result in \S \ref{sec:arc_obs} at all since galaxies at $z\sim 1$ are the
main sources of lensed arcs for our sample of clusters at $z>0.2$ (\S
\ref{sec:arc_clusteremss}). 

Except for HDF1, the Schechter parameters were derived assuming the
Einstein-de Sitter (EdS) model ($\Omega_M=1$, $\Omega_\Lambda=0$) in the
original references. We convert them into the counterparts in the
Lambda-dominated universe ($\Omega_M=0.3$, $\Omega_\Lambda=0.7$) as
follows. 

Since the number of galaxies in the redshift interval $[z_{\rm S},z_{\rm
S}+dz_{\rm S}]$,
\begin{equation}
 dN_{\rm g}(z_{\rm S})\propto D^2_{\rm OS}\frac{c\,dt}{dz_{\rm S}}dz_{\rm S}n_{\rm g}(L,z_{\rm S})dL, 
\end{equation}
is observable, it should be invariant. Thus the luminosity function in
the Lambda-dominated universe is related to that in the EdS as:
\begin{equation}
 \left[n_{\rm g}(L',z_{\rm S})dL'\right]_{\rm Lambda}
=\frac{\left[D^2_{\rm OS}(c\,dt/dz_{\rm S})\right]_{\rm EdS}}
{\left[D^2_{\rm OS}(c\,dt/dz_{\rm S})\right]_{\rm Lambda}}
\left[n_{\rm g}(L,z_{\rm S})dL\right]_{\rm EdS}, 
\end{equation}
where 
\begin{equation}
 L' \equiv \frac{\left[D^2_{\rm OS}\right]_{\rm Lambda}}
{\left[D^2_{\rm OS}\right]_{\rm EdS}}L.
\end{equation}

The resulting luminosity functions in terms of the absolute
magnitude $M$:
\begin{equation}
 \phi_M(M,z)dM=0.921\phi^*10^{-0.4(\alpha_{\rm s}+1)(M-M^*)}
\exp\left(-10^{-0.4(M-M^*)}\right)dM,
\label{arc_schechter_m}
\end{equation}
at $z=0.3$, $0.6$, $0.9$, and $1.2$ are plotted in Figure \ref{fig:arc_lf}.
Clearly the uncertainty increases at fainter luminosities at $z>1$,
which may significantly change the predicted number of arcs.  Therefore,
while we adopt HDF1 as our fiducial model, we also attempt to evaluate
the uncertainty due to the different choice of luminosity functions using
the other three.

\subsection{Predicted Cross Sections and Numbers of Arcs}

\begin{figure}[tb]
\begin{center}
 \includegraphics[width=0.7\hsize]{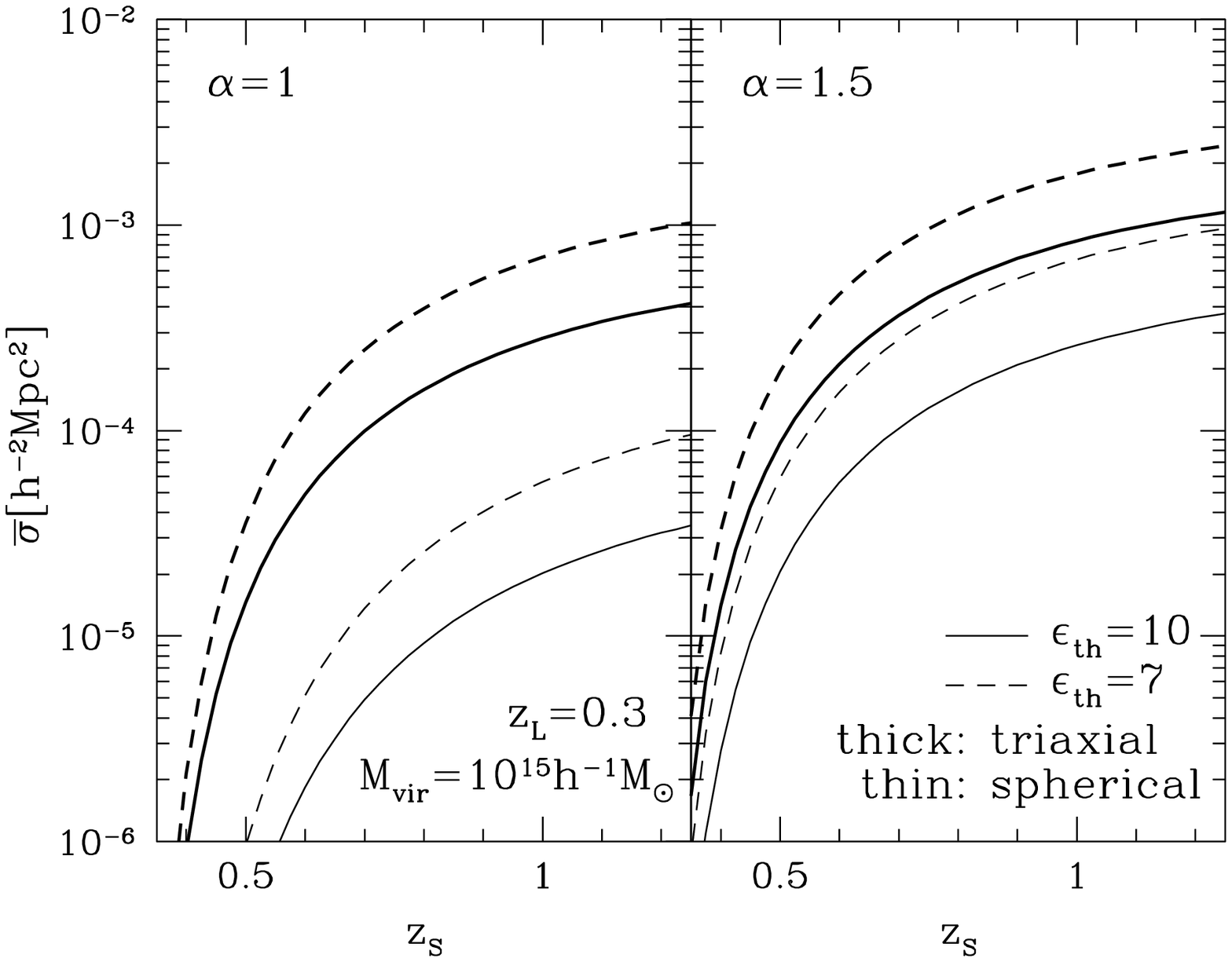}
\caption{Average cross sections (eq. [\ref{arc_cs_av}]) for triaxial and
 spherical dark matter halo models as a function of source redshift
 $z_{\rm S}$ for both $\alpha=1$ ({\it left}) and $1.5$ ({\it right}),
 where $\alpha$ is the inner slope of dark matter halo density profile. 
The lens cluster has a mass $M_{\rm vir}=10^{15}h^{-1}M_\odot$ and is
 placed at $z_{\rm L}=0.3$. For the threshold axis ratio of arcs, we
 adopt both $\epsilon_{\rm th}=10$ ({\it solid}) and $7$ ({\it dashed}). 
\label{fig:arc_cross_zs}}
\end{center}
\end{figure}

\begin{figure}[p]
\begin{center}
 \includegraphics[width=0.7\hsize]{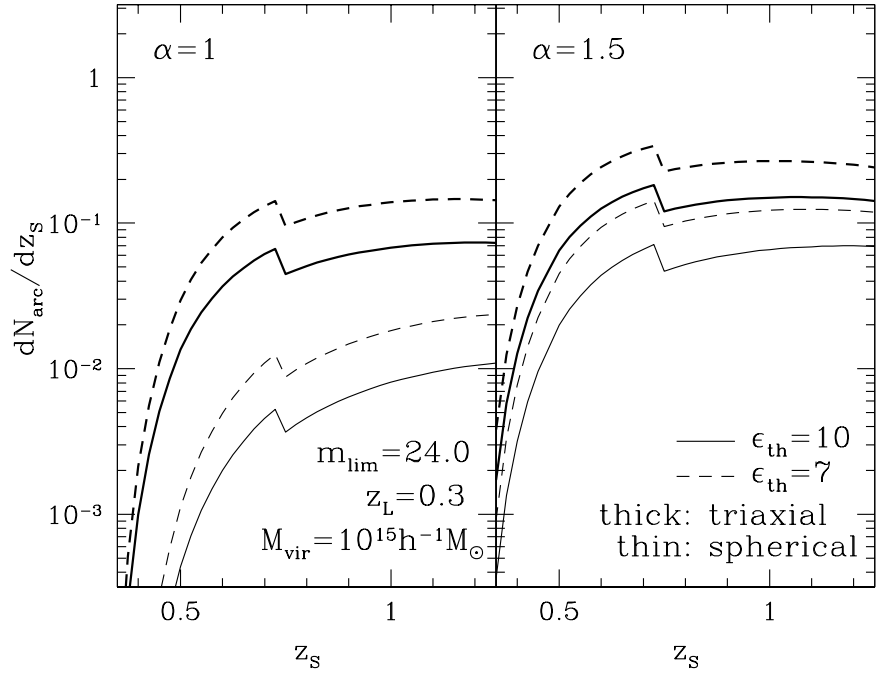}
\caption{Number distributions of arcs (eq. [\ref{arc_dnds}]) for triaxial and
 spherical dark matter halo models. The B-band magnitude limit for
 arcs is set to $m_{\rm lim}=24$. The distributions are discontinuous
 at $z_{\rm S}=0.75$ because we adopt binned luminosity function (see
 Table \ref{table:arc_lf}).
\label{fig:arc_dndzs}}
\end{center}
\end{figure}
\begin{figure}[p]
\begin{center}
 \includegraphics[width=0.7\hsize]{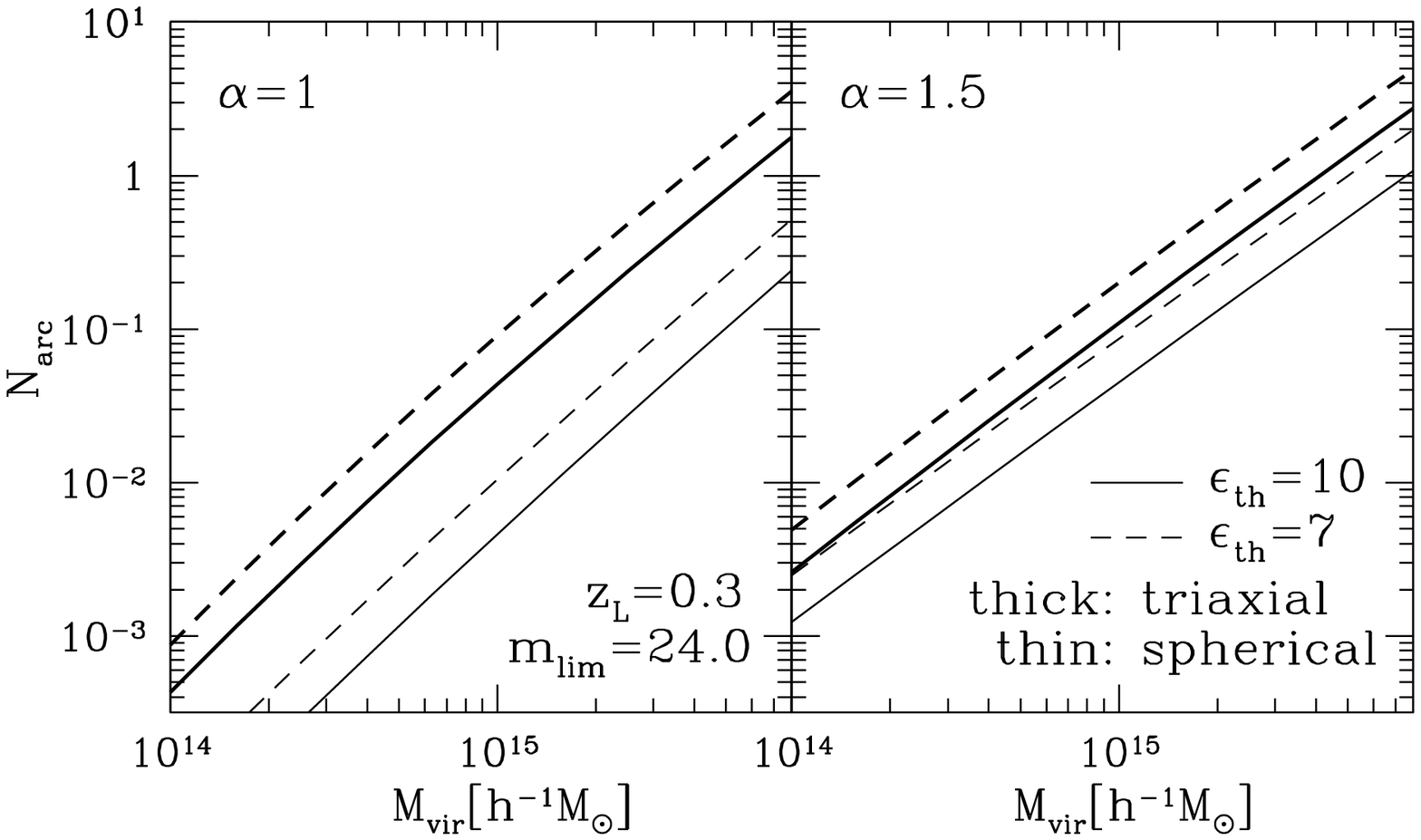}
\caption{Predicted numbers of arcs (eq. [\ref{arc_n_arc}]) as a function of
 halo mass $M_{\rm vir}$. The redshift of the dark halo is still fixed
 to $z_{\rm L}=0.3$. The B-band magnitude limit for
 arcs is set to $m_{\rm lim}=24$. 
\label{fig:arc_num_mvir}}
\end{center}
\end{figure}

Figure \ref{fig:arc_cross_zs} shows the average cross sections (eq.
[\ref{arc_cs_av}]) of a dark matter halo of $M_{\rm
vir}=10^{15}h^{-1}M_\odot$ at $z_{\rm L}=0.3$.  The cross section for
the triaxial model is larger by a factor of 10 ($\alpha=1$) and of 4
($\alpha=1.5$) than that for the spherical counterpart.  Since the
magnification factor is always larger for smaller cross sections (see
Figures \ref{fig:arc_cross_a10} and \ref{fig:arc_cross_a15}), the magnification
bias further reduces the difference between $\alpha=1$ and 1.5 for the
triaxial model. This explains the behavior of Figure \ref{fig:arc_dndzs}
where the source redshift distribution of arcs (eq. [\ref{arc_dnds}]) is
plotted. Actually the figure indicates that the non-spherical effect
even exceeds that of the difference due to the inner slope.

\begin{figure}[p]
\begin{center}
 \includegraphics[width=0.7\hsize]{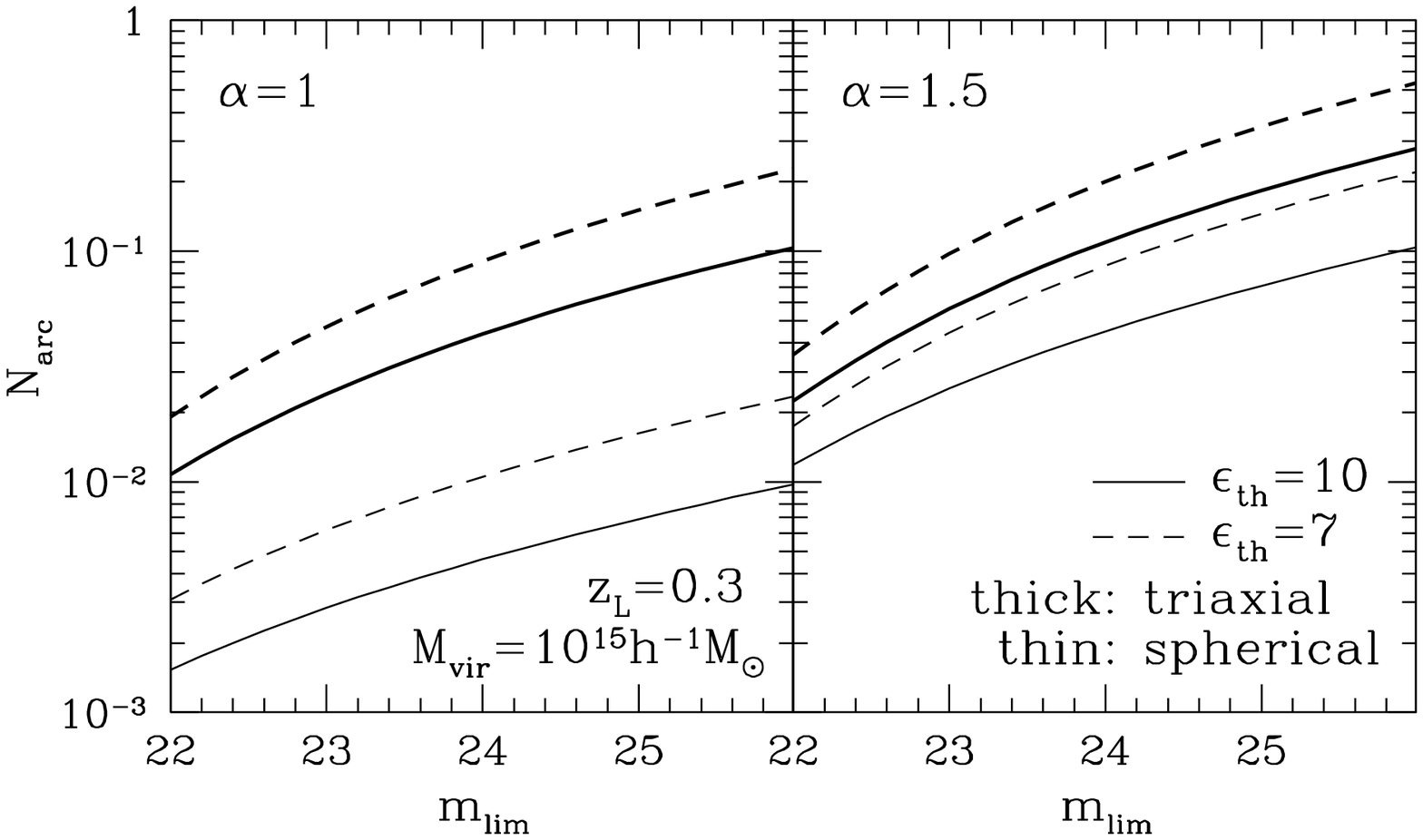}
\caption{Predicted numbers of arcs as a function of B-band magnitude
 limit $m_{\rm lim}$. The mass of lens cluster is 
 $M_{\rm vir}=10^{15}h^{-1}M_\odot$.  
\label{fig:arc_num_mlim}}
\end{center}
\end{figure}

\begin{figure}[p]
\begin{center}
 \includegraphics[width=0.7\hsize]{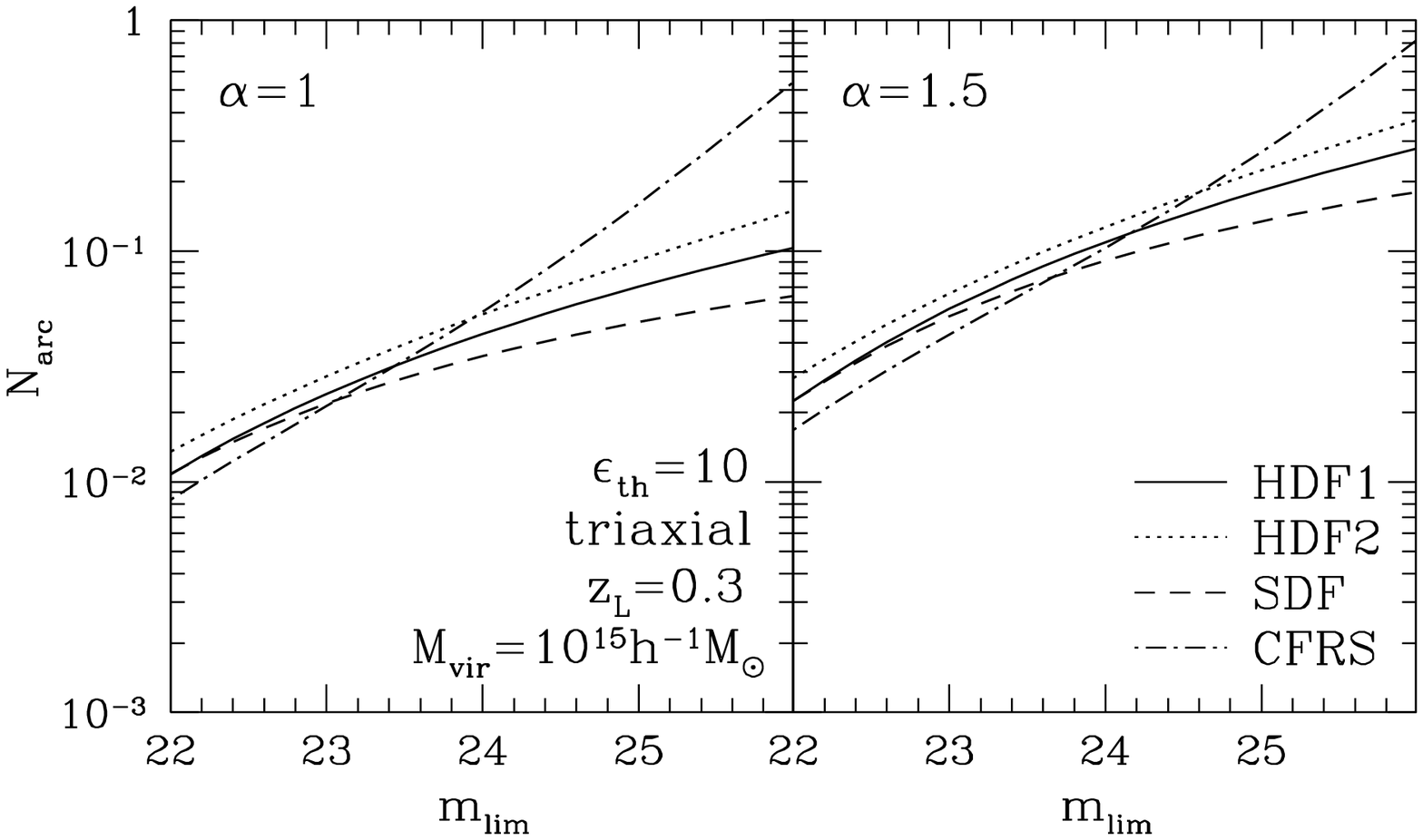}
\caption{Predicted numbers of arcs for different luminosity functions of
 source galaxies as a function of B-band magnitude limit
 $m_{\rm lim}$. Parameters of luminosity functions are given in Table
 \ref{table:arc_lf}. Only triaxial dark matter halo model and threshold
 axis ratio $\epsilon_{\rm th}=10$ are considered.
\label{fig:arc_num_mlim_lf}}
\end{center}
\end{figure}

Figures \ref{fig:arc_num_mvir}, \ref{fig:arc_num_mlim} and 
\ref{fig:arc_num_mlim_lf} show how the predicted number of arcs depends
on the mass of a lensing halo, the limiting magnitude of the survey, and
the adopted luminosity function of source galaxies.  Figure
\ref{fig:arc_num_mvir} shows that the number of arcs is sensitive to the
mass of halo, implying the estimate of the mass of the target cluster is
essential in interpreting the data. In addition, the difference between
$\alpha=1$ and 1.5 becomes smaller for the triaxial model of $M_{\rm
vir}>10^{15}M_\odot$. Thus in order to distinguish the inner slope
clearly as well, one needs a sample of less massive clusters that have
lensed arcs. 


Figure \ref{fig:arc_num_mlim} indicates that the number of arcs is also
sensitive to the magnitude limit, suggesting that the well-controlled
selection function for the arc survey is quite important.  On the other
hand, the uncertainty of the luminosity function of source galaxies
seems to be less critical, at least for arcs of galaxies at $z_{\rm
s}<1.25$ that we consider in this paper (Figure \ref{fig:arc_num_mlim_lf}).
The difference among the four luminosity functions (see Table
\ref{table:arc_lf}) is merely up to 50 \% for $m_{\rm lim}< 24$, and is
within a factor of 2 even at $m_{\rm lim}< 26$ except CFRS.  The
predictions based on HDF1 approximately correspond to the median among
the four and this is why we choose this as our fiducial model in what
follows.

\section{Comparison with the Observed Number of Arcs}
\label{sec:arc_obs}
\markboth{CHAPTER \thechapter.
{\MakeUppercase\mychapheadname}}{\thesection.
\MakeUppercase{Comparison with the Observed Number of Arcs}}

\subsection{Cluster Data}
\label{sec:arc_clusteremss}

We use a sample of $38$ X-ray selected clusters compiled by
\citet{luppino99}. The clusters are selected from the {\it Einstein
Observatory} Extended Medium Sensitivity Survey (EMSS). For all the
clusters, deep imaging observations with B-band limiting magnitude
$m_{\rm lim}\sim 26.0$ were carried out to search for arcs.

As we remarked in the previous section, the mass estimate of those
clusters is important in understanding the implications from the
observed arcs statistics. For this purpose, we first construct a gas
temperature -- X-ray luminosity (in the Einstein band) relation from a
subset of the above clusters whose temperature is determined.  Then we
estimate the temperature of the remaining clusters using the temperature
-- luminosity relation. Finally we estimate the mass of each cluster
employing the virial mass -- gas temperature relation of
\citet*{finoguenov01}. 

\begin{figure}[tb]
\begin{center}
 \includegraphics[width=0.5\hsize]{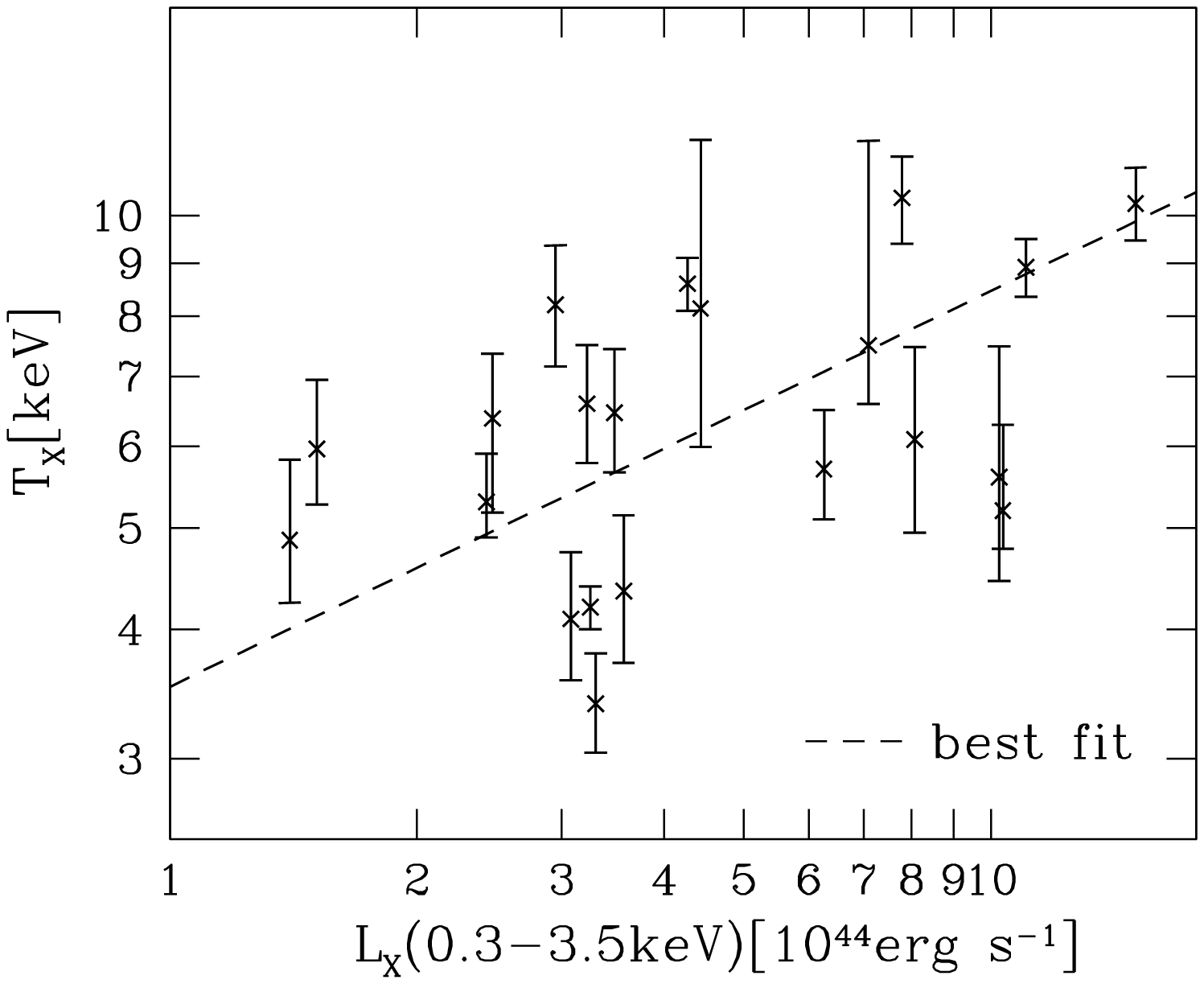}
\caption{The luminosity -- temperature relation for the EMSS cluster
 sample. Among 38 clusters, we use 21 clusters with measured temperature 
 to derive luminosity -- temperature relation. The best-fit
 luminosity -- temperature relation is shown in equation
 (\ref{arc_ltrelation}). 
\label{fig:arc_lt}}
\end{center}
\end{figure}

\begin{table}[p]
 \begin{center}
  \begin{tabular}{lcccccl}\hline\hline
 & & $L_X$(EdS)${}^{\rm a}$ & $L_X$(Lambda)${}^{\rm b}$ & 
 $T_X$ & $M_{\rm vir}$ & \\  
 Name & $z_{\rm L}$ & [$10^{44}{\rm erg\, s^{-1}}$] &
 [$10^{44}{\rm erg\, s^{-1}}$] & [${\rm keV}$] &
   [$10^{14}h^{-1}M_\odot$] & Ref.\\ \hline 
MS 0011.7$+$0837 & $0.163$ &  $3.77$ & $2.24$ & $4.79^{+0.48}_{-0.44}{}^{\rm c}$ & $3.89^{+0.72}_{-0.66}$ & 1 \\
MS 0015.9$+$1609 & $0.546$ & $14.64$ &$11.03$ & $8.92^{+0.57}_{-0.56}$ & $12.30^{+1.46}_{-1.43}$ & 1, 2\\
MS 0302.5$+$1717 & $0.425$ &  $2.88$ & $2.04$ & $4.62^{+0.45}_{-0.41}{}^{\rm c}$ & $3.64^{+0.66}_{-0.60}$ & 1 \\
MS 0302.7$+$1658 & $0.426$ &  $5.04$ & $3.57$ & $4.35^{+0.80}_{-0.64}$ & $3.25^{+1.11}_{-0.89}$ & 1, 2\\
MS 0353.6$-$3642 & $0.320$ &  $5.24$ & $3.48$ & $6.46^{+0.98}_{-0.80}$ & $6.77^{+1.90}_{-1.55}$ & 1, 2\\
MS 0433.9$+$0957 & $0.159$ &  $4.34$ & $2.57$ & $5.04^{+0.53}_{-0.48}{}^{\rm c}$ & $4.27^{+0.83}_{-0.75}$ & 1 \\
MS 0440.5$+$0204 & $0.190$ &  $4.01$ & $2.43$ & $5.30^{+0.60}_{-0.40}$ & $4.69^{+0.98}_{-0.66}$ & 1, 3\\
MS 0451.5$+$0250 & $0.202$ &  $6.98$ & $4.27$ & $8.60^{+0.50}_{-0.50}$ & $11.50^{+1.24}_{-1.24}$ & 1, 3\\
MS 0451.6$-$0305 & $0.539$ & $19.98$ &$15.00$ & $10.27^{+0.85}_{-0.80}$ & $15.97^{+2.45}_{-2.30}$ & 1, 2\\
MS 0735.6$+$7421 & $0.216$ &  $6.12$ & $3.79$ & $5.85^{+0.68}_{-0.61}{}^{\rm c}$ & $5.63^{+1.21}_{-1.09}$ & 1 \\
MS 0811.6$+$6301 & $0.312$ &  $2.10$ & $1.40$ & $4.87^{+0.95}_{-0.63}$ & $4.01^{+1.45}_{-0.96}$ & 1, 2\\
MS 0839.8$+$2938 & $0.194$ &  $5.35$ & $3.25$ & $4.20^{+0.20}_{-0.20}$ & $3.05^{+0.27}_{-0.27}$ & 1, 3\\
MS 0906.5$+$1110 & $0.180$ &  $5.77$ & $3.47$ & $5.65^{+0.64}_{-0.58}{}^{\rm c}$ & $5.28^{+1.11}_{-1.00}$ & 1 \\
MS 1006.0$+$1201 & $0.221$ &  $4.82$ & $2.99$ & $5.34^{+0.58}_{-0.52}{}^{\rm c}$ & $4.76^{+0.96}_{-0.86}$ & 1 \\
MS 1008.1$-$1224 & $0.301$ &  $4.49$ & $2.95$ & $8.21^{+1.15}_{-1.05}$ & $10.55^{+2.74}_{-2.50}$ & 1, 2\\
MS 1054.5$-$0321 & $0.823$ &  $9.28$ & $7.79$ & $10.4^{+1.00}_{-1.00}$ & $16.35^{+2.91}_{-2.91}$ & 1, 4\\
MS 1137.5$+$6625 & $0.782$ &  $7.56$ & $6.26$ & $5.70^{+0.80}_{-0.60}$ & $5.37^{+1.40}_{-1.05}$ & 1, 5\\
MS 1147.3$+$1103 & $0.303$ &  $2.30$ & $1.51$ & $5.96^{+0.99}_{-0.69}$ & $5.83^{+1.79}_{-1.25}$ & 1, 2\\
MS 1201.5$+$2824 & $0.167$ &  $2.03$ & $1.21$ & $3.78^{+0.34}_{-0.32}{}^{\rm c}$ & $2.51^{+0.42}_{-0.39}$ & 1 \\
MS 1208.7$+$3928 & $0.340$ &  $2.03$ & $1.37$ & $3.97^{+0.36}_{-0.33}{}^{\rm c}$ & $2.75^{+0.46}_{-0.42}$ & 1 \\
MS 1224.7$+$2007 & $0.327$ &  $4.61$ & $3.08$ & $4.09^{+0.65}_{-0.52}$ & $2.90^{+0.85}_{-0.68}$ & 1, 2\\
MS 1231.3$+$1542 & $0.238$ &  $2.88$ & $1.81$ & $4.41^{+0.42}_{-0.39}{}^{\rm c}$ & $3.34^{+0.59}_{-0.55}$ & 1 \\
MS 1241.5$+$1710 & $0.549$ & $10.70$ & $8.07$ & $6.09^{+1.38}_{-1.14}$ & $6.07^{+2.55}_{-2.10}$ & 1, 2\\
MS 1244.2$+$7114 & $0.225$ &  $3.84$ & $2.39$ & $4.90^{+0.50}_{-0.46}{}^{\rm c}$ & $4.06^{+0.77}_{-0.71}$ & 1 \\
MS 1253.9$+$0456 & $0.230$ &  $3.14$ & $1.96$ & $4.55^{+0.44}_{-0.40}{}^{\rm c}$ & $3.54^{+0.63}_{-0.58}$ & 1 \\
MS 1358.4$+$6245 & $0.327$ & $10.62$ & $7.09$ & $7.50^{+4.30}_{-0.91}{}^{\rm d}$ & $8.93^{+9.48}_{-2.01}$ & 1, 2, 6\\
MS 1426.4$+$0158 & $0.320$ &  $3.71$ & $2.47$ & $6.38^{+0.98}_{-1.20}$ & $6.62^{+1.88}_{-2.30}$ & 1, 2\\
MS 1455.0$+$2232 & $0.259$ & $16.03$ &$10.23$ & $5.60^{+1.88}_{-1.15}{}^{\rm d}$ & $5.20^{+3.23}_{-1.98}$ & 1, 3, 6\\
MS 1512.4$+$3647 & $0.372$ &  $4.81$ & $3.30$ & $3.39^{+0.40}_{-0.35}$ & $2.05^{+0.45}_{-0.39}$ & 1, 2\\
MS 1546.8$+$1132 & $0.226$ &  $2.94$ & $1.83$ & $4.43^{+0.43}_{-0.39}{}^{\rm c}$ & $3.37^{+0.61}_{-0.55}$ & 1 \\
MS 1618.9$+$2552 & $0.161$ &  $2.24$ & $1.33$ & $3.92^{+0.36}_{-0.33}{}^{\rm c}$ & $2.68^{+0.46}_{-0.42}$ & 1 \\
MS 1621.5$+$2640 & $0.426$ &  $4.55$ & $3.22$ & $6.59^{+0.92}_{-0.81}$ & $7.02^{+1.82}_{-1.60}$ & 1, 2\\
MS 1910.5$+$6736 & $0.246$ &  $4.39$ & $2.78$ & $5.20^{+0.55}_{-0.50}{}^{\rm c}$ & $4.53^{+0.89}_{-0.81}$ & 1 \\
MS 2053.7$-$0449 & $0.583$ &  $5.78$ & $4.43$ & $8.14^{+3.68}_{-2.15}$ & $10.39^{+8.70}_{-5.08}$ & 1, 2\\
MS 2137.3$-$2353 & $0.313$ & $15.62$ &$10.34$ & $5.20^{+1.09}_{-0.42}{}^{\rm d}$  & $4.53^{+1.76}_{-0.68}$ & 1, 2, 6\\ 
MS 2255.7$+$2039 & $0.288$ &  $2.04$ & $1.33$ & $3.92^{+0.36}_{-0.33}{}^{\rm c}$ & $2.68^{+0.46}_{-0.42}$ & 1 \\
MS 2301.3$+$1506 & $0.247$ &  $3.29$ & $2.08$ & $4.65^{+0.46}_{-0.42}{}^{\rm c}$ & $3.68^{+0.67}_{-0.62}$ & 1 \\
MS 2318.7$-$2328 & $0.187$ &  $6.84$ & $4.14$ & $6.05^{+0.73}_{-0.65}{}^{\rm c}$ & $6.00^{+1.34}_{-1.19}$ & 1 \\ 
\hline
\end{tabular}
\caption{Properties of clusters in the $38$ EMSS distant cluster
 sample. Errors are at $68$\% C.L.\protect\\
\footnotesize{\hspace*{5mm}${}^{\rm a}$ X-ray luminosity in the $0.3-3.5{\rm keV}$ band for
 $\Omega_M=1$, $\Omega_\Lambda=0$, and $h=0.5$ universe.\hspace{10mm}\protect\\
\hspace*{5mm}${}^{\rm b}$ X-ray luminosity in the $0.3-3.5{\rm keV}$ band for
 $\Omega_M=0.3$, $\Omega_\Lambda=0.7$, and $h=0.7$ universe.\hspace{70mm}\protect\\
\hspace*{5mm}${}^{\rm c}$ Estimated from $L_X-T_X$ relation (eq.
 [\ref{arc_ltrelation}]).\hspace{70mm}\protect\\
\hspace*{5mm}${}^{\rm d}$ The effects of cooling flows are corrected.\hspace{70mm}\protect\\
\hspace*{5mm}Ref. --- (1) \citealt{luppino99}; (2) \citealt*{novicki02};
  (3) \citealt{mushotzky97}; (4) \citealt{jeltema01}; (5)
 \citealt{borgani01}; (6) \citealt{allen98}}
}
\label{table:arc_emss}
 \end{center}
\end{table}

More specifically, our best-fit luminosity -- temperature relation from
Figure \ref{fig:arc_lt} is
\begin{equation}
 T_X=T_{X,0}\left(\frac{L_X(0.3-3.5{\rm keV})}
{10^{44}{\rm erg\, s^{-1}}}\right)^\gamma,
\label{arc_ltrelation}
\end{equation}
where $\gamma=0.381\pm 0.052$ and $T_{X,0}=3.52^{+0.32}_{-0.29}{\rm
keV}$. The derived luminosity -- temperature relation is consistent with 
recent other estimations \citep[e.g.,][]{ikebe02}. Neglecting the
possible redshift evolution for the luminosity -- temperature relation
\citep[e.g.,][]{mushotzky97}, we estimate the temperature of those
clusters without spectroscopic data as shown in Table
\ref{table:arc_emss}. The mass -- temperature relation that we adopt is
\begin{equation}
 T_X=2.3{\rm keV}\left(\frac{M_{\rm vir}}{10^{14}h^{-1}M_\odot}\right)^{0.54}.
\end{equation}
This relation is derived by \citet{shimizu03} who converted the result
of \citet{finoguenov01} in terms of $M_{\rm vir}$ assuming the density
profile; the difference between $\alpha=1$ and $1.5$ turned out to be
negligible. 

\subsection{Observed Number of Arcs}
\label{sec:arc_obs_num}

\begin{table}[tb]
 \begin{center}
  \begin{tabular}{lccccccl}\hline\hline
 Cluster & $z_{\rm L}$ & Arc & $z_{\rm S}$ & $l/w$ & $m_{\rm arc}$ &
 Notes & Ref. \\ \hline
 MS 0302.7$+$1658 & $0.426$ & A1 & $\sim 0.8{}^{\rm a}$ & $>18$ &
 $B=23.8$ & $\cdots$ & 1, 2\\
               &           & A1W& $\cdots$ & $>12$ & $B=24.9$ & $\cdots$ & \\
MS 0440.5$+$0204 & $0.190$ & A1 & $0.532$ & $>10$ & $B=22.9$ & $\cdots$ &
 1, 3, 4\\
                 &         & A3 & $\cdots$ & $>20$ & $B=24.0$ & $\cdots$ & \\
MS 0451.6$-$0305 & $0.539$ & A1 & $\cdots$ & $10$  & $V=24.6$ & $\cdots$ & 1\\
MS 1006.0$+$1201 & $0.221$ & A2$+$A3  & $\cdots$ & $>20$& $V<22.1$ &
 Candidate & 1, 5\\
                 &         & A4 & $\cdots$ & $12.9$ & $V=21.4$ & Candidate & \\
MS 1008.1$-$1224 & $0.301$ & A2 & $\cdots$ & $10.0$ & $B=23.4$ & Candidate & 1\\
MS 1358.4$+$6245 & $0.328$ & A1 & $4.92$  & $>21$  & $\cdots$  & $\cdots$
 & 1, 6\\
MS 1621.5$+$2640 & $0.426$ & A1 & $\cdots$ & $>18$  & $B=23.1$ & $\cdots$
 & 1, 7\\
MS 1910.5$+$6736 & $0.246$ & A1 & $\cdots$ & $10.5$ & $R=20.6$ &
 Candidate & 1, 5\\
MS 2053.7$-$0449 & $0.583$ & AB & $\cdots$ & $>22$  & $V=22.4$ & $\cdots$
 & 1, 7\\
MS 2137.3$-$2353 & $0.313$ & A1 & $1.501$ & $18.1$ & $B=22.0$ & $\cdots$
 & 1, 8, 9, 10\\
\hline
\end{tabular}
\caption{Giant arcs ($l/w>10$) in the $38$ EMSS distant cluster
  sample.\protect\\ 
\footnotesize{\hspace*{5mm}${}^{\rm a}$ Estimated from color of the
  arc.\hspace{70mm}\protect\\ 
\hspace*{5mm}Ref. --- (1) \citealt{luppino99}; (2) \citealt{mathez92}; (3)
 \citealt{luppino93}; (4) \citealt{gioia98} (5); \citealt{lefevre94};
 (6) \citealt{franx97}; (7) \citealt{luppino92}; (8) \citealt{fort92};
 (9) \citealt{hammer97}; (10) \citealt{sand02}}
}
\label{table:arc_arc}
 \end{center}
\end{table}

The observed giant arcs ($\epsilon_{\rm th}=10$) in the $38$ EMSS
cluster sample are listed in Table \ref{table:arc_arc}. The number of
arcs in this sample is roughly consistent with more recent data from
different cluster samples \citep{zaritsky03,gladders03}. In order to be
consistent with our adopted luminosity functions and K-correction of
source galaxies, we need to select the arcs with $z<1.25$.  In reality,
this is quite difficult; most of the observed arcs do not have a
measured redshift, while uncertainties of source redshifts may
systematically change lensing probabilities. For instance,
\citet{wambsganss04} explicitly showed that it is important to take
correctly account of the source redshift which can change cross sections
by an order of magnitude. Moreover four in the list labeled
``Candidate'' in Table \ref{table:arc_arc} are even controversial and
may not be real lensed arcs. Thus we consider the two extreme cases; one
is to select only the two arcs with measured redshifts less than 1.25,
and the other is to assume that all the arcs without measured redshifts
in the list (including the candidates) are located at $z<1.25$. Of
course the reality should be somewhere in between, and thus we assume
that the range between the two cases well represents the current
observational error. This means that the observational error can be
greatly reduced if redshifts of all arcs are measured in the future
observations.  

\subsection{Comparison of Theoretical Predictions with Observations}

\begin{figure}[tb]
\begin{center}
 \includegraphics[width=0.6\hsize]{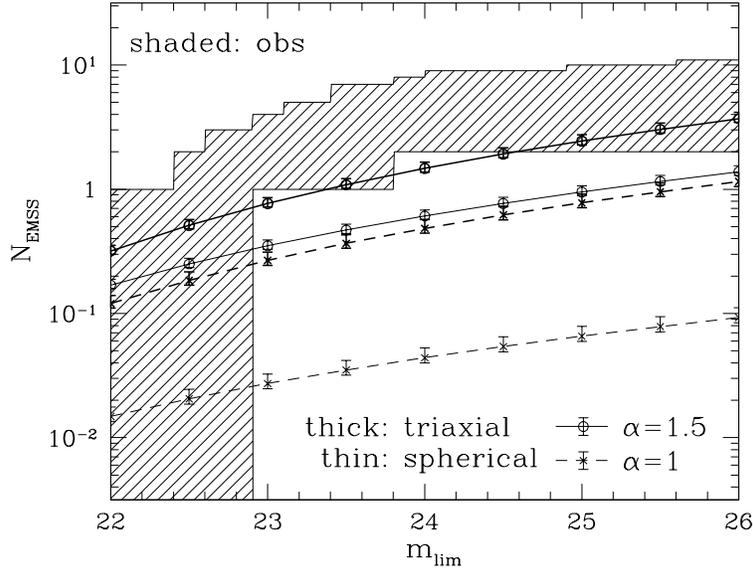}
\caption{The number of arcs in the 38 EMSS cluster sample (eq.
 [\ref{arc_n_emss}]) as a function of B-band limiting magnitude $m_{\rm lim}$. 
 The threshold axis ratio is $\epsilon_{\rm th}=10$. The observed number 
 of arcs taking account of several uncertainties, which is shown by the
 shaded region, is discussed in \S \ref{sec:arc_obs_num}. 
\label{fig:arc_num_obs_mlim}}
\end{center}
\end{figure}

Finally let us compare our theoretical predictions with the data in
detail. Our prediction of the number of arcs is the sum of equation
(\ref{arc_n_arc}) over all the 38 EMSS clusters:
\begin{equation}
 N_{\rm EMSS}\equiv\sum^{38}_{i=1}N_{\rm arc}(M_{{\rm vir},i},z_{{\rm L},i}).
\label{arc_n_emss}
\end{equation}
We also compute the error of the predicted number of arcs by propagating
the mass uncertainty for each cluster (Table \ref{table:arc_emss}).  Figure
\ref{fig:arc_num_obs_mlim} shows the number of arcs in the 38 EMSS cluster
sample as a function of the B-band limiting magnitude $m_{\rm lim}$.
When the B-band magnitude of an arc is not available, we convert its
corresponding V- or R-band magnitude into the B-band assuming typical
colors of spiral galaxies at $z\sim 1$, $B-V=V-R=1$ \citep{fukugita95}.

The important conclusion that we draw from Figure \ref{fig:arc_num_obs_mlim}
is that the triaxial model in the Lambda-dominated CDM universe with the
inner slope of $\alpha=1.5$ successfully reproduces the observed number of
arcs, and that the spherical model prediction with $\alpha=1$ fails by a
wide margin. Both the triaxial model with $\alpha=1$ and the spherical
model with $\alpha=1.5$ are marginal in a sense that the presence of
substructure in the dark halo which we ignore in the current method
should systematically increase our predicted number of arcs. Indeed
\citet*{meneghetti03a} reported that the substructure enhances the number
of arcs with $\epsilon_{\rm th}=10$ typically by a factor 2 or 3. This
is exactly the amount of enhancement that is required to reconcile those
two models with the observation.

We note here that the additional contribution due to galaxies inside a
cluster is generally small; \citet*{flores00} and \citet{meneghetti00}
found that galaxies increase the number of arcs merely by $\sim$10\%.
Even a central cD galaxy produces the number of arcs by not more than
$\sim$50\% \citep*{meneghetti03b}.

\section{Discussion}
\label{sec:arc_discussion}
\markboth{CHAPTER \thechapter.
{\MakeUppercase\mychapheadname}}{\thesection.
\MakeUppercase{Discussion}}

\subsection{Comparison with the previous result}

Our result that the halos in a Lambda-dominated CDM universe reproduces
the observed number of arcs seems inconsistent with the previous result
of \citet{bartelmann98} who claimed that only open CDM models can
reproduce the observation. One possibility to explain the apparent
discrepancy is the difference of the inner profile of halos; we showed
that the slope of $\alpha=1.5$ is required to reproduce the observation.
This implies that N-body simulations may underestimate the real number
of arcs unless they have sufficient spatial resolution. On the other hand,
cluster-scale halos may indeed have a shallower inner profile 
\citep{jing00a}. Therefore this is closely related to the well known
problem of the inner slope of CDM dark matter halos
\citep{navarro96,navarro97,fukushige97,fukushige01,fukushige03,moore99b,
ghigna00,jing00a,klypin01,power03,fukushige04,hayashi04,navarro04},
and would need further investigation.  Moreover, in reality, the mass
estimate for each cluster, the limiting magnitude of source galaxies,
and the adopted luminosity function would also affect the prediction in
a more complicated fashion, and the further quantitative comparison is
not easy at this point.

\begin{figure}[tb]
\begin{center}
 \includegraphics[width=0.6\hsize]{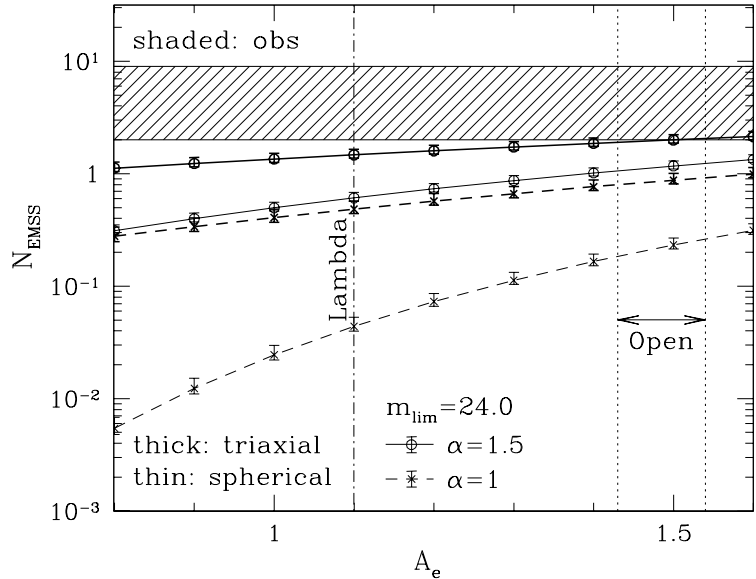}
\caption{The number of arcs in the 38 EMSS cluster sample as a function
 of $A_e$ for $m_{\rm lim}=24$. Dash-dotted line indicates the fiducial value
 for $A_e$, $A_e=1.1$, in a Lambda-dominated CDM model. Dotted lines suggest
 possible range of $A_e$ with taking account of the enhancement of the
 concentration parameter in an open CDM model (see text for details).
\label{fig:arc_num_obs_ce}}
\end{center}
\end{figure}

Nevertheless we can point out the general tendency that open CDM models
produce more arcs than Lambda-dominated CDM models because of the larger
value of the concentration parameter in the former. Thus it is unlikely
that difference between open and Lambda-dominated CDM models results
from the ``global'' effect of the cosmological parameters.  In order to
show this, we compute the number of arcs as a function of  $A_e$ still
assuming the Lambda-dominated CDM model. Figure \ref{fig:arc_num_obs_ce}
plots $N_{\rm EMSS}$ for $m_{\rm lim}=24$ as a function of $A_e$. While
\citet{jing02} found $A_e=1.1$ in a Lambda-dominated CDM models, their
fitting formula \citep[see also][]{bartelmann98} tend to predict $\sim
30-40\%$ larger concentration parameter in open CDM models. This
enhancement of the concentration parameter corresponds to
$A_e=1.43\sim1.53$ if we still assume Lambda-dominated CDM models as a
background cosmology. Thus the effect of $A_e$ alone increases the
number of arc by $\sim 50\%-100\%$ even for triaxial cases, which is
qualitatively consistent with the result of \citet{bartelmann98}.

\subsection{Required Non-sphericity of Lensing Halos}

\begin{figure}[tb]
\begin{center}
 \includegraphics[width=0.7\hsize]{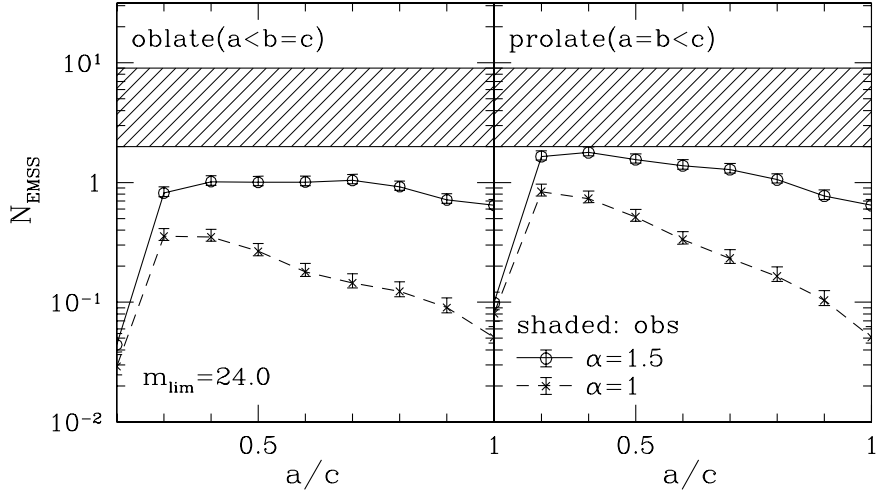}
\caption{The number of arcs in the 38 EMSS cluster sample for fixed axis
 ratios of dark matter halos. The B-band limiting magnitude is set to
 $m_{\rm lim}=24$. Left panel plots the oblate case ($a<b=c$) while
 right panel is the prolate case ($a=b<c$).
\label{fig:arc_num_obs_def}}
\end{center}
\end{figure}

Although we showed that the triaxial halos predicted in the
Lambda-dominated CDM model reproduce the observed number of arcs, the
analysis employed a series of fairly complicated PDFs for the axial
ratios, and it is not so clear what degree of non-sphericity for lensing
halos is required to account for the observation.  Thus we rather
simplify the situation and consider that all halos consist of oblate
($a<b=c$) or prolate ($a=b<c$) halos with a fixed axial ratio. This is
equivalent to replacing $p(a/c)$ (eq. [\ref{tri_p_a}]) or $p(a/b)$ (eq.
[\ref{tri_p_ab}]) by the corresponding $\delta$-functions.  Figure
\ref{fig:arc_num_obs_def} plots the result of this exercise.

The predicted number of arcs is indeed sensitive to the axis ratios of
dark matter halos, and prolate halos of $a/c\lesssim0.5$ in the
$\alpha=1.5$ case reproduce the observation. This is basically
consistent with the finding of \citet{jing02} for halo properties.

The reason why prolate halos tend to produce the larger number of arcs
than oblate halos is explained as follows.  Notice first that to keep
the mass of dark matter halo invariant with the change of the axial
ratio, $b_{\rm TNFW}$ should be approximately proportional to
$(ab/c^2)^{-1}$. Suppose that oblate and prolate halos are projected
onto their axisymmetric direction ($x$ for oblate and $z$ for
prolate). Then their lensing cross sections should scale as
\begin{eqnarray}
 \sigma({\rm oblate})
&\propto & \tilde{\sigma}((a/c)(a/c)^{-1}b_{\rm TNFW},1)\nonumber\\
&=&\tilde{\sigma}(b_{\rm TNFW},1),\\
 \sigma({\rm prolate})
&\propto & \left(\frac{a}{c}\right)^2\tilde{\sigma}((a/c)^{-2}b_{\rm TNFW},1)\nonumber\\  
&=&\left(\frac{a}{c}\right)^{2-2\delta}\tilde{\sigma}(b_{\rm TNFW},1),
\end{eqnarray}
where we assume 
\begin{equation}
 \tilde{\sigma}(b_{\rm TNFW},q)\propto b_{\rm TNFW}^\delta.
\end{equation}
Since Figures \ref{fig:arc_cross_a10} and \ref{fig:arc_cross_a15}
suggest $\delta\gtrsim 2$, we find that $\sigma({\rm
prolate})\gg\sigma({\rm oblate})$ for $a/c <1$.  If those halos are
projected along the $y$-direction, on the other hand, their cross
sections are almost the same: 
\begin{equation}
 \sigma({\rm oblate})\sim\sigma({\rm prolate})
\propto \left(\frac{a}{c}\right)^{-\delta}\tilde{\sigma}(b_{\rm TNFW},a/c).
\end{equation}
The above consideration explains the qualitative difference between
oblate and prolate halos, and points out that the elongation along the
line-of-sight is also important in the arc statistics as well as the
asymmetry of the projected mass density.

\subsection{Are Clusters Equilibrium Dark Matter Halos?}

So far we have assumed the one-to-one correspondence between dark matter
halos and X-ray clusters.  This assumption, however, is definitely
over-simplified \citep{suto01,suto03}.  If ``dark clusters'' which are
often reported from recent weak lensing analyses
\citep{hattori97a,wittman01,miyazaki02a} are real, the one-to-one
correspondence approximation may be unexpectedly inaccurate.  As an
extreme possibility, let us suppose that observed X-ray clusters
preferentially correspond to halos in equilibrium.  According to
\citet{jing00b}, such halos have generally larger concentration
parameters and their scatter is small.  In order to imitate this
situation, we repeat the computation using $A_e=1.3$ and the scatter of
$0.18$ \citep{jing00b,jing02}.  We find that this modified model
increases the number of arcs merely by 10\%$-$20\%. Thus our conclusion
remains the same. 

\subsection{Sample Variance}

\begin{figure}[tb]
\begin{center}
 \includegraphics[width=0.6\hsize]{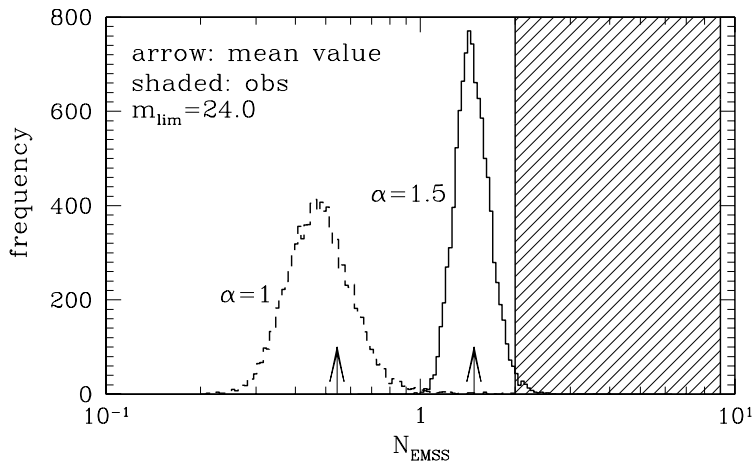}
\caption{The effect of the sample variance. The number of arcs in the 38
 EMSS cluster sample is calculated, not being averaged over axis ratios,
 orientation, and concentration parameters, but using fixed values for
 each clusters.  Axis ratios, orientation, and concentration parameters
 for each cluster are randomly chosen according to their corresponding 
 PDFs (\S \ref{sec:tri_tnfw}).  We calculate 10000 realizations and plot
 the histogram of frequency. Averaged values are shown by arrows. 
\label{fig:arc_samvar}}
\end{center}
\end{figure}

The predicted number of arcs for the EMSS cluster that we have presented
so far is based on the {\it averaged} cross section. This is a
reasonably good approximation in the situation that the number of sample
clusters is large enough, but in the current sample, its validity is not
clear.  To examine the sample variance, we re-compute the number of arcs
in the 38 EMSS cluster sample without using the average statistics.
Instead, we first randomly choose values of the axis ratios, the
orientation angles, and the concentration parameters for each cluster
according to their corresponding PDFs (\S \ref{sec:tri_tnfw}). Then we
sum up the number of arcs for the entire cluster sample.  We repeat the
procedure 10000 times each for $\alpha=1$ and 1.5, and construct a
distribution function of $N_{\rm EMSS}$ as plotted in Figure
\ref{fig:arc_samvar}. The resulting 1$\sigma$ sample variance is
$\sim$30\% for $\alpha=1$ and $\sim$15\% for $\alpha=1.5$.  Therefore we
confirm that the effect of the sample variance does not  change our
overall conclusion. 

\section{Summary}
\label{sec:arc_conclusion}
\markboth{CHAPTER \thechapter.
{\MakeUppercase\mychapheadname}}{\thesection.
\MakeUppercase{Summary}}

We have presented a semi-analytic method to predict the number of lensed
arcs, for the first time taking proper account of the triaxiality of
lensing halos. We found that Lambda-dominated concordance CDM models
successfully reproduce the observed number of arcs of X-ray-selected
clusters \citep{luppino99} if the inner slope of the density profile is
close to $\alpha=1.5$.  Since the spherical models significantly
underestimate the expected number of arcs, we conclude that the observed
number of arcs indeed requires the non-sphericity of the lensing halos.
In fact, the number of arcs is sensitive to the axis ratios of those
halos, and the non-sphericity that reproduces the observed number
corresponds to the minor to major axis ratios of $\sim 0.5$. This value
is perfectly consistent with the findings of \citet{jing02} in the
Lambda-dominated CDM models. In this sense, we may even argue that the
arc statistics lend strong support for the collisionless CDM paradigm at
the mass scale of clusters.  As discussed in \citet{meneghetti01},
self-interacting dark matter models \citep{spergel00} for instance, are
inconsistent with the observed number of arcs not only because they
erase the central cusp but because they produce much rounder dark matter
halos (see Figure \ref{fig:halo_ny}). Since we have exhibited that even
the current arc surveys have a great impact in testing the collisionless
CDM paradigm, larger surveys with well-controlled systematics in near
future will unveil the nature of dark matter more precisely. 

\chapter{Theoretical Predictions for Large-Separation Lensed Quasars with
Triaxial Dark Halos}\label{chap:lat} 
\def\mychapheadname{Theoretical Predictions for Large-Sep Lensed Quasars}
\markboth{CHAPTER \thechapter.
{\MakeUppercase\mychapheadname}}{}

\section{Introduction}
\markboth{CHAPTER \thechapter.
{\MakeUppercase\mychapheadname}}{\thesection.
\MakeUppercase{Introduction}}

Since the discovery of the first gravitationally lensed quasar Q0957+561  
\citep{walsh79}, about 80 strong lens systems have been found so far.
All of the lensed quasars have image separations smaller than $7''$,
and they are lensed by massive galaxies (sometimes with small boosts
from surrounding groups or clusters of galaxies). The probability that
distant quasars are lensed by intervening galaxies was originally
estimated by \citet{turner84} to be 0.1\%--1\%, assuming that galaxies
can be modeled as singular isothermal spheres (SIS). This prediction
has been verified by several optical and radio lens surveys, such as the
{\it HST} Snapshot Survey \citep{bahcall92}, the Jodrell Bank/Very Large
Array Astrometric Survey \citep[JVAS;][]{patnaik92}, and the Cosmic Lens
All Sky Survey \citep[CLASS;][]{myers95}.  The lensing probability is
sensitive to the volume of the universe, so it can be used to place
interesting constraints on the cosmological constant $\Omega_\Lambda$
(see Chapter \ref{chap:conc}). 

In contrast, lenses with larger image separations should probe a
different deflector population: massive dark matter halos that host
groups and clusters of galaxies. Such lenses therefore offer valuable
and complementary information on structure formation in the universe,
including tests of the CDM paradigm
\citep{narayan88,cen94,wambsganss95,kochanek95b,flores96,nakamura97}. 
So far the observed lack of large-separation lensed quasars has been
used to infer that, unlike galaxies, cluster-scale halos cannot be
modeled as singular isothermal spheres \citep*{keeton98a,porciani00,
kochanek01,keeton01a,sarbu01,li02,li03,oguri02c,ma03}.  The difference
can probably be ascribed to baryonic processes: baryonic infall and
cooling have significantly modified the total mass distribution in
galaxies but not in clusters \citep[e.g.,][]{rees77,blumenthal86,kochanek01}.
As a result, large-separation lenses may constrain the density profiles
of dark matter halos of cluster more directly than small-separation
lenses \citep*{maoz97,keeton01b,wyithe01,takahashi01,li02,oguri02a,
oguri03c,huterer04a,kuhlen04}. Alternatively, large-separation lensed
quasars may be used to place limits on the abundance of massive halos if
the density profiles are specified \citep{narayan88,wambsganss95,
kochanek95b,nakamura97,mortlock00,oguri03c,chen03a,chen04,lopes04}.
Better yet, the full distribution of lens image separations may provide
a systematic diagnostic of baryonic effects from small to large scales
in the CDM scenario. 

\begin{figure}[tb]
\begin{center}
 \includegraphics[width=0.32\hsize]{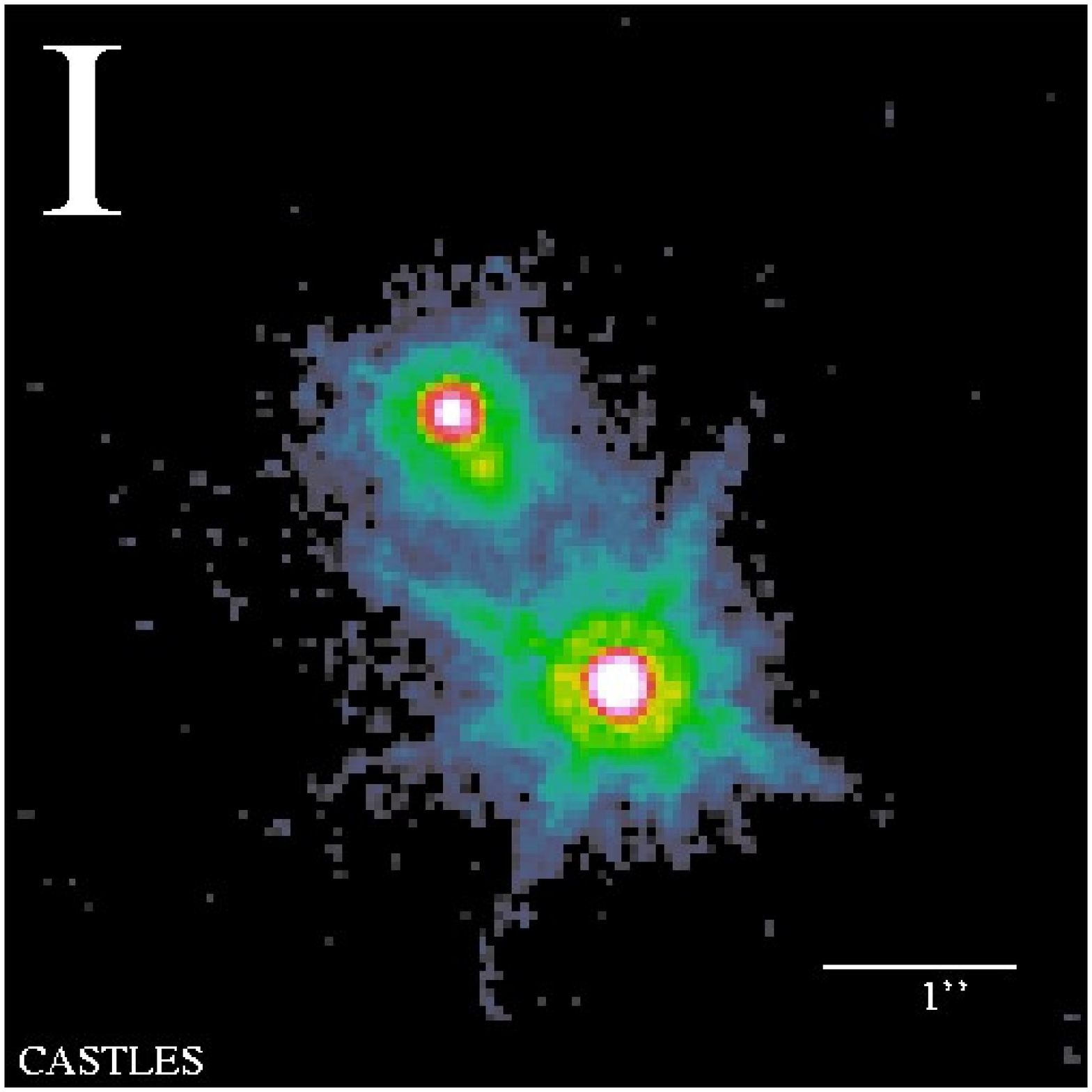}
 \includegraphics[width=0.32\hsize]{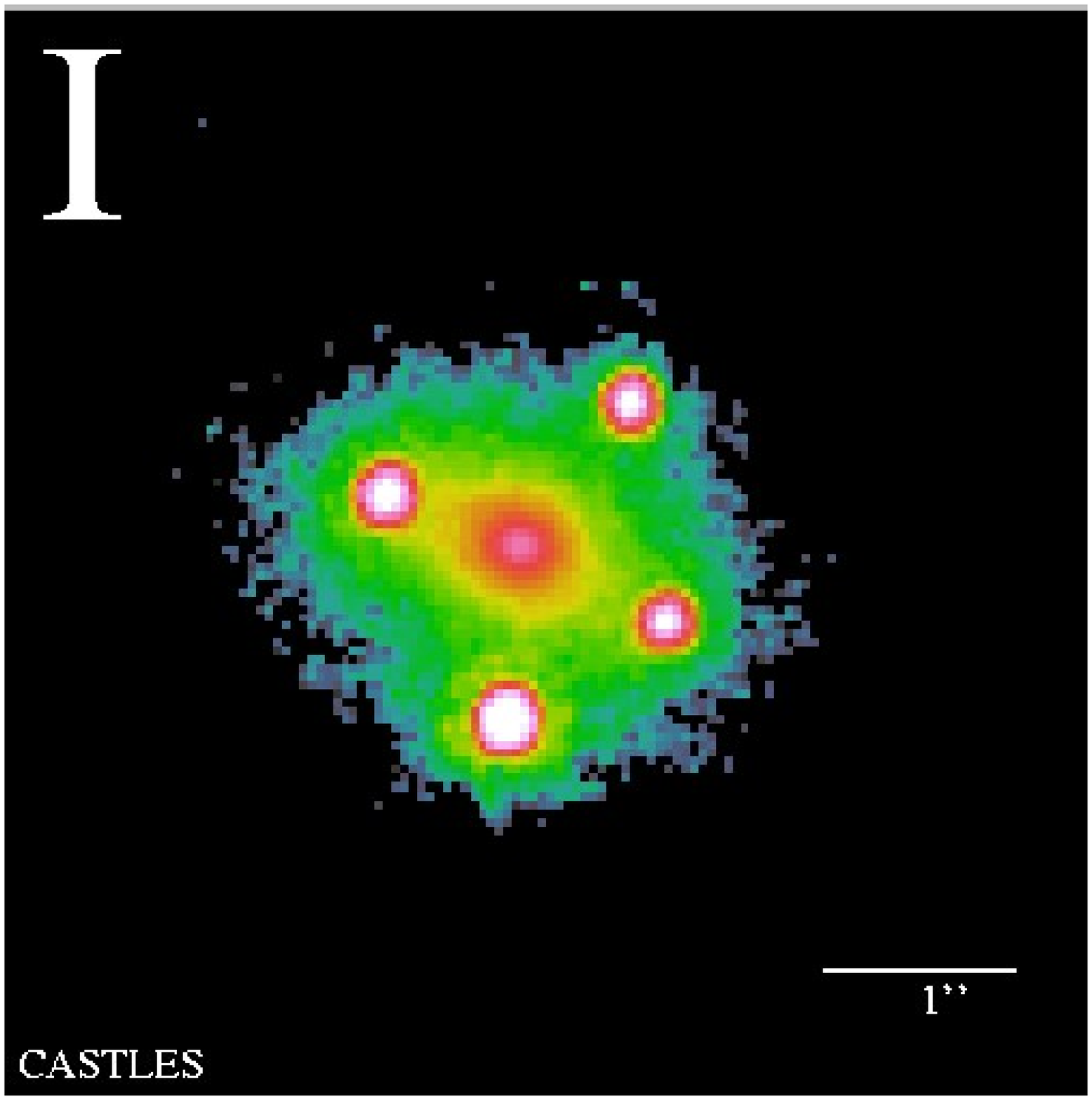}
 \includegraphics[width=0.32\hsize]{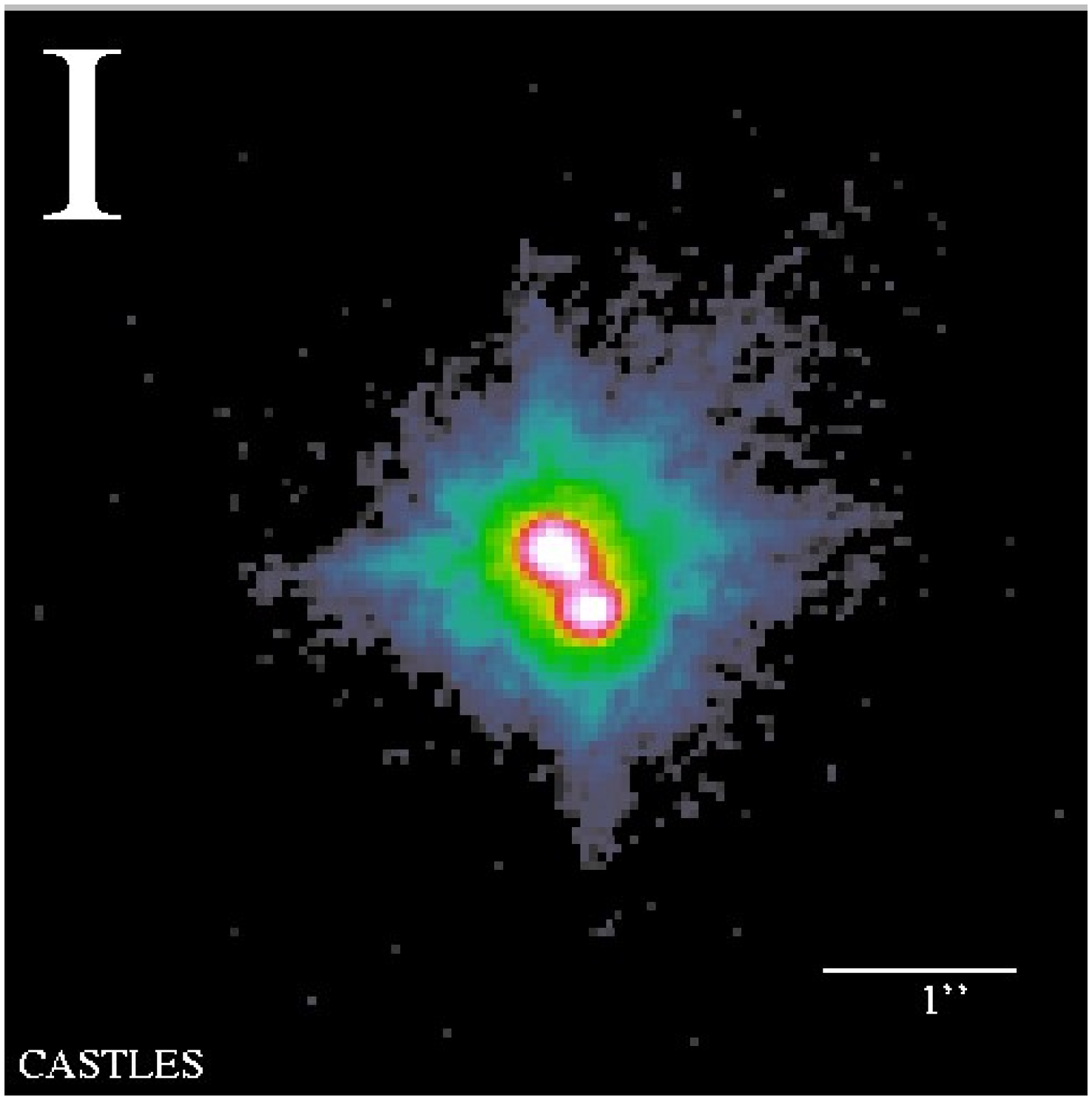}
\caption{Examples of small-separation lenses with different image
 multiplicities; double ({\it left}), quadruple ({\it center}), and
 naked cusp ({\it right}). These Figures are taken from
 \citet{kochanek04b}. 
\label{fig:lat_ex}}
\end{center}
\end{figure}

In all previous analytic work on the statistics of large-separation
lensed quasars, the lens objects were assumed to be spherical. However,
in the CDM model dark halos are not spherical at all but triaxial
(see Chapter \ref{chap:tri}).  It is already known that triaxiality has
a significant effect on the statistics of lensed arcs, from both
analytic (see Chapter \ref{chap:arc}) and numerical
\citep{meneghetti03a,dalal04a} points of view.  In the statistics of
normal lensed quasars, triaxiality (or ellipticity) has been thought to
mainly affect the image multiplicities, with only small changes to the 
total lensing probability \citep*{kochanek96,keeton97,evans02,chae03,
huterer04b}. However, that conclusion is based on nearly-singular
isothermal lens models, and the situation may be quite different for the
less concentrated mass distributions of the massive halos that create
large-separation lenses.  Moreover, only triaxial modeling allows us to
study image multiplicities. For the simple mass distribution (i.e., not
merging two galaxies, etc.), there are mainly three types of image
configurations; double, quadruple, and naked cusp. From the observations
of small-separation lensed quasars,it is found that the fraction of
double and quadruple lenses is 2:1, and naked cusp lenses are quite rare
(see Figure \ref{fig:lat_ex}). Since image multiplicities depend both on
the shape and central concentration of lens objects, and since they are
measured easily from observations, they can be a new simple test of the
CDM paradigm.   

\section{Small versus Large Separation Lenses}
\markboth{CHAPTER \thechapter.
{\MakeUppercase\mychapheadname}}{\thesection.
\MakeUppercase{Small versus Large Separation Lenses}}

Before we study large-separation lensed quasars, we review the
qualitative differences between
 small- and large-separation lenses from the
theoretical point of view, and see why we concentrate on
large-separation lenses.

\begin{figure}[tb]
\begin{center}
 \includegraphics[width=0.6\hsize]{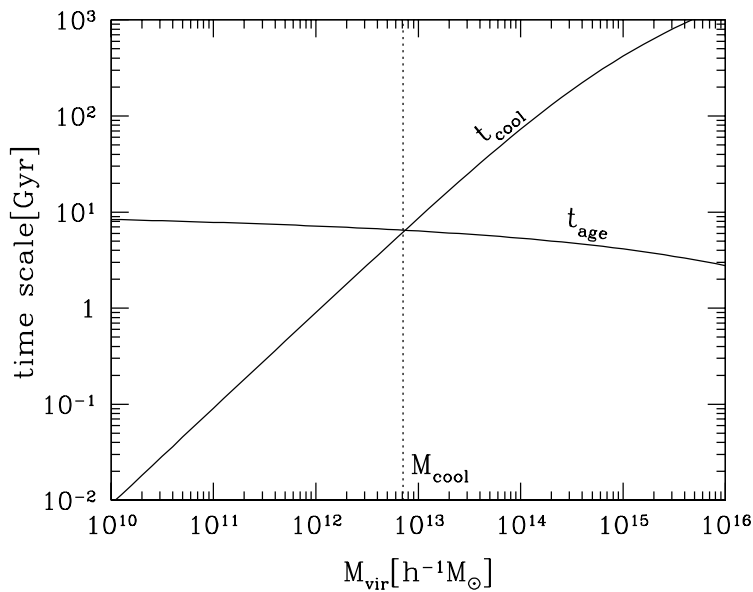}
\caption{Comparison of two timescales, $t_{\rm cool}$ (eq.
 [\ref{lat_tcol}]) and $t_{\rm age}$ (eq. [\ref{lat_tage}]), as a
 function of the virial mass of dark halos $M_{\rm vir}$. Two timescales
 become equal at $M_{\rm vir}=M_{\rm cool}\sim 10^{13}h^{-1}M_\odot$.
\label{fig:lat_time}}
\end{center}
\end{figure}

First, we see the simple picture of galaxy formation inside dark halos
\citep[e.g.,][]{rees77,blumenthal86,kochanek01}. For the gas to collapse
and form stars, it must release its internal energy; thus the galaxy
formation is governed by the radiative cooling timescale. From the
cooling function $\Lambda(T)$ for a plasma
\citep[e.g.,][]{sutherland93}, the timescale in the concordance model 
can be approximated as \citep{peacock99}
\begin{eqnarray}
 t_{\rm cool}&=&\frac{3kT}{2\Lambda(T)n_b}\nonumber\\
&\sim& 500 h^{-2}\left(T_8^{-\frac{1}{2}}+0.5T_8^{-\frac{3}{2}}\right)^{-1}[{\rm Gyr}],
\label{lat_tcol}
\end{eqnarray}
here we assumed the self-similar collapse. The first term
($\Lambda(T)\propto T^{-1/2}$) represents bremsstrahlung, and the second
term ($\Lambda(T)\propto T^{-3/2}$) denotes (effectively) atomic line
cooling. If we assume that the gas temperature is heated to the virial
temperature of dark halos via shocks, $T_8\equiv T/(10^8K)$ reduces to
\begin{equation}
 T_8\sim 6\times 10^{-3}\left(\frac{M_{\rm vir}}{10^{12}h^{-1}M_\odot}\right)^{2/3}.
\end{equation}
Another important timescale is the age of dark halos (i.e., the age of
the universe $t_0$ minus the formation epoch of dark halos).
This can be easily estimated from the formation epoch distribution
$dp/dt$ \citep[see, e.g.,][]{lacey93,kitayama96}:
\begin{equation}
 t_{\rm age}=t_0-\int t\frac{dp}{dt}dt.
\label{lat_tage}
\end{equation}
Figure \ref{fig:lat_time} shows these timescales, $t_{\rm cool}$ and
$t_{\rm age}$, as a function of the virial mass of dark halos $M_{\rm
vir}$. It is found that $t_{\rm cool}$ rapidly increases as the virial
mass increases.  The age $t_{\rm age}$ does not depend on $M_{\rm vir}$
so much, but slightly deceases as the virial mass increases. Thus we can
define the characteristic mass $M_{\rm cool}\sim 10^{13}h^{-1}M_\odot$,
by equating these two time scales, $t_{\rm cool}=t_{\rm age}$: At $M_{\rm
vir}<M_{\rm cool}$, baryon cooling takes place efficiently, and galaxies
are formed successfully. On the other hand, at $M_{\rm vir}>M_{\rm
cool}$ the cooling timescale is so large that baryons are expected to
remain mostly in the form of hot gas. This picture naturally explains
the qualitative differences between galaxies and clusters. Since the
baryon cooling modifies the mass distributions of dark halos
significantly and makes them more centrally concentrated (so that it is
well approximated by the singular isothermal mass distribution,
$\rho(r)\propto r^{-2}$) than original NFW density profile, it does have
a great impact also on strong gravitational lensing. 

\begin{figure}[tb]
\begin{center}
 \includegraphics[width=0.4\hsize]{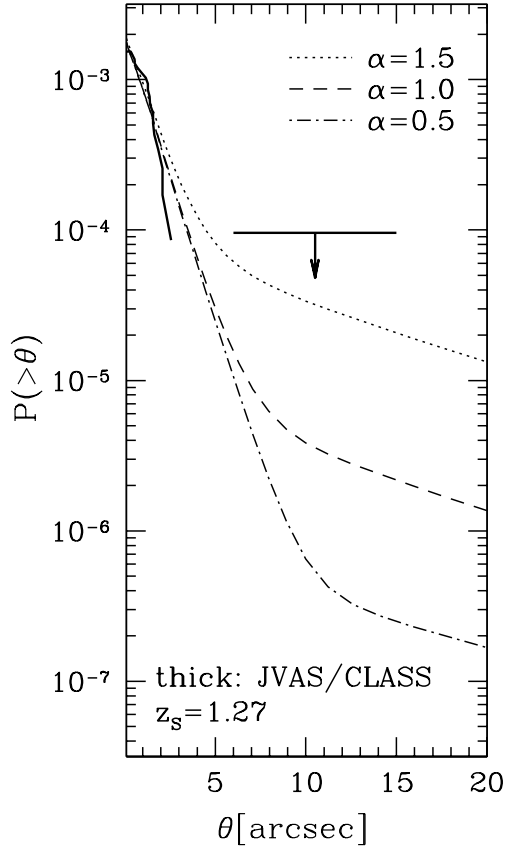}
\caption{Cumulative distributions of image separation $\theta$, on the
 basis of the model described by \citet{oguri02c}. In this model, we
 adopt the spherical lens objects. For more details of the model, please
 see the text. All the probabilities are normalized by the CLASS
 observation at small image separations \citep{myers03,browne03}, which
 is shown by the solid line. The lack of lenses from the explicit search at
 $6''<\theta<15''$ in CLASS \citep{phillips01b} is also shown by the
 horizontal line with an arrow.
 \label{fig:lat_prob}} 
\end{center}
\end{figure}

Bearing the picture of galaxy formation in mind, one can make a rough
prediction for the image separation distributions of strongly lensed
quasars \citep*{keeton98a,porciani00,kochanek01,keeton01a,sarbu01,li02,
li03,oguri02c,ma03}. In Figure \ref{fig:lat_prob}, we plot an example of
image separation distributions, calculated from the model described by
\citet{oguri02c}.  The model comprises the galaxy formation probability
$p_{\rm g}(M)$ (which has a sharp cutoff at $M\sim M_{\rm cool}$) and
the ratio of circular velocities of galaxies to virial velocities of
dark halos $\gamma\equiv v_{\rm c}/v_{\rm vir}$.\footnote{This parameter
is similar to the cooled baryon fraction in another papers
\citep[e.g.,][]{kochanek01,keeton01b}, because $\gamma$ becomes larger
when more and more baryons are cooled and turn into stars inside the
halos.} In this plot, we choose $\gamma=2$ and adjust $p_{\rm g}(M)$ so
as to reproduce the velocity function of galaxies. We simply assume the
spherical halos, and adopt the generalized NFW profile with
$\alpha=0.5$, $1$, and $1.5$ (see \S \ref{sec:lat_sph}). We approximate
the mass distribution of {\it galaxies} by the singular isothermal
sphere:
\begin{equation}
 \rho(r)=\frac{\sigma^2}{2\pi G r^2},
\end{equation}
where one-dimensional velocity dispersion $\sigma$ is related with the
circular velocity as $v_{\rm c}=\sqrt{2}\sigma$. All the probabilities
are normalized by the CLASS observation at small image separations
\citep{myers03,browne03}.  

It is clear from this Figure that the distributions of image separation
$\theta$ have a ``break'' at around $\theta\sim 7''$ which roughly
corresponds to $M_{\rm cool}$. The probabilities for large-separation
lenses ($\theta\gtrsim 7''$) are much smaller than those for
small-separation lenses, and depend strongly on the inner slope $\alpha$.
Therefore, statistics of large-separation lenses can be a useful tool to
probe directly the mass distributions of dark halos, unlike
small-separation lenses for which mass distributions of lens objects
were significantly modified due to baryon cooling. This argument also
justifies lensed arcs in clusters as a direct test of the CDM model.

We note several caveats in calculating image separation distribution
from this simple picture of galaxy formation. First of all, the relation
between visible galaxies and the hosting dark halos is still uncertain;
it has been often assumed that galaxies velocity dispersions are equal
to the virial velocities of dark halos (i.e., $\gamma=1$ in the above
model), but this is not trivial at all. Actually, some theoretical
models, semi-analytic models, and observations do favor $\gamma>1$
\citep[][ and references therein]{oguri02c}, and this larger $\gamma$
significantly increases the (relative) probabilities of small-separation
lenses. Next, the model of image separation distributions based on dark
halos can account for central galaxies (galaxies that lie at the center
of dark halos) only; but in reality satellite galaxies (i.e., galaxies
in clusters), which are associated with substructures in dark halos,
act as lens objects, too. Apparently, these problems come from our poor
understanding of galaxy formation, and only affect the relative
probabilities of small- and large-separation lenses. Therefore, while
full distribution of image separations offer us invaluable information on
galaxy formation, theoretical predictions for large separation lenses are 
robust because abundances and mass distributions of dark halos in the
CDM model are accurately known from $N$-body simulations.
  
\section{Modeling Lens Probabilities: Spherical Case}
\label{sec:lat_sph}
\markboth{CHAPTER \thechapter.
{\MakeUppercase\mychapheadname}}{\thesection.
\MakeUppercase{Modeling Lens Probabilities: Spherical Case}}

First we briefly review the modeling of lens probabilities with the
spherical dark halos which can be done much more easily than the case of
the triaxial lens model. 

\subsection{Lens Probabilities}
 
Let the physical image position in the lens plane and physical source
position in the source plane as $\xi$ and $\eta$, respectively.
Consider the probability that a quasar at $z_{\rm S}$ with luminosity
$L$ is strongly lensed. The probability of lensing with image separation
larger than $\theta$ is given by 
\begin{equation}
P(>\!\theta; z_{\rm S}, L) =
\int_{0}^{z_{\rm S}} dz_{\rm L}\,(1+z_{\rm L})^3\,\frac{c\,dt}{dz_{\rm L}}
\int_{M(\theta)}^{\infty} dM\, \frac{dn}{dM}\
\sigma_{\rm lens}\ B(z_{\rm S}, L)
\label{lat_cpd_bias}
\end{equation}
where 
\begin{equation}
 \sigma_{\rm lens}=\pi\eta_{\rm r}^2\frac{D_{\rm OL}^2}{D_{\rm OS}^2}
\end{equation}
is the cross section for lensing, with $\eta_{\rm r}$ being the physical 
radius of the radial caustic in the source plane.  The lower limit of
the mass integral is the mass $M(\theta)$ that corresponds to the
image separation $\theta$; this can be computed once the density profile
of the lens object is specified. The mass function of dark halos,
$dn/dM$, is given in \S \ref{sec:cosmo_mf}. The magnification bias
$B(z_{\rm S}, L)$ is \citep{turner80,turner84}  
\begin{equation}
B(z_{\rm S}, L) = \frac{2}{\eta_{\rm r}^2\,\phi_L(z_{\rm S}, L)}
\int_{0}^{\eta_{\rm r}} d\eta\ \eta\ \phi_L(z_{\rm S}, L/\mu(\eta))\,
\frac{1}{\mu(\eta)},
\label{lat_biasfactor}
\end{equation}
where $\phi_L(z_{\rm S}, L)$ is the luminosity function of source quasars. 
Note that the magnification factor $\mu(\eta)$ may be interpreted as the
total magnification or the magnification of the brighter or fainter
image, depending on the observational selection criteria
\citep{sasaki93,cen94}. We adopt the magnification of the fainter image,
because we concentrate on the large-separation lenses for which the
images are completely deblended. Finally, differential distribution can
be obtained simply by differentiating equation (\ref{lat_cpd_bias}):
\begin{equation}
\frac{dP}{d\theta}(\theta; z_{\rm S}, L) =
\int_{0}^{z_{\rm S}} dz_{\rm L}\,(1+z_{\rm L})^3\,\frac{c\,dt}{dz_{\rm L}}
\left[\frac{dM}{d\theta} \frac{dn}{dM}\
\sigma_{\rm lens}\ B(z_{\rm S}, L)\right]_{M(\theta)}.
\label{lat_dpd_bias}
\end{equation}

\subsection{Generalized NFW Profile}
\label{sec:lat_gnfw}

As discussed, the lensing probability distribution at large-separation
reflects the properties of dark halos, rather than galaxies. For the
statistics calculation, the debate over the inner slope of the density
profile seen in $N$-body simulations leads us to consider the
generalized version \citep{zhao96,jing00a} of the NFW density profile:
\begin{equation}
 \rho(r)=\frac{\rho_{\rm crit}(z)\delta_{\rm c}(z)}
{\left(r/r_{\rm s}\right)^\alpha\left(1+r/r_{\rm s}\right)^{3-\alpha}}.
\end{equation}
While the correct value of $\alpha$ is still unclear, the existence of
cusps with $1\lesssim\alpha\lesssim1.5$ has been established in recent
$N$-body simulations \citep{navarro96,navarro97,moore99b,ghigna00,jing00a,
klypin01,fukushige97,fukushige01,fukushige03,power03,fukushige04,hayashi04}.
The case $\alpha=1$ corresponds to the original NFW profile, while the
case $\alpha=1.5$ resembles the profile proposed by \citet{moore99b}.
The scale radius $r_{\rm s}$ is related to the concentration parameter
as  
\begin{equation}
 c_{\rm vir}=\frac{r_{\rm vir}}{r_{\rm s}}.
\end{equation}
Then the characteristic density $\delta_{\rm c}(z)$ is given in terms of
the concentration parameter: 
\begin{equation}
\delta_{\rm c}(z)=\frac{\Delta_{\rm vir}(z)\Omega(z)}{3}
\frac{c_{\rm vir}^3}{m(c_{\rm vir})},
\end{equation}
where $m(c_{\rm vir})$ is given by equation (\ref{tri_mce}).
The {\it mean\/} overdensity $\Delta_{\rm vir}(z)$ can be computed using
the nonlinear spherical collapse model (see \S \ref{sec:cosmo_mf}).

We define 
\begin{eqnarray}
\tilde{\xi}&\equiv&\xi/r_{\rm s}\\
\tilde{\eta}&\equiv&\eta D_{\rm OL}/r_{\rm s}D_{\rm OS}.
\end{eqnarray}
Then the lensing deflection angle $\alpha(\tilde{\xi})$ is related to the
dark halo profile as follows: 
\begin{equation}
\alpha(\tilde{\xi})=\frac{b_{\rm NFW}}{\tilde{\xi}}\int_0^\infty dz
\int_0^{\tilde{\xi}} dx \frac{x}{\left(\sqrt{x^2+z^2}\right)^\alpha
\left(1+\sqrt{x^2+z^2}\right)^{3-\alpha}}.
\end{equation}
The lensing strength parameter $b_{\rm NFW}$ is defined as
\begin{equation}
b_{\rm NFW}\equiv\frac{4\rho_{\rm crit}(z)\delta_{\rm c}(z)r_{\rm s}}{\Sigma_{\rm crit}}.
\label{lat_bnfw}
\end{equation}
For sources inside the caustic
($\eta<\eta_{\rm r}$), the lens equation has three solutions
$\tilde{\xi}_1>\tilde{\xi}_2>\tilde{\xi}_3$, where image \#1 is on
the same side of the lens as the source and images \#2 and \#3 are
on the opposite side.\footnote{The third image is usually predicted
to be very faint, so in practice just two images are actually
observed.} The lens image separation is then
\begin{equation}
 \theta=\frac{r_{\rm s}(\tilde{\xi}_1+\tilde{\xi}_2)}{D_{\rm OL}}\simeq\frac{2r_{\rm s}\tilde{\xi}_{\rm t}}{D_{\rm OL}},
\label{lat__sep}
\end{equation}
where $\tilde{\xi}_{\rm t}$ is a radius of the tangential critical
curve \citep{hinshaw87,oguri02a}. The magnification of the fainter
image may be approximated by \citep{oguri02a}
\begin{equation}
 \mu_{\rm faint}(\eta)\simeq\frac{\tilde{\xi}_{\rm t}}{\tilde{\eta}(1-\alpha'(\tilde{\xi}_{\rm t}))}.
\label{lat_approxfaint}
\end{equation}
These approximations are sufficiently accurate over the range of
interest here \citep[see][]{oguri02a}. Although it is often adopted in
searching for lensed quasars that the flux ratios should be smaller than,
e.g., $10:1$, this condition does not affect our theoretical predictions
because the flux ratios of strong lensing by NFW halos are typically
much smaller than $10:1$ \citep{oguri02a,rusin02}.

The concentration parameter $c_{\rm vir}$ depends on a halo's mass
and redshift. Moreover, even halos with the same mass and redshift
show significant scatter in the concentration which reflects the
difference in formation epoch \citep{wechsler02}, and which is well
described by a log-normal distribution.  For the median of this
distribution, we adopt the mass and redshift dependence reported by
\citet{bullock01} as a canonical model:
\begin{equation}
 c_{\rm Bullock}(M, z)=\frac{10}{1+z}\left(\frac{M}{M_*(0)}\right)^{-0.13},
\label{lat_conmed_bullock}
\end{equation}
where $M_*(z)$ is the mass collapsing at redshift $z$ (defined by
$\sigma_M(z) = \delta_{\rm c} \equiv 1.68$). To study uncertainties
related to the concentration distribution we also consider other mass
and redshift dependences, e.g.,
\begin{equation}
 c_{\rm CHM}(M, z)=10.3(1+z)^{-0.3}\left(\frac{M}{M_*(z)}\right)^{-0.24(1+z)^{-0.3}},
\label{lat_conmed_chm}
\end{equation}
from \citet*{cooray00}, and
\begin{equation}
 c_{\rm JS}(M, z)=2.44\sqrt{\frac{\Delta_{\rm vir}(z_c)}{\Delta_{\rm vir}(z)}}\left(\frac{1+z_c}{1+z}\right)^{3/2},
\label{lat_conmed_js}
\end{equation}
from \citet{jing02}, with $z_c$ being the collapse redshift of the
halo of mass $M_{\rm vir}$. Note that these relations were derived
under the assumption of $\alpha=1$. We can extend them to
$\alpha\neq 1$ by multiplying the concentration by a factor $2-\alpha$
\citep{keeton01a,jing02}, because it has turned out that dark halos can
be fitted reasonably well for $\alpha\neq 1$ only if we apply this
translation.  

The statistics of large-separation lenses are highly sensitive to the 
degree of scatter in the concentration
\citep{keeton01a,wyithe01,kuhlen04}. \citet[][ see also
\citealp{wechsler02}]{bullock01} found $\sigma_c \sim 0.32$ in their
simulations. \citet{jing00b} found a smaller scatter $\sigma_c\sim 0.18$
among well relaxed halos, but \citet{jing02} found $\sigma_c\sim 0.3$ if
all halos are considered. From these results, it is reasonable to use
$\sigma_c=0.3$, which is adopted throughout the thesis. 

\subsection{Examples}

\begin{figure}[tb]
\begin{center}
 \includegraphics[width=0.8\hsize]{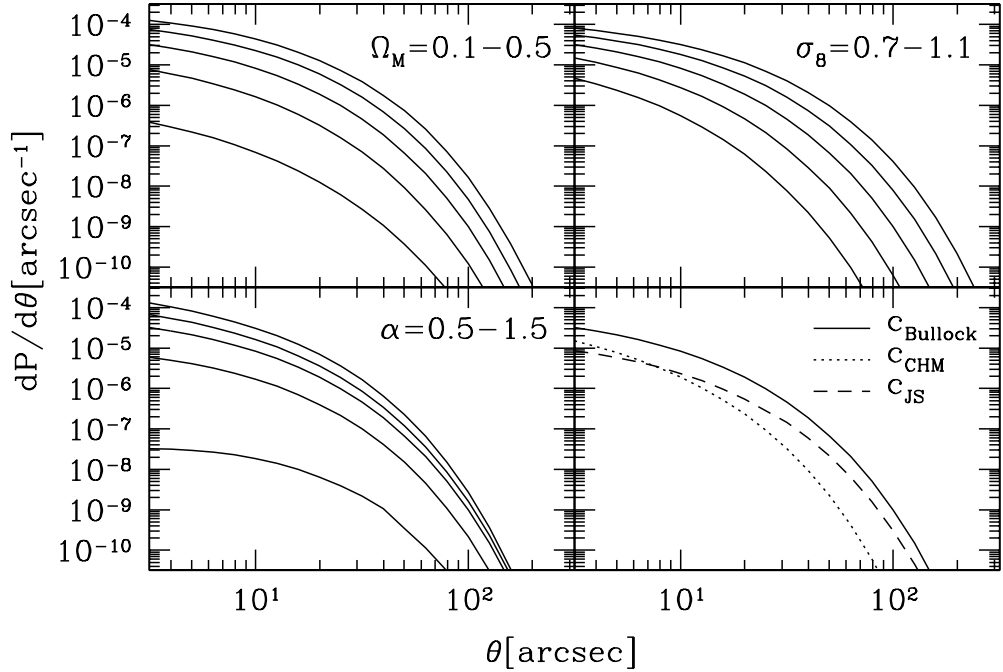}
\caption{Differential distributions of image separation distributions
 (eq. [\ref{lat_dpd_bias}]). We fix the source at $z=2$ and adopt a
 power-law luminosity function $\phi_L\propto L^{-2.5}$. We use
 $\alpha=1$ for the fiducial value. Dependences of several parameters
 are shown; $\Omega_M$ ({\it upper left}), $\sigma_8$ ({\it upper
 right}), and $\alpha$ ({\it lower left}). It is also demonstrated how
 distributions change by adopting different models of concentration
 parameters ({\it lower right}). 
\label{fig:lat_tdis}}
\end{center}
\end{figure}

Here we show examples of image separation distributions for
large-separation lenses, computed from the spherical model described
above. In Figure \ref{fig:lat_tdis}, we show image separation
distributions and their parameter dependences. We assumed a source at
$z=2$ and a power-law luminosity function $\phi_L\propto L^{-2.5}$, just
for simplicity. As a fiducial model, we consider the spherical halo with
$\alpha=1$. Fiducial cosmological parameters are those in Table
\ref{table:conc_para}. In each panel, we distribute the parameter we
concentrate on, while the other parameters are fixed to fiducial values.

It is found that distributions are sensitive to the inner slope $\alpha$
as well as $\Omega_M$ and $\sigma_8$. The parameter $\alpha$ determines
the cross sections of lensing, while $\Omega_M$ and $\sigma_8$ changes
the abundance of lensing clusters. Probabilities at very large $\theta$
are particularly sensitive to $\sigma_8$. We also use three different
models of concentration parameters (eqs. [\ref{lat_conmed_bullock}],
[\ref{lat_conmed_chm}], and [\ref{lat_conmed_js}]), and show that the
uncertainties of concentration parameters result in large differences of
lensing probabilities. These are why there have been many attempts to
probe the mass distributions and/or abundances of clusters using
large-separation lensed quasars. 

\section{Modeling Lens Probabilities: Triaxial Case}\label{sec:lat_theory}
\markboth{CHAPTER \thechapter.
{\MakeUppercase\mychapheadname}}{\thesection.
\MakeUppercase{Modeling Lens Probabilities: Triaxial Case}}

\subsection{Cross sections and image separation
  distributions}\label{sec:lat_cross} 

\begin{figure}[tb]
\begin{center}
 \includegraphics[width=0.8\hsize]{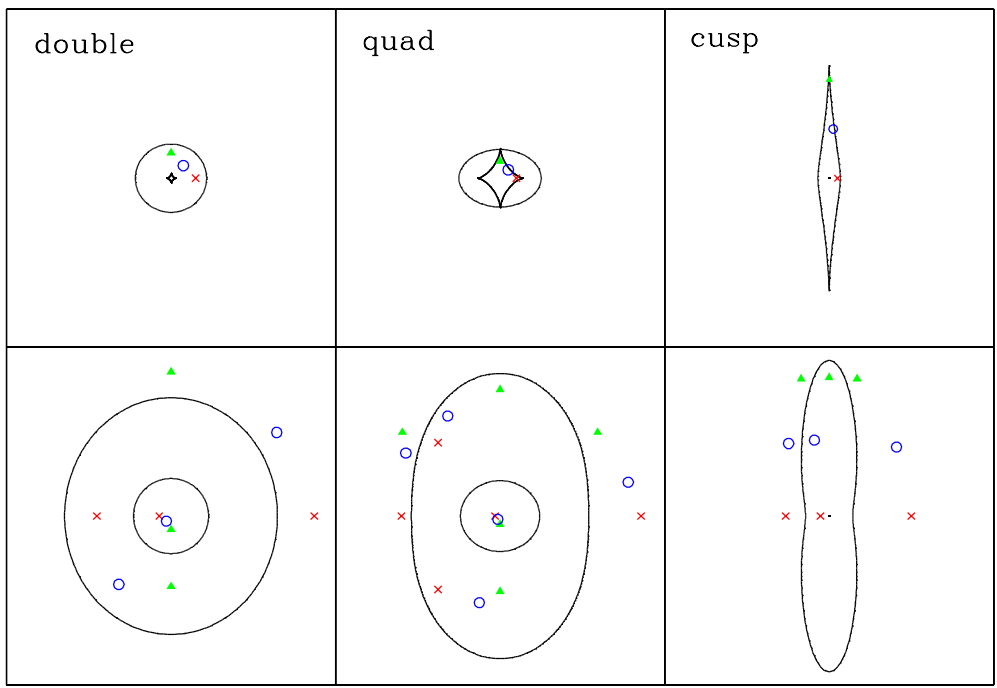}
\caption{Sample image configurations. The top panels show the source
planes, and the bottom panels show the corresponding image planes.
The solid lines indicate the caustics and critical curves.  We show
three sources (denoted by triangles, circles, crosses), and their
corresponding images.  From left to right, the lenses are doubles,
quadruples, and cusps. Specific values of $(b_{\rm TNFW}, q)$ for each
 example are $(2,0.95)$, $(2,0.75)$, and $(0.6,0.25)$ for doubles,
 quadruples, and cusps, respectively. Doubles and cusps are
 distinguished by the image parities: doubles have one positive-parity
 image and one negative-parity image, plus a central double-negative
 image that is usually too faint to be observed; while cusps have two
 positive-parity images and one negative parity image, all of comparable
 brightnesses. 
\label{fig:lat_multi}}
\end{center}
\end{figure}

\begin{figure}[p]
\begin{center}
 \includegraphics[width=0.85\hsize]{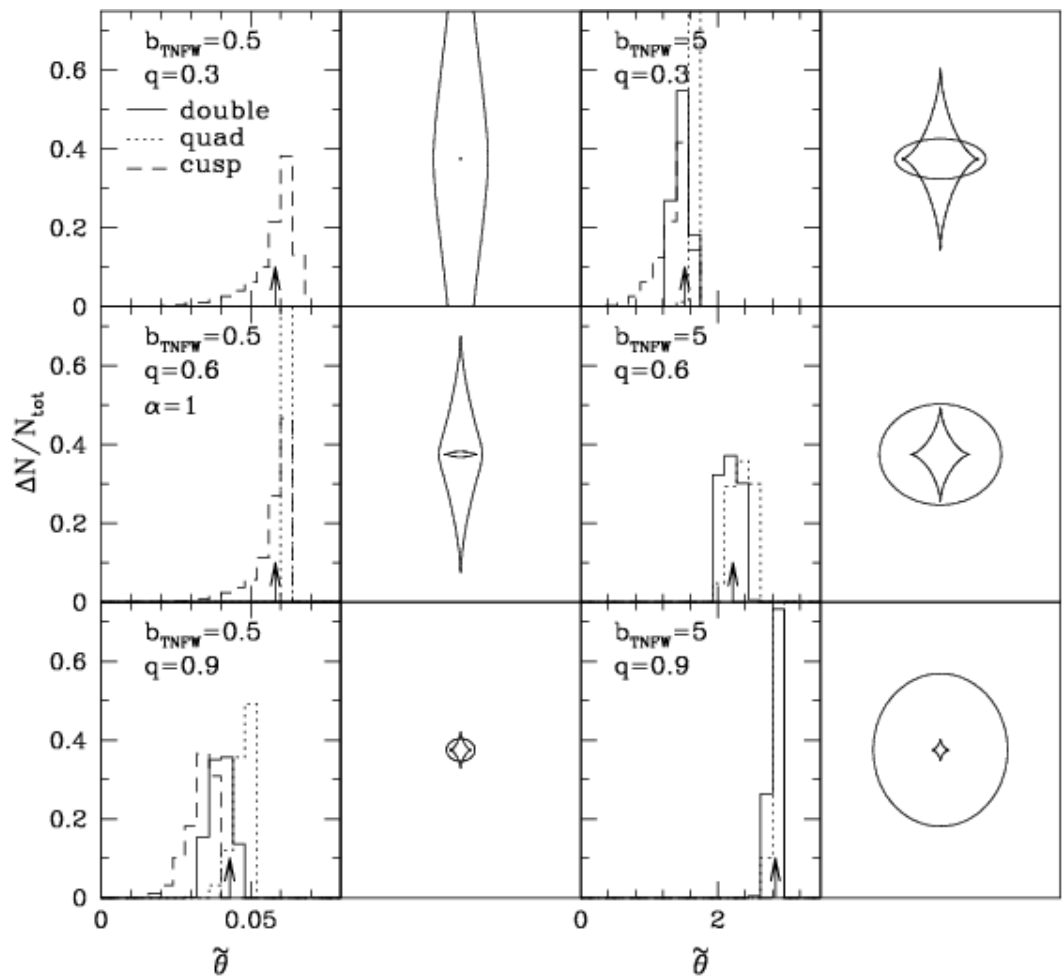}
\caption{Image separation distributions for sample lenses with
$\alpha=1$.  Arrows indicate the average separations.  The
corresponding caustics are shown for reference.  For each
$b_{\rm TNFW}$, the caustics are all plotted on the same scale.
\label{fig:lat_dist_theta_a100}}
\end{center}
\end{figure}

\begin{figure}[p]
\begin{center}
 \includegraphics[width=0.85\hsize]{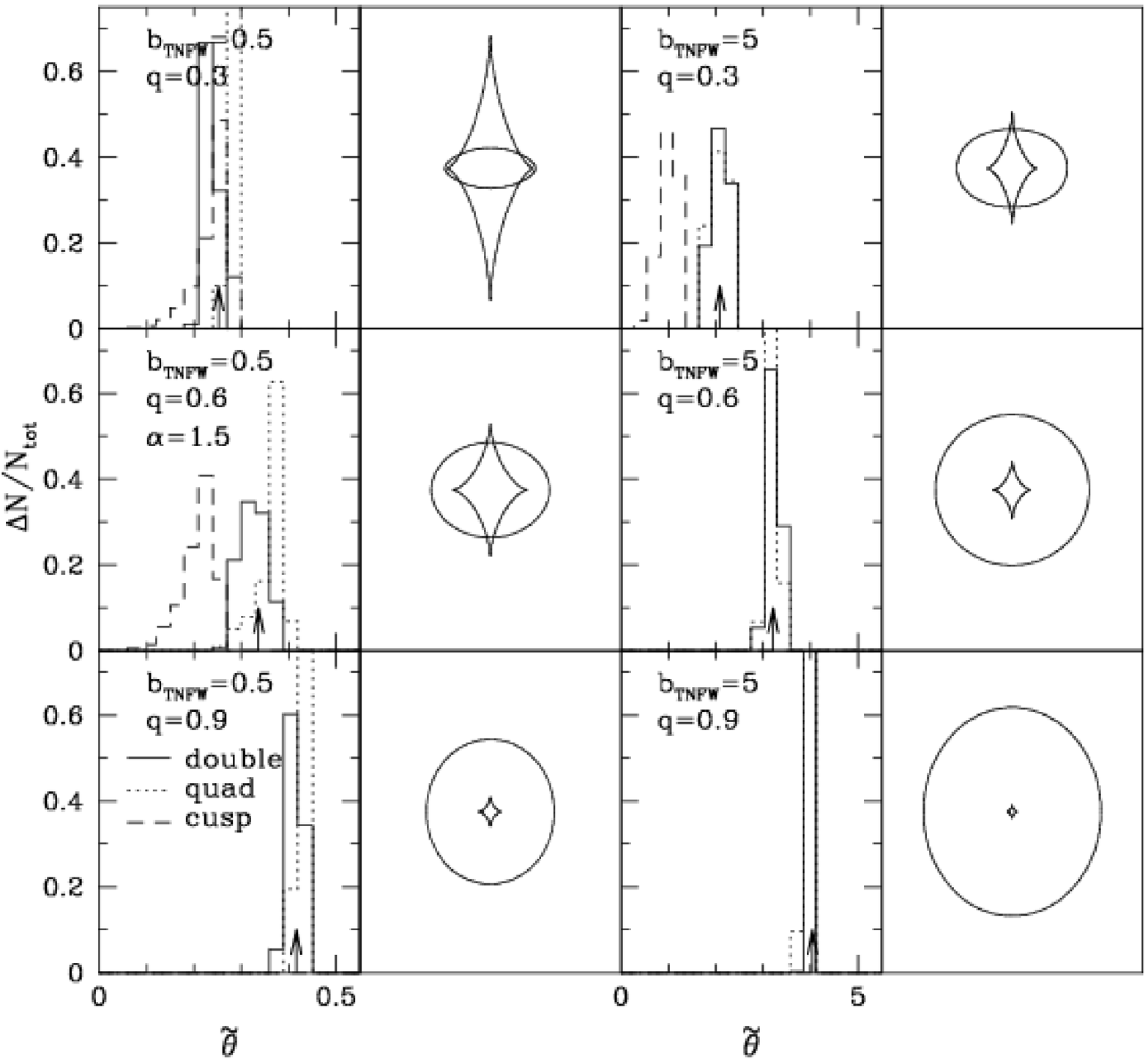}
\caption{Similar to Figure~\ref{fig:lat_dist_theta_a100}, but for 
 $\alpha=1.5$.
\label{fig:lat_dist_theta_a150}}
\end{center}
\end{figure}

\begin{figure}[p]
\begin{center}
 \includegraphics[width=0.85\hsize]{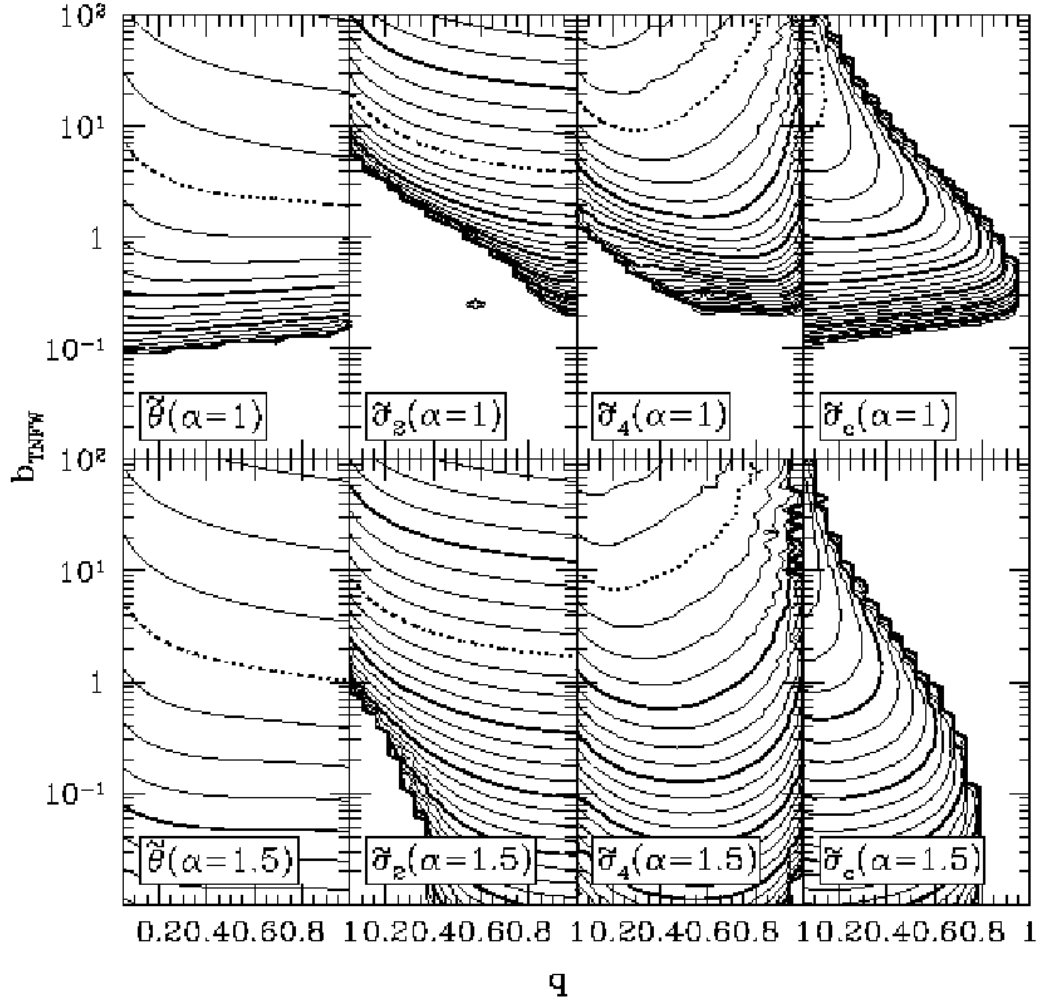}
\caption{Dimensionless image separations $\tilde{\theta}$ and cross
 sections $\tilde{\sigma}$ in the $b_{\rm TNFW}$-$q$ plane. The second,
 third, and fourth panels from left denote cross sections for double
 lenses, quadruple lenses, and naked cusp lenses, respectively. Dotted
 lines are contours for $=1$, and thick (thin) solid lines are drawn at
 $10^{2n}$ ($10^{0.5n}$), where $n$ is integer.  
 \label{fig:cross}}
\end{center}
\end{figure}

We compute lensing cross sections using Monte Carlo methods. Working in
dimensionless coordinates $X \equiv x'/L_0$ and $Y \equiv y'/L_0$, we
pick random sources and use the {\it gravlens} software by
\citet{keeton01d} to solve the lens equation.
Figure~\ref{fig:lat_multi} shows examples of the three different kinds
of image configurations: double, quadruple, and naked cusp
lenses.\footnote{We use the terms ``double'' and ``quadruple'' because
the third and fifth images are usually too faint to be observed,
although with the density profiles we use here they are probably not as
faint as for nearly-isothermal lenses \citep[see][]{rusin02}.}  We count
the number of sources that produce lenses of different image
multiplicities to determine the dimensionless cross sections
$\tilde{\sigma}_2$, $\tilde{\sigma}_4$, and $\tilde{\sigma}_{\rm c}$ for
doubles, quadruples, and cusps, respectively.  For each set of images,
we define the dimensionless image separation $\tilde{\theta}$ to be the
maximum separation between any pair of images; this is a convenient
definition that depends only on observable quantities and is well
defined for all image configurations (no matter how many images there
are).  We bin the sources by the image separations they produce to
derive image separation distributions, as shown in
Figures~\ref{fig:lat_dist_theta_a100} and \ref{fig:lat_dist_theta_a150}.
For a given halo there is a range of separations, but it tends to be
fairly narrow ($\lesssim$20\%); the main exception is for cusp
configurations, which show a tail to small separations that corresponds
to sources near the cusp in the caustic. We also plot dimensionless
image separations and cross sections in the $b_{\rm TNFW}$-$q$ plane in
Figure \ref{fig:cross}. Hereafter we use the cross sections and image
separations tabulated in the range $0.1\leq q\leq 1$ and $10^{-2}\leq
b_{\rm TNFW} \leq 10^{2}$.  

If we not only count the sources but also weight them appropriately, we
can compute the magnification bias. Specifically, if the sources have a
simple power law luminosity function $\phi_L(L)\propto L^{-\beta}$ then
the ``biased cross section'' can be written as
\begin{eqnarray}
 B\tilde{\sigma} &=& \int dX dY \frac{\phi_L(L/\mu)/\mu}{\phi_L(L)}\nonumber\\
 &=& \int dX dY \mu^{\beta-1},
\end{eqnarray}
where the integral is over the multiply-imaged region of the source
plane.  We can compute the biased cross sections for doubles,
quadruples, and cusps similarly.  Each source is to be weighted by
$\mu^{\beta-1}$, where we take $\mu$ to be the magnification of the
second brightest image to reflect the popular method of searching for
large-separation lenses in observational data such as the SDSS (see
Chapter \ref{chap:sdss}).  

An important qualitative result is already apparent from
Figure~\ref{fig:lat_multi}.  CDM-type dark matter halos are very
sensitive to departures from spherical symmetry, in the sense that even
small projected ellipticities lead to large tangential caustics and
hence large quadruple cross sections.  When the ellipticity is large,
the tangential caustic is much larger than the radial caustic and nearly
all the images correspond to cusp configurations.  This situation is
notably different from what happens in lensing by galaxies that have
concentrated, roughly isothermal mass distributions.  In that case, the
ellipticity must approach unity before cusp configurations become common 
\citep[see][]{keeton97,rusin01b}.  Such large ellipticities are uncommon,
and cusp configurations are correspondingly rare among observed
galaxy-scale lenses: among $\sim$80 known lenses there is only one
candidate \citep[APM~08279+5255;][]{lewis02}.  The incidence of cusp
configurations therefore appears to be a significant distinction between
normal and large-separation lenses.

\subsection{Lensing probabilities}

The probability that a source at redshift $z_{\rm S}$ is lensed into a
system with image separation $\theta$ is computed by summing the biased
cross section over an appropriate population of lens halos:
\begin{eqnarray}
 \frac{dP}{d\theta}(\theta,z_{\rm S})&=&
   \int dz_{\rm L}\frac{c\,dt}{dz_{\rm L}}(1+z_{\rm L})^3\int d(a/c)
   \int dc_e \int d(a/b) \int d\theta \int d\phi \nonumber\\
 &&\times \left[ p(a/c)\,p(c_e)\,p(a/b|a/c)\,p(\theta)\,p(\phi)\,B\sigma
   \,\frac{dn}{dM} \right]_{M(\theta)}.
   \label{lat_prob}
\end{eqnarray}
The first integral is over the volume between the observer and the
source.  The next three integrals are over the structural parameters of
the lens halos, while the last two integrals cover the different
orientations.  The mass function of dark halos is represented by
$dn/dM$.  Finally, $M(\theta)$ is the mass of a halo that produces image
separation $\theta$ (for given redshift and other parameters), which is
given by the solution of  
\begin{equation}
 \theta = R_0\,q_x\,\tilde{\theta}(b_{\rm TNFW},q).
\end{equation}
The square brackets in equation (\ref{lat_prob}) indicate that the
integrand is to be evaluated only for parameter sets that produce
the desired image separation.  Equation (\ref{lat_prob}) gives the
total lensing probability, but we can simply replace the total
biased cross section $B \sigma$ with $B_2 \sigma_2$, $B_4 \sigma_4$,
or $B_c \sigma_c$ to compute the probability for doubles, quadruples,
or cusps.

\section{Lensing Probabilities and Image Multiplicities in the
Triaxial Halo Model}\label{sec:lat_prob}
\markboth{CHAPTER \thechapter.
{\MakeUppercase\mychapheadname}}{\thesection.
\MakeUppercase{Lensing Probabilities and Image Multiplicities}}

\subsection{Dependence of the triaxiality}

\begin{figure}
\begin{center}
 \includegraphics[width=0.7\hsize]{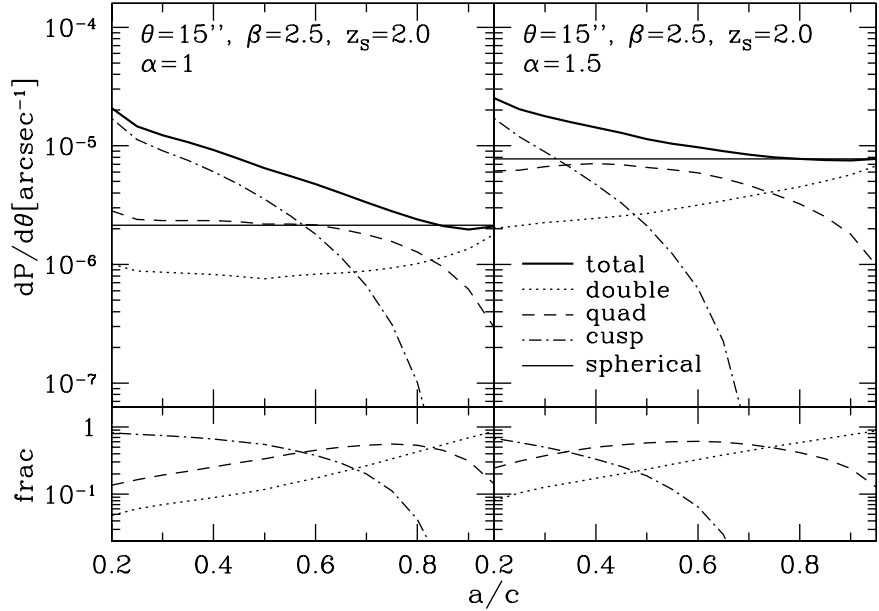}
\caption{Lensing probabilities and image multiplicities as a function
of $a/c$.  The left panels are for inner slope $\alpha=1$, and the
right panels for $\alpha=1.5$.  We adopt $\theta=15''$, $\beta=2.5$,
and $z_{\rm S}=2.0$.  We show the total lensing probability (thick
solid line), as well as the probabilities for double (dotted),
quadruple (dashed), and cusp (dash-dotted) lenses.  For comparison,
the lensing probability of spherical halos is shown by the thin solid
line.
\label{fig:lat_afix}}
\end{center}
\end{figure}

We begin by examining how the lensing probabilities and image
multiplicities vary when we change the degree of triaxiality. We remove
the integral over $a/c$ in equation (\ref{lat_prob}) to compute the
lensing probabilities at fixed triaxiality (we still integrate over the
intermediate axis ratio $a/b$ and over random orientations).  We can
then plot the probabilities as a function of $a/c$, as shown in
Figure~\ref{fig:lat_afix}.  In this example, we place the source at
redshift $z_{\rm S}=2.0$, and we use a source luminosity function with
slope $\beta=2.5$.  We compute the probabilities for an image separation
of $\theta=15''$, for concreteness.

For $a/c \to 1$ we recover the spherical case.  As $a/c$ decreases (the
triaxiality increases), at first the total lensing probability stays
roughly constant but the fraction of quadruples rises; this is similar
to the effects of ellipticity on isothermal lenses
\citep[see][]{keeton97,rusin01b}.  Then the probability for naked cusp
image configurations begins to rise dramatically, and they come to
dominate the total probability.  Interestingly, the sum of the
probabilities for quadruple and double lenses is roughly equal to the
probability for spherical halos for most values of $a/c$, at least for
this example with image separation $\theta=15''$ (also see
Figures~\ref{fig:lat_sep}--\ref{fig:lat_zs} below).  This suggests that
the enhancement in the total lensing probability is mainly driven by
naked cusp configurations.  The $\alpha=1$ and 1.5 cases both have these
qualitative features, and they differ only in the quantitative details.
Since the typical triaxiality in CDM simulations is $a/c \sim 0.5$
(see Figure \ref{fig:tri_pdf}), it appears that triaxiality can have a
significant effect on the statistics of large-separation lenses.

\subsection{Full results}

\begin{figure}
\begin{center}
 \includegraphics[width=0.7\hsize]{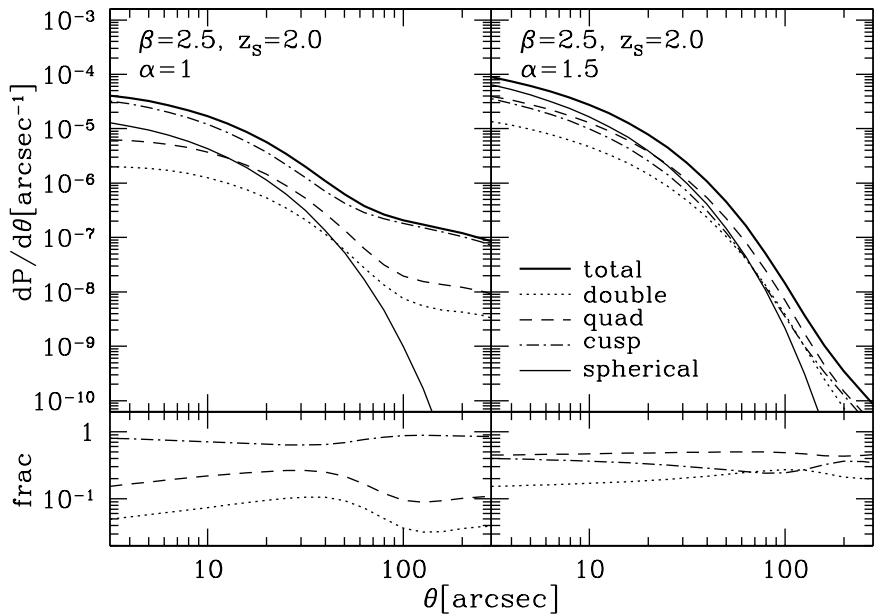}
\caption{Lensing probabilities and image multiplicities with triaxial
dark halos as a function of image separation $\theta$.  The source is
placed at $z_{\rm S}=2.0$, and slope of the source luminosity function
is fixed to $\beta=2.5$.
\label{fig:lat_sep}}
\end{center}
\end{figure}

To compute the full impact of triaxiality on lens statistics, we must
integrate over an appropriate triaxiality distribution (as in equation
\ref{lat_prob}).  Figure~\ref{fig:lat_sep} shows the resulting lensing
probabilities and image multiplicities as a function of the image
separation $\theta$.  Again, we place the source quasar at $z_{\rm
S}=2.0$, and fix the slope of the source luminosity function to
$\beta=2.5$.  The first important result is that the triaxial model
predicts larger lensing probabilities than the spherical model for all
image separations.  The enhancement is a factor of $\sim$4 for
$\alpha=1$, and a factor of $\sim$2 for $\alpha=1.5$, if the image
separation is not so large ($\theta\lesssim 30''$).  At larger
separations it seems that the effect of triaxiality is even more
significant, especially for $\alpha=1$; we will discuss this issue in
\S\ref{sec:lat_largesep}. 

There are several interesting results in the image multiplicities. The
$\alpha=1$ and $1.5$ cases have very different multiplicities: with
$\alpha=1$ the lensing probability is dominated by cusp configurations;
while with $\alpha=1.5$ quadruple lenses are somewhat more common than
cusps.  Neither result is very sensitive to the image separation.  In
both cases double lenses are fairly uncommon, which is very different
from the situation with normal arcsecond-scale lenses produced by
nearly-isothermal galaxies. This result is consistent with previous
theoretical conclusions that image multiplicities depend on the central
concentration of the lens mass distribution, such that less concentrated
profiles tend to produce more quadruple and cusp lenses
(\citealt{kassiola93}; \citealt*{kormann94}; \citealt{rusin01b};
\citealt{evans02}; \citealt{dalal04a}). The important point for
observations is that if dark halos have inner profiles with $\alpha
\lesssim 1.5$, then many or even most large-separation lenses should be
quadruples or cusps rather than doubles.  Another point is that the
image multiplicities are sensitive to the inner density profile, so they
offer a new method for probing dark matter density profiles that is
qualitatively different from methods discussed before. 

\begin{figure}
\begin{center}
 \includegraphics[width=0.7\hsize]{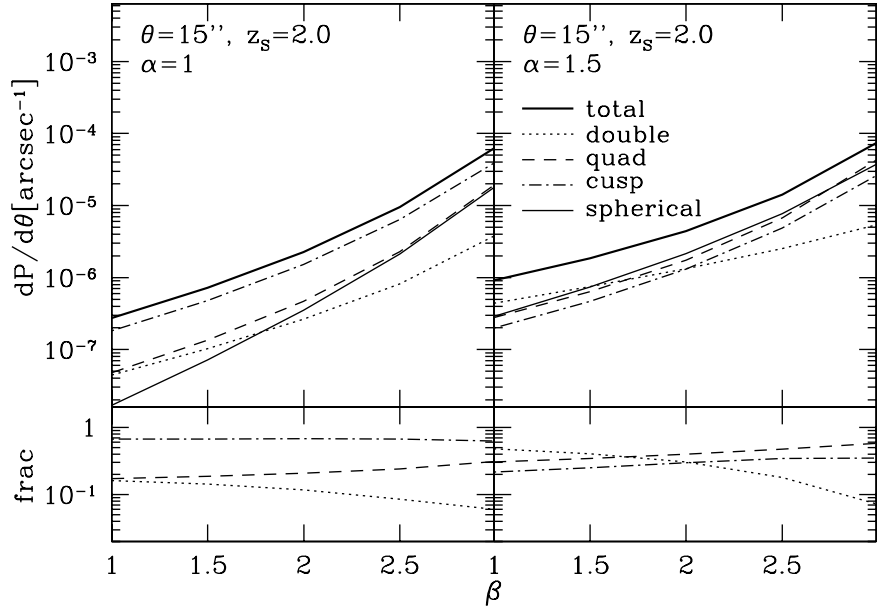}
\caption{Lensing probabilities and image multiplicities with triaxial
dark halos as a function of the slope of the source luminosity function
$\beta$.  We consider an image separation of $\theta=15''$, and we
place the source at $z_{\rm S}=2.0$.
 \label{fig:lat_bias}}
\end{center}
\end{figure}

Magnification bias is important in lens statistics, particularly in
image multiplicities, because it gives more weight to quadruple and cusp
configurations (which tend to have large magnifications) than to
doubles.  We should therefore understand what happens when we modify the
magnification bias by varying the slope $\beta$ of the source luminosity
function.  The results are shown in Figure~\ref{fig:lat_bias} (for image
separation $\theta=15''$ and source redshift $z_{\rm S}=2.0$).  The
lensing probabilities increase as $\beta$ increases, because as the
source luminosity function becomes steeper magnification bias becomes
stronger.\footnote{Note that the magnification bias diverges if the
luminosity function is a pure power law with $\beta \ge 3$, so we are
restricted to shallower cases.}  Interestingly, the increase in the
total probability due to triaxiality weakens as $\beta$ increases,
although the effect is not strong.  As for the image multiplicities,
$\alpha=1$ halos are always dominated by cusp lenses, although for
sufficiently steep luminosity functions quadruples become fairly common.
With $\alpha=1.5$ halos, when magnification bias is weak ($\beta \sim
1$) doubles are the most probable, but as magnification bias strengthens
($\beta$ increases) quadruples receive more weight and become the most
likely.  In practice, the effective values of $\beta$ are larger than
$\sim$1.5 for both optical \citep[e.g.,][]{boyle00} and radio surveys
\citep[e.g.,][]{rusin01b}, so we expect cusps to dominate for $\alpha=1$
and quadruples to be the most common for $\alpha=1.5$.

\begin{figure}
\begin{center}
 \includegraphics[width=0.7\hsize]{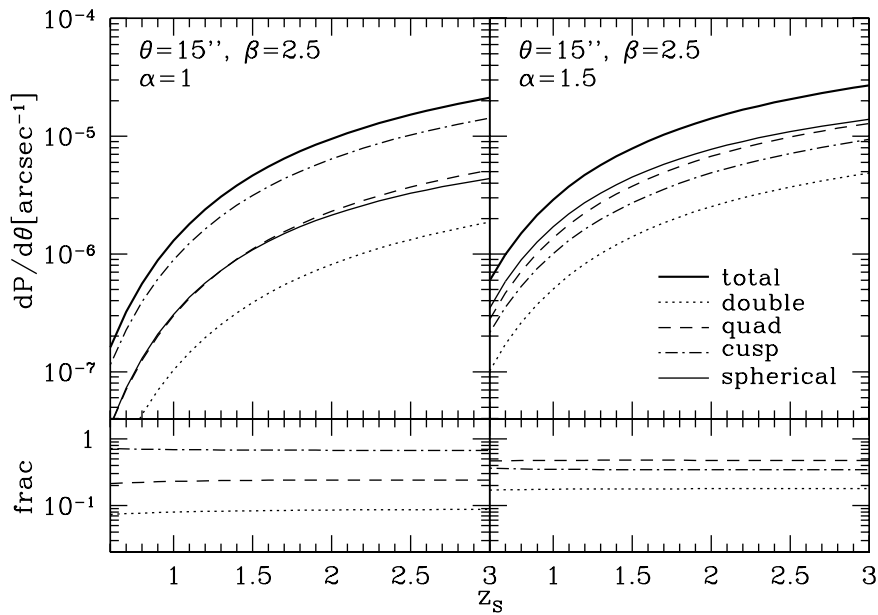}
\caption{Lensing probabilities and image multiplicities as a function
of source redshift $z_{\rm S}$.  We fix the image separation to
$\theta=15''$ and the of the source luminosity function to $\beta=2.5$.
\label{fig:lat_zs}}
\end{center}
\end{figure}

Finally, we consider whether the results depend on the source redshift,
as shown in Figure~\ref{fig:lat_zs}.  The lensing probabilities rise
with $z_{\rm S}$, because there is more volume and hence more deflectors
between the observer and the source.  In addition, geometric effects
mean that a given halo can produce a larger image separation when the
source is more distant, so the halo mass required to produce a given
image separation goes down and the abundance of relevant deflectors goes
up.  However, the probability increase affects the different image
configurations in basically the same way, so the image multiplicities
are quite insensitive to the source redshift.  Therefore, we conclude
that details of the source redshift distribution are not so important
for image multiplicities, at least when the source luminosity function
has a power law shape. 

\subsection{Statistics at larger image separations}\label{sec:lat_largesep}

In Figure~\ref{fig:lat_sep}, at very large image separations
($\theta\gtrsim100''$) the lensing probabilities in the triaxial halo
model are orders of magnitude larger than those in the spherical halo
model. In addition, at these very large separations the $\alpha=1$ case
produces higher probabilities than the more concentrated $\alpha=1.5$
case.  Both features are puzzling and invite careful consideration.

\begin{figure}
\begin{center}
 \includegraphics[width=0.45\hsize]{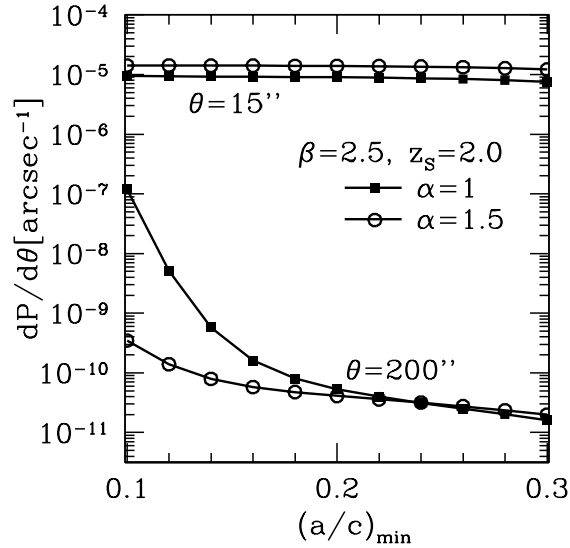}
\caption{Dependence of the lensing probability on the cutoff in $a/c$.
Both $\theta=15''$ and $200''$ are shown.  Filled squares and open
circles denote $\alpha=1$ and $1.5$, respectively.  The slope of the
luminosity function is fixed to $\beta=2.5$.
\label{fig:lat_amin}}
\end{center}
\end{figure}

Figure~\ref{fig:lat_amin} shows the dependence of total lensing
probability on the lower limit of the integral over $a/c$.  In the
previous calculations, we assumed $(a/c)_{\rm min}=0.1$. This figure
shows that for $\theta=15''$ the results are quite insensitive to
$(a/c)_{\rm min}$, suggesting that the contribution from extremely
triaxial halos is negligible. For $\theta=200''$, however, the lensing
probability rapidly decreases as $(a/c)_{\rm min}$ increases. In other
words, the lensing probability at very large image separations seems to
be dominated by very small $a/c$, or very large triaxialities.

Results that are dominated by such extreme halos are probably not very
reliable. They depend sensitively on both the assumed PDF for the axis
ratio $a/c$ (eq.~[\ref{tri_p_a}]) and the correlation between $a/c$ and
the concentration $c_e$ (eq.~[\ref{tri_median-ce}]) at very small axis
ratios. The fitting forms presented by \citet{jing02} were intended to
reproduce the PDF and correlation at $a/c\gtrsim 0.3$ (see Figures
\ref{fig:tri_pdf} and \ref{fig:tri_c_ac}), and it is unclear whether
they are still accurate at $a/c \sim 0.1$. In addition, even if we know
accurate fitting forms, such a situation implies that sample variance
(i.e., the effect of the finite number of lensing clusters) may be quite
large.  

Another ambiguity is the projection effect. Here, we assumed that the
density profile (eq.~[\ref{tri_gnfw}]) extends beyond the virial radius,
and in projecting along the line of sight we integrated the profile to
infinity. Although it is not clear whether we should cut off the profile
at the virial radius or not \citep[e.g.,][]{takada03}, the effect of the
extended profile on the gravitational lensing is not so large for normal
dark halos. However, when $a/c$ is small enough, equation
(\ref{tri_median-ce}) indicates that the concentration parameter $c_e$
becomes smaller than unity, so the effect of the extended profile
outside the virial radius is quite significant. The puzzling feature
that the $\alpha=1$ case produces higher probabilities than the
$\alpha=1.5$ case at very large separations can be ascribed to the
projection effect, because the effect is more significant for shallower
density profiles. 

Thus, lenses with extremely large image separations are associated with
the most extreme dark matter halos, and it may be difficult to make
reliable predictions about them.  We emphasize, though, that these
issues do not apply to lenses with separations $\theta \lesssim 30''$,
and on these scales we believe our results to be robust.

\section{Summary}\label{sec:lat_sum}
\markboth{CHAPTER \thechapter.
{\MakeUppercase\mychapheadname}}{\thesection.
\MakeUppercase{Summary}}

The dark matter halos predicted by the CDM model are triaxial rather
than spherical, which has a significant effect on the statistics of
large-separation gravitational lenses.  Triaxiality systematically
enhances the lensing probability by a factor of $\sim$4 if dark halos
have an inner density profile with $\alpha=1$, or a factor of $\sim$2 if
$\alpha=1.5$.  The effects may be even more dramatic at very large image
separations ($\theta \gtrsim 100''$), although such lenses are very
sensitive to the most triaxial halos and so the predictions are not as
reliable.  Thus, triaxiality must be added to the list of important
systematic effects that need to be included in calculations of
large-separation lens statistics.  (Some of the other effects are the
inner density profile and the shape of the distribution of concentration
parameters, as found in \S \ref{sec:lat_sph}.) 

Triaxial modeling allows us to predict the image multiplicities for
large-separation lenses.  We found that the multiplicities depend
strongly on the density profile: for $\alpha=1$, lenses are dominated by
naked cusp image configurations; while for $\alpha=1.5$, quadruple
configurations are the most probable. Double lenses, which are dominant
among normal arcsecond-scale lenses, are subdominant in both cases. Note
that cusp lenses can be distinguished from doubles by the presence of a
third image comparable in brightness to the other two, and by the
configuration of image positions.  The differences can be ascribed to
the different mass density profiles, and they indicate that the
multiplicities of large-separation lenses will provide a qualitatively
new probe of the central density profiles of massive dark matter halos
(and hence a new test of CDM). 

The prediction that triaxial halos produce significant fractions of
quadruple and cusp lenses should be kept in mind when considering
samples of candidate large-separation lenses.  For example,
\citet{miller04} found six large-separation double lens candidates in
the Two-degree Field (2dF) Quasar Redshift Survey, but no quadruple or
cusp lens candidates.  Even accounting for small number statistics, our
results suggest that such a high fraction of doubles would be
inconsistent with CDM at more than 3$\sigma$,\footnote{If the predicted
double fraction is $f_2$, then the Poisson probability of having $N$
doubles and no quadruples or cusps is $\mathcal{L}(f_2) \propto
(f_2)^N$.} and that it would be surprising if many of the six candidates
are genuine lens systems.  

While our theoretical model is much more realistic than the simple
spherical model, we have still made several simplifying assumptions.
One is that we have neglected substructure in dark halos.  The galaxies
in massive cluster halos do not have a large effect on the statistics of
lensed arcs \citep{meneghetti00}, but it is not obvious whether or not
they would affect large-separation lenses.  Substructure can affect the
image multiplicities for isothermal lenses \citep{cohn04}, so it should
be considered for CDM halos as well.  Another effect we have neglected
is the presence of a massive central galaxy in a cluster.
\citet{meneghetti03b} claim that central galaxies do not have a large
effect on arc statistics.  However, because our results depend on the
inner slope of the density profile, and a central galaxy effectively
increases the concentration, this effect should be considered. A third
phenomenon we have neglected is cluster merger events. Indeed, mergers
can change the shapes of critical curves and caustics substantially, and
thus have a great impact on lensing cross sections \citep{torri04}.  To
estimate the effect on large-separation lens statistics, we would need a
realistic model of the cluster merger event rate and the physical
conditions of merger events.  Addressing these various issues to improve
the accuracy of the theoretical predictions is beyond the scope of this
thesis, but is certainly of interest for future work.

\chapter{Discovery of the Large-Separation Lensed Quasar SDSS
J1004+4112}
\label{chap:sdss}
\def\mychapheadname{Discovery of the Large-Separation Lensed Quasar}
\markboth{CHAPTER \thechapter.
{\MakeUppercase\mychapheadname}}{}

\section{Introduction}
\markboth{CHAPTER \thechapter.
{\MakeUppercase\mychapheadname}}{\thesection.
\MakeUppercase{Introduction}}

The CDM model naturally predicts the existence of strong gravitational
lens systems with image separations of $\sim\!10''$ or even larger, as
shown in Chapter \ref{chap:lat}. Observations of massive clusters of
galaxies have revealed many systems of ``giant arcs'' representing
lensed images of background galaxies (see Chapter \ref{chap:arc}).
However, until recently all lensed quasars and radio sources had image
separations $<\!7''$ corresponding to lensing by galaxies, despite some
explicit searches for lenses with larger separations. 

\begin{figure}[tb]
\begin{center}
\includegraphics[width=0.5\hsize]{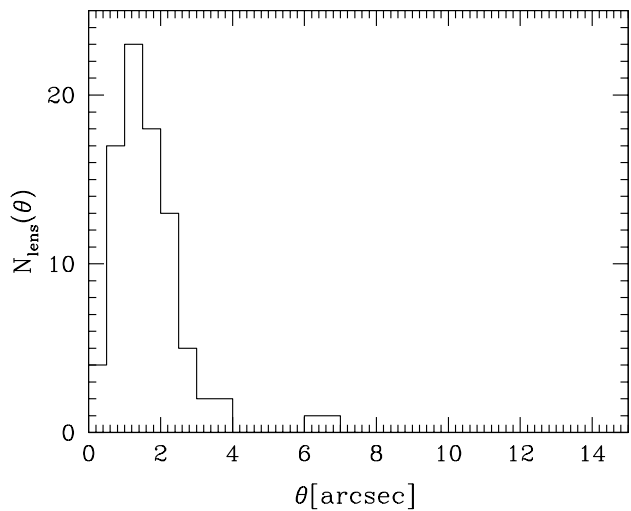}
\caption{Number distribution of strongly lensed quasars known so far. No
 confirmed lens system has image separation $\theta>7''$. The data are
 taken from \citet{kochanek04b}.
 \label{fig:sdss_sep}}
\end{center}
\end{figure}

The fact that clusters have less concentrated mass distributions than
galaxies implies that large-separation lensed quasars should be less
abundant than small-separation lensed quasars by one or two orders of
magnitude (see, e.g., Figure \ref{fig:lat_prob}). This explains why past
surveys have failed to unambiguously identify large-separation lensed
quasars
\citep*{kochanek95a,maoz97,hewett98,phillips01a,zhdanov01,ofek01,ofek02}.
For instance, CLASS found 22 small-separation lenses but no
large-separation lenses among $\sim$11,000 radio sources
\citep{phillips01b}. Although several large separation lensed quasar
candidates have been found \citep*[e.g.,][]{mortlock99}, they are
thought to be physical (unlensed) 
pairs on the basis of individual observations \citep[e.g.,][]{green02}
or statistical arguments \citep*{kochanek99,rusin02}. Recently
\citet{miller04} found 6 candidate lens systems with image separations
$\theta > 30''$ among $\sim$22,000 quasars in the Two-degree Field (2dF)
quasar sample.  Given the lack of high-resolution spectra and deep
imaging for the systems, however, it seems premature to conclude that
they are true lens systems. We note that both because the expected
number of lenses with such large image separations in the 2dF sample is
much less than unity \citep{oguri03c} and because all these candidates
are doubles unlike the theoretical prediction (see Chapter
\ref{chap:lat}), these systems would present a severe challenge to
standard models if confirmed as lenses. In Figure \ref{fig:sdss_sep}, we
show image separation distribution of known lensed quasars so far. There
is a cutoff at $\theta\sim 3''$, and no lens system has image separation
larger than $7''$. We summarize the previous large-separation lens
searches in Table \ref{table:sdss_survey}.

To find a first unambiguous large-separation lensed quasar, we started a
project to search for large-separation lenses in the quasar sample of
the Sloan Digital Sky Survey \citep[SDSS;][]{york00}.  This project
complements ongoing searches for small-separation lenses in SDSS
\citep*{inada03a,inada04a,inada04b,pindor03,pindor04,johnston03,morgan03,oguri04b}.
The SDSS has completed less than   
half of its planned observations, but already it contains more than
30,000 quasars and is superior to previous large-separation lens surveys
in several ways.  The full SDSS sample will comprise $\sim$100,000
quasars, so we ultimately expect to find several large-separation lensed
quasars \citep{keeton01a,takahashi01,li02,kuhlen04}.  One of the most
important advantages of the SDSS in searching for large-separation
lensed quasars is that imaging in five broad optical bands allows us
to select lens candidates quite efficiently (see Appendix \ref{chap:appsdss}).

\begin{table}[tb]
 \begin{center}
  \begin{tabular}{ccccccl}\hline\hline
Name & Wavelength & $\theta$ Range & \# of Quasars & Median $z$ & \# of
   Lenses & Ref. \\
\hline
{\it HST} Snapshot & optical & $7''-50''$ & $\sim 80$ & $\sim 2$ & $0$ & 1 \\
LBQS & optical & $3''-10''$ & $\sim 1,000$ & $\sim 1.3$ & $0$ & 2, 3 \\
ARCS & radio & $15''-60''$ & $\sim 1,000$ & $\sim 1.3$ & $0$ & 4\\
CLASS & radio & $6''-15''$ & $\sim 11,000$ & $\sim 1.3$ & $0$ & 5\\
FIRST & radio & $5''-30''$ & $\sim 9,000$ & $\sim 1.0$ & $0$ & 6\\
2dF & optical & $30''-200''$ & $\sim 22,000$ & $\sim 1.5$ & $>6$ ?? & 7\\
\hline
\end{tabular}
\caption{Surveys for large-separation lensed quasars. \hspace{70mm}\protect\\
\footnotesize{\hspace*{5mm}Ref. --- (1) \citealt{maoz97}; 
  (2) \citealt{hewett98}; (3) \citealt{mortlock00}; 
  (4) \citealt{phillips01a}; (5) \citealt{phillips01b}; 
  (6) \citealt{ofek01}; (7) \citealt{miller04} }
} 
\label{table:sdss_survey}
 \end{center}
\end{table}

We search for large-separation lensed quasars in a sample of
$\sim$30,000 spectroscopically-confirmed SDSS quasars at redshifts $z$
of $0.6-2.3$, a sample larger than those used in any  previous searches.
Even with this large sample, the expected number of large-separation
lensed quasars is of the order of unity. Then we report the discovery of
the first large-separation lensed quasar SDSS~J1004+4112 at
$z=1.73$\footnote{The discovery itself was also reported in
\citet{inada03b} and \citet{inada04b}; in this thesis we describe
details of how we searched and identified it as a gravitationally lensed
quasar, and discussed theoretical implications of the discovery.}.
The quasar itself turned out to be previously identified in the {\it
ROSAT} All Sky Survey \citep*{cao99} and the Two-Micron All-Sky Survey
\citep{barkhouse01}, but was not recognized as a lensed system. In this
Chapter, we describe lens search in the SDSS data, and also photometric
and spectroscopic follow-up  observations of SDSS~J1004+4112 in detail.
We discuss the spectra of lensed quasar components, including puzzling
differences between emission lines seen in the different images. We
analyze deep multicolor imaging data to show the existence of a lensing
cluster robustly. We also discuss the implications of this system for
the statistics of large-separation lenses, using both spherical and
triaxial lens models. The mass modeling of this system is of great
interest, and we describe it in Appendix \ref{chap:1004}. 

\section{Candidate Selection from the SDSS Object Catalog}
\label{sec:sdss_select}
\markboth{CHAPTER \thechapter.
{\MakeUppercase\mychapheadname}}{\thesection.
\MakeUppercase{Candidate Selection from the SDSS Object Catalog}}

The SDSS is a survey to image a quarter of the Celestial Sphere at high
Galactic latitude and to measure spectra of galaxies and quasars found
in the imaging data \citep{blanton03}. The dedicated 2.5-meter telescope
at Apache Point Observatory (APO) is equipped with a multi-CCD
camera \citep{gunn98} with five broad bands centered at $3561$,
$4676$, $6176$, $7494$, and $8873${\,\AA} \citep{fukugita96}. The
imaging data are automatically reduced by a photometric pipeline
\citep{lupton01}. The astrometric positions are accurate to about
$0\farcs1$ for sources brighter than $r=20.5$ \citep{pier03}. The
photometric errors are typically less than 0.03 magnitude
\citep{hogg01,smith02}. The SDSS quasar selection algorithm is presented
in \citet{richards02}. The SDSS spectrographs are used to obtain spectra,
covering $3800$--$9200${\,\AA} at a resolution of $1800$--$2100$, for
the quasar candidates. The public data releases of the SDSS are
described in \citet{stoughton02} and \citet{abazajian03,abazajian04}.
Please refer Appendix \ref{chap:appsdss} for more details.

\begin{figure}[tb]
\begin{center}
\includegraphics[width=0.5\hsize]{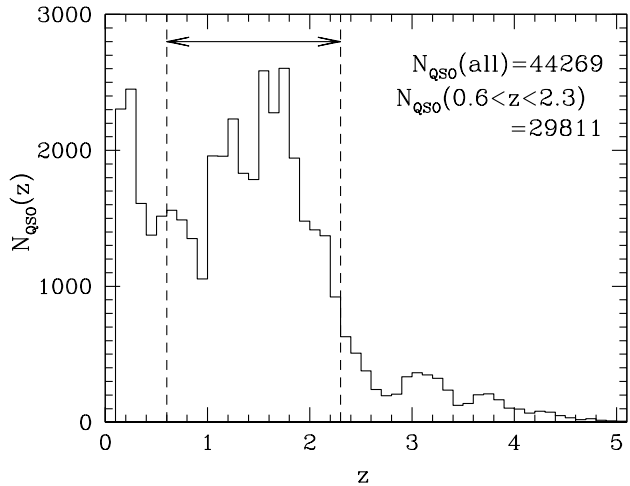}
\caption{Redshift distribution of quasars identified by the
 spectroscopic pipeline in the SDSS.  Dashed vertical lines show the
 redshift cut $0.6<z<2.3$ used for the statistical analysis.
 \label{fig:sdss_z_qso}}
\end{center}
\end{figure}

Large-separation lens candidates can be identified from the SDSS data
as follows. First, we select objects that were initially identified as
quasars by the spectroscopic pipeline. Specifically, among SDSS
spectroscopic targets we select all objects that have spectral
classification of {\tt SPEC\_QSO} or {\tt SPEC\_HIZ\_QSO} with
confidence {\tt z\_conf} larger than $0.9$ \citep[see][for details of
the SDSS spectral codes]{stoughton02}. Next we check the colors of
nearby unresolved sources to see if any of those sources could be an
additional quasar image, restricting the lens search to separations
$\theta<60''$.  We define a large-separation lens by $\theta>7''$ so
that it exceeds the largest image separation lenses found so far:
Q0957+561 with $\theta=6\farcs26$ \citep{walsh79} and RX~J0921+4529 with
$\theta=6\farcs97$ \citep{munoz01}, both of which are produced by
galaxies in small clusters. We regard the stellar object as a candidate
companion image if the following color conditions are satisfied:
\begin{equation}
 \left|\Delta(j-k)\right|<3\sigma_{\Delta(j-k)}=3\sqrt{\left(\sigma_{j,{\rm err}}^2+\sigma_{k,{\rm err}}^2\right)_{\rm quasar}+\left(\sigma_{j,{\rm err}}^2+\sigma_{k,{\rm err}}^2\right)_{\rm stellar}},
\end{equation}
\begin{equation}
\left|\Delta(j-k)\right|<0.1,
\end{equation}
\begin{equation}
\left|\Delta i^*\right|<2.5,
\end{equation}
where $\{j,k\}=\{u^*,g^*\}$, $\{g^*,r^*\}$, $\{r^*,i^*\}$, and
$\{i^*,z^*\}$,\footnote{The starred magnitudes ($u^*g^*r^*i^*z^*$)
are used to denote still-preliminary 2.5m-based photometry
\citep[see][]{stoughton02}.}
and $\Delta$ denotes the difference between the spectroscopically
identified quasar and the nearby stellar object.  Note that this
selection criterion is tentative; we still do not know much about
large-separation lenses, so selection criteria may evolve as we learn more. 

\begin{table}[tb]
 \begin{center}
  \begin{tabular}{ccccccc}\hline\hline
Object & $i^*$ & $u^*-g^*$ & $g^*-r^*$ & $r^*-i^*$ & $i^*-z^*$ &
   Redshift \\
\hline
A & $18.46\pm0.02$ & $0.15\pm0.05$ & $-0.03\pm0.04$ &
 $0.24\pm0.03$ & \hspace*{3mm}$0.02\pm0.05$ & $1.7339\pm0.0001$\\
B & $18.86\pm0.06$ & $0.18\pm0.08$ & $-0.05\pm0.08$ &
 $0.23\pm0.08$ & $-0.03\pm0.09$ & $1.7335\pm0.0001$\\
C & $19.36\pm0.03$ & $0.03\pm0.05$ & $-0.03\pm0.04$ &
 $0.38\pm0.04$ & \hspace*{3mm}$0.05\pm0.08$ & $1.7341\pm0.0002$\\
D & $20.05\pm0.04$ & $0.15\pm0.09$ & \hspace*{3mm}$0.15\pm0.05$&
 $0.46\pm0.05$ & \hspace*{3mm}$0.09\pm0.13$ & $1.7334\pm0.0003$\\
\hline
\end{tabular}
\caption{Magnitudes and colors for the four quasar images, taken from
  the SDSS photometric data. Redshifts are derived from Ly$\alpha$ lines
  in the Keck LRIS spectra (see Figure~\ref{fig:sdss_spec}). } 
\label{table:sdss_sdssphoto}
 \end{center}
\end{table}

\begin{table}[tb]
 \begin{center}
  \begin{tabular}{cccccc}\hline\hline
Object & R.A.(J2000) & Dec.(J2000) & $\Delta$R.A.[arcsec]${}^{\rm a}$ &
 $\Delta$Dec.[arcsec]${}^{\rm a}$ \\
\hline
A & 10 04 34.794 & $+$41 12 39.29 & \hspace*{5mm}$0.000\pm0.012$   & \hspace*{3mm}$0.000\pm0.012$ \\
B & 10 04 34.910 & $+$41 12 42.79 & \hspace*{5mm}$1.301\pm0.011$   & \hspace*{3mm}$3.500\pm0.011$\\
C & 10 04 33.823 & $+$41 12 34.82 & $-10.961\pm0.012$ & $-4.466\pm0.012$\\
D & 10 04 34.056 & $+$41 12 48.95 & \hspace*{2mm}$-8.329\pm0.007$  & \hspace*{3mm}$9.668\pm0.007$\\
G1   & 10 04 34.170 & $+$41 12 43.66 & \hspace*{2mm}$-7.047\pm0.053$  & \hspace*{3mm}$4.374\pm0.053$\\
\hline
\end{tabular}
\caption{Astrometry of SDSS~J1004+4112 from the deep imaging data taken
  with Suprime-Cam (see \S\ref{sec:sdss_imaging}). The absolute
  coordinates are calibrated using the SDSS data. \protect\\
\footnotesize{\hspace*{5mm}${}^{\rm a}$ Positions relative to component A.}  
\label{table:sdss_astrometry}
}
 \end{center}
\end{table}

\begin{figure}[tbh]
\begin{center}
\includegraphics[width=0.6\hsize]{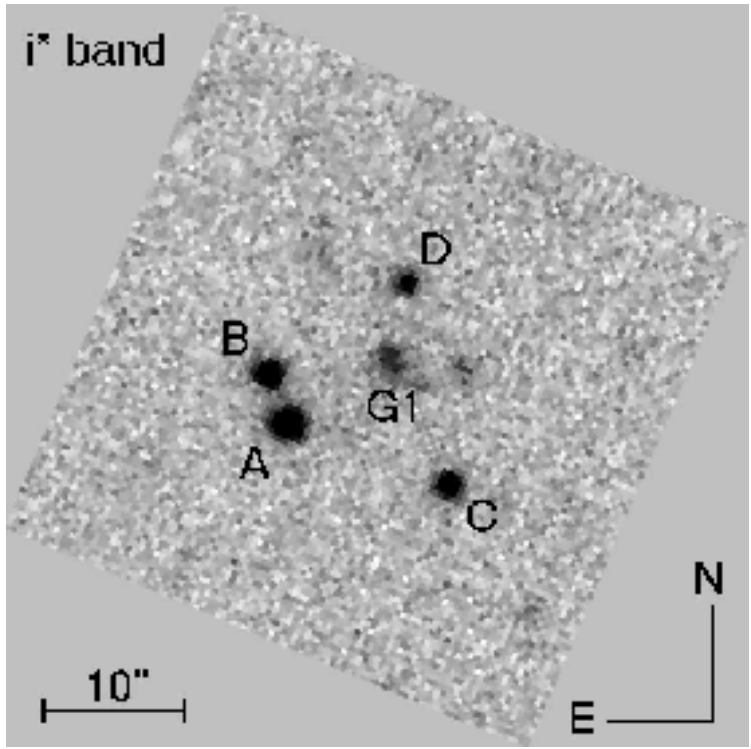}
\caption{SDSS $i^*$-band image of SDSS~J1004+4112. Components A, B,
C, and D are lensed images while component G1 is the brightest galaxy
in the lensing cluster.
 \label{fig:sdss_sdss1004_i}}
\end{center}
\end{figure}

Our full sample contains 44,269 quasars with the redshift distribution
shown in Figure~\ref{fig:sdss_z_qso}.  For the lens search we select the 
subset of 29,811 quasars with $0.6<z<2.3$, making the redshift cuts
for four reasons: (1) at $z<0.6$ quasars are often extended, which can
complicated both lens searches and also lens statistics analyses;
(2) at $z<0.6$ the sample is contaminated by narrow emission line
galaxies; (3) at $z>2.3$ we may miss a number of quasar candidates
because of large color errors; and (4) lens statistics calculations for
high redshift quasars are not very reliable because of uncertainties in
the quasar luminosity function \citep{wyithe02a,wyithe02b}.  Lens
surveys of high-redshift quasars are of course very interesting for
insights into the abundance and formation of distant quasars; a search
for lenses among high-redshift SDSS quasars is the subject of a separate
analysis by \citet{richards04a}.  

SDSS~J1004+4112 was first selected as a lens candidate based on a
pair of components, A and B (see Figure~\ref{fig:sdss_sdss1004_i}),
where B is the SDSS spectroscopic target. Components C and D were
identified by visual inspection and found to have colors similar
to those of A and B (even though they do not match the above color
criteria).  Table~\ref{table:sdss_sdssphoto} summarizes the photometry
for the four components, and Table~\ref{table:sdss_astrometry} gives the 
astrometry for the four components as well as the galaxy G1 seen in
Figure~\ref{fig:sdss_sdss1004_i}.  The reason that components C and D
have somewhat different colors from B is still unclear, but it must be
understood in order to discuss the completeness of the lens survey. The
difference might be ascribed to differential absorption or extinction by
intervening material \citep{falco99}, or to variability in the source on
time scales smaller than the time delays between the images
\citep*[e.g.,][]{devries03}, both of which are effects that become more
important as the image separation grows.

\section{Data Analysis}
\label{sec:data}
\markboth{CHAPTER \thechapter.
{\MakeUppercase\mychapheadname}}{\thesection.
\MakeUppercase{Data Analysis}}

\subsection{Spectroscopic Follow-up Observations}

\subsubsection{Quasar Images}

\begin{figure}[p]
\begin{center}
\includegraphics[width=0.6\hsize]{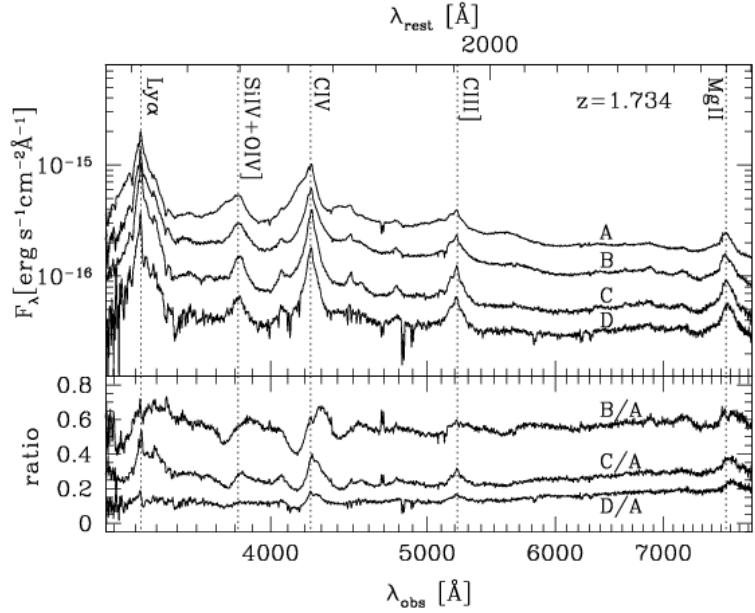}
\caption{Spectra ({\it top}) and flux ratios ({\it bottom}) of
 SDSS~J1004+4112 components A, B, C, and D taken with LRIS on Keck
 I. In the upper panel, we can confirm that all components have
 Ly$\alpha$, \ion{Si}{4}, \ion{C}{4}, \ion{C}{3]}, and \ion{Mg}{2}
 emission lines at $z=1.734$. The flux ratios shown in the lower panel
 are almost constant for a wide range of wavelength. Several absorption
 lines are also seen in the spectra (see text for details).
 \label{fig:sdss_spec}}
\end{center}
\end{figure}
\begin{figure}[p]
\begin{center}
\includegraphics[width=0.5\hsize]{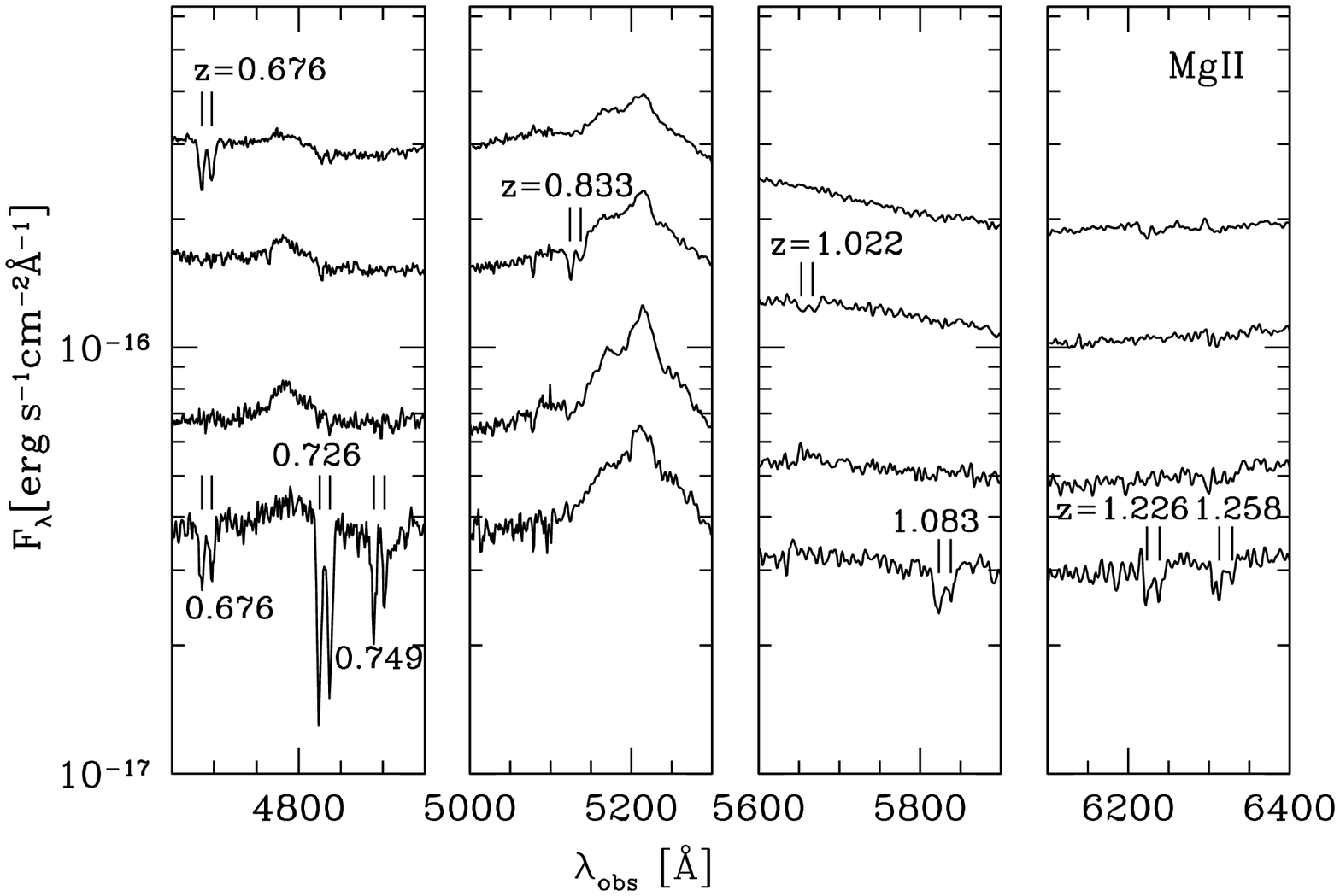}
\includegraphics[width=0.36\hsize]{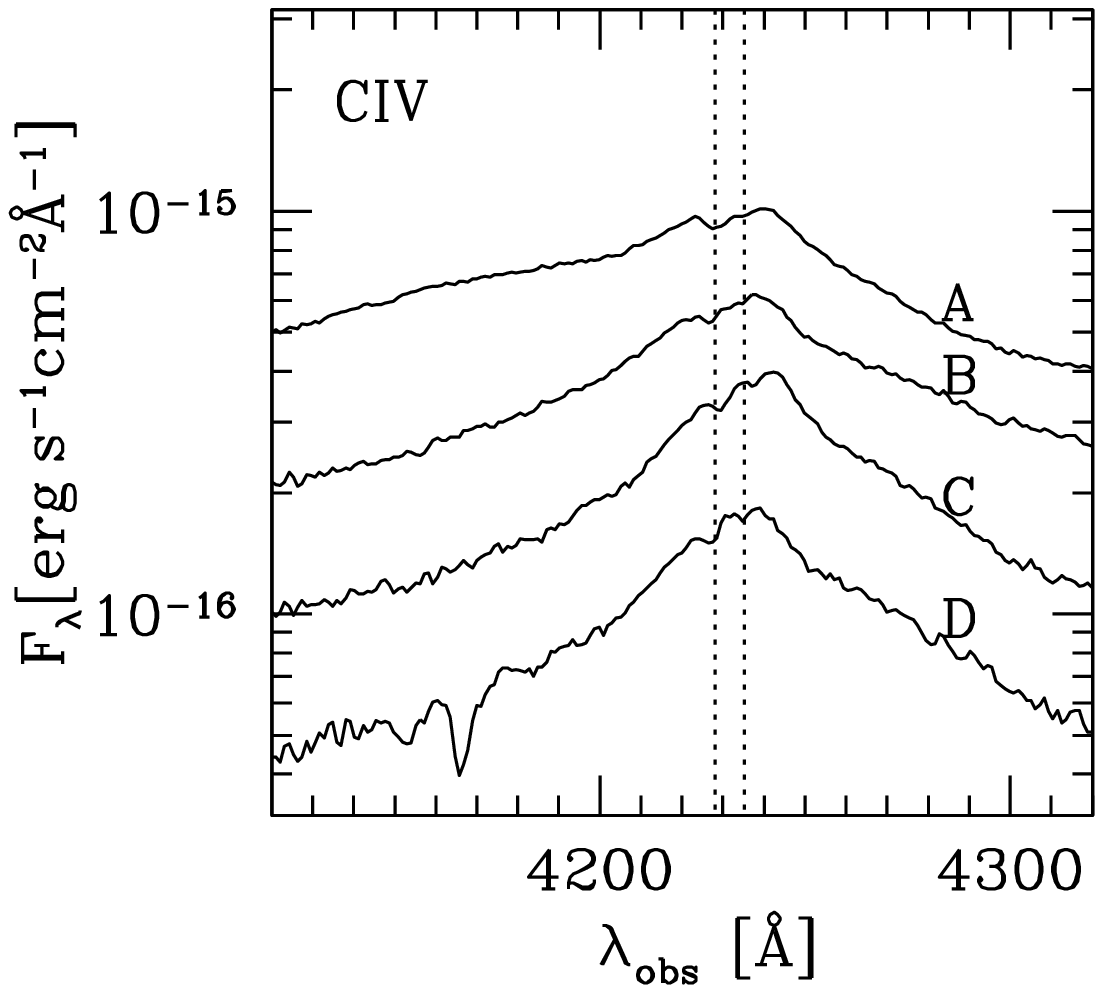}
\caption{Left: The \ion{Mg}{2} doublet absorption lines (rest frame
 wavelengths of $2795.5${\,\AA} and $2802.7${\,\AA}, rest frame
 equivalent width $W_{\rm r}\gtrsim 0.5${\,\AA}) of SDSS~J1004+4112
 components A, B, C, and D at various wavelengths. The absorption lines
 are indicated by vertical lines. Right: The \ion{C}{4} lines of
 SDSS~J1004+4112 components A, B, C, and D taken with LRIS. The
 associated \ion{C}{4} doublet absorption lines (rest frame wavelengths
 $1548.2${\,\AA} and $1550.8${\,\AA}, denoted by dotted lines) are seen
 in all four components. \label{fig:sdss_spec_abs}}
\end{center}
\end{figure}

Since only component B has a spectrum from SDSS, we obtained spectra of
the other components to investigate the lensing hypothesis. The first
spectroscopic follow-up observations were done on 2003 May 2 and 5 with
the Double Imaging Spectrograph of the Astrophysical Research Consortium
(ARC) 3.5-m telescope at APO. All four components have a prominent
\ion{C}{4} emission line ($1549.06${\,\AA}) at $\lambda_{\rm obs} \sim
4230${\,\AA}, indicating that they are quasars with very similar
redshifts. Spectra with higher resolution and longer wavelength range
were taken on 2003 May 30 with the Low-Resolution Imaging Spectrometer
\citep[LRIS;][]{oke95} of the Keck~I telescope at the W.~M.~Keck
Observatory on Mauna Kea, Hawaii, USA. The blue grism is 400 line
mm$^{-1}$, blazed at $3400${\,\AA}, $1.09${\,\AA} pixel$^{-1}$, covering
$3000${\,\AA} to $5000${\,\AA}. The red grating is 300 line mm$^{-1}$,
blazed at $5000${\,\AA}, $1.09${\,\AA} pixel$^{-1}$, covering
$5000${\,\AA} to the red limit of the detector. The spectra were
obtained with 900 sec exposures and a $1''$ slit in $0\farcs9$ seeing.
The data were reduced in a standard way using IRAF.\footnote{IRAF is
distributed by the National Optical Astronomy Observatories, which are
operated by the Association of Universities for Research in Astronomy,
Inc., under cooperative agreement with the National Science Foundation.}
The  Keck/LRIS spectra are shown in Figure~\ref{fig:sdss_spec}. All four
components show emission lines of Ly$\alpha$, \ion{Si}{4}, \ion{C}{4},
\ion{C}{3]}, and \ion{Mg}{2}.  They have nearly identical redshifts of
$z=1.734$, with velocity differences less than $50\,{\rm km\,s^{-1}}$
(see Table~\ref{table:sdss_sdssphoto}).  The flux ratios between the
images (see Figure~\ref{fig:sdss_spec}) are almost constant over the
wavelength range 3000--8000{\,\AA}, indicating that these are actual
lensed images. From the spectra, we conclude that the color differences
found in the SDSS images are mainly caused by differences in the
emission lines (discussed below) and by slightly different continuum
slopes. 

\begin{figure}[tb]
\begin{center}
\includegraphics[width=0.7\hsize]{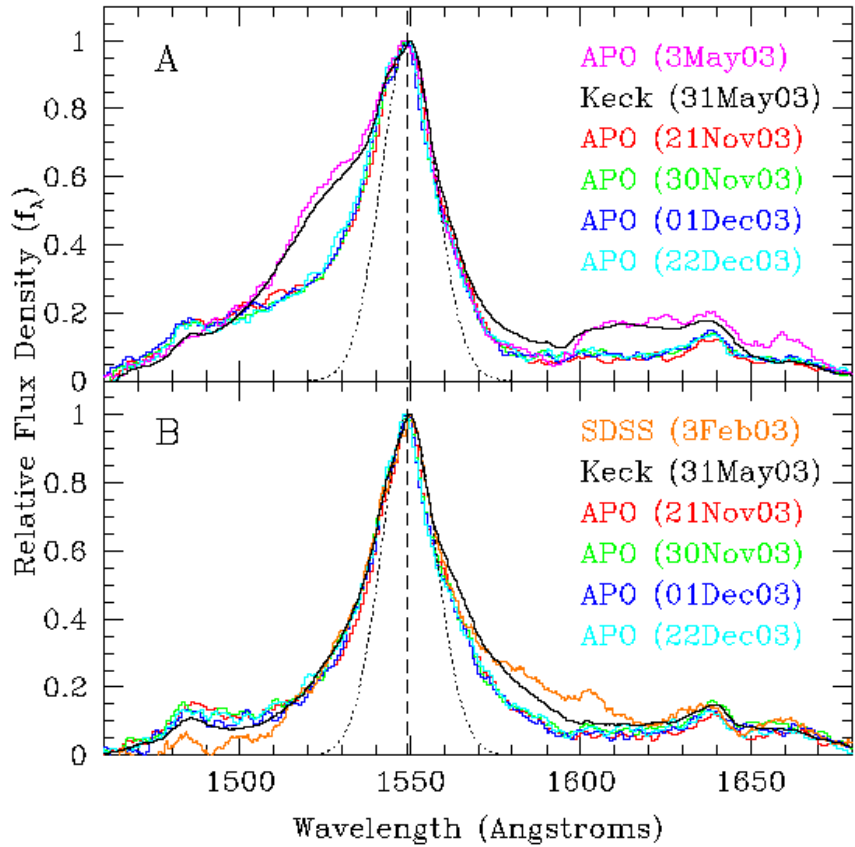}
\caption{Seven epoch data of \ion{C}{4} emission lines of components A
 and B. A power-law continuum is subtracted from each spectrum. The
 spectra are normalized to the peak of \ion{C}{4} emission lines. 
See \citet{richards04b} for more details.
 \label{fig:sdss_micro}}
\end{center}
\end{figure}

Several absorption line systems are seen in the spectra. Components A
and D have intervening \ion{Mg}{2}/\ion{Fe}{2} absorption systems at
$z=0.676$; this redshift is similar to that of the foreground galaxies
(\S\ref{sec:sdss_spec_gal}), suggesting that this absorption system is
associated with the lensing galaxies. Component D has additional
\ion{Mg}{2} absorption systems at $z=0.726$, $0.749$, $1.083$, $1.226$,
and $1.258$. Figure~\ref{fig:sdss_spec_abs} left identifies the various
\ion{Mg}{2} absorbers. We also note that all four components have
\ion{C}{4} absorption lines just blueward of \ion{C}{4} emission lines
(see Figure~\ref{fig:sdss_spec_abs} right). The velocity difference
between the emission and absorption lines is $\sim\!500\,{\rm
km\,s^{-1}}$, so the absorption system is likely to be associated with
the quasar.  The fact that all four components have \ion{C}{4}
absorption lines offers further evidence that SDSS~J1004+4112 is indeed
a gravitational lens.  

Figure~\ref{fig:sdss_spec_abs} right shows notable differences in the
\ion{C}{4} emission line profiles in the different components.  One
possible explanation is the time delay between the lensed images; at any 
given observed epoch, the images represent different epochs in the
source frame. However, the fact that the \ion{C}{4} emission lines in
components A and B differ seems to rule out the time delay explanation:
the expected delay is shorter than the month or year time scale on
which \ion{C}{4} emission lines typically vary
\citep[e.g.,][]{vandenberk04}. Other possible explanations include
differences between the viewing angles probed by the images,
microlensing amplification of part of the quasar emission region,
significant errors in the predicted time delay between A and B, or just
that the quasar is extremely unusual.  

To resolve this puzzling differences, we obtained additional spectra at
different epochs. Figure \ref{fig:sdss_micro} shows how the strong
enhancement in the blue wings of the \ion{C}{4} emission line was faded.
The enhancement lasted at least 28 days, and the predicted time delay
between A and B is $\lesssim 30$ days, thus the event is unlikely to be
intrinsic to the quasar. The enhancement is seen also in other emission
lines, and the amount of the enhancement depends on ionization states so
that higher ionization lines tend to be enhanced more. Since the broad
emission line region is stratified by ionization and higher ionization
lines are found closer to the center \citep{peterson99}, it is
consistent with the interpretation that the enhancement was caused by
microlening of part of broad emission line region of the quasar
\citep[see][ for more details]{richards04b}.

\subsubsection{Galaxies}
\label{sec:sdss_spec_gal}

\begin{figure}[p]
\begin{center}
\includegraphics[width=0.6\hsize]{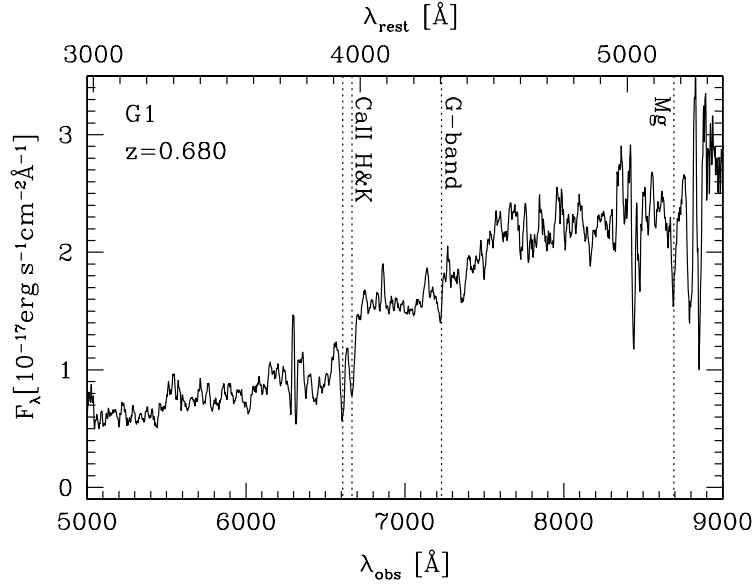}
\caption{Spectrum of the galaxy G1 taken with LRIS on Keck
 I. The break, \ion{Ca}{2} H\&K absorption lines, and Mg absorption line
 are consistent with redshift $z=0.680$ ($z=0.6799\pm0.0001$ from
 the \ion{Ca}{2} H line). The G-band also appears in the spectrum. 
 \label{fig:sdss_spec_g}}
\end{center}
\end{figure}
\begin{figure}[p]
\begin{center}
\includegraphics[width=0.6\hsize]{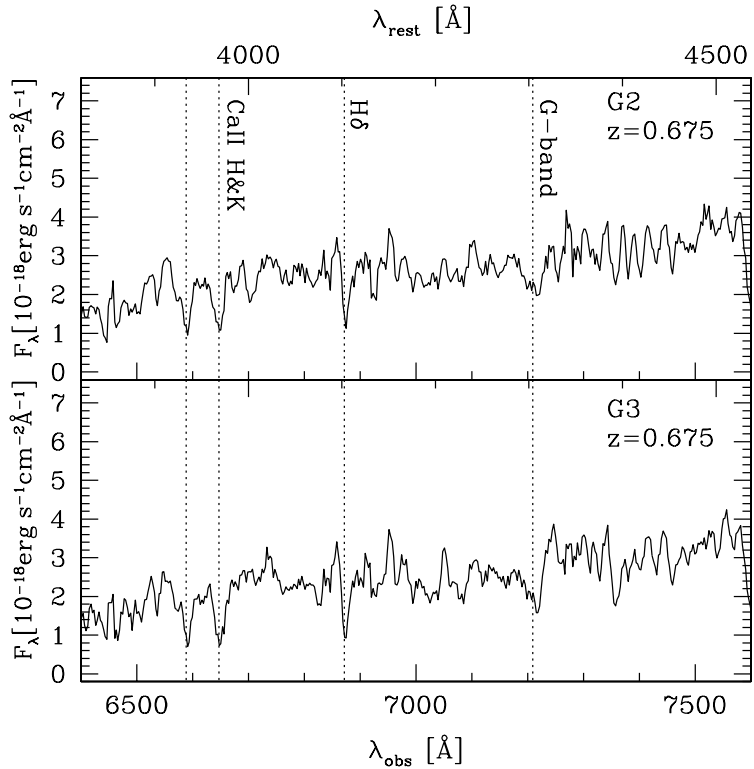}
\caption{Spectra of galaxies G2 and G3 taken with FOCAS on the Subaru
 8.2-m telescope. From the absorption lines \ion{Ca}{2} H\&K, H$\delta$,
 and G-band, we find that the redshifts of both galaxies are $z=0.675$
 ($z=0.6751\pm0.0001$ from the H$\delta$ lines).
 \label{fig:sdss_spec_focas}}
\end{center}
\end{figure}

The spectrum of the galaxy G1, the brightest object near the center of 
the quasar configuration (see Figure~\ref{fig:sdss_sdss1004_i}), was
acquired on 2003 May 30 with LRIS.  The spectrum measured from a 900 sec
exposure is shown in Figure~\ref{fig:sdss_spec_g}. We confirm the break
and \ion{Ca}{2} H\&K lines at $\lambda_{\rm obs}\sim6700${\,\AA}. The
G-band also appears in the spectrum. From the \ion{Ca}{2} H\&K and Mg
lines we derive the redshift of G1 as $z=0.680$.  

The spectra of two additional galaxies near G1 (see
\S\ref{sec:sdss_imaging}) were taken on 2003 June 20 with the Faint
Object Camera and Spectrograph \citep[FOCAS;][]{kashikawa02} on the
Subaru 8.2-m telescope of the National Astronomical Observatory of Japan
on Mauna Kea, Hawaii, USA. We used the 300B grism together with the SY47
filter, and took optical spectra covering $4100${\,\AA} to
$10000${\,\AA} with resolution $2.84${\,\AA} pixel$^{-1}$. The seeing
was $0\farcs7$, and the exposure time was 1740 sec for both galaxies.
The data were reduced in a standard way using IRAF. The spectra are
shown in Figure~\ref{fig:sdss_spec_focas}.  Both galaxies, denoted as G2
and G3, have $z = 0.675$, only $\sim\!700\,{\rm km\,s^{-1}}$ from the
redshift of G1.  

\subsection{Imaging Follow-up Observations}
\label{sec:sdss_imaging}

\subsubsection{Observations}

\begin{figure}[p]
\begin{center}
\includegraphics[width=0.9\hsize]{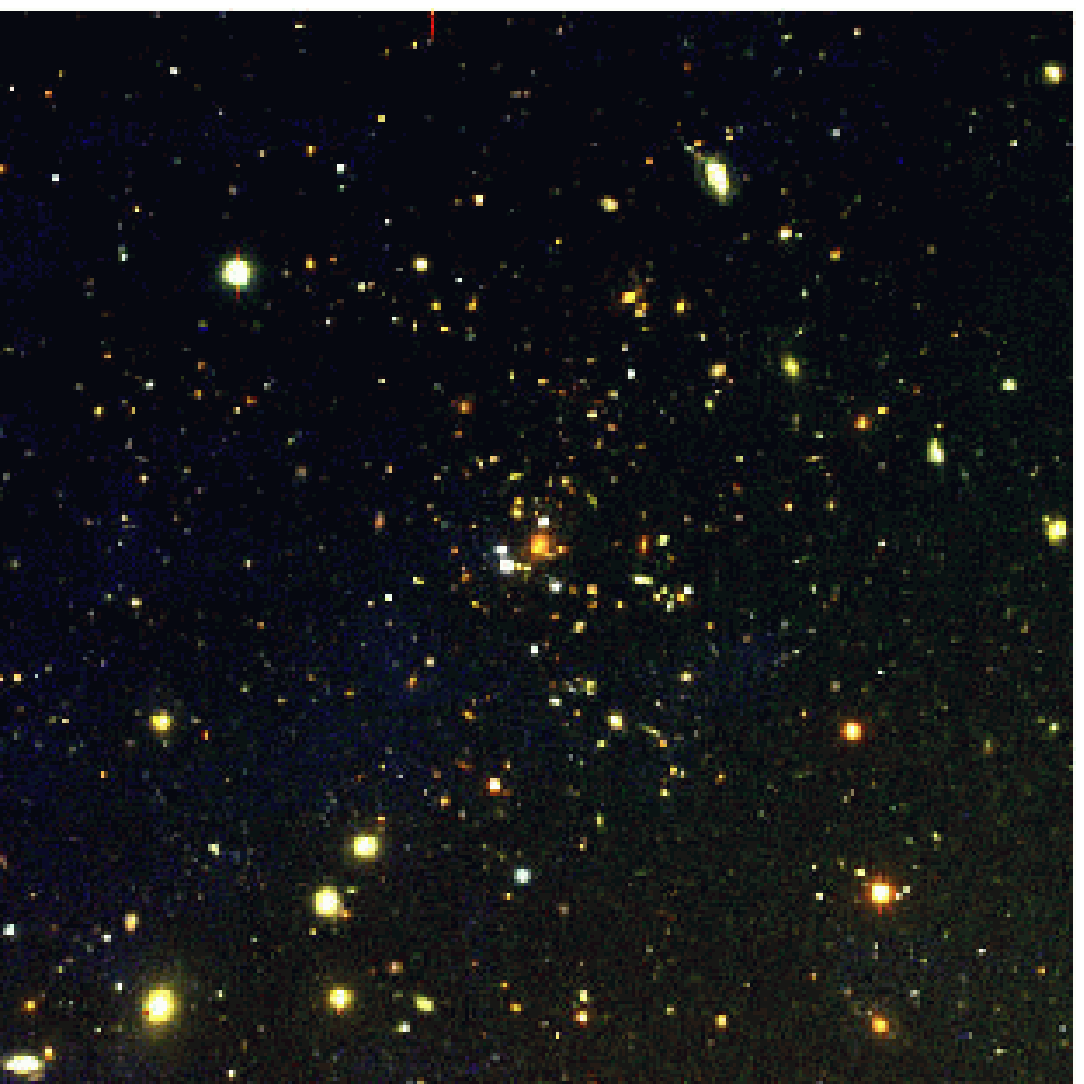}
\caption{The $gri$ composite Subaru image of the field around SDSS
J1004+4112. Many faint galaxies can be seen --- their positions
and colors are consistent with being members of a cluster ($z=0.68$)
 centered on component G1. 
\label{fig:sdss_subaru1004color}}
\end{center}
\end{figure}

\begin{figure}[tbh]
\begin{center}
\includegraphics[width=0.6\hsize]{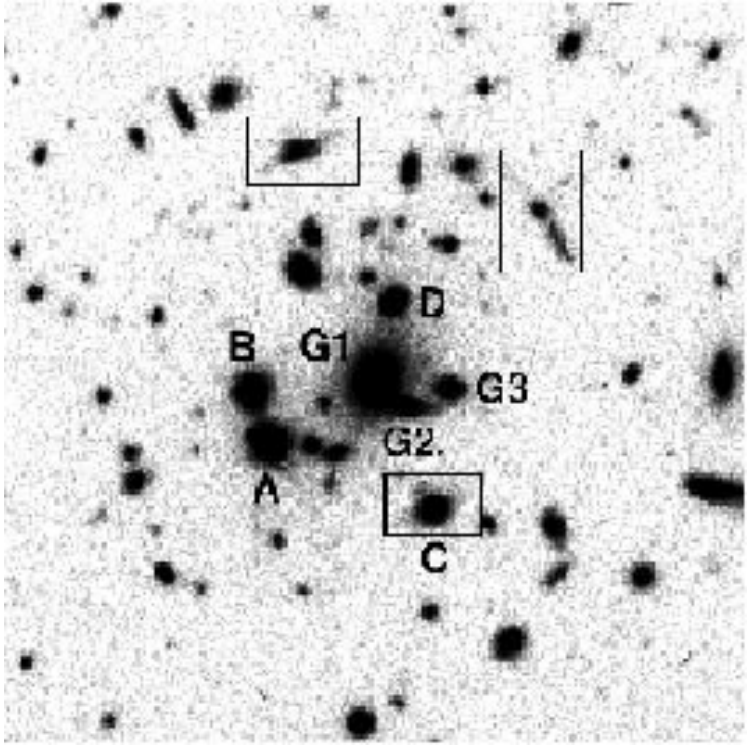}
\caption{The central region of the Suprime-Cam $i$-band image.
 The galaxies with measured redshifts (G1 from LRIS and G2 and G3 from
 FOCAS) as well as the four lensed images are labeled. The possible
 lensed arclets are marked with rectangles.
 \label{fig:sdss_subaru_center}}
\end{center}
\end{figure}

\begin{table}[b]
 \begin{center}
  \begin{tabular}{ccc}\hline\hline
Band & Exptime & $m_{\rm lim}$ ${}^{\rm a}$\\
\hline
$g$ & 810  & 27.0 \\
$r$ & 1210 & 26.9 \\
$i$ & 1340 & 26.2 \\
$z$ & 180  & 24.0 \\
\hline
\end{tabular}
\caption{Total exposure time in seconds (exptime) and limiting
magnitude ($m_{\rm lim}$) for the Subaru deep imaging
  observations.\protect\\
\footnotesize{\hspace*{5mm}${}^{\rm a}$ Defined by $S/N>5$ for point
  sources. \hspace{70mm}}
\label{table:sdss_subaruobs}}
 \end{center}
\end{table}

A deep $r$-band image of SDSS~J1004+4112 was taken on 2003 May 5 with
the Seaver Prototype Imaging camera of the ARC 3.5-m telescope at APO.
The image shows rich structure, with many galaxies between and around
the quasar components suggesting a possible galaxy cluster in the field.
For a further check, we obtained deeper multi-color ($griz$) images on
2003 May 28 with the Subaru Prime Focus Camera
\citep[Suprime-Cam;][]{miyazaki02b} on the Subaru 8.2-m telescope. The
exposure times and limiting magnitudes of the Suprime-Cam images are
given in Table~\ref{table:sdss_subaruobs}. Suprime-Cam has a pixel scale
of $0\farcs2\,{\rm pixel^{-1}}$, and the seeing was
$0\farcs5$--$0\farcs6$.  The frames were reduced (bias-subtracted and
flat-field corrected) in a standard way. The resulting images are shown
in Figure \ref{fig:sdss_subaru1004color}. It is clear that there are
many red galaxies around the four images. Moreover, we find three
possible lensed arclets (distorted images of galaxies behind the
cluster), which are shown in Figure~\ref{fig:sdss_subaru_center}. The
fact that the arclets are relatively blue compared with the brighter
galaxies in the field  (see Figure \ref{fig:sdss_subaru1004color})
suggests that the arclets may be images of distant galaxies
\citep*[e.g.,][]{colley96}. Confirming that they are lensed images will
require higher resolution images and measurements of the arclets'
redshifts. If the hypothesis is confirmed, the arclets will provide
important additional constraints on the lens mass distribution.

\subsubsection{Colors of Nearby Galaxies}

The colors of galaxies in the vicinity of SDSS~J1004+4112 can help us
search for the signature of a cluster.  The central regions of clusters
are dominated by early-type galaxies \citep[e.g.,][]{dressler80} that
show tight correlations among their photometric properties
\citep*{bower92}. These correlations make it possible to search for
clusters using color-magnitude and/or color-color diagrams
\citep{dressler92,gladders00,goto02}. 

We measure the colors of galaxies using the deep Suprime-Cam $griz$
images. Object identifications are performed using the Source Extractor
algorithm \citep[SExtractor;][]{bertin96}; we identify objects with
SExtractor parameter {\tt CLASS\_STAR} smaller than 0.6 in the $i$ band
image as galaxies. Note that this star/galaxy separation criterion is
successful only for objects with $i \lesssim 24$. The magnitudes in the
images are calibrated using nearby stars whose magnitudes are taken from
the SDSS photometric data. 

\begin{figure}[tb]
\begin{center}
\includegraphics[width=0.6\hsize]{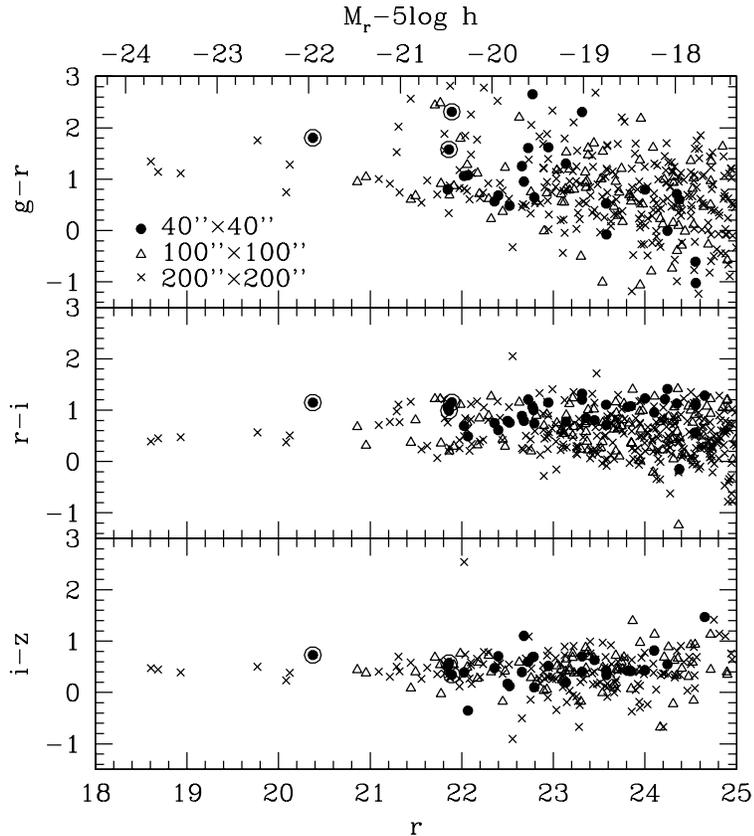}
\caption{Color-magnitude diagrams for the SDSS~J1004+4112 field taken
 with Suprime-Cam. We divide the galaxies into three categories
 according to their positions: filled circles denote galaxies inside a
 $40''\times40''$ box centered on G1, open triangles denote galaxies
 inside a $100''\times100''$ box, and crosses denote galaxies inside a
 $200''\times200''$ box. These box sizes correspond to 
 $0.2h^{-1}\,{\rm Mpc}\times0.2h^{-1}{\rm Mpc}$, 
 $0.5h^{-1}{\rm Mpc}\times0.5h^{-1}{\rm Mpc}$, and 
 $1.0h^{-1}{\rm Mpc}\times1.0h^{-1}{\rm Mpc}$ at $z=0.68$, respectively. 
 Three spectroscopically confirmed member galaxies are marked with open
 circles. The corresponding $r$-band absolute magnitudes at $z=0.68$
 (without K-correction) are given at the top of the frame. 
\label{fig:sdss_colormag}}
\end{center}
\end{figure}

\begin{figure}[tb]
\begin{center}
\includegraphics[width=0.45\hsize]{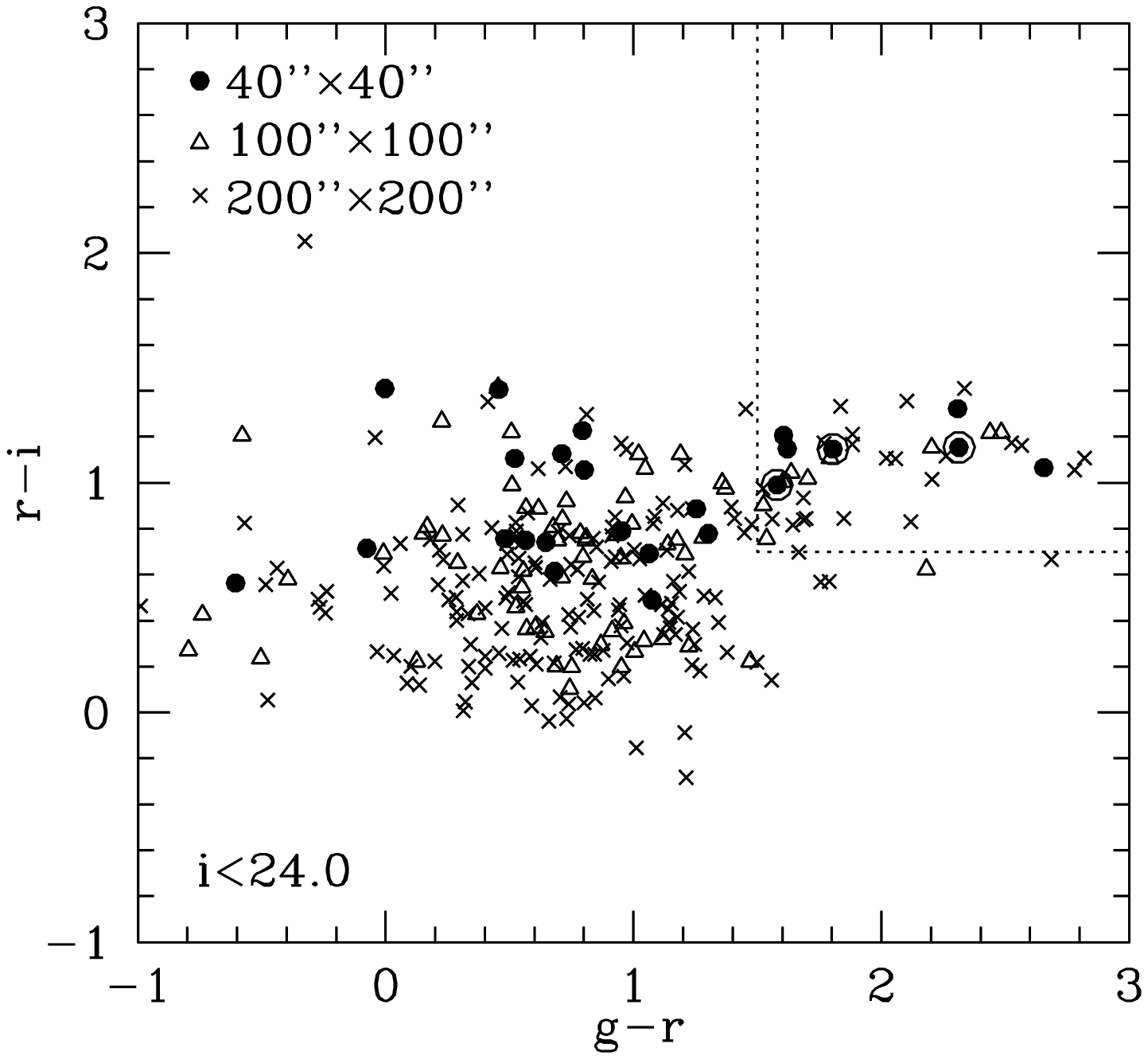}
\includegraphics[width=0.45\hsize]{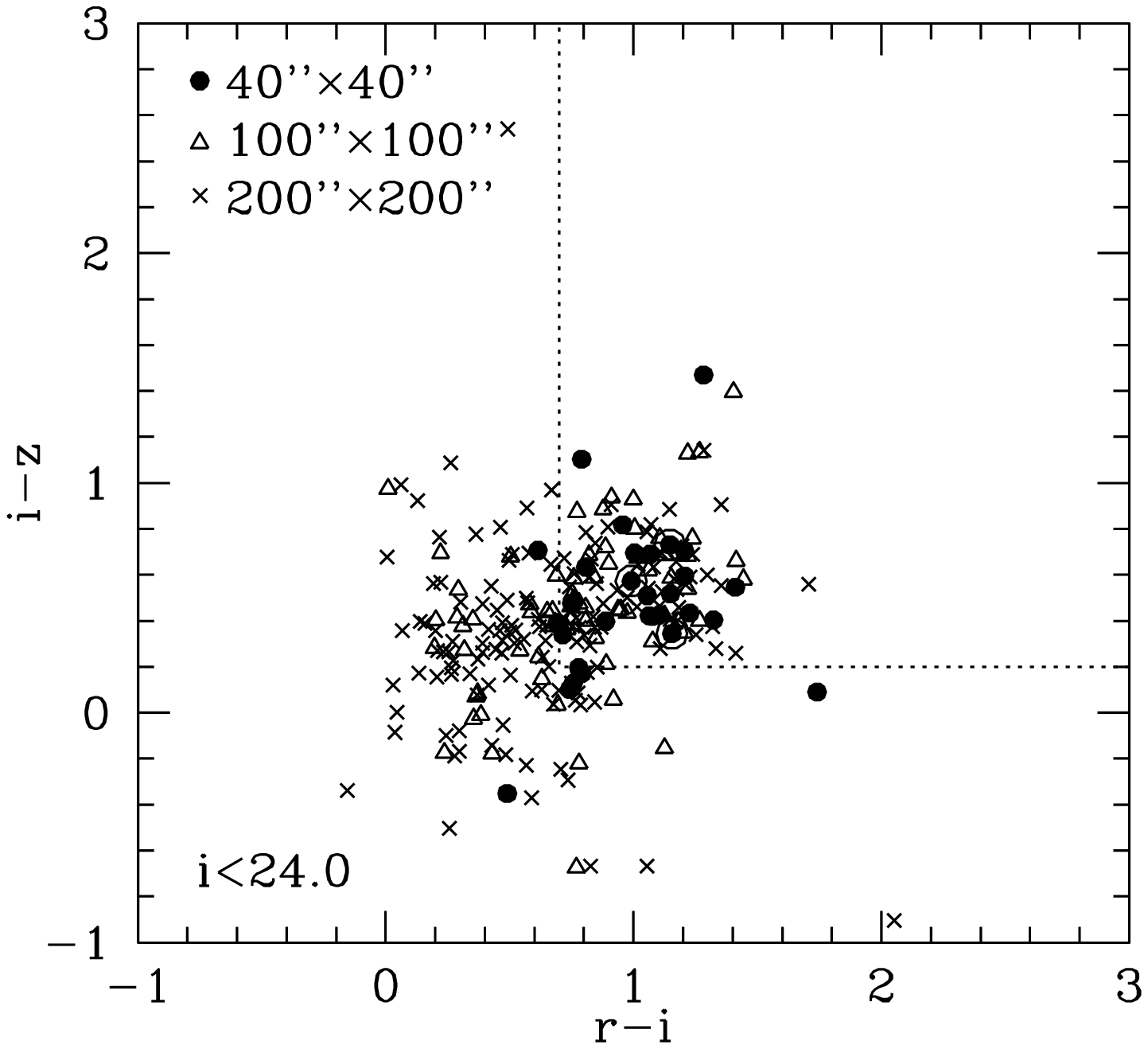}
\caption{The $g-r-i$ and $r-i-z$ color-color diagrams of galaxies
 brighter than $i=24$. Symbols are the same as in
 Figure~\ref{fig:sdss_colormag}. Dotted lines indicate color cuts to
 find cluster members. 
 \label{fig:sdss_cc_gri}}
\end{center}
\end{figure}

Since the red galaxies in clusters are dominant in the central regions,
and the center of the cluster is thought to be near G1, we divide the
galaxies in the field into three categories: galaxies inside a
$40''\times40''$ (corresponding to $0.2h^{-1}{\rm
Mpc}\times0.2h^{-1}{\rm Mpc}$ at $z=0.68$) box centered on G1; galaxies
inside a $100''\times100''$ ($0.5h^{-1}{\rm Mpc}\times0.5h^{-1}{\rm
Mpc}$) box (except for those in the first category); and galaxies inside
a $200''\times200''$ ($1.0h^{-1}{\rm Mpc}\times1.0h^{-1}{\rm Mpc}$) box
(except for those in the first two categories).
Figure~\ref{fig:sdss_colormag} shows color-magnitude diagrams for the
three categories. It is clear that the color-magnitude relations,
particularly $r-i$ and $i-z$, show tight correlations for galaxies
inside the $40''\times40''$ box. Ridge lines at $r-i \sim 1.1$ and $i-z
\sim 0.5$ strongly suggest a cluster of galaxies at $z \sim 0.6$
\citep{goto02}. The result is consistent with the Keck and Subaru
spectroscopic results showing that the redshifts of galaxies G1, G2, and
G3 are all $z\sim0.68$. 

\begin{figure}[tbh]
\begin{center}
\includegraphics[width=0.7\hsize]{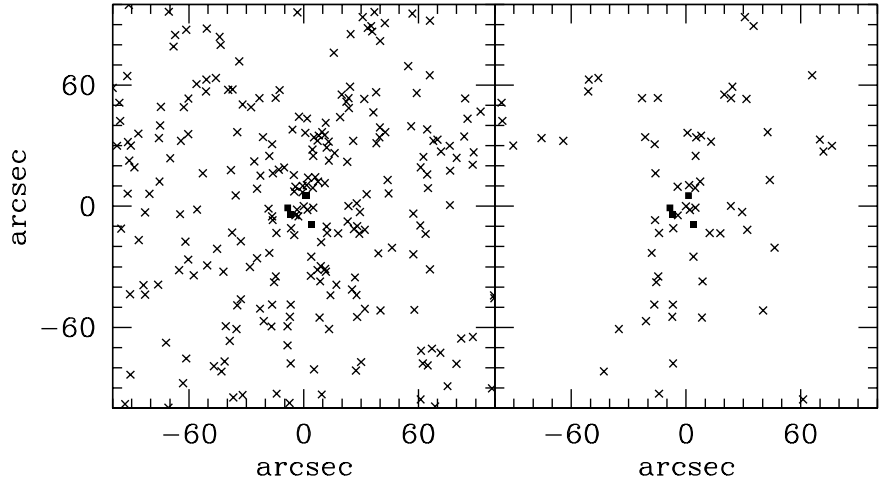}
\caption{Distributions of galaxies brighter than $i=24$ with  ({\it
 right}) and without ({\it left}) the color cut. The origin $(0,0)$ is
 set to the position of the central galaxy G1. Filled squares denote the
 four lensed images. 
 \label{fig:sdss_colorcut}}
\end{center}
\end{figure}

We identify cluster members by their location in color-color space
\citep{dressler92,goto02}. We show $g-r-i$ and $r-i-z$ color-color
diagrams in Figure \ref{fig:sdss_cc_gri}. We restrict the plots to
galaxies brighter than $i=24$ because of the limitation of the
star/galaxy separation.  We make color-color cuts based on the colors
expected of elliptical galaxies \citep{fukugita95}: $g-r>1.5$,
$r-i>0.7$, and $i-z>0.2$ for elliptical galaxies at $z \gtrsim 0.5$. The
galaxy distributions with and without the color cuts are shown in
Figure~\ref{fig:sdss_colorcut}.  The galaxies that survive the color
cuts are concentrated around G1, so we conclude that they are candidate
members of a cluster of galaxies at $z=0.68$ whose center is near G1. We
note that the distribution of candidate cluster members is not spherical
and appears to be elongated North--South.  

\section{Lens Statistics with Spherical Dark Halos}
\label{sec:sdss_stat}
\markboth{CHAPTER \thechapter.
{\MakeUppercase\mychapheadname}}{\thesection.
\MakeUppercase{Lens Statistics with Spherical Dark Halos}}

In this section, we calculate the expected rate of large-separation
lensing in the SDSS quasar sample.  The discovery of SDSS~J1004+4112
allows us to move past the upper limits obtained from previous
large-separation lens searches, although at present we focus on testing
whether the detection of one large-separation lens in the current sample
is consistent with standard theoretical models in the CDM scenario. In
this section, we perform a traditional analysis with the spherical lens
model.  

\subsection{Number of Lensed Quasars in the SDSS}
\label{sec:sdss_sdsscalc}

Because the lensing probability depends on the source redshift and
luminosity, we compute the predicted number of lenses in redshift and
luminosity bins and then sum the bins.  Specifically, let $N(z_j,i^*_k)$
by the number of quasars in a redshift range $z_j-\Delta
z/2<z<z_j+\Delta z/2$ that have a magnitude in the range $i^*_k-\Delta
i^*/2<i^*<i^*_k+\Delta i^*/2$. Then the predicted total number of lensed
quasars is  
\begin{equation}
N_{\rm lens}(>\!\theta)=\sum_{z_j}\sum_{i^*_k} N(z_j,i^*_k)\,
P(>\!\theta; z_j, L(i^*_k)).
\end{equation}
We adopt bins of width $\Delta z=0.1$ and $\Delta i^*=0.2$. The quasar
sample we used comprises 29,811 quasars with mean redshift $\langle
z\rangle=1.45$ (see Figure~\ref{fig:sdss_z_qso}). 

\begin{figure}[th]
\begin{center}
\includegraphics[width=0.5\hsize]{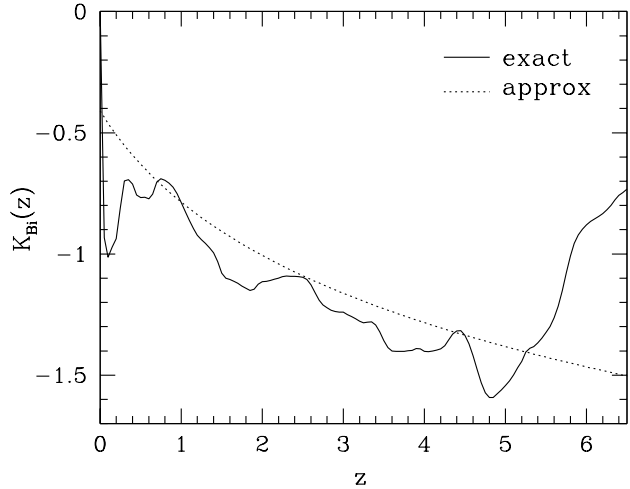}
\caption{The cross-filter K-correction, computed from the SDSS composite
 quasar spectrum created by \citet{vandenberk01}. Dotted line indicate
 the approximation (eq.~[\ref{sdss_kcor_approx}]) with $\alpha_{\rm s}=0.5$.
\label{fig:sdss_kcor}} 
\end{center}
\end{figure}

To calculate the $B$-band absolute luminosity $L(i^*)$ corresponding to
observed magnitude $i^*$, we must estimate the cross-filter K-correction
$K_{Bi}(z)$. The K-correction calculated from the composite quasar
spectrum created from the SDSS sample by \citet{vandenberk01} is shown
in Figure~\ref{fig:sdss_kcor}. As a simplification, one might use the
following approximation: 
\begin{equation}
 K_{Bi}(z)=-2.5(1-\alpha_{\rm s})\log(1+z)-2.5\alpha_{\rm s}\log\left(\frac{7500}{4400}\right)-0.12,
\label{sdss_kcor_approx}
\end{equation}
where the offset 0.12 mainly arises from the difference between
$AB(4400)$ and $B$ magnitudes \citep*[calculated assuming $\alpha_{\rm
s}=0.5$;][]{schmidt95}. Here we use the K-correction directly calculated
from composite quasar spectrum. 

The luminosity function of quasars is needed to compute magnification
bias. We adopt the standard double power law $B$-band luminosity
function \citep*{boyle88} 
\begin{equation}
 \phi_L(z_{\rm S},L)dL=\frac{\phi_*}{[L/L_*(z_{\rm S})]^{\beta_l}+[L/L_*(z_{\rm S})]^{\beta_h}}\frac{dL}{L_*(z_{\rm S})}.
\end{equation}
As a fiducial model of the evolution of the break luminosity, we assume
the form proposed by \citet*{madau99}, 
\begin{equation}
 L_*(z_{\rm S})=L_{*}(0)(1+z_{\rm S})^{\alpha_{\rm s}-1}\frac{e^{\zeta z_{\rm S}}(1+e^{\xi z_*})}{e^{\xi z_{\rm S}}+e^{\xi z_*}},
\end{equation}
where a power-law spectral distribution for quasar spectrum has been
assumed, $f_\nu\propto \nu^{-\alpha_{\rm s}}$. \citet{wyithe02b}
determined the parameters so as to reproduce the low-redshift luminosity
function as well as the space density of high-redshift quasars for a
model with $\beta_h=3.43$ below $z_{\rm S}=3$, $\beta_h=2.58$ above
$z_{\rm S}=3$, and $\beta_l=1.64$. The resulting parameters are
$\phi_*=624\,{\rm Gpc^{-3}}$, $L_{*}(0)=1.50\times 10^{11}~L_\odot$,
$z_*=1.60$, $\zeta=2.65$, and $\xi=3.30$. We call this model LF1. To
estimate the systematic effect, we also use another quasar luminosity
function (LF2) derived by \citet{boyle00}: $\beta_h=3.41$,
$\beta_l=1.58$ and an evolution of the break luminosity $L_*(z_{\rm
S})=L_*(0)10^{k_1z_{\rm S}+k_2z_{\rm S}^2}$ with $k_1=1.36$,
$k_2=-0.27$, and $M_*=-21.15+5\log h$. 

\subsection{Results}

\begin{figure}[tbh]
\begin{center}
\includegraphics[width=0.45\hsize]{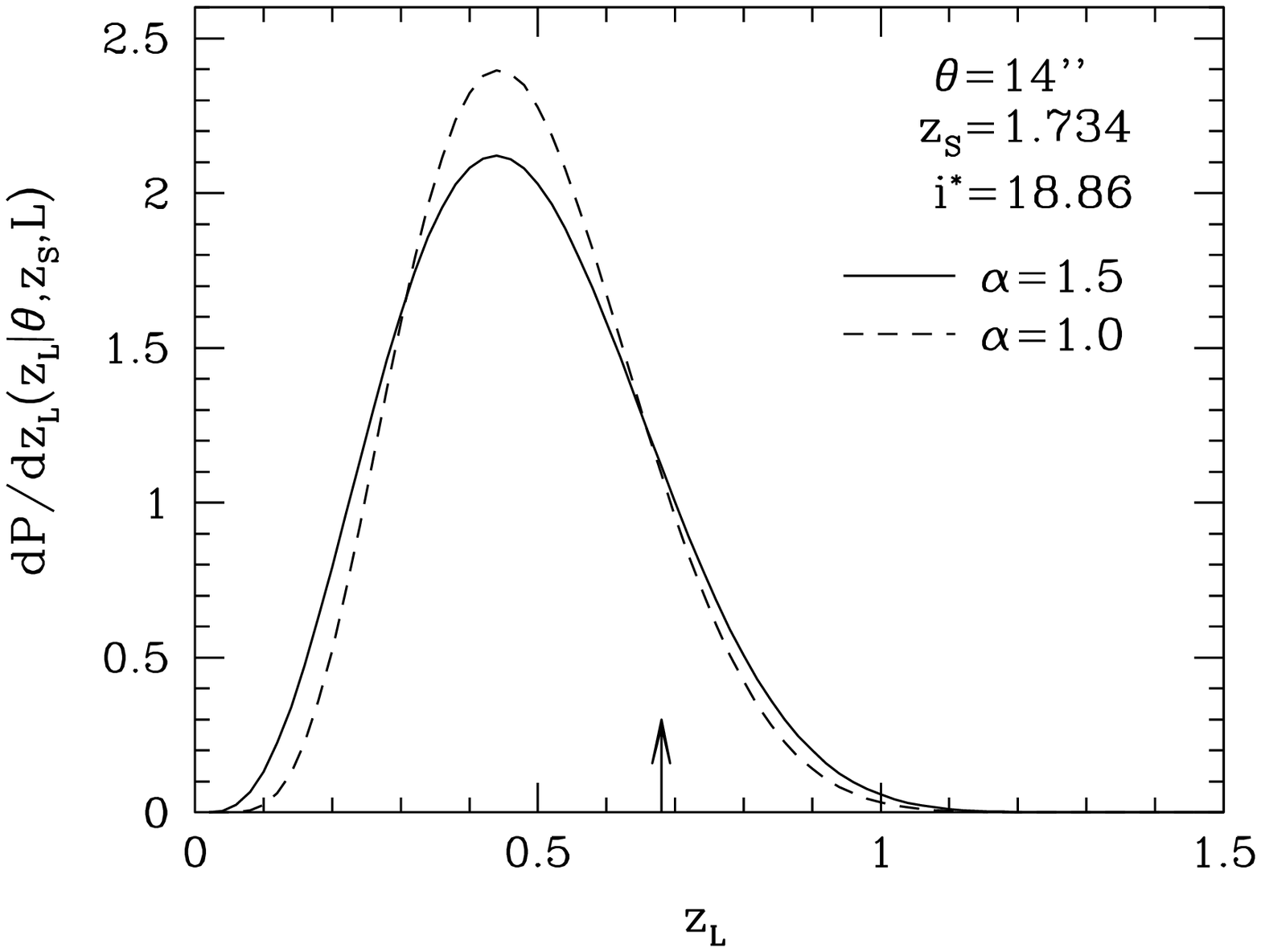}
\includegraphics[width=0.45\hsize]{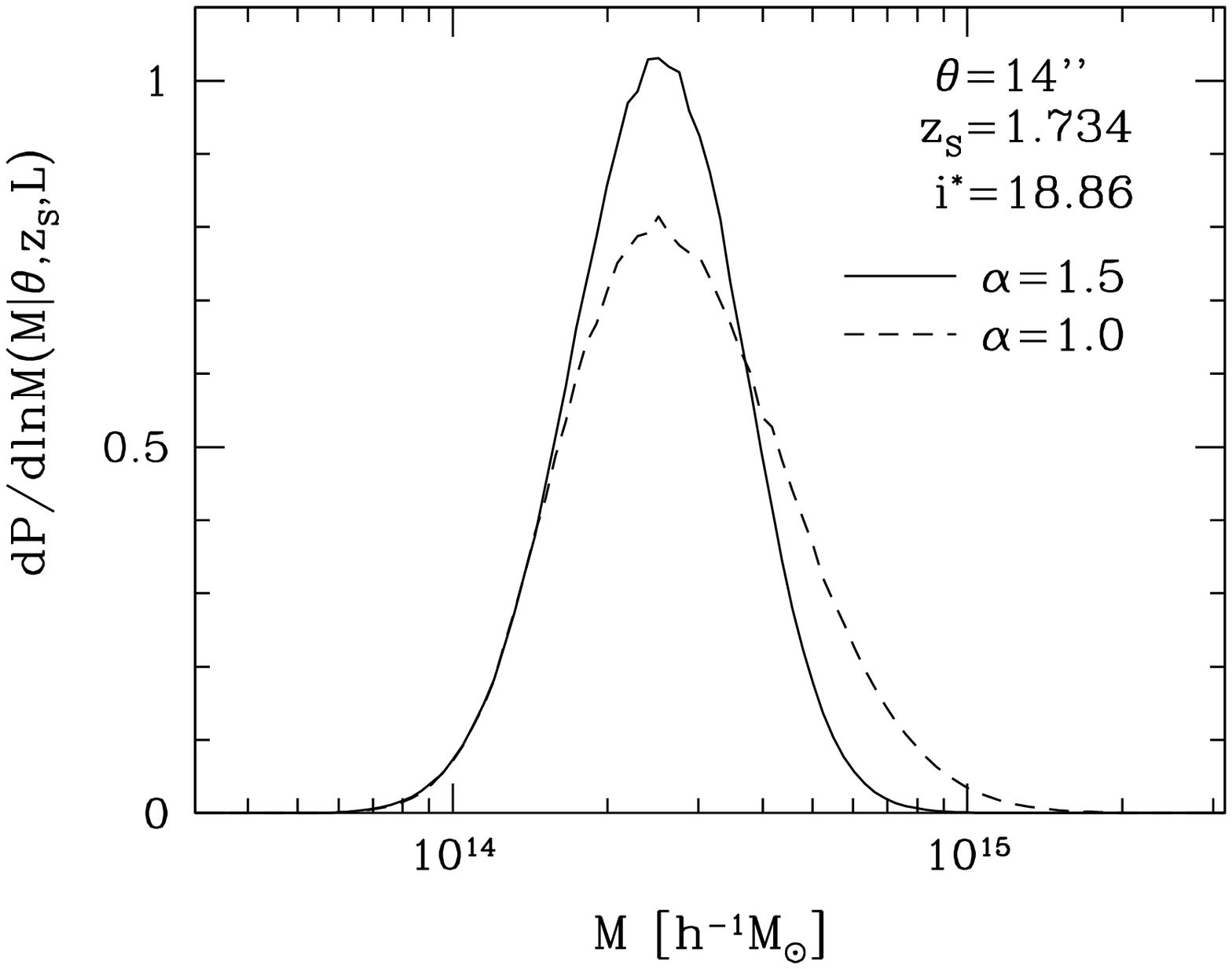}
\caption{Conditional probability distributions for the lens redshift and
 mass in SDSS~J1004+4112, given the image separation $\sim14''$, source
 redshift $z_{\rm S}=1.739$, and apparent magnitude $i^*=18.86$. Solid
 and dashed lines show the probability distributions with $\alpha=1.5$
 and $1.0$, respectively. The arrow shows the measured redshift of the
 lensing cluster.   
 \label{fig:sdss_dist_z}}
\end{center}
\end{figure}

First we show the conditional probability distributions
\begin{equation}
 \frac{dP}{dz_{\rm L}}(z_{\rm L}|\theta,z_{\rm S},L)\equiv\left|\frac{d^2P/dz_{\rm L}d\theta}{dP/d\theta}\right|,
\end{equation}
\begin{equation}
 \frac{dP}{d\ln M}(M|\theta,z_{\rm S},L)\equiv\left|\frac{d^2P/d\ln Md\theta}{dP/d\theta}\right|,
\end{equation}
in order to identify the statistically reasonable ranges of redshift and
mass for the lensing cluster. Figure~\ref{fig:sdss_dist_z} shows the
conditional probability distributions for the lens redshift and lens
mass, given that the gravitational lens system SDSS~J1004+4112 has image
separation $\sim\!14''$, source redshift $z_{\rm S}=1.734$, and apparent
magnitude $i^*=18.86$. We find the most probable lens redshift to be
$z_{\rm L} \sim 0.5$, but the distribution is broad and the measured
redshift $z_{\rm L}=0.68$ is fully consistent with the distribution. We
also find a cluster mass $M\sim 2$--$3\times 10^{14}\,h^{-1}\,M_\odot$
to be most probable for this system. Note that we do not include
information on the measured redshift $z_{\rm L}=0.68$ in the conditional
probability distributions for the lens mass in
Figure~\ref{fig:sdss_dist_z}, which might cause a slight underestimate 
of the lens mass. 

Next we consider the statistical implications of SDSS~J1004+4112.
Although our large-separation lens search is still preliminary, and we
have several other candidates from the current SDSS sample that still
need follow-up observations, we can say that the current sample contains
{\it at least\/} one large-separation lens system. This is enough for
useful constraints because of the complementary constraints available
from the lack of large-separation lenses in previous lens surveys.
Among the previous large-separation lens surveys, we adopt the CLASS
$6''<\theta<15''$ survey comprising a statistically complete sample of
9,284 flat-spectrum radio sources \citep{phillips01b}. For the CLASS
sample, we use a source redshift $z_{\rm S}=1.3$ \citep{marlow00} and a
flux distribution $N(S) dS \propto S^{-2.1} dS$ \citep{phillips01b} to
compute the expected number of large-separation lenses.

\begin{figure}[tbh]
\begin{center}
\includegraphics[width=0.45\hsize]{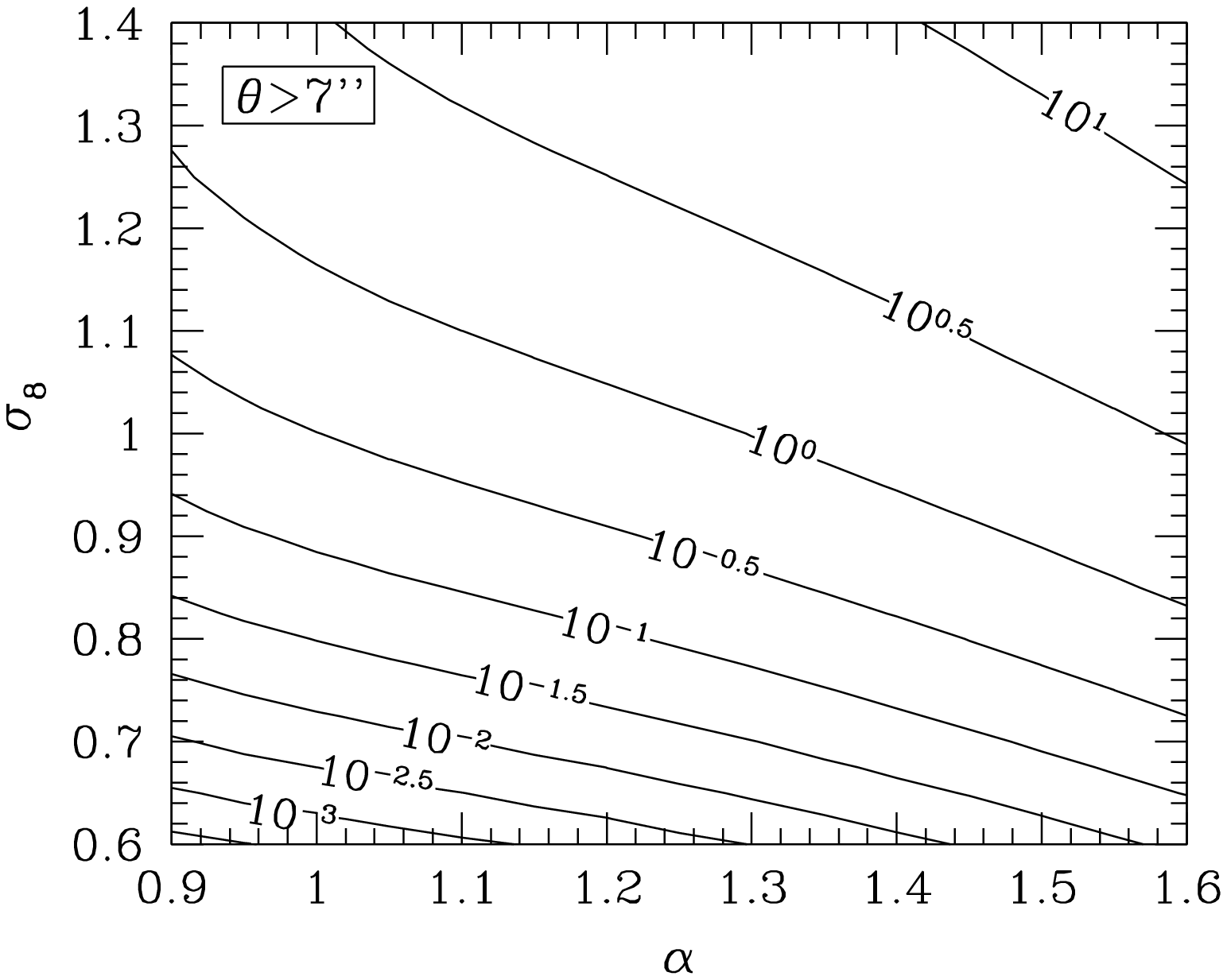}
\includegraphics[width=0.45\hsize]{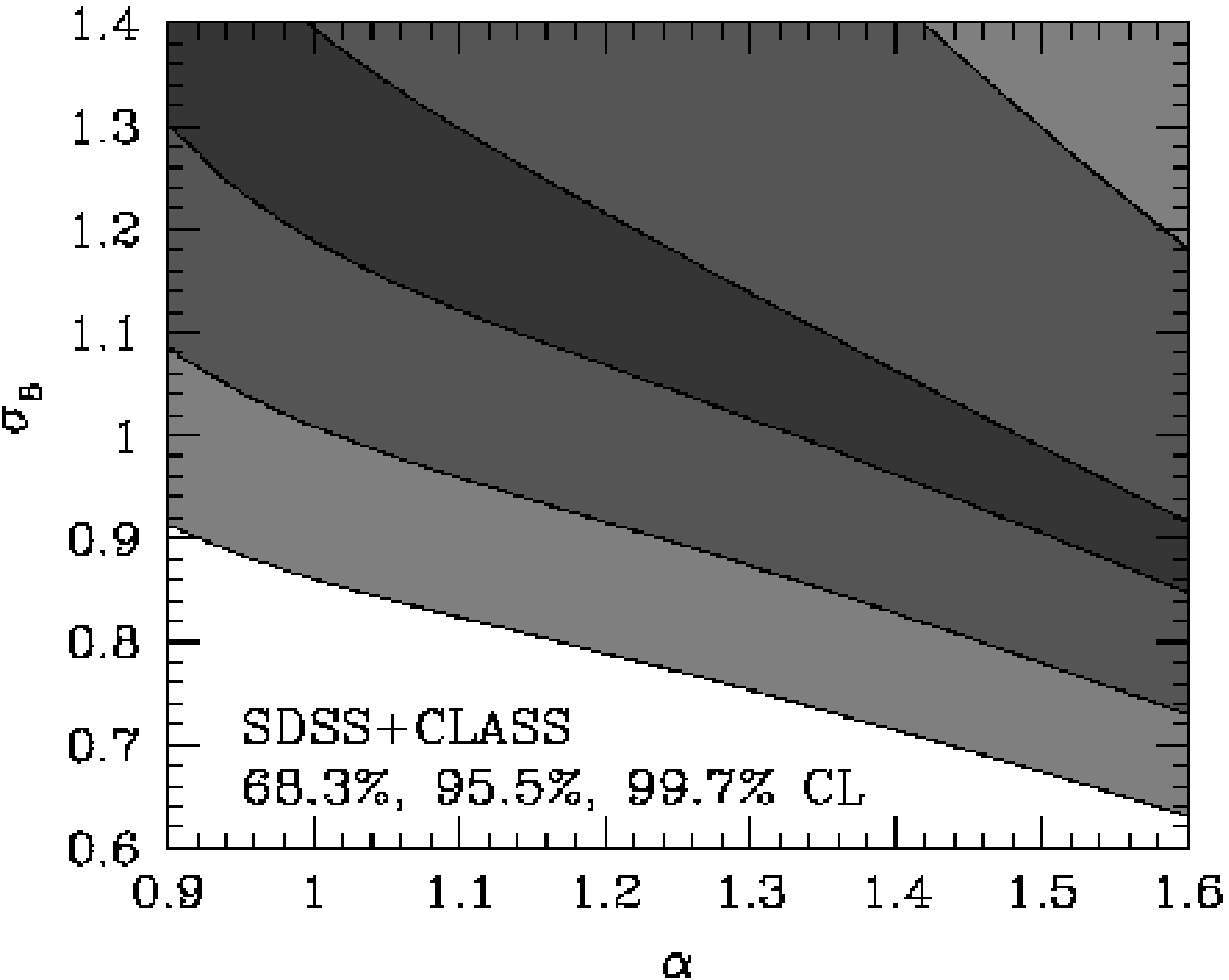}
\caption{Left: Contours of the predicted number of large-separation
 ($\theta>7''$) lenses in the current SDSS sample in the
 $(\alpha,\sigma_8)$ plane. Right: Constraints from both SDSS and CLASS
 in the $(\alpha,\sigma_8)$ plane. The discovery of one large-separation
 ($\theta>7''$) lens in SDSS provides lower limits on $\alpha$ and
 $\sigma_8$, while the lack of large-separation lenses
 ($6''<\theta<15''$) in CLASS yields the upper limit. The regions in
 which both SDSS and CLASS limits are satisfied are shown by the
 shadings. The confidence levels are $68.3\%$, $95.5\%$, and $99.7\%$ in
 the dark, medium, and light shaded regions, respectively.  Both plots
 are based on the spherical lens model.
 \label{fig:sdss_cont}}
\end{center}
\end{figure}

Figure~\ref{fig:sdss_cont} shows contours of the predicted number of
large-separation lenses with $\theta>7''$ in the SDSS quasar sample.
Since  the number of lenses is very sensitive to both the inner slope of
the density profile $\alpha$ and the mass fluctuation normalization
$\sigma_8$ \citep[e.g.,][]{oguri03c}, we draw contours in the
$(\alpha,\sigma_8)$ plane. Constraints from the existence of
SDSS~J1004+4112 together with the lack of large-separation lenses in the
CLASS sample are also shown in Figure~\ref{fig:sdss_cont}. To explain
both observations, we need a relatively large $\alpha$ or $\sigma_8$,
such as $\sigma_8=0.95^{+0.35}_{-0.2}$ (95\% confidence) for
$\alpha=1.5$. This value is fully consistent with other observations
(see Table \ref{table:conc_sigms8}). By contrast, if we adopt $\alpha=1$ then
the required value of $\sigma_8$ is quite large, $\sigma_8 \gtrsim 1.2$.
Thus, our result might be interpreted as implying that dark matter halos
have cusps steeper than $\alpha=1$. Alternatives to collisionless CDM,
such as self-interacting dark matter \citep{spergel00} or warm dark
matter \citep*{colin00,bode01}, tend to produce less concentrated mass
distributions which are effectively expressed by low $\alpha$; such
models would fail to explain the discovery of SDSS~J1004+4112 unless
$\sigma_8$ is unexpectedly large. This result is consistent with results
from strong lensing of galaxies by clusters (i.e., giant arcs), which
also favors the collisionless CDM model (see Chapter \ref{chap:arc}). 
We note that the abundance of large-separation lenses produces a
degeneracy between $\alpha$ and $\sigma_8$ seen in
Figure~\ref{fig:sdss_cont}, but additional statistics such as the
distribution of time delays can break the degeneracy \citep{oguri02a}.   

\begin{table}[bt]
 \begin{center}
  \begin{tabular}{cc}\hline\hline
Models & 
\multicolumn{1}{@{}c@{}}{\begin{tabular}{cccc}
\multicolumn{4}{c}{$N_{\rm lens}(>7'')$ for $(\alpha,\sigma_8)$}\\ 
\hline $(1.0,0.7)$ & $(1.5,0.7)$ & $(1.0,1.1)$ & $(1.5,1.1)$ \end{tabular}} \\
\hline
fiducial model & 
\multicolumn{1}{@{}l@{}}{\begin{tabular}{llll} 
\hspace*{2mm}$0.0055$ & \hspace*{5mm}$0.114$ & \hspace*{7mm}$0.66$ & \hspace*{8mm}$3.9$ \end{tabular}}\\
$c_{\rm Bullock}\rightarrow c_{\rm CHM}$ & 
\multicolumn{1}{@{}l@{}}{\begin{tabular}{llll} 
\hspace*{2mm}$0.0001$ & \hspace*{5mm}$0.019$ & \hspace*{7mm}$0.27$ & \hspace*{8mm}$2.5$ \end{tabular}}\\
$c_{\rm Bullock}\rightarrow c_{\rm JS}$ & 
\multicolumn{1}{@{}l@{}}{\begin{tabular}{llll}
\hspace*{2mm}$0.0008$ & \hspace*{5mm}$0.094$ & \hspace*{7mm}$0.23$ & \hspace*{8mm}$3.1$ \end{tabular}}\\
$dn_{\rm Jenkins}/dM\rightarrow dn_{\rm Evrard}/dM$ & 
\multicolumn{1}{@{}l@{}}{\begin{tabular}{llll}
\hspace*{2mm}$0.0012$ & \hspace*{5mm}$0.033$ & \hspace*{7mm}$0.27$ & $\hspace*{8mm}1.9$ \end{tabular}}\\
$dn_{\rm Jenkins}/dM\rightarrow dn_{\rm STW}/dM$ & 
\multicolumn{1}{@{}l@{}}{\begin{tabular}{llll}
\hspace*{2mm}$0.0083$ & \hspace*{5mm}$0.163$ & \hspace*{7mm}$0.72$ & \hspace*{8mm}$4.1$ \end{tabular}}\\
LF1$\rightarrow$LF2 & 
\multicolumn{1}{@{}l@{}}{\begin{tabular}{llll}
\hspace*{2mm}$0.0048$ & \hspace*{5mm}$0.106$ & \hspace*{7mm}$0.59$ & \hspace*{8mm}$3.7$ \end{tabular}}\\
$\Omega_M=0.3\rightarrow 0.25$ & 
\multicolumn{1}{@{}l@{}}{\begin{tabular}{llll}
\hspace*{2mm}$0.0016$ & \hspace*{5mm}$0.050$ & \hspace*{7mm}$0.37$ & \hspace*{8mm}$2.5$ \end{tabular}}\\
$\Omega_M=0.3\rightarrow 0.35$ & 
\multicolumn{1}{@{}l@{}}{\begin{tabular}{llll}
\hspace*{2mm}$0.0144$ & \hspace*{5mm}$0.227$ & \hspace*{7mm}$1.06$ & \hspace*{8mm}$5.7$ \end{tabular}}\\
\hline
\end{tabular}
\caption{Sensitivity of the predicted number of large-separation lensed
  quasars in the SDSS quasar sample to various changes in the statistics
  calculations. The numbers are computed with the spherical lens model.}  
\label{table:sdss_lens}
 \end{center}
\end{table}

Table~\ref{table:sdss_lens} summarizes the sensitivity of our
predictions to various model parameters.  The uncertainties in our
predictions are no more than a factor of 2--3, dominated by
uncertainties in the concentration parameter and the matter density
$\Omega_M$.  This error roughly corresponds to $\Delta\sigma_8\sim 0.1$,
and so does not significantly change our main results. 

\section{Lens Statistics with Triaxial Dark Halos}
\label{sec:sdss_stat_tri}
\markboth{CHAPTER \thechapter.
{\MakeUppercase\mychapheadname}}{\thesection.
\MakeUppercase{Lens Statistics with Triaxial Dark Halos}}

Next we move on to the triaxial lens model and see whether the discovery
is still consistent with more accurate predictions based on the triaxial
lens model. We also consider whether it is statistically natural that
the first discovered large-separation lens is a quadruple.

We adopt LF1 for the luminosity function of source quasars. In practice
we actually use the cumulative luminosity function
\begin{equation}
\Phi_L(z_{\rm S},L)=\int^\infty_L\phi_L(z_{\rm S},L)dL,
\end{equation}
to calculate the biased cross section (see \S\ref{sec:lat_cross})
\begin{equation}
 B\tilde{\sigma}=\int dXdY \frac{\Phi_L(L/\mu)}{\Phi_L(L)}.
\end{equation}
We approximate that the SDSS quasar sample is a sample with a flux limit
of $i^*=19.7$.\footnote{The SDSS quasar target selection is aimed to
choose quasars with $i^*\lesssim 19.1$ \citep{richards02}. However, we
assume a flux limit of $i^*=19.7$ because there are quasars with
$i^*>19.1$ in the SDSS quasar sample which are, for example, first
targeted as different objects but revealed to be quasars.}  One needs a
cross-filter K-correction to convert observed $i^*$ magnitudes to
absolute $B$-band luminosity.  Here we adopt the approximation
(\ref{sdss_kcor_approx}) with $\alpha_{\rm s}=0.5$. Finally, we
approximate the redshift distribution (see Figure \ref{fig:sdss_z_qso})
with the following Gaussian distribution:
\begin{equation}
p(z_{\rm S})dz_{\rm S}=\frac{1}{1.21}\exp\left\{-\frac{(z-1.45)^2}{2(0.55)^2}\right\}dz_{\rm S}\;\;\;\mbox{($0.6<z_{\rm S}<2.3$)},
\label{sdss_zdist}
\end{equation}
to reduce computational efforts. We have confirmed that the results
using these approximations agree well with those obtained by fully
taking account of the observed redshift and magnitude distributions.

\begin{figure}[th]
\begin{center}
\includegraphics[width=0.7\hsize]{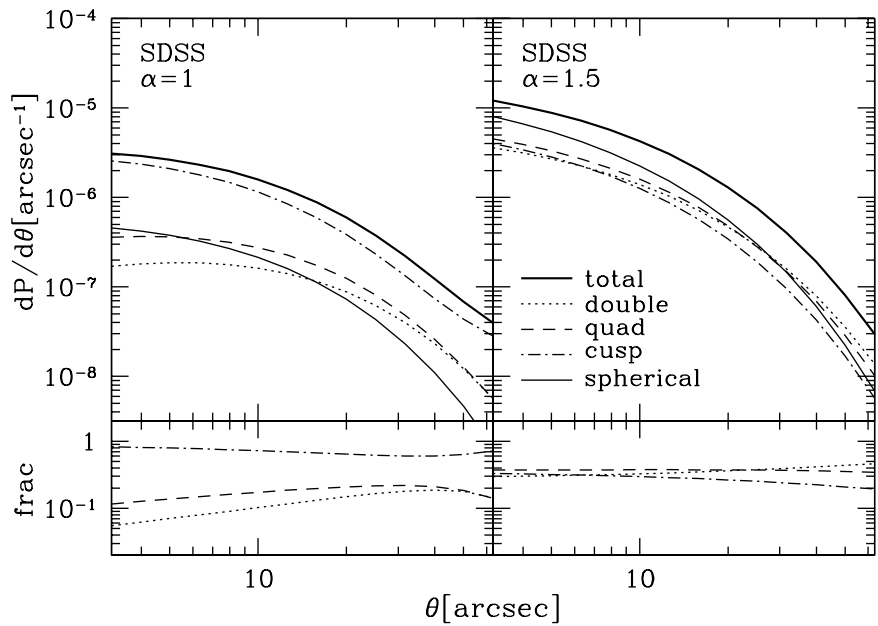}
\caption{Lensing probabilities and image multiplicities for SDSS quasars
 at redshifts $0.6<z_{\rm S}<2.3$. We adopted the triaxial lens model to
 compute the probabilities,  which was described in Chapter \ref{chap:lat}.
\label{fig:sdss_sdss_sep}}
\end{center}
\end{figure}

Figure~\ref{fig:sdss_sdss_sep} shows the lensing probabilities and image
multiplicities as a function of image separation $\theta$ for the SDSS
quasar sample.  Again large-separation lenses are dominated by cusp
configurations for $\alpha=1$, but all three configurations are almost
equally likely for $\alpha=1.5$. Therefore, we confirm that the image
multiplicities in SDSS large-separation lenses will offer interesting
information on the density profile of dark halos.  We can now consider
whether it is statistically natural that the first large-separation lens
in the SDSS is a quadruple lens.  We find that for an image separation
$\theta=15''$ the fractions of quadruple lenses are $\sim$0.2 and
$\sim$0.4 for $\alpha=1$ and $1.5$, respectively. Thus $\alpha=1.5$
could explain the discovery of the quadruple lens somewhat better, but
$\alpha=1$ is also not unnatural. 

Finally, we can use the discovery of SDSS J1004+4112, together with the
lack of large-separation lenses in the CLASS, to constrain the
cosmological parameter $\sigma_8$ describing the normalization of the
density fluctuation power spectrum.  In the previous section, we found
that the discovery of SDSS J1004+4112 required rather large values of
either $\alpha$ or $\sigma_8$, but given the importance of triaxiality
we should revisit this question. For the SDSS, we compute the expected
number of large-separation lenses with $7''<\theta<60''$ among the 29,811
SDSS quasars.  For CLASS, we adopt a power law source luminosity
function with $\beta=2.1$ \citep[see][]{rusin01b}, fix the source
redshift to $z_{\rm S}=1.3$ \citep[see][]{marlow00}, and again calculate the
expected number of lenses with $6''<\theta<15''$ among 9,284
flat-spectrum radio sources \citep{phillips01b}.  We then compute the
likelihood 
\begin{equation}
\mathcal{L}\propto \left(1-e^{-N_{\rm SDSS}}\right) e^{-N_{\rm CLASS}},
\end{equation}
which represents the Poisson probability of observing no
large-separation lenses in CLASS when $N_{\rm CLASS}$ are expected, and
at least one large-separation lens in SDSS when $N_{\rm SDSS}$ are
expected.  (There may be other large-separation lenses in the SDSS
sample that have not yet been identified.)  We consider two
possibilities for the expected number of lenses in the SDSS: (1) the
total number of lenses $N_{\rm tot}$ is used as $N_{\rm SDSS}$; (2) only
the number of quadruple lenses $N_{\rm quad}$ is because the discovered
lens is quadruple. 

\begin{figure}[th]
\begin{center}
\includegraphics[width=0.45\hsize]{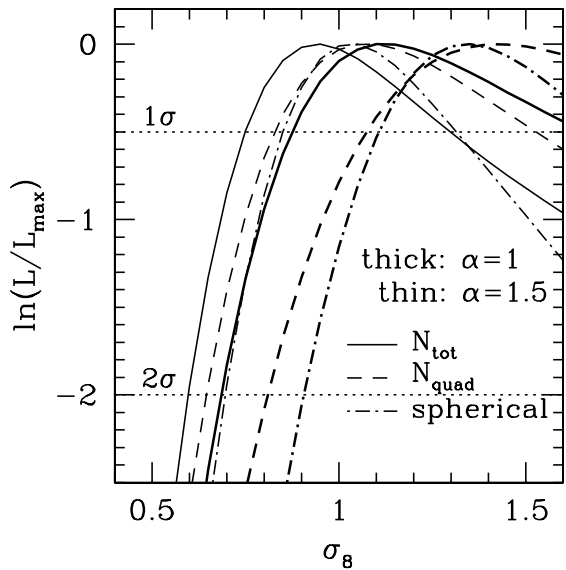}
\caption{Maximum likelihood estimates for $\sigma_8$ with the triaxial
 model, obtained by combining the discovery of SDSS J1004+4112 in SDSS
 with the lack of large-separation lenses in CLASS.  In making
 predictions for SDSS, we consider two cases: the appropriate prediction
 could be the total number of lenses ({\it solid}); or since SDSS
 J1004+4112 is a quad the appropriate quantity could be the number of
 quadruples ({\it dashed}).  The likelihoods for $\alpha=1$ and $1.5$
 are shown by thick and thin lines, respectively.  Results for spherical
 halos are also shown by dash-dotted lines for reference. We note that
 this time we used  $c_{\rm JS}$ as a model for the concentration
 parameter, and we adopt several approximations, thus the result with
 the spherical model is slightly different from the result in \S
 \ref{sec:sdss_stat}. 
\label{fig:sdss_s8}}
\end{center}
\end{figure}

Figure~\ref{fig:sdss_s8} shows the resulting maximum likelihood
constraints on $\sigma_8$.  We find that $\sigma_8 \sim 1$ explains the
data well, although the details depend on the value of $\alpha$ and the
choice of $N_{\rm SDSS}$. Only the case with $\alpha=1$ and $N_{\rm
SDSS}=N_{\rm quad}$ prefers relatively large $\sigma_8$, but
$\sigma_8=1$ is still allowed at the 2$\sigma$ level. At present the
data do not allow particularly strong constraints on $\sigma_8$.
Nevertheless, we can conclude that the status large-separation lenses is
quite consistent with the predictions of CDM given $\sigma_8 \sim 1$.

For comparison, Figure~\ref{fig:sdss_s8} also shows results for
spherical halos.  It turns out that the spherical model overestimates
the value of $\sigma_8$ by $\sim$0.1 for $\alpha=1$ and $\sim$0.2 for
$\alpha=1.5$, compared with cases where we take $N_{\rm SDSS} = N_{\rm
tot}$. Triaxiality is therefore an important systematic effect in these
cases.  Interestingly, the best-fit values of $\sigma_8$ from spherical
models are quite similar to those from triaxial models with $N_{\rm
SDSS} = N_{\rm quad}$.  In both cases, the likelihood function for the
spherical model is narrower than for the triaxial model, indicating that
the spherical model would underestimate the statistical uncertainties in
$\sigma_8$. 

\section{Summary}
\label{sec:sdss_discussion}
\markboth{CHAPTER \thechapter.
{\MakeUppercase\mychapheadname}}{\thesection.
\MakeUppercase{Summary}}

We have searched for large-separation lensed quasars in a sample of
$\sim$30,000 spectroscopically-confirmed SDSS quasars. From the quasar
catalog, we have discovered an excellent quadruple lens candidates
SDSS~J1004+4112. The system consists of four components with image
separation $\theta \sim 14''$. The spectroscopic and photometric
follow-up observations confirm SDSS~J1004+4112 to be a lens system;
spectroscopic observations of four components showed that they have
nearly identical spectra with  $z=1.734$. Deep images and spectroscopy
of nearby galaxies indicate that there is a cluster of galaxies with
$z=0.68$, whose center is likely to be among the four components.  We
conclude that the cluster is responsible for this large-separation lens.
Puzzling differences between the \ion{C}{4} emission line profiles in the
four images are likely to be due to microlensing of part of broad
emission line region, concluded from 7 epoch spectroscopic
observations. 

Although the large-separation lens search in the SDSS is still underway,
we can already constrain model parameters from the discovery of
SDSS~J1004+4112.  The existence of at least one large separation lens in
SDSS places a lower limit on the lensing probability that complements
the upper limits from previous surveys. Both results can be explained if
clusters have the density profiles predicted in the CDM
scenario and moderate values of the mass fluctuation parameter,
$\sigma_8\sim 1$.  Various systematic errors are estimated to be
$\Delta\sigma_8 \sim 0.1$, dominated by uncertainties in the
distribution of the concentration parameter $c_{\rm vir}$ and in the
matter density parameter $\Omega_M$. Still, our overall conclusion is that
the discovery of SDSS~J1004+4112 is fully consistent with the standard
model of structure formation.

We have computed image multiplicities for the SDSS quasar sample from
the triaxial lens model developed in Chapter \ref{chap:lat}.  We predict
that for both $\alpha=1$ and $1.5$ most of the large-separation lenses
should be quadruples or cusps.  The fractions of quadruple lenses at
separations of  $\theta=15''$ are $\sim$0.2 and $\sim$0.4 for $\alpha=1$
and $1.5$, respectively.  This means that it is not surprising that the
first large-separation lens discovered is a quadruple.  Thus, the
discovery of SDSS J1004+4112 can be interpreted as additional support
for CDM on non-linear cluster scales also from this point of view.

In summary, SDSS~J1004+4112 is a fascinating new lens system that
illustrates how large-separation lenses can be used to probe the
properties of clusters and test models of structure formation.  The full
SDSS sample is expected to contain several more large-separation lenses.
The complete sample of lenses, and the distribution of their image
separations, will be extremely useful for understanding the assembly of
structures from galaxies to clusters.  More immediately, the discovery
of a quasar lensed by a cluster of galaxies fulfills long-established
theoretical predictions and resolves uncertainties left by previously
unsuccessful searches. 

\chapter{Conclusion}\label{chap:sum}
\def\mychapheadname{Conclusion}
\markboth{CHAPTER \thechapter.
{\MakeUppercase\mychapheadname}}{}

In this thesis, we have studied the statistics of strong gravitational
lenses, as a test of the CDM model on non-linear scales. Specifically, we
used two complementary statistics, strongly lensed arcs and quasars, to
probe the mass distributions of dark halos. 

Although there have been many analytic studies on cluster-scale lens
statistics, all they adopted the spherical lens model. However, lensing
cross sections are quite sensitive to the deviation from the spherical
symmetry. Since dark halos in the CDM universe are not spherical at all,
it is required to include the non-sphericity of dark halos statistically and
systematically. In addition, the non-spherical modeling enables us to
apply new statistics -- image multiplicities.  Image multiplicities are
significantly affected by both the central concentration and
non-sphericity of dark halos, thus they can be a new powerful test of
the CDM model. 

In order to include the non-sphericity in a systematic manner, we have
adopted the triaxial dark halo model presented by \citet{jing02}, and
studied its lensing properties in detail. We have shown that the
triaxiality has a great impact on cross sections for strong lensing, and 
also on lensing probabilities.

First, we studied arc statistics. We have developed a semi-analytic
method to predict the number of lensed arcs, for the first time taking
proper account of the triaxiality of lensing halos. We have found that
triaxial dark matter halos significantly increase the number of arcs
relative to spherical models; the difference amounts to more than one
order of magnitude while the value of enhancement depends on the
specific properties of density profiles. Then we have compared our
theoretical predictions with the observed number of arcs from $38$ X-ray
selected clusters. In contrast to the previous claims, our triaxial dark
matter halos with inner density profile $\alpha=1.5$ in a
Lambda-dominated CDM universe reproduces well the observation. 
Since both the central concentration and large triaxiality of dark halos
are required to account for the observation, our result may be
interpreted to lend strong support for the CDM paradigm.

Next we developed a model to predict probabilities and image
multiplicities of large-separation lensed quasars with the triaxial lens
model. We have found that the triaxiality significantly enhances lensing
probabilities by a factor of $\sim$2--4, so it cannot be ignored.   
We have pointed out that a significant fraction ($\gtrsim$20\%) of
large-separation lenses should have naked cusp image configurations,
if CDM halos have central density slopes $\alpha \lesssim 1.5$; this
contrasts with lensing by isothermal ($\alpha \approx 2$) galaxies where
naked cusp configurations are rare.  The image multiplicities depend
strongly on the inner density slope $\alpha$: for $\alpha=1$, the naked
cusp fraction is $\gtrsim$60\%; while for $\alpha=1.5$, quadruple lenses
are actually the most probable.  Thus, the image multiplicities in
large-separation lenses offer a simple new probe of the mass
distributions of dark matter halos. 

The main disadvantage of large-separation lensed quasars is that such
lenses have not been discovered so far, despite several explicit searches.
To find first large-separation lens, we searched from the imaging and
spectroscopic data of the SDSS. The imaging data around $\sim 30,000$
quasars, which are more than those used in any previous surveys, are
used to find large separation lenses. From the quasar sample, we have
found the first large-separation lensed quasar SDSS~J1004+4112; the
system consists of four components with image separation $\theta \sim
14''$. The extensive spectroscopic and photometric follow-up
observations unambiguously confirmed that SDSS~J1004+4112 is a true lens
system; spectroscopic observations of four components have showed that
they have nearly identical spectra with  $z=1.734$. Deep images and
spectroscopy of nearby galaxies have succeeded in detecting a lensing
cluster at $z=0.68$, which is responsible for this large-separation
lens. We have also computed the expected probabilities and image
multiplicities for lensed quasars in the SDSS, and argued that the
discovery of the large-separation quadruple lens SDSS J1004+4112
is consistent with expectations for the CDM model. 

Our results both indicate that strong gravitational lenses represent
strong support for the CDM model at small non-linear scales. It is
surprising that the CDM model, which is based on the simple assumption, 
works very well at such non-linear regimes as well as at large scales
where linear theory is applicable. However, we still believe it is quite
important to do as many tests on the CDM model as possible in order to
really understand what the dark matter is; this will lead us to better
understandings of the universe, and if we come to find phenomena which
cannot be explained by the CDM model unambiguously, they will offer a
clue to unveil the dark side of the universe. Although the concordance
model plus the collisionless CDM paradigm seems to be good enough to
explain the evolution and contents of the universe both at large and
small scales, we are just {\it at the end of the beginning} of our journey
to understand the universe.  

\appendix

\chapter{Cosmology Fundamentals}
\label{chap:cosmo}
\def\mychapheadname{Cosmology Fundamentals}
\markboth{CHAPTER \thechapter.
{\MakeUppercase\mychapheadname}}{}

\section{The Dynamics of the Universe}
\markboth{CHAPTER \thechapter.
{\MakeUppercase\mychapheadname}}{\thesection.
\MakeUppercase{The Dynamics of the Universe}}

We adopt the hypothesis that all positions and directions in the
universe is equivalent. This is sometimes called cosmological principle.
Then dynamics of the universe is described by the following Robertson-Walker
metric:  
\begin{equation}
 ds^2=-dt^2+a^2(t)\left\{\frac{dr^2}{1-kr^2}+r^2\left(d\theta^2+\sin^2\theta\,d\varphi^2\right)\right\},
\label{cosmo_rwmetric}
\end{equation}
where $a(t)$ is the scale factor and $k$ is the spatial curvature. The
scale factor $a(t)$ is normalized to unity at the present:
\begin{equation}
 a_0=1,
\end{equation}
where the subscript $0$ means the present value. The scale factor is also
related to the redshift $z=(\lambda_0-\lambda_{\rm s})/\lambda_{\rm s}$
due to the expansion of the universe, where 
$\lambda_{\rm s}$ is the wavelength of a particular emission line at the
source and $\lambda_0$ at the observer, as follows
\begin{equation}
1+z=\frac{1}{a}.
\end{equation}

If we transform the radial coordinate $r$ to $\chi$:
\begin{equation}
 d\chi=\frac{dr}{\sqrt{1-kr^2}},
\end{equation}
then Robertson-Walker metric (\ref{cosmo_rwmetric}) is rewritten as
\begin{equation}
 ds^2=-dt^2+a^2(t)\left\{d\chi^2+f^2(\chi)\left(d\theta^2+\sin^2\theta d\varphi^2\right)\right\},
\end{equation}
where
\begin{eqnarray}
 f(\chi)=
\begin{cases}
 {\displaystyle \frac{1}{\sqrt{k}}\sin\left(\sqrt{k}\chi\right)} & (k>0)\\
 \chi     & (k=0)\\
 {\displaystyle \frac{1}{\sqrt{-k}}\sinh\left(\sqrt{-k}\chi\right)} & (k<0).
\end{cases}
\label{cosmo_f_chi}
\end{eqnarray}
Spaces with $k>0$, $k=0$, and $k<0$ are called closed, flat, and open,
respectively. One may also define a conformal time:
\begin{equation}
 d\eta=\frac{dt}{a(t)}.
\end{equation}
In terms of $\eta$, the metric reduces to
\begin{gather}
 ds^2=a^2(\eta)d\tilde{s}^2=a^2(\eta)\tilde{g}_{\mu\nu}dx^\mu dx^\nu, 
\label{cosmo_rwmetric_ano}\\
 d\tilde{s}^2=-d\eta^2+d\chi^2+f^2(\chi)\left(d\theta^2+\sin^2\theta
 d\varphi^2\right).
\end{gather}
That is, $g_{\mu\nu}$ and $\tilde{g}_{\mu\nu}$ are conformally related.

The energy-momentum tensor $T_{\mu\nu}$ in this spacetime takes the same 
form as the perfect fluid:
\begin{equation}
 T_{\mu\nu}=(\rho+p)u_\mu u_\nu+p g_{\mu\nu},
\label{cosmo_rwenergymomentum}
\end{equation}
where $u_\mu$ denotes the 4-velocity in units of $c$ and $\rho$ and $p$
are the density and the pressure of the universe, respectively. From
equations (\ref{cosmo_rwmetric}) and (\ref{cosmo_rwenergymomentum}), the
Einstein equation reduces to the following two
independent equations: 
\begin{gather}
 \left(\frac{\dot{a}}{a}\right)^2+\frac{k}{a^2}-\frac{\Lambda}{3}=\frac{8\pi G}{3}\rho,
\label{cosmo_friedmann1}\\
 \frac{\ddot{a}}{a}-\frac{\Lambda}{3}=-\frac{4\pi G}{3}\left(\rho+3p\right),
\label{cosmo_friedmann2}
\end{gather}
where a dot denotes the time derivative. To close the equations, one needs
equation of state which is often assumed to be described by the following form:
\begin{equation}
 p=w\rho,
\end{equation}
with $w$ being a dimensionless constant. For instance, relativistic
particles have  $w=1/3$ while non-relativistic particles have $w=0$.
This expression, with equations (\ref{cosmo_friedmann1}) and
(\ref{cosmo_friedmann2}), allows us to describe the density $\rho$ as a
function of scale factor,  
\begin{equation}
 \rho=\rho_0a^{-3(1+w)}.
\end{equation}

Here we define the following cosmological parameters:
\begin{alignat}{2}
 H(a) &\equiv \frac{\dot{a}}{a} & \hspace{5mm} &: \mbox{Hubble parameter,}\\
 \Omega(a) &\equiv \frac{\rho}{\rho_{\rm crit}} \equiv \frac{8\pi G\rho}{3H^2} & &: \mbox{density parameter,} \\
 \Omega_\Lambda(a) &\equiv \frac{\Lambda}{3H^2} & &: \mbox{dimensionless
 cosmological constant,}\\
 \Omega_K(a) &\equiv \frac{k}{a^2H^2} & &: \mbox{curvature parameter,}\\
 q(a) &\equiv -\frac{\ddot{a}a}{\dot{a}^2} & &: \mbox{deceleration parameter.}
\end{alignat}
With these parameters, equations (\ref{cosmo_friedmann1}) and
(\ref{cosmo_friedmann2}) reduce to 
\begin{gather}
\Omega(a)+\Omega_\Lambda(a)-\Omega_K(a)=1,\\
q(a)=\frac{1}{2}\left(1+3w\right)\Omega(a)-\Omega_\Lambda(a).
\end{gather}
In particular, equation (\ref{cosmo_friedmann1}) can be rewritten in terms of
the above cosmological parameters at present and the scale
factor\footnote{Here we neglect the contribution of relativistic
particles (including photons), since their energy fraction in the
universe is sufficiently small at present, i.e., $\Omega_{\rm rela}(a=1)\ll 1$.}: 
\begin{equation}
 H^2(a)=H_0^2\left[\Omega_Ma^{-3}-\Omega_Ka^{-2}+\Omega_\Lambda\right],
\label{cosmo_hubble-scalefactor}
\end{equation}
where we wrote $H(a=1)=H_0$ and $\Omega_M(a=1)=\Omega_M$, etc.
The present value of the Hubble parameter $H_0$ is sometimes expressed as
\begin{equation}
 h=\frac{H_0}{100{\rm km\,s^{-1}\,Mpc^{-1}}}.
\end{equation}
This is the dimensionless Hubble parameter and observationally turns out
to be of order unity.

By solving equation (\ref{cosmo_hubble-scalefactor}), we obtain the
time-dependence of scale factor, $a(t)$. Below we show the results for
three representative cases:
\begin{itemize}
 \item[(a)] $\Omega_M=1$, $\Omega_\Lambda=0$
\begin{equation}
 a(t)=\left(\frac{3}{2}H_0t\right)^{2/3}.
\end{equation}
 \item[(b)] $\Omega_M<1$, $\Omega_\Lambda=0$
\begin{equation}
 a(\theta)=\frac{\Omega_M}{2(1-\Omega_M)}(\cosh\theta-\theta),\;\;\;\; H_0t(\theta)=\frac{\Omega_M}{2(1-\Omega_M)^{3/2}}(\sinh\theta-\theta).
\end{equation}
 \item[(c)] $\Omega_M<1$, $\Omega_\Lambda=1-\Omega_0$
\begin{equation}
 a(t)=\left(\frac{\Omega_M}{1-\Omega_M}\right)^{1/3}\left[\sinh\left(\frac{3\sqrt{1-\Omega_M}}{2}H_0t\right)\right]^{2/3}.
\end{equation}
\end{itemize}

\section{Structure Formation: Linear Perturbation Theory}
\label{sec:cosmo_lin}
\markboth{CHAPTER \thechapter.
{\MakeUppercase\mychapheadname}}{\thesection.
\MakeUppercase{Structure Formation: Linear Perturbation Theory}}
Consider next the evolution of mass fluctuations in the universe. The
evolution of cosmological perturbations should be described in the
framework of general relativity. We, however, focus on the fluctuations
of sub-horizon scale where the Newtonian approach is applicable. In this
scale, the fluid dynamics is governed by the following three equations: 
\begin{eqnarray}
 \mbox{Continuity equation:}&&\displaystyle{\frac{\partial\rho}{\partial t}+\vec{\nabla}\cdot(\rho\vec{u})=0,}\label{cosmo_fluid11}\\
 \mbox{Poisson's equation:}&&\displaystyle{\triangle\Phi=4\pi G\rho,}\\
 \mbox{Euler's equation:}&&\displaystyle{\frac{\partial\vec{u}}{\partial t}+(\vec{u}\cdot\vec{\nabla})\vec{u}=-\frac{1}{\rho}\vec{\nabla}p-\vec{\nabla}\Phi,}\label{cosmo_fluid13}
\end{eqnarray}
where symbols have their usual meanings. In the cosmological situations,
it is useful to rewrite the above in terms of the comoving quantities:
\begin{eqnarray}
 \vec{x}&\equiv&\frac{\vec{r}}{a(t)},\\
 \vec{v}&\equiv&a(t)\dot{\vec{x}},\\
 \delta(\vec{x}, t)&\equiv&\frac{\rho(\vec{x}, t)}{\bar{\rho}(t)}-1,\\
 \phi(\vec{x}, t)&\equiv&\Phi(\vec{x}, t)+\frac{1}{2}a(t)\ddot{a}(t)x^2.
\end{eqnarray}
Then equations (\ref{cosmo_fluid11})-(\ref{cosmo_fluid13}) reduce to
\begin{eqnarray}
 \mbox{Continuity equation:}&&\frac{\partial\delta}{\partial t}+\frac{1}{a}\vec{\nabla}\cdot\left[(1+\delta)\vec{v}\right]=0,\\
 \mbox{Poisson's equation:}&&\triangle\phi=4\pi G\bar{\rho}\delta a^2,\label{cosmo_def_potential}\\
 \mbox{Euler's equation:}&&\frac{\partial\vec{v}}{\partial t}+\frac{1}{a}(\vec{v}\cdot\vec{\nabla})\vec{v}+\frac{\dot{a}}{a}\vec{v}=-\frac{1}{\rho a}\vec{\nabla}p-\frac{1}{a}\vec{\nabla}\phi,
\end{eqnarray}
Now we explore ``linear'' fluctuations, where the amplitude is
sufficiently small ($\delta\ll1$). In practice, this is a good
approximation for fluctuations at scales larger than a few Mpc. In this
case, the first order terms of above equations yield 
\begin{eqnarray}
 \mbox{Continuity equation:}&&\frac{\partial\delta}{\partial t}+\frac{1}{a}\vec{\nabla}\cdot\vec{v}=0,\\
 \mbox{Poisson's equation:}&&\triangle\phi=4\pi G\bar{\rho}\delta a^2,\\
 \mbox{Euler's equation:}&&\frac{\partial\vec{v}}{\partial t}+\frac{\dot{a}}{a}\vec{v}=-\frac{c_{\rm s}^2}{a}\vec{\nabla}\delta-\frac{1}{a}\vec{\nabla}\phi.\label{cosmo_fluid33}
\end{eqnarray}
In equation (\ref{cosmo_fluid33}), $c_{\rm s}$ is the sound velocity $c_{\rm
s}\equiv\sqrt{\partial p/\partial\rho}$. Eliminating $\vec{v}$ and
$\phi$ from the above equations, one obtains the evolution equation of
density fluctuations in linear theory: 
\begin{equation}
 \ddot{\delta}+2\frac{\dot{a}}{a}\dot{\delta}-\left(\frac{c_{\rm s}^2}{a^2}\triangle\delta+4\pi G\bar{\rho}\delta\right)=0.
\label{cosmo_delta_evo_real}
\end{equation}
To solve this equation, it is convenient to consider the Fourier
decomposition of $\delta$: 
\begin{equation}
 \delta_k(t)=\int\delta(\vec{x}, t)e^{-i\vec{k}\cdot\vec{x}}d^3x.
\end{equation}
Then equation (\ref{cosmo_delta_evo_real}) becomes
\begin{equation}
 \ddot{\delta_k}+2\frac{\dot{a}}{a}\dot{\delta_k}+\left(\frac{c_{\rm s}^2k^2}{a^2}-4\pi G\bar{\rho}\right)\delta_k=0.
\label{cosmo_delta_evo_k}
\end{equation}
This is the equation governing the time evolution of linear density
fluctuations. As easily seen from the above equation, $\delta_k$ can
have the growing solution if the wavelength $\lambda$ is longer than
some critical value $\lambda_{\rm J}$, the Jeans length:
\begin{equation}
 \lambda=\frac{2\pi a}{k}>\lambda_{\rm J}\equiv c_{\rm s}\sqrt{\frac{\pi}{G\bar{\rho}}}.
\end{equation}
Equation (\ref{cosmo_delta_evo_k}) indicates that fluctuations with the
wavelength smaller than the Jeans length oscillate because the pressure
gradient balances the gravitational infall.

Actually we are mainly interested in the matter dominated universe when
the pressure is negligible, then equation (\ref{cosmo_delta_evo_k})
reduces to 
\begin{equation}
 \ddot{\delta_k}+2\frac{\dot{a}}{a}\dot{\delta_k}-4\pi G\bar{\rho}\delta_k=0.
\label{cosmo_delta_evo_k_md}
\end{equation}
This differential equation has two independent solutions, a growing
solution denoted by $D_+(t)$ and a decaying solution denoted by
$D_-(t)$. The general solution is expressed as their combination:
\begin{equation}
 \delta_k(t)=C_+(\vec{k})D_+(t)+C_-(\vec{k})D_-(t),
\end{equation}
where $C_+(\vec{k})$ and $C_-(\vec{k})$ are arbitrary
time-independent functions. It is known that $D_+$ and $D_-$ satisfy the
following relations \citep[e.g.,][]{peebles80}:  
\begin{eqnarray}
 D_+(a)&=&H(a)\int_0^a\frac{da}{\left[aH(a)\right]^3},\\
 D_-(a)&=&H(a),
\end{eqnarray}
in terms of the Hubble parameter $H(a)$.

Below we summarize the growing solution $D_+$ for the previous three
representative models.
\begin{itemize}
 \item[(a)] $\Omega_M=1$, $\Omega_\Lambda=0$
\begin{equation}
 D_+(z)=\frac{1}{1+z}.
\end{equation}
 \item[(b)] $\Omega_M<1$, $\Omega_\Lambda=0$
\begin{eqnarray}
D_+(z)&=&\frac{1}{1+z}{}_2F_1(1,2;7/2;1-\Omega_M^{-1}(z))\nonumber\\
&=&1+\frac{3}{x}+3\sqrt{\frac{1+x}{x^3}}\ln\left(\sqrt{1+x}-\sqrt{x}\right),\;\;\;\;x\equiv\frac{1-\Omega_M}{\Omega_M(1+z)},
\end{eqnarray}
where ${}_2F_1(a,b;c;x)$ is the hypergeometric function.
 \item[(c)] $\Omega_M<1$, $\Omega_\Lambda=1-\Omega_M$
\begin{eqnarray}
D_+(z)&=&\frac{1}{1+z}{}_2F_1(1/3,1;11/6;1-\Omega_M^{-1}(z))\nonumber\\
&=&\sqrt{1+\frac{2}{x^3}}\int_0^x\left(\frac{y}{2+y^3}\right)^{3/2}dy,\;\;\;\;x\equiv\frac{\left\{2\left(\Omega_M^{-1}-1\right)\right\}^{1/3}}{1+z}.
\end{eqnarray}
Incidentally $D_+$ is well approximated by the following empirical
fitting function \citep*{carroll92}:
\begin{eqnarray}
 D_+(z)&=&\frac{g(z)}{1+z},\\
 g(z)&=&\frac{5\Omega_M(z)}{2}\left[\Omega_M^{\frac{7}{4}}(z)-\Omega_\Lambda(z)+\left\{1+\frac{\Omega_M(z)}{2}\right\}\left\{1+\frac{\Omega_\Lambda(z)}{70}\right\}\right]^{-1}.\label{cosmo_g_z}
\end{eqnarray}
\end{itemize}

\section{Cosmological Distances}
\markboth{CHAPTER \thechapter.
{\MakeUppercase\mychapheadname}}{\thesection.
\MakeUppercase{Cosmological Distances}}
The meaning of distance is no longer unique in the dynamical universe. 
Therefore we must define the distances according to the situations we 
consider. 

\subsection*{Comoving Distance}
The comoving distance $D_{\rm c}(z)$ is defined by the distance for
which light propagates from $z$ to present in the comoving coordinate.
Since light run along with the light cone (considering the radial
propagation, $d\theta=d\varphi=0$),
\begin{equation}
ds^2=-dt^2+a^2(t)\frac{dr^2}{1-kr^2}=0,
\end{equation}
which reduces to
\begin{equation}
\int_0^{D_{\rm c}(z)}\frac{dr}{\sqrt{1-kr^2}}=\frac{1}{H_0}\int_0^z\frac{dz}{\sqrt{\Omega_M(1+z)^3+(1-\Omega_M-\Omega_\Lambda)(1+z)^2+\Omega_\Lambda}}.
\end{equation}
For the special case, $\Omega_M<1$ and $\Omega_\Lambda=0$, this can be
calculated  analytically \citep{matting58}:
\begin{equation}
D_{\rm c}(z)=\frac{2}{H_0\Omega_M^2(1+z)}\left[2-\Omega_M+\Omega_Mz-\left(2-\Omega_M\right)\sqrt{1+\Omega_Mz}\right].
\end{equation}

\subsection*{Proper Distance}
The light traveling distance is called the proper distance $D_{\rm prop}(z)$,
\begin{equation}
D_{\rm prop}(z)\equiv\int^{t_0}_{t(z)}dt.
\end{equation}
This becomes
\begin{equation}
D_{\rm prop}(z)=\frac{1}{H_0}\int_0^z\frac{dz}{(1+z)\sqrt{\Omega_M(1+z)^3+(1-\Omega_M-\Omega_\Lambda)(1+z)^2+\Omega_\Lambda}}.
\end{equation}
Although this definition of the distance may seem to be the most 
straightforward and reasonable, the proper distance is hardly used in the 
study of cosmology, because it is not related directly with the
observable quantities.

\subsection*{Luminosity Distance}
The luminosity distance $D_{\rm L}(z)$ is defined so as to reproduce the 
relation between the source luminosity $L$ and the observed flux $S$:
\begin{equation}
S\equiv\frac{L}{4\pi D_{\rm L}^2}.
\end{equation}
From this definition, the luminosity distance can be calculated. Firstly, the 
source luminosity $L$ is written in terms of the energy $\delta E$ emitted 
in the time interval $\delta t$,
\begin{equation}
L=\frac{\delta E_{\rm S}}{\delta t_{\rm S}}=(1+z)^2\frac{\delta E_{\rm O}}{\delta t_{\rm O}},
\end{equation}
where subscripts S and O denote the quantities at the source and the 
observer, respectively. On the other hand, the observed flux $S$ becomes
\begin{equation}
S=\frac{\delta E_{\rm O}}{4\pi[D_{\rm c}(z)]^2\delta t_{\rm O}},
\end{equation}
since at present the physical distance from the observer to the source is 
given by $D_{\rm c}(z)$. Combining these equations, the luminosity distance 
is written as
\begin{equation}
D_{\rm L}(z)=(1+z)D_{\rm c}(z).
\end{equation}

\subsection*{Angular Diameter Distance}
The distance most used in the study of the gravitational lensing is the
angular diameter distance $D_{\rm A}(z)$ which is defined by
\begin{equation}
D_{\rm A}(z)\equiv\frac{\delta\ell}{\delta\theta},
\end{equation}
where $\delta\ell$ is the {\it proper} length of some distant object and 
$\delta\theta$ is the angle subtended by that object. From the 
Robertson-Walker metric (\ref{cosmo_rwmetric}), $\delta\ell$ is written as
\begin{equation}
(\delta\ell)^2=a^2(z)[D_{\rm c}(z)]^2(\delta\theta)^2.
\end{equation}
Therefore the angular diameter distance becomes
\begin{equation}
D_{\rm A}(z)=\frac{D_{\rm c}(z)}{1+z}.
\label{cosmo_distance_ad}
\end{equation}

\subsection*{Comparison of the Distances}

\begin{figure}[t]
\begin{center}
\includegraphics[width=0.5\hsize]{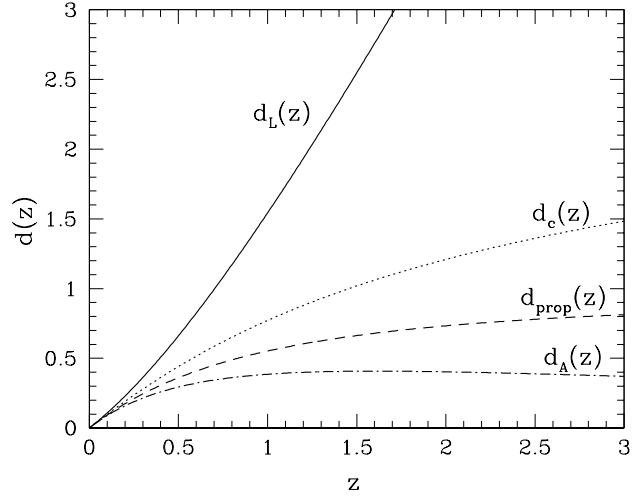}
\caption{Distances used in cosmology. The dimensionless distances
 (eq. [\ref{cosmo_dimensionlessdistance}]) are shown.
\label{fig:cosmo_distance}}
\end{center}
\end{figure}

We compare these distances in Figure \ref{fig:cosmo_distance}. We use
the dimensionless distances $d(z)$ defined by
\begin{equation}
 D(z)\equiv\frac{1}{H_0}d(z)=(2997.9h^{-1}{\rm Mpc})d(z),
\label{cosmo_dimensionlessdistance}
\end{equation}
instead of dimensional distances $D(z)$. The difference of distances in
Figure \ref{fig:cosmo_distance} is also seen by expanding $d(z)$ in $z$
assuming $z\ll 1$:
\begin{eqnarray}
 d_{\rm c}(z)&\sim&z-(q_0+1)\frac{z^2}{2},\\
 d_{\rm prop}(z)&\sim&z-(q_0+2)\frac{z^2}{2},\\
 d_{\rm L}(z)&\sim&z-(q_0-1)\frac{z^2}{2},\\
 d_{\rm A}(z)&\sim&z-(q_0+3)\frac{z^2}{2},
\end{eqnarray}
where $q_0\equiv q(a=1)$ denotes the deceleration parameter at present.
From the above equations, one obtains $d_{\rm L}(z)>d_{\rm c}(z)>d_{\rm
prop}(z)>d_{\rm A}(z)$. One also finds from this expansion that the
observation of the distance-redshift relation at $z\lesssim 1$ well
constrains the deceleration parameter $q_0=\Omega_M/2-\Omega_\Lambda$. 

\section{Mass Functions of Dark Halos}
\label{sec:cosmo_mf}
\markboth{CHAPTER \thechapter.
{\MakeUppercase\mychapheadname}}{\thesection.
\MakeUppercase{Mass Functions of Dark Halos}}

\subsection{Spherical Collapse Model}
\label{sec:cosmo_sc}

Consider a local spherical region of radius $r$ and mass $M$. Then its
equation of motion is 
\begin{equation}
 \frac{d^2r}{dt^2}=-\frac{GM}{r^2}.
\end{equation}
The solution of this equation is described in a parametric form:
\begin{equation}
 r=\frac{GM}{C}(1-\cos\theta),\;\;\;\;\;t=\frac{GM}{C^{3/2}}(\theta-\sin\theta),
\end{equation}
where $C$ is some positive constant. At $\theta\ll 1$, a mean density of
this region becomes 
\begin{equation}
 \bar{\rho}(<r; t)=\frac{1}{6\pi G t^2}\left[1+\frac{3C}{20}\left(\frac{6t}{GM}\right)^{2/3}+\cdots\right],
\end{equation}
which reproduces the behavior of the mean density and the linear
perturbation in the EdS (Einstein de-Sitter; $\Omega_M=1$, $\Omega_\Lambda=0$)
universe. As $\theta$ approaches unity, it starts to deviate from the
linear theory prediction. There are two interesting phases
characterizing the above solution. 

\begin{enumerate}
 \item Turn-around.\\
       The spherical region reaches the maximum radius at $\theta=\pi$. At this
       point, $r$ and $t$ are
\begin{gather}
 r_{\rm ta}=\frac{2GM}{C},\\
 t_{\rm ta}=\frac{\pi GM}{C^{3/2}}.
\end{gather}
       Thus the mean overdensity of this region, 
       $\bar{\rho}(<r_{\rm ta}; t_{\rm ta})/\bar{\rho}(t_{\rm ta})$ and
       the {\it extrapolation} of linear density fluctuation 
       $\delta_{\rm linear}(<r_{\rm ta}; t_{\rm ta})$ are written as 
\begin{gather}
 \frac{\bar{\rho}(<r_{\rm ta}; t_{\rm ta})}{\bar{\rho}(t_{\rm ta})}=\frac{9\pi^2}{16}\sim5.55,\\
 \delta_{\rm linear}(<r_{\rm ta}; t_{\rm ta})=\frac{3\left(6\pi\right)^{2/3}}{20}\sim1.06.
\end{gather}
       It should be noted that they do not depend on either $M$ or $C$.
 \item Virialization.\\
       At $\theta=2\pi$, the spherical region ``collapses'' to a point and
       produces a singularity formally. But in practice the collapse to
       a point never occurs, and after a while the region is supposed to
       be in a virial equilibrium. In this case, $r$ and $t$ are
\begin{gather}
 r_{\rm vir}=\frac{r_{\rm ta}}{2}=\frac{GM}{C},\\
 t_{\rm vir}=2t_{\rm ta}=\frac{2\pi GM}{C^{3/2}},
\end{gather}
       using the virial theorem. The mean density and the linear
       fluctuation are also given as 
\begin{gather}
 \frac{\bar{\rho}(<r_{\rm vir}; t_{\rm vir})}{\bar{\rho}(t_{\rm vir})}\equiv\Delta_{\rm c}=18\pi^2\sim177.7,\\
 \delta_{\rm linear}(<r_{\rm vir}; t_{\rm vir})\equiv\delta_{\rm
 c}=\frac{3\left(12\pi\right)^{2/3}}{20}\sim1.69.
\label{cosmo_criticaldelta}
\end{gather}
      Again these do not depend on $M$ or $C$.
\end{enumerate}

Although the above discussion assures the EdS model, one can generalize
this result  to other models \citep[e.g.,][]{kitayama96}. 

\begin{itemize}
 \item[(a)] $\Omega_M=1$, $\Omega_\Lambda=0$
\begin{gather}
 \Delta_{\rm c}=18\pi^2\sim177.7,\\
\delta_{\rm c}=\frac{3\left(12\pi\right)^{2/3}}{20}\sim1.69.
\end{gather}
 \item[(b)] $\Omega_M<1$, $\Omega_\Lambda=0$
\begin{gather}
 \Delta_{\rm c}=4\pi^2\frac{\left(\cosh\eta_{\rm vir}-\eta_{\rm
 vir}\right)^3}{\left(\sinh\eta_{\rm vir}-\eta_{\rm vir}\right)^2},\\
 \delta_{\rm c}=\frac{3}{2}\left[\frac{3\sinh\eta_{\rm
 vir}\left(\sinh\eta_{\rm vir}-\eta_{\rm
 vir}\right)}{\left(\cosh\eta_{\rm
 vir}-1\right)^2}-2\right]\left[1+\left(\frac{2\pi}{\sinh\eta_{\rm
 vir}-\eta_{\rm vir}}\right)^{2/3}\right],
\end{gather}
	    where $\eta_{\rm vir}\equiv\cosh^{-1}\left(2/\Omega_M(z_{\rm vir})-1\right)$. 
 \item[(c)] $\Omega_M<1$, $\Omega_\Lambda=1-\Omega_M$
\begin{gather}
 \Delta_{\rm c}\simeq 18\pi^2\left(1+0.40929w_{\rm vir}^{0.90524}\right),\\
 \delta_{\rm c}\simeq
 \frac{3\left(12\pi\right)^{2/3}}{20}\left(1+0.012299\log_{10}\Omega_M(z_{\rm
 vir})\right),
\end{gather}
where $w_{\rm vir}\equiv 1/\Omega_M(z_{\rm vir})-1$. These
	    approximations were given by \citet{nakamura97}.  
\end{itemize}

\subsection{Press-Schechter Theory}

The Press-Schechter theory \citep{press74} predicts the abundance of
dark halos in the universe. It is based on simple assumptions, linear 
perturbation theory (\S \ref{sec:cosmo_lin}) and the spherical collapse
model (\S \ref{sec:cosmo_sc}), but agrees well with numerical simulations. 

Consider an initial density field $\delta(\vec{x}, M, z_i)$ smoothed
over the region containing mass $M$. If the initial density field is
random Gaussian, the PDF of $\delta$ at any point is given by  
\begin{equation}
P\left[\delta(M, z_i)\right]=\frac{1}{\left(2\pi\right)^{1/2}\sigma_M(z_i)} \exp\left[-\frac{\delta^2(M, z_i)}{2\sigma^2_M(z_i)}\right],
\end{equation}
where $\sigma_M(z_i)=\sigma(R_M, z_i)$ is the mass variance. From the
discussion of \S \ref{sec:cosmo_sc}, we know that the region is already
virialized at $z$ if the linearly extrapolated density contrast
$\delta_{\rm linear}(M, z)$ exceeds the critical value $\delta_{\rm c}$
(eq. [\ref{cosmo_criticaldelta}]): 
\begin{equation}
 \delta_{\rm linear}(M, z)=\delta(M, z_i)\frac{D_+(z)}{D_+(z_i)}>\delta_{\rm c},
\end{equation}
which reduces to
\begin{equation}
 \delta(M, z_i)>\delta_{\rm c}\frac{D_+(z_i)}{D_+(z)}\equiv\delta_{\rm c}(z, z_i).
\end{equation}
Therefore the probability that the region with mass $M$ is already
virialized is given by
\begin{eqnarray}
 f(M, t)&=&\int^\infty_{\delta_{\rm c}(z, z_i)}P\left[\delta(M, z_i)\right]d\delta\nonumber\\
&=&\frac{1}{2}{\rm erfc}\left(\frac{\delta_{\rm c}(z, z_i)}{\sqrt{2}\sigma_M(z_i)}\right)\nonumber\\
&=&\frac{1}{2}{\rm erfc}\left(\frac{\delta_{\rm c}(z)}{\sqrt{2}\sigma_M}\right),
\label{cosmo_fractionfunction}
\end{eqnarray}
where ${\rm erfc}(x)$ is the complementary error function:
\begin{equation}
 {\rm erfc}(x)\equiv\frac{2}{\sqrt{\pi}}\int_x^\infty e^{-y^2}dy,
\end{equation}
$\delta_{\rm c}(z)\equiv\delta_{\rm c}D_+(z=0)/D_+(z)$, and
$\sigma_M\equiv\sigma_M(z=0)$. This result indicates that the mass function
does not depend on our choice of $z_i$.

From equation (\ref{cosmo_fractionfunction}), we finally obtain the comoving
number density of halos of mass $M$ at time $z$, the mass function:
\begin{eqnarray}
\frac{dn_{\rm PS}}{dM}(M, z)&=&2\frac{\Omega_M\rho_{\rm crit}(0)}{M}\left|\frac{\partial f}{\partial M}\right|\nonumber\\
&=&\sqrt{\frac{2}{\pi}}\frac{\Omega_M\rho_{\rm crit}(0)}{M}\frac{\delta_{\rm c}(z)}{\sigma_M^2}\left|\frac{d\sigma_M}{dM}\right|\exp\left[-\frac{\delta_{\rm c}^2(z)}{2\sigma_M^2}\right].
\label{cosmo_pressschechter}
\end{eqnarray}
The above expression is called the Press-Schechter mass function
\citep{press74}. The extra factor 2 is introduced to match the
normalization: 
\begin{equation}
 \int_0^\infty M\,\frac{dn_{\rm PS}}{dM}(M, z)dM=\rho_0.
\end{equation}
Although the meaning of the factor 2 had been unclear for a long time,
this problem was partially solved by taking account of the region
$\delta>\delta_c$ for some mass but $\delta<\delta_c$ for smaller mass 
\citep{peacock90,bower91,bond91,lacey93}.

\begin{figure}[t]
\begin{center}
\includegraphics[width=0.6\hsize]{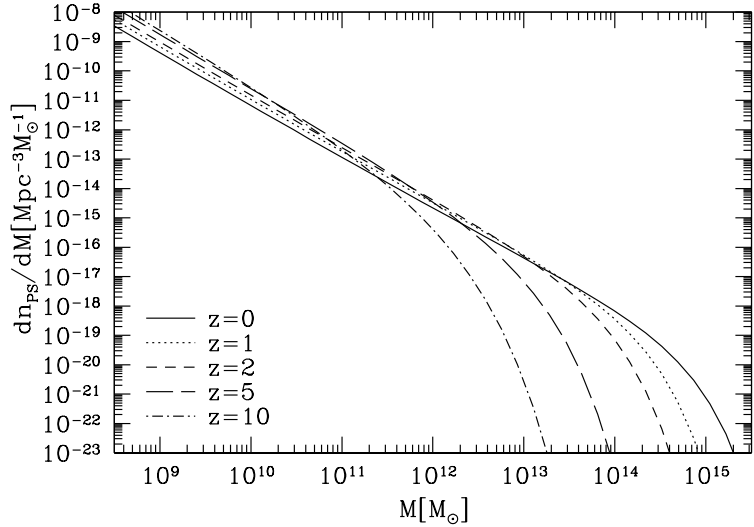}
\caption{The Press-Schechter mass function (eq.
 [\ref{cosmo_pressschechter}]) for different redshifts.
\label{fig:cosmo_ps}}
\end{center}
\end{figure}

We plotted the mass function in Figure \ref{fig:cosmo_ps}. The
number of massive objects monotonically increases as $z$ decreases. 
This means that structure formation in the CDM model is hierarchical,
i.e., massive halos are formed through merging of small halos.

\subsection{Accurate Mass Functions}

Although the Press-Schechter mass function has been tested against
several N-body simulation and shown to be in reasonable agreement
\citep[e.g.,][]{efstathiou88, lacey94}, \citet{jenkins01} reported
the small disagreement against N-body simulation, underpredictions for
the massive halos and overpredictions for the less massive halos. Instead
they derived more accurate fitting formula. Unless otherwise specified,
we adopt equation (B3) of \citet{jenkins01}: 
\begin{equation}
 \frac{dn_{\rm Jenkins}}{dM}=A\frac{\Omega_M\rho_{\rm crit}(0)}{M}\frac{d\ln\sigma_M^{-1}}{dM}\exp\left(-|\ln\sigma_M^{-1}(z)+B|^\epsilon\right),
\label{cosmo_mf_jenkins}
\end{equation}
where $A=0.301$, $B=0.64$, and $\epsilon=3.82$. We use the approximation
of $\sigma_M$ given by \citet{kitayama96}, and the shape parameter
presented by \citet{sugiyama95}. Note that this mass function is given
in terms of the mean overdensity $\Delta_{\rm c}=180$ instead of
$\Delta_{\rm vir}(z)$. Therefore, the mass function should be converted
correctly \citep[e.g.,][]{komatsu02}, when one needs mass functions in
terms of $M_{\rm vir}$. To study uncertainties related to the mass
function we also consider two other possibilities: the mass function
derived in the Hubble volume simulations, $dn_{\rm Evrard}/dM$, which is
given by equation (\ref{cosmo_mf_jenkins}) with $A=0.22$, $B=0.73$,
$\epsilon=3.86$ in terms of the mean overdensity $\Delta_{\rm
c}=200/\Omega(z)$ \citep{evrard02}; and the mass function given by
\citet{sheth99}  
\begin{equation}
 \frac{dn_{\rm STW}}{dM}=A\frac{\Omega_M\rho_{\rm crit}(0)}{M}\left[1+\left(\frac{\sigma_M^2(z)}{a\delta_{\rm c}^2}\right)^p\right]\sqrt{\frac{2a}{\pi}}\frac{\delta_{\rm c}}{\sigma_M(z)}\frac{d\ln\sigma_M^{-1}}{dM}\exp\left(-\frac{a\delta_{\rm c}^2}{2\sigma_M^2(z)}\right),
\end{equation}
with $A=0.29$, $a=0.66$, $p=0.33$ in terms of the mean overdensity
$\Delta_{\rm c}=180$ \citep{white02}.

\chapter{Power Spectrum}
\label{chap:pk}
\def\mychapheadname{Power Spectrum}
\markboth{CHAPTER \thechapter.
{\MakeUppercase\mychapheadname}}{}

\section{Power Spectra in Various Dark Matter Models}
\label{sec:pk_dm}
\markboth{CHAPTER \thechapter.
{\MakeUppercase\mychapheadname}}{\thesection.
\MakeUppercase{Power Spectra in Various Dark Matter Models}}

The power spectrum of density fluctuations is defined by
\begin{equation}
 P(k)\equiv\langle\left|\delta_k^2\right|\rangle.
\end{equation}
If the density field is random-Gaussian, its statistical properties are
completely specified by the power spectrum. In a homogeneous and
isotropic universe, the power spectrum does not depend on the direction
of $\vec{k}$. 

The primordial power spectrum is often assumed to have the scale-free
form:
\begin{equation}
 P_i(k)\propto k^{n_s}.
\end{equation}
The power spectrum with $n_s=1$ is sometimes called the
Harrison-Zel'dovich spectrum \citep{harrison70,zeldovich72} which has an
interesting feature that fluctuations for all wavelengths come into
horizon with the same amplitude. Furthermore, inflationary scenarios
also predict the power spectrum with $n_s\sim 1$.

The power spectrum at the present epoch differs from the simple
power-law due to many physical processes.  Such a modification is
encapsulated in the transfer function defined by
\begin{equation}
 T^2(k, z)\equiv\frac{|\delta_k|^2(z=0)}{|\delta_k|^2(z)D_+^2(z)},
\label{pk_trans}
\end{equation} 
where $D_+(z)$ is the linear growth rate (see \S \ref{sec:cosmo_lin}). With
this definition, present power spectrum is written as
\begin{equation}
 P(k)=A\left\{D_+(z_i)\right\}^2T^2(k,z_i)P_i(k),
\end{equation}
where $A$ is the normalization constant and should be determined from
observations. Below we give results for transfer functions of three
non-baryonic dark matter model, obtained by assuming
$\Omega_b\ll\Omega_M$ and considering linear perturbations
\citep{bardeen86}. 

\begin{itemize}
 \item Hot Dark Matter (HDM)
\begin{gather}
 T_{{\rm HDM}}(k)=\exp\left(-3.9q-2.1q^2\right),\\
 q\equiv\frac{k/\left(h {\rm Mpc^{-1}}\right)}{\Omega_{\nu 0} h},
\end{gather}
 \item Warm Dark Matter (WDM)
\begin{gather}
 T_{{\rm WDM}}(k)=\exp\left\{-\frac{kR_{\rm fw}}{2}-\frac{\left(kR_{\rm
 fw}\right)^2}{2}\right\}\left[1+1.7q+\left(4.3q\right)^{3/2}+q^2\right]^{-1},\\
q\equiv\frac{k/\left(h {\rm Mpc^{-1}}\right)}{\Omega_Mh},\\
R_{\rm fw}=0.2\left(\frac{g_{\rm
 dec}}{100}\right)^{-4/3}\left(\Omega_Mh^2\right)^{-1}{\rm Mpc},
\end{gather}
       where $g_{\rm dec}$ is the effective number of degrees of freedom
       for dark matter particles decoupling, typically 
       $g_{\rm dec}={\cal O}(100)$.  
 \item Cold Dark Matter (CDM)
\begin{gather}
T_{{\rm
 CDM}}(k)=\frac{\ln(1+2.34q)}{2.34q}\left[1+3.89q+\left(16.1q\right)^2+\left(5.46q\right)^3+\left(6.71q\right)^4\right]^{-1/4},
\label{pk_bbks}\\
q\equiv\frac{k/\left(h {\rm Mpc^{-1}}\right)}{\Gamma},
\end{gather}
       where $\Gamma$ is called a shape parameter. A general form for
       the shape parameter was obtained by \citet{sugiyama95} as
\begin{equation}
 \Gamma=\Omega_Mh\exp\left[-\Omega_b\left(1+\frac{\sqrt{2h}}{\Omega_M}\right)\right].
\label{pk_shape}
\end{equation}
\end{itemize}

\begin{figure}[t]
\begin{center}
\includegraphics[width=0.7\hsize]{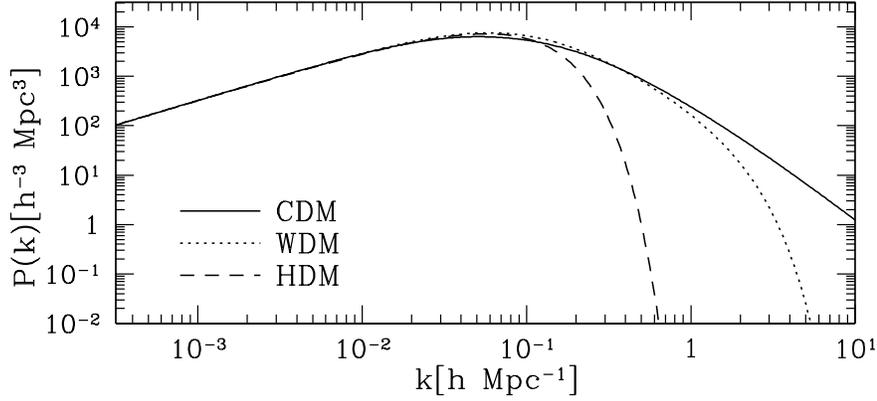}
\caption{Power spectra for three dark matter models; CDM, WDM, and HDM. 
\label{fig:pk_dm}}
\end{center}
\end{figure}

Power spectra for three dark matter models are plotted in Figure
\ref{fig:pk_dm}. As clearly exhibited in these plots, there is a strong
cut-off at large $k$ (small scale) in WDM and HDM models. That
wavenumber corresponds to their free streaming scale. In CDM model,
there is a characteristic scale at $\sim 10 h^{-1}$Mpc which is the
horizon size at matter-radiation equality, $\lambda_{\rm eq}$. This is
understood as follows. Fluctuations with scales larger than
$\lambda_{\rm eq}$ enter the horizon when the universe is matter
dominant. Such fluctuations grow as soon as they enter the horizon.
Therefore the power spectrum at those scales does not change the shape;
$P(k)\sim k^{n_s}$. On the other hand, fluctuations with scales less
than $\lambda_{\rm eq}$ enter the horizon when the universe is still
radiation dominant. Fluctuations in that epoch do not grow in practice
because the Jeans length of the radiation is very large (almost same as
the horizon size) and this suppresses fluctuations of dark matter
(sometimes called ``stagspansion''). Smaller fluctuations suffer from
the longer period of the stagspansion, and result in the modification of
spectrum as $P(k)\sim k^{n_s-4}$. Therefore the power spectrum becomes 
\begin{eqnarray}
 P(k)=\begin{cases}
       k^{n_s}   & (k\ll k_{\rm eq})\\
       k^{n_s-4} & (k\gg k_{\rm eq}),
      \end{cases}
\end{eqnarray}
where $k_{\rm eq}=2\pi/\lambda_{\rm eq}$. This explains the asymptotic
behavior of the fitting formula (\ref{pk_bbks}).

\section{Effects of Baryon}
\markboth{CHAPTER \thechapter.
{\MakeUppercase\mychapheadname}}{\thesection.
\MakeUppercase{Effects of Baryon}}

\begin{figure}[p]
\begin{center}
\includegraphics[width=0.95\hsize]{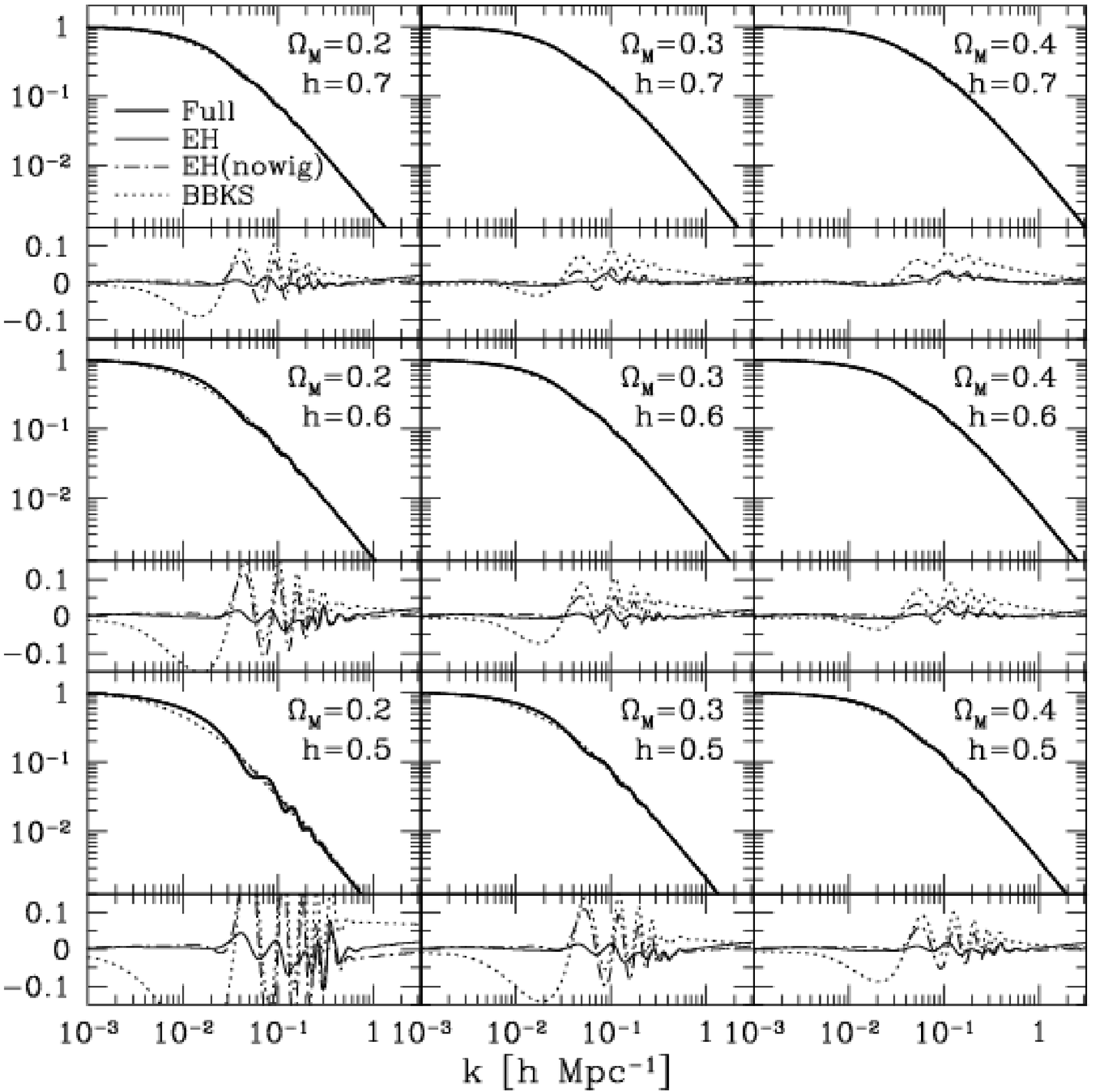}
\caption{Comparison of transfer functions (eq. [\ref{pk_trans}]) for
 different values of $\Omega_M$ and $h$ (we fix $\Omega_bh^2=0.022$). 
 We compare transfer functions obtained from the Boltzmann equation ({\it
 Full}), fitting formula of \citet{eisenstein98} including baryon wiggle
 ({\it EH}), fitting formula of \citet{eisenstein98} without baryon
 wiggle ({\it EH(nowig)}), and fitting formula of \citet{bardeen86} with
 a shape parameter of equation (\ref{pk_shape}) ({\it BBKS}). For each
 set of cosmological parameters, we show transfer functions ({\it
 upper}) and the fractional deviations from the ``Full'' transfer
 function ({\it lower}).
\label{fig:pk_trans}}
\end{center}
\end{figure}

Although the effect of baryon can be included as a simple modification
of the shape parameter, (eq. [\ref{pk_shape}]), in reality including
baryon yields rather complicated features in the power spectrum, such as
oscillations and a sharp suppression in the transfer function below
sound horizon.  The fitting formula including these effects was
obtained by \citet{eisenstein98}. In Figure \ref{fig:pk_trans}, we
compare several fitting formulae of transfer functions with direct
integrals of the Boltzmann equations with CMBEASY \citep{doran03}, in
order to see how accurate these fitting formulae are. The amount of
errors depends on adopted cosmological parameters, but around the
concordance model the accuracy of the transfer function of
\citet{bardeen86} plus equation (\ref{pk_shape}) is not so good,
$\lesssim 10\%$. Fitting formulae of \citet{eisenstein98} are much
accurate, but still contains errors of a few \%. In using power spectra
in research, it should be kept in mind that the approximations sometimes
contain larger errors than accuracies of results.  

\section{Mass Variance}
\markboth{CHAPTER \thechapter.
{\MakeUppercase\mychapheadname}}{\thesection.
\MakeUppercase{Mass Variance}}

The amplitude of density fluctuations is characterized by the mass
variance, $\sigma^2$. First we consider a density fluctuation averaged
over volume $V=3M/4\pi\bar{\rho}=R_M^3$: 
\begin{eqnarray}
 \delta\left(\vec{x}, R_M, z\right)&=&\int\delta\left(\vec{y}, z\right)W_M\left(\left|\vec{x}-\vec{y}\right|; R_M\right)d^3y,\\
&=&\frac{1}{\left(2\pi\right)^3}\int\delta_k\left(z\right)\tilde{W}_M\left(kR_M\right) e^{i\vec{k}\cdot\vec{x}}d^3k,
\end{eqnarray}
where $W_M(r; R_M)$ and $\tilde{W}_M(kR_M)$ are the spatial window
function and its Fourier transform, respectively. Typical choices for
the window functions include
\begin{enumerate}
 \item Top-hat:
\begin{eqnarray}
 W_M(r; R_M)&=&\frac{3}{4\pi R_M^3}\Theta(R_M-r),\\
 \tilde{W}_M(kR_M)&=&\frac{3}{\left(kR_M\right)^3}\left[\sin\left(kR_M\right)-kR_M\cos\left(kR_M\right)\right],
\end{eqnarray}
with $\Theta(r)$ being the Heaviside step function.
 \item Gaussian:
\begin{eqnarray}
 W_M(r; R_M)&=&\frac{1}{\left(2\pi\right)^{3/2}R_M^3}\exp\left(-\frac{r^2}{2R_M^2}\right),\\
 \tilde{W}_M(kR_M)&=&\exp\left(-\frac{\left(kR_M\right)^2}{2}\right).
\end{eqnarray}
\end{enumerate}
Then mass variance is given by
\begin{equation}
 \sigma_M(z)\equiv\sigma^2(R_M, z)\equiv\langle\left|\delta\left(\vec{x}, R_M, z\right)\right|^2\rangle=\frac{1}{\left(2\pi\right)^3}\int P(k, t)\tilde{W}_M^2(kR_M)d^3k.
\end{equation}

In the CDM cosmology with $n_s=1$, its useful fitting formula is
\citep{kitayama96}
\begin{equation}
 \sigma_M\propto\left[1+2.208m^p-0.7668m^{2p}+0.7949m^{3p}\right]^{-2/9p},
\end{equation}
where $p=0.0873$ and $m\equiv M(\Gamma
h)^3/\left(\Omega_Mh^2\right)/10^{12}M_\odot$. The above approximation
and its derivative give accurate fits in the range $10^{-7}\lesssim 
m\lesssim 10^5$.

It is conventional to use $\sigma_8$:
\begin{equation}
 \sigma_8\equiv\sigma(R_M=8h^{-1}{\rm Mpc}, z=0),
\end{equation}
as the parameter characterizing the normalization of the power spectrum.
The reason for using the mass variance of $R_M=8h^{-1}{\rm Mpc}$ is that
the observed two-point correlation function of galaxies translates to
$\sigma_8=1$ if galaxies perfectly trace mass. 

\chapter{Models for Dark Matter}
\label{chap:dm}
\def\mychapheadname{Models for Dark Matter}
\markboth{CHAPTER \thechapter.
{\MakeUppercase\mychapheadname}}{}

\section{Candidates of Cold Dark Matter}
\markboth{CHAPTER \thechapter.
{\MakeUppercase\mychapheadname}}{\thesection.
\MakeUppercase{Candidates of Cold Dark Matter}}

In this section, we review several specific candidates of cold dark matter.

\subsection*{Axion}
Axions were proposed to solve the strong CP problem, i.e., large
CP-violation due to degenerate vacua of $SU(3)_{\rm color}$ gauge group
\citep{peccei77}. Although the mass of axions is very small, axions can
be cold dark matter because they are generated non-thermally (they form
a Bose-Einstein condensate) with small momentum, $\ll{\rm keV}$. The
mass of axions has been constrained by making use of the coupling with
photon-photon, but still axions with mass $\mu{\rm eV}\lesssim m\lesssim
1{\rm meV}$ are good candidate of cold dark matter.

\subsection*{Neutralino}

Neutralinos are particles predicted in the framework of supersymmetry
(SUSY); the SUSY scenario solves the hierarchy problem by assuming a boson
partner for each fermion (and vise versa). Now it is assumed that the
SUSY is broken, thus SUSY partners are much heavier than standard
particles. Although ``massive'' sometimes means unstable, the lightest
superpartners are stable because of  the $R$-parity 
conservation which require that superpartners are produced/destroyed
only in pairs. Thus the (lightest) neutralinos, which are the superpartner
of neutrinos, are ideal candidate of cold dark matter. 

An important implication of neutralino dark matter is that it
annihilates, though the cross section is thought to be very small,
$\sigma\lesssim 10^{-35}{\rm cm^2}$. Therefore, neutralino dark matter
predicts several astrophysically interesting phenomena such as
high-energy neutrinos from the Sun/Earth \citep[e.g.,][]{kamionkowski95}
and gamma-rays from the galactic centers \citep[e.g.,][]{gondolo99,boehm04}.  

\subsection*{WIMPZILLA}

WIMPZILLAs are very massive relic particles produced gravitationally at
the end of inflation \citep*[e.g.,][]{chung98}. Gravitational
production of particles is caused by the change of vacuum states from
the inflationary (de-Sitter) phase to the matter-dominant phase. 
They can be cold dark matter if their mass is  $\sim 10^{13}{\rm GeV}$. 

\subsection*{Soliton}

The SUSY model permits non-topological solitions, ($Q$-balls), which is
stable and carry a large number of $U(1)$ charge. $Q$-balls may be
stable or unstable depending on the situation, and stable $Q$-balls can
be cold dark matter \citep{kusenko98}. 

\section{Alternatives to Cold Dark Matter}
\markboth{CHAPTER \thechapter.
{\MakeUppercase\mychapheadname}}{\thesection.
\MakeUppercase{Alternatives to Cold Dark Matter}}

In this section, we review alternatives to the CDM model, which are
(mainly) aimed to solve the possible problems in the CDM model at small
scales (see \S \ref{sec:halo_crisis}). This section is partly based on the
review of \citet{ostriker03}.

\subsection*{Self-Interacting Dark Matter}

In this model, dark matter particles have a significant self-scattering
cross section comparable to the nucleon-nucleon cross section,
$\sigma_{XX}/m_X\sim 10^{-24}{\rm cm^2GeV^{-1}}$ \citep{spergel00}.
Therefore at the high-density regions (e.g., cores of dark halos),
collisions between dark matter particles significantly modify the
structure. Basically, collisions make the dark halos much less centrally
concentrated and much rounder. In addition, collisions help to reduce
the number of substructures in dark halos.  

\subsection*{Warm Dark Matter}

As shown in \S \ref{sec:pk_dm}, warm dark matter has a larger velocity
dispersion, and hence induces a cutoff in the power spectrum. Thus,
while the cluster scale objects are not affected so much, it has a
significant effect on small-scale structure; it lowers the number of
small halos, and also makes small halos less centrally concentrated
\citep{colin00,bode01}. Examples of warm dark matter are gravitinos and
keV-scale sterile neutrinos.

\subsection*{Repulsive Dark Matter}

If we consider a condensate of massive bosons interacting via a
repulsive interacting potential, this results in a minimum length-scale
for bound objects, and to super-fluidity. From these properties, both
less central concentration and smaller substructures of galactic dark
halos can be achieved \citep{goodman00}. 

\subsection*{Fuzzy Dark Matter}

This model assumes that dark matter takes the form of ultra-light scalar
particles ($m\sim 10^{-22}{\rm eV}$). The Compton wavelength becomes so
large ($\sim$ the size of the galaxy core) that it provides soft cores
and suppressed small-scale structures \citep*{hu00}.

\subsection*{Self-Annihilating Dark Matter}

Instead of self-scattering, this model considers the significant
self-annihilating cross section, $\sigma/m \sim 10^{-29}{\rm
cm^2}{\rm GeV^{-1}}$ \citep*{kaplinghat00}. This annihilation removes
central cusps of halos nearly independently with their mass. As an
example of such model, \citet{riotto00} considered a self-interacting
Bose-field. 

\subsection*{Fluid Dark Matter}

In this mode, dark matter is regarded as a classical scalar field that
interacts only with gravity and with itself. The dark matter behaves
like an ideal fluid with pressure that is a function only of the mass
density. This model could have effects on the core radii and  on the
abundance of low-mass objects \citep{peebles00}.

\subsection*{Decaying Dark Matter}

\begin{figure}[t]
\begin{center}
\includegraphics[width=0.50\hsize]{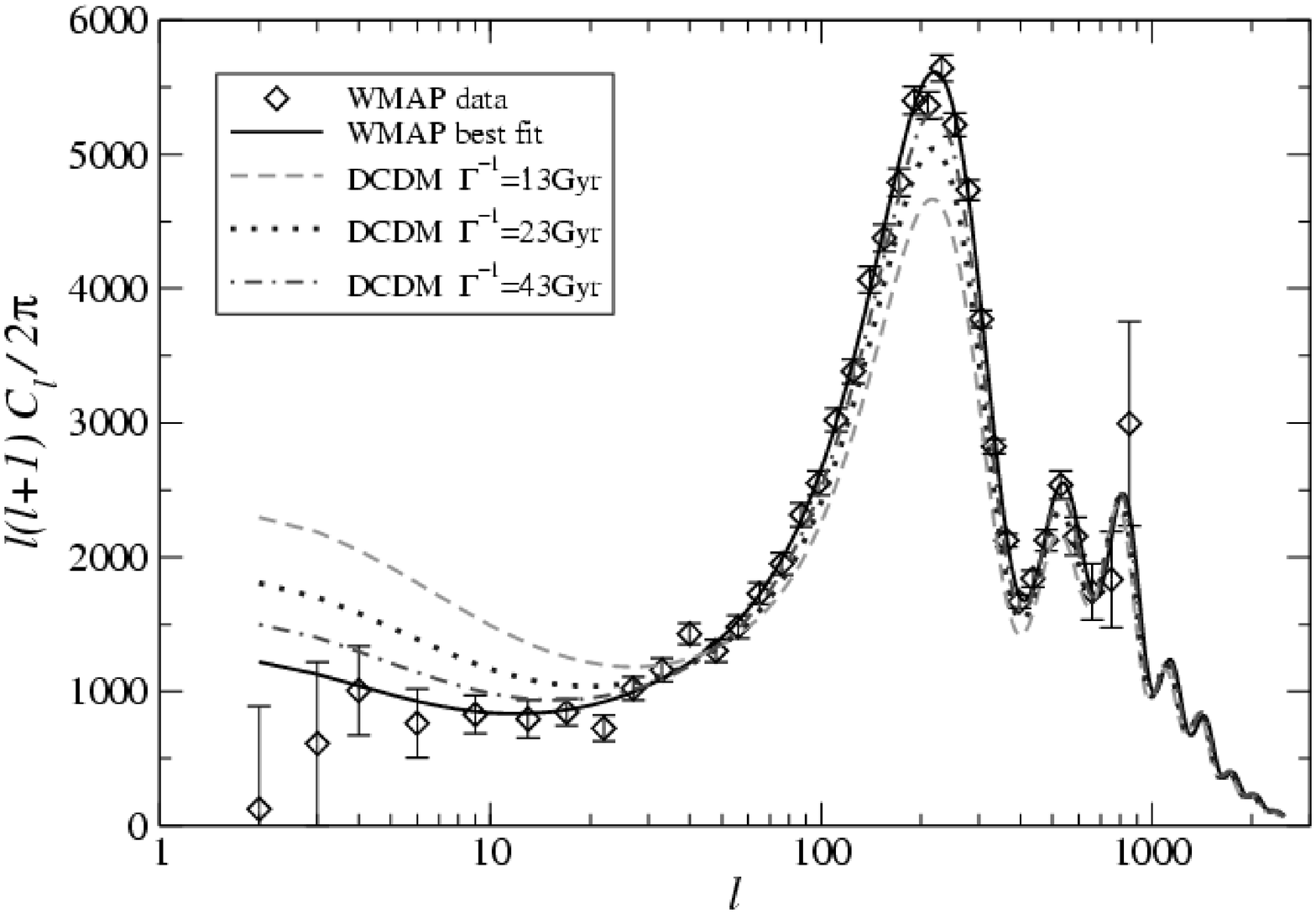}
\includegraphics[width=0.4\hsize]{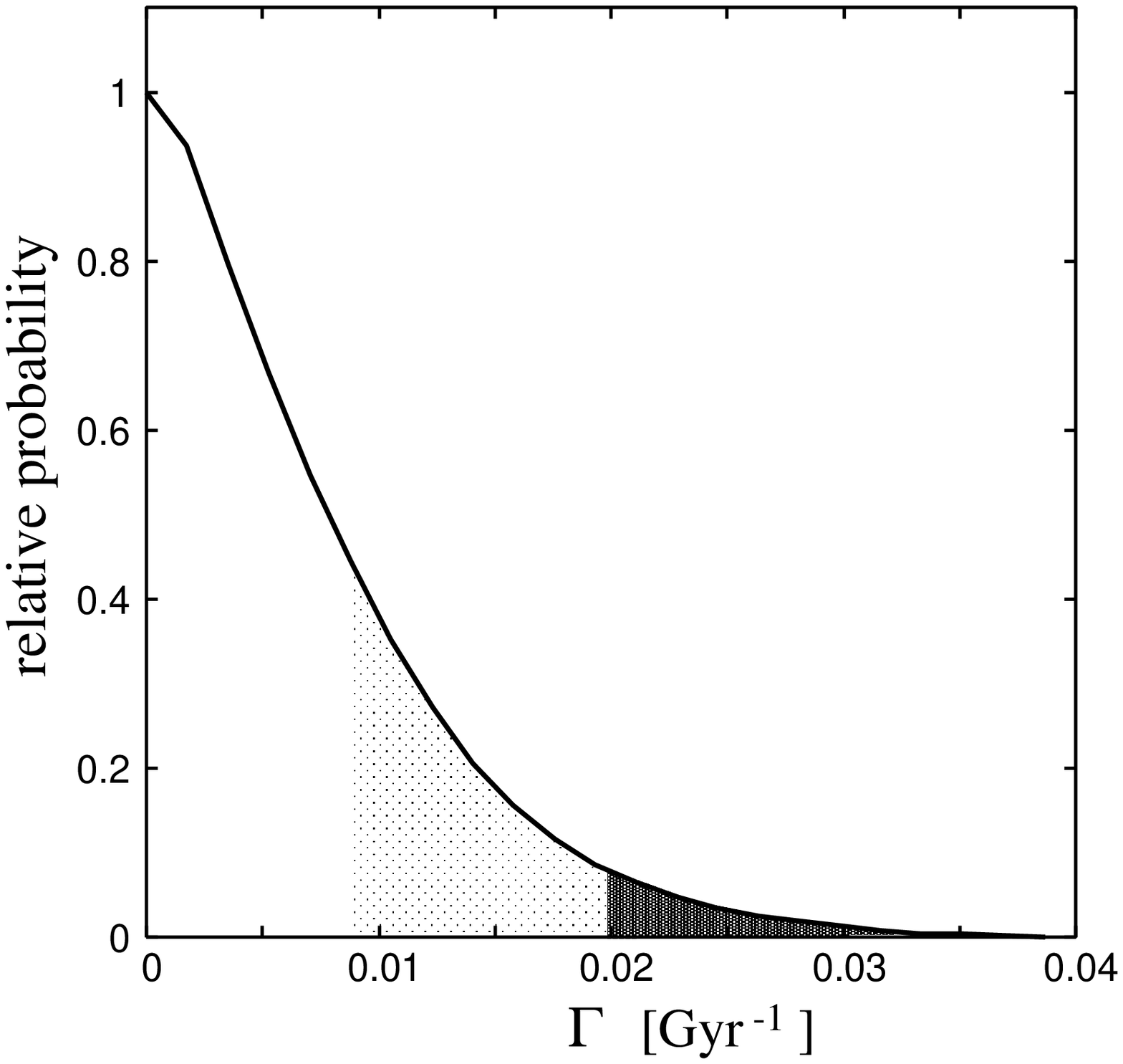}
\caption{Left: The CMB angular power spectrum with and without decays of
 CDM particles. The decays basically enhance $C_l$ at low-$l$ through
 late ISW effect, and also lower first peak because of larger
 $\Omega_Mh^2$ at early epochs. Right: The constraint on the decay rate
 $\Gamma$, marginalized over the other parameters. Confidence limits of 68\%
 and 95.4\% are shown by shaded regions. These Figures are taken from
 \citet*{ichiki04}. \label{fig:dm_ddm}}
\end{center}
\end{figure}

\citet{cen01} proposed decaying dark matter as a solution to the
possible CDM problems; if dark matter particles in dark halos decay into
relativistic particles, then the central densities are lowered due to the
re-expansion of halos. In addition, \citet*{ichiki03} and
\citet{oguri03d} showed that introducing decays of dark matter can
improve the fits of observational data sets of type-Ia supernovae,
mass-to-light ratios and X-ray gas fraction of clusters, and the
evolution of the cluster abundance \citep*[see also][]{takahashi04a}.
However, it turned out that the large amount of late-time decays of dark 
matter is not allowed by the {\it WMAP} data; the lifetime should be
$>123{\rm Gyr}$ (68\% C.L.) if cold dark matter consists only of such
decaying particles \citep*[][ see Figure \ref{fig:dm_ddm}]{ichiki04}. 

\subsection*{Massive Black Holes}

It is possible to consider massive black holes as dark matter. Indeed,
black hole dark matter may explain several observations better, such as 
dynamics in our galaxy \citep{lacey85} and anomalous flux ratios in
lensed quasar systems \citep{mao04}. One of the powerful methods to
detect such compact dark matter is gravitational lensing
\citep[e.g.,][]{inoue03}. However, the observations of wide
binaries seem to be inconsistent with the dark matter model consists of
$M\gtrsim 50M_\odot$ compact objects \citep*{yoo04}.

\chapter{Gravitational Lens Theory}
\label{chap:lens}
\def\mychapheadname{Gravitational Lens Theory}
\markboth{CHAPTER \thechapter.
{\MakeUppercase\mychapheadname}}{}

\section{The Lens Equation}
\markboth{CHAPTER \thechapter.
{\MakeUppercase\mychapheadname}}{\thesection.
\MakeUppercase{The Lens Equation}}

In this section we derive the lens equation which is the fundamental
equation in studying the gravitational lensing. There are many novel
approaches to derive the lens equation
\citep*[e.g.,][]{schneider85,schneider92,sasaki93b,seitz94,futamase95,surpi96}.
In this Appendix, we take more ``intuitive'' approach to derive the lens
equation.  

\begin{figure}[t]
\begin{center}
\includegraphics[width=0.7\hsize]{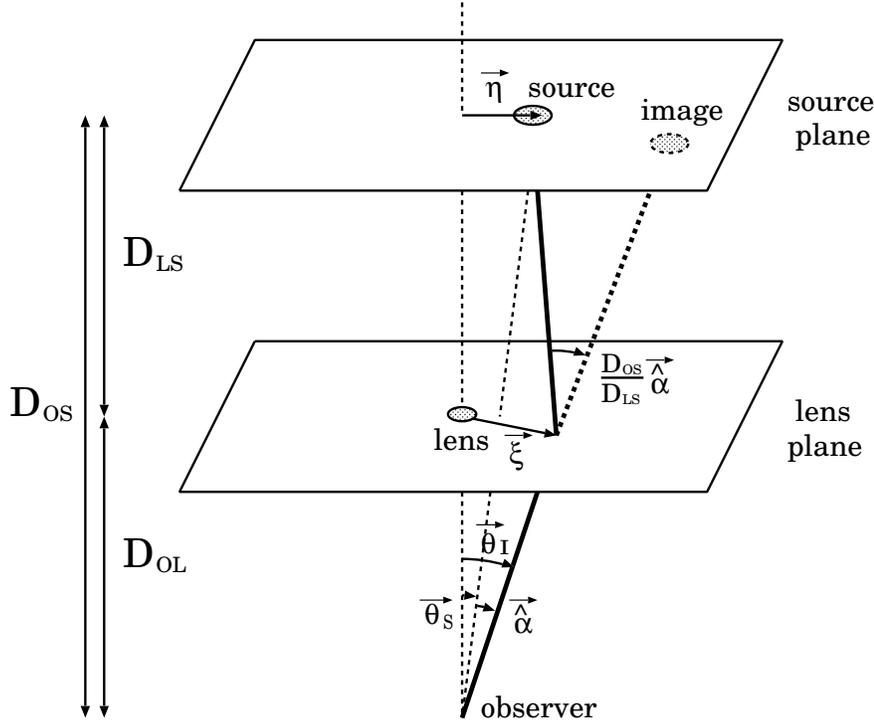}
\caption{A schematic diagram of the lensing system. The image and source
 positions in each plane are denoted by $\vec{\xi}$ and  $\vec{\eta}$.
 Angular diameter distances should be used for the distances shown in
 this Figure.  
\label{fig:lens_lenseq}}
\end{center}
\end{figure}

In most cases of strong gravitational lens studies, we can apply an
approximation called {\it thin lens approximation}; that is, the light
deflection takes place suddenly within a small distance. In this approximation, 
the deflection geometry is described in Figure \ref{fig:lens_lenseq}.
From the geometrical consideration, one can write down the lens equation
as
\begin{equation}
 \vec{\theta}_{\rm S}=\vec{\theta}_{\rm I}-\vec{\hat{\alpha}},
\end{equation}
or equivalently
\begin{equation}
 \frac{D_{\rm OL}}{D_{\rm OS}}\vec{\eta}=\vec{\xi}-D_{\rm OL}\vec{\hat{\alpha}}. 
\end{equation}
The deflection angle $\vec{\hat{\alpha}}$ can be computed as
\begin{eqnarray}
 \vec{\hat{\alpha}}&=&\frac{4G}{c^2}\frac{D_{\rm LS}}{D_{\rm OS}}\int d\chi \vec{\nabla}_\perp \phi_{\rm G}(\chi)\nonumber\\
&=&\frac{1}{\pi}\int d^2\theta\frac{\vec{\theta}_{\rm I}-\vec{\theta}}{|\vec{\theta}_{\rm I}-\vec{\theta}|^2}\hat{\kappa}(D_{\rm OL}\vec{\theta}).
\end{eqnarray}
where $\phi_{\rm G}$ is the gravitational potential and $d\chi$ denotes
the integral along the line of sight. We defined the dimensionless
surface density as
\begin{gather}
 \hat{\kappa}(D_{\rm OL}\vec{\theta})=\frac{\Sigma(D_{\rm OL}\vec{\theta})}{\Sigma_{\rm crit}},\\
 \Sigma_{\rm crit}\equiv\frac{c^2}{4\pi G}\frac{D_{\rm OS}}{D_{\rm
 OL}D_{\rm LS}}.\label{lens_critsurf}
\end{gather}
Here we defined the critical surface mass density $\Sigma_{\rm crit}$.

It is useful to rewrite the above equations in a dimensionless form.
We define a characteristic length $\xi_0$ in the lens plane and a
corresponding length $\eta_0=\xi_0D_{\rm OS}/D_{\rm OL}$ in the source
plane. Then we introduce the dimensionless vectors: 
\begin{eqnarray}
 \vec{x}&=&\frac{\vec{\xi}}{\xi_0},\\
 \vec{y}&=&\frac{\vec{\eta}}{\eta_0},
\end{eqnarray}
in each plane. With this definitions, the dimensionless lens equation becomes
\begin{gather}
 \vec{y}=\vec{x}-\vec{\alpha}(\vec{x}),
\label{lens_lenseq}\\
 \vec{\alpha}(\vec{x})\equiv \frac{D_{\rm OL}}{\xi_0}\vec{\hat{\alpha}}=\frac{1}{\pi}\int
 d^2x'\,\kappa(\vec{x'})\frac{\vec{x}-\vec{x'}}{|\vec{x}-\vec{x'}|^2},
\label{lens_definition_deflection}\\
 \kappa(\vec{x})\equiv\hat{\kappa}(\vec{\xi})=\frac{\Sigma(\xi_0\vec{x})}{\Sigma_{\rm crit}}=\frac{1}{\Sigma_{\rm crit}}\int_{-\infty}^\infty \rho(\vec{r})\,dz.
\label{lens_definition_kappa}
\end{gather}
In the following discussion, we mainly adopt this dimensionless lens
equation (eqs. [\ref{lens_lenseq}], [\ref{lens_definition_deflection}], and
[\ref{lens_definition_kappa}]). If the lens object is axially symmetric,
$\kappa(\vec{x})=\kappa(|\vec{x}|)$, then the lens 
equation reduces to the scalar equation: 
\begin{gather}
 y=x-\alpha(x),\\
 \alpha(x)=\frac{2}{x}\int_0^x x'\kappa(x')dx',
\end{gather}
because $\vec{\alpha}(\vec{x})\parallel \vec{x}$.

The expression of the scaled deflection angle $\vec{\alpha}$ (eq.
[\ref{lens_definition_deflection}]) shows that the scaled deflection angle is a
gradient of some scalar function $\psi$:
\begin{equation}
 \vec{\alpha}(\vec{x})=\vec{\nabla}\psi(\vec{x}),
 \label{lens_lenspotfromalpha}
\end{equation}
where
\begin{equation}
 \psi(\vec{x})\equiv\frac{1}{\pi}\int d^2x'\kappa(\vec{x}')\ln\left|\vec{x}-\vec{x}'\right|.
\label{lens_definition_lenspot}
\end{equation}
This is called the lens potential. From the definition
(\ref{lens_definition_lenspot}), it is easily seen that the Laplacian of
$\psi$ reduces to $2\kappa$:
\begin{equation}
 \triangle\psi(\vec{x})=2\kappa(\vec{x}).
\label{lens_psi-kappa}
\end{equation}

\section{Magnification and Image Distortion}
\markboth{CHAPTER \thechapter.
{\MakeUppercase\mychapheadname}}{\thesection.
\MakeUppercase{Magnification and Image Distortion}}

\subsection{Magnification}

Gravitational lensing changes not only the light path but also the area
and shape of a light bundle. This leads to the amplification and
distortion of images. Since the surface brightness $I_\nu$ is invariant
under the gravitational deflection of light \citep[e.g.,][]{schneider92},  
the change of flux is determined only by the change of the area of a
light bundle. Thus magnification $\mu$, the ratio of flux $S_\nu$ with
and without gravitational lensing, is simply given by
\begin{equation}
 \mu=\frac{S_\nu}{S_{\nu0}}=\frac{I_\nu\Delta\Omega}{I_\nu\Delta\Omega_0}=\frac{\Delta\Omega}{\Delta\Omega_0},
\end{equation}
where the subscript $0$ means the quantities in the absence of lensing,
and $\Delta\Omega$ is the solid angle of the image. Note that the
magnification is independent of the frequency. 
 
Therefore, if we define the Jacobi matrix from the lens equation,
\begin{equation}
 A_{ij}(\vec{x})\equiv\frac{\partial y_i}{\partial x_j}=\delta_{ij}-\psi_{,x_ix_j},
\label{lens_jacobiandef}
\end{equation}
where $\delta_{ij}$ is the Kronecker delta, $\vec{y}=(y_1, y_2)$, 
$\vec{x}=(x_1, x_2)$, and $\psi_{,x_ix_j}\equiv \partial^2\psi/\partial
x_i\partial x_j$, then magnification of the image is 
\begin{equation}
\mu(\vec{x})=\frac{1}{\det A(\vec{x})}. 
\end{equation}
That is, an image at $\vec{x}$ is magnified by a factor
$|\mu(\vec{x})|$. The magnification factor $\mu(\vec{x})$ can take
either positive or negative value, corresponding to the image with
positive or negative parity (see also \S \ref{sec:lens_math}). The
magnification factor formally diverges\footnote{This does not mean that
the image is actually infinitely bright; in reality the magnification is
saturated due to the finite size of source and/or the break down of the
approximation of geometrical optics.} when $\det A(\vec{x})=0$; this
point is called critical point. In many cases, a set of critical points
makes one closed curve; this curve is called critical curve.  

\subsection{Convergence and Shear}

Using the identity (\ref{lens_psi-kappa}), we can write down the Jacobi
matrix (\ref{lens_jacobiandef}) as 
\begin{equation}
 A(\vec{x})=\left(\begin{array}{cc}
    1-\kappa-\gamma_1 & -\gamma_2 \\
    -\gamma_2 & 1-\kappa+\gamma_1
	 \end{array}\right),
\label{lens_jacobian}
\end{equation}
where $\kappa$ is defined by equation (\ref{lens_definition_kappa}), and
$\gamma$ is 
\begin{eqnarray}
 \gamma_1 & = & \frac{1}{2}\left(\psi_{,x_1x_1}-\psi_{,x_2x_2}\right)\\
\gamma_2 & = & \psi_{,x_1x_2}.
\end{eqnarray}
We note that the Jacobi matrix is the mapping which transforms the
shape of the image to that of the source. This can be easily seen from
\begin{equation}
 \delta\vec{y}=A(\vec{x})\delta\vec{x},
\end{equation}
which is obtained from the perturbation of the lens equation
(\ref{lens_lenseq}).  

\begin{figure}[t]
\begin{center}
\includegraphics[width=0.7\hsize]{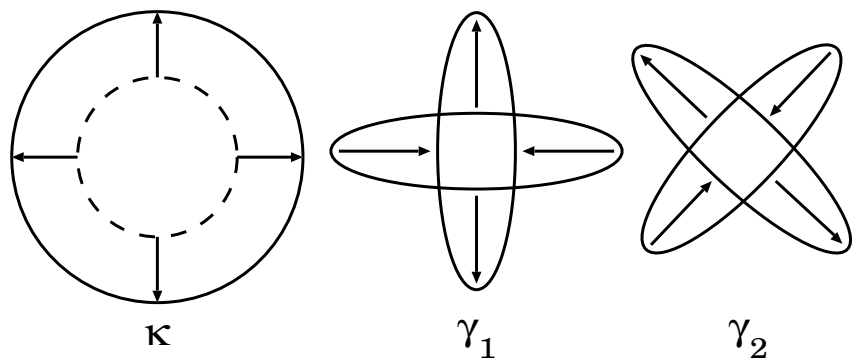}
\caption{The schematic meaning of $\kappa$, $\gamma_1$, and $\gamma_2$
 in the Jacobi matrix (\ref{lens_jacobian}). The convergence $\kappa$
 expands a light bundle, while the shear $\gamma$ changes the shape.
\label{fig:lens_convshear}}
\end{center}
\end{figure}

Figure \ref{fig:lens_convshear} shows the meaning of each quantity in the
Jacobi matrix (\ref{lens_jacobian}). The trace part of the Jacobi matrix,
$\kappa$, is called convergence\footnote{It is also called the Ricci
focusing.}. On the other hand, the traceless part, $\gamma_1$ and
$\gamma_2$, is called shear. 

\subsection{Critical Curves and Caustics}
\label{sec:lens_cri}

The determinant of the Jacobi matrix can be calculated from equation
(\ref{lens_jacobian}) as
\begin{equation}
 \det A=\left(1-\kappa\right)^2-\gamma^2,
\end{equation}
where $\gamma$ is defined by
\begin{equation}
 \gamma=\sqrt{\gamma_1^2+\gamma_2^2}.
\end{equation}
The eigenvalues of the Jacobi matrix are
\begin{equation}
 \lambda_{\pm}=1-\kappa\pm\gamma.
\end{equation}
Curves defined by $\lambda_\pm(\vec{x})=0$ are called the critical
curves. Corresponding curves in the source plane are called caustics. As
the source approaches caustics, images are highly magnified. Moreover it
has been shown that the number of images changes by two if, and only if,
the source crosses a caustic \citep[e.g.,][]{schneider92}. Therefore the 
caustics play a central role in the study of strong gravitational
lensing. This also suggests\footnote{Here we consider the usual case
that $\kappa$ and $\gamma$ are sufficiently small at
$|\vec{x}|\rightarrow\infty$.} that the multiple images must occur if
$\kappa(\vec{x})>1$, or equivalently $\Sigma(\vec{\xi})>\Sigma_{\rm
crit}$, for some point $\vec{x}$, because in this case the curve defined by
$\lambda_{-}=0$ should exist at the outside of the point $\vec{x}$
(i.e., $\kappa(\vec{x})>1$ is a sufficient condition for multiple images).

In the case of axially symmetric lenses, eigenvectors of the Jacobi
matrix point to the tangential and radial directions of the lens object.
Then corresponding eigenvalues are
\begin{eqnarray}
 \lambda_{\rm t}(x) &=& 1-\kappa(x)-\gamma(x)\;\;=\;\;1-\frac{\alpha(x)}{x},
\label{lens_lambda_t}\\
 \lambda_{\rm r}(x) &=& 1-\kappa(x)+\gamma(x)\;\;=\;\;1-\frac{d}{dx}\alpha(x),
\label{lens_lambda_r}
\end{eqnarray}
where the subscripts t and r mean tangential and radial, respectively.
We also define the magnification for each direction by
\begin{eqnarray}
 \mu_{\rm t}(x) &=& \frac{1}{\lambda_{\rm t}(x)},
\label{lens_mu_t}\\ 
 \mu_{\rm r}(x) &=& \frac{1}{\lambda_{\rm r}(x)}.
\label{lens_mu_r}
\end{eqnarray}
The magnification factor then becomes
\begin{equation}
 \mu(x)=\mu_{\rm t}(x)\mu_{\rm r}(x).
\end{equation}
The above expressions also indicate that the image near the curve
$\lambda_{\rm t}=0$ ($\lambda_{\rm r}=0$) is highly elongated  in the
tangential (radial) direction. Thus tangential (radial) arcs are
understood as the images near the critical line defined by $\lambda_{\rm
t}=0$ ($\lambda_{\rm r}=0$).  

\section{Differential Time Delays}
 \label{sec:timedelay}
\markboth{CHAPTER \thechapter.
{\MakeUppercase\mychapheadname}}{\thesection.
\MakeUppercase{Differential Time Delays}}

When multiple images of a single source are formed by the lens, the
traveling time along the each path will, in general, be different. 
This fact is known as time delay
\citep{refsdal64a,refsdal64b,refsdal66}, which may be ascribed to the
following two effects \citep{cooke75}. First, the curved rays are
geometrically longer than straight rays, which results in a geometrical
time delay $\Delta t_{\rm geom}$. Second, when the light rays travel in
the gravitational field, they experience the relativistic time dilation,
which results in a potential time delay $\Delta t_{\rm pot}$. Although
there are many ways to derive the time delay effect
\citep{refsdal66,cooke75,borgeest83,kayser83,schneider85,schneider92}, 
we calculate both geometrical and potential contribution separately in
order to clarify the physical meaning. Again we adopt a simple argument to
derive gravitational lens time delay.

The time delay of a deflected light ray relative to an unlensed
light ray is
\begin{eqnarray}
 c\Delta t&\simeq&\Delta\chi-\frac{2}{c^2}\int\phi_{\rm G} d\chi\nonumber\\
&\simeq&(1+z_{\rm L})\Delta\ell-\frac{2(1+z_{\rm L})}{c^2}\int\phi d\ell\nonumber\\
&\equiv&c\Delta t_{\rm geom}+c\Delta t_{\rm pot},
\end{eqnarray}
where $z_{\rm L}$ means the redshift at the lens and
$\ell=\chi/(1+z_{\rm L})$ is the physical length at the lens. The above
substitution holds because the deviation from the unlensed rays occurs  
only near the lens. Below we calculate each term in turn.

\begin{figure}[t]
\begin{center}
\includegraphics[width=0.75\hsize]{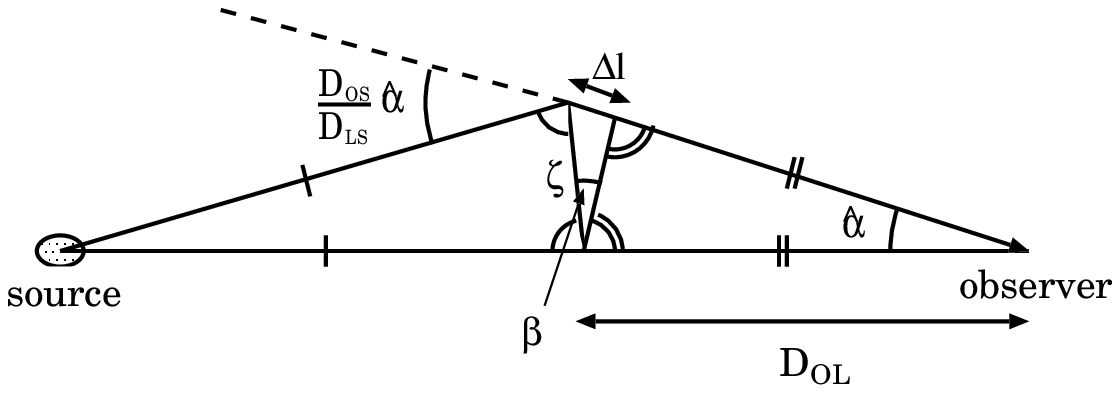}
\caption{Calculation of the geometrical time delay. We consider the
 plane on which the source, observer, and deflection point (image)
 exist. From the purely geometrical consideration, one finds
 $\Delta\ell\simeq\beta\zeta$ and $\beta=D_{\rm OS}\hat{\alpha}/2D_{\rm LS}$. 
\label{fig:lens_lenseq_td}}
\end{center}
\end{figure}

\begin{itemize}
 \item Geometrical time delay $\Delta t_{\rm geom}$\\
From Figure \ref{fig:lens_lenseq_td}, it is found
\begin{equation}
 \Delta\ell\simeq\beta\zeta\simeq\left(\frac{D_{\rm OS}\hat{\alpha}}{2D_{\rm LS}}\right)(D_{\rm OL}\hat{\alpha})=\frac{D_{\rm OL}D_{\rm OS}}{D_{\rm LS}}\frac{\hat{\alpha}^2}{2}.
\end{equation}
From the lens equation (\ref{lens_lenseq}), $\hat{\alpha}$ is written as
\begin{equation}
 \hat{\alpha}^2=\frac{\xi_0^2}{D_{\rm OL}^2}(\vec{x}-\vec{y})^2.
\end{equation}
Then the geometrical time delay $\Delta t_{\rm geom}$ becomes
\begin{equation}
 c\Delta t_{\rm geom}=\frac{\xi_0^2D_{\rm OS}}{D_{\rm OL}D_{\rm LS}}(1+z_{\rm L})\frac{(\vec{x}-\vec{y})^2}{2}.
\end{equation}
\item  Potential time delay $\Delta t_{\rm pot}$\\
The potential time delay is obtained by performing the integral:
\begin{eqnarray}
c\Delta t_{\rm pot}&=&-\frac{2(1+z_{\rm L})}{c^2}\int\phi_{\rm G} d\ell\nonumber\\
&\simeq&-\frac{4G(1+z_{\rm L})}{c^2}\int d^2\xi' \Sigma(\vec{\xi'})\ln\frac{|\vec{\xi}-\vec{\xi}'|}{\xi_0}\nonumber\\
&=&-\frac{\xi_0^2D_{\rm OS}}{D_{\rm OL}D_{\rm LS}}(1+z_{\rm L})\psi(\vec{x}).
\end{eqnarray}
\end{itemize}

What we observe is the total time delay, not individual terms. The
total time delay of a deflected light ray $\Delta t$ is derived simply
by summing up these two effects:  
\begin{eqnarray}
 c\Delta t&=&c\Delta t_{\rm geom}+c\Delta t_{\rm pot}\nonumber\\
&=&\frac{\xi_0^2D_{\rm OS}}{D_{\rm OL}D_{\rm LS}}(1+z_{\rm L})\phi(\vec{x}, \vec{y}),
\end{eqnarray}
where we defined the Fermat potential:
\begin{equation}
 \phi(\vec{x}, \vec{y})\equiv\frac{(\vec{x}-\vec{y})^2}{2}-\psi(\vec{x}).
\label{lens_fermat}
\end{equation}
The Fermat potential has several interesting properties. For example, the lens
equation (\ref{lens_lenseq}) is rewritten in terms of the Fermat potential:
\begin{equation}
 \vec{\nabla}\phi(\vec{x}, \vec{y})=0.
\end{equation}
This means that the lensed images are formed at the points where the
Fermat potential is stationary. This is certainly Fermat's principle in
geometrical optics, which says that the actual light trajectory is such
that the light-travel time is stationary on it under the first order
variations.

Since the time delay is, in practice, observed between two images
produced from one source, we write this as
\begin{equation}
 c\Delta t(\vec{y})=\frac{\xi_0^2D_{\rm OS}}{D_{\rm OL}D_{\rm LS}}(1+z_{\rm L})\left[\phi(\vec{x}^{(1)}, \vec{y})-\phi(\vec{x}^{(2)}, \vec{y})\right],
\label{lens_timedelay}
\end{equation}
where $\vec{x}^{(i)}$, $i=1,2$, are two image positions, produced by a
source at $\vec{y}$. 

\section{Mathematical Aspects}
\label{sec:lens_math}
\markboth{CHAPTER \thechapter.
{\MakeUppercase\mychapheadname}}{\thesection.
\MakeUppercase{Mathematical Aspects}}

In this section, we briefly review some mathematical aspects of
strong gravitational lensing, which may be useful to understand image
position/magnification patterns in strong lens systems. Basically we
follow the discussion of \citet{schneider92}; please refer it for more
comprehensive discussions.

\subsection{Classification of Images}

It has been shown in the previous section that lensed images are formed
at the points where the derivative of the Fermat potential $\phi$ (eq.
[\ref{lens_fermat}]) vanishes. Thus images are located at local minima
and saddle points of the constant Fermat potential surface (i.e., the
arrival time surface). The properties of such stationary point is
characterized by the second derivative of the Fermat potential, i.e., 
the Jacobi matrix (eq. [\ref{lens_jacobiandef}]). Specifically, images
are classified into the following three types:

\begin{itemize}
 \item Type I: Minimum of $\phi$, defined by ${\rm det}A>0$ and ${\rm tr}A>0$.
 \item Type II: Saddle point of $\phi$, defined by ${\rm det}A<0$.
 \item Type III: Maximum of $\phi$, defined by ${\rm det}A>0$ and ${\rm tr}A<0$.
\end{itemize}

Since the magnification factor is $\mu=({\rm det}A)^{-1}$, it is
said that images of types I and III have positive parities, while images
of type II have negative parity.

We denote the number of images of each type as $n_{\rm I}$, etc. The total
number of images is denoted by $n=n_{\rm I}+n_{\rm II}+n_{\rm III}$. 
We consider a single, thin lens system whose surface mass density
$\kappa(\vec{x})$ is smooth and decrease faster than $|\vec{x}|^{-2}$
at $|\vec{x}|\rightarrow\infty$. The deflection angle $\vec{\alpha}(\vec{x})$ is
continuous, bounded ($|\vec{\alpha}(\vec{x})|\leq a$), and
$|\vec{\alpha}(\vec{x})|\rightarrow 0$ for both $|\vec{x}|\rightarrow0$ and
$\infty$.\footnote{A point mass lens does not satisfy this condition
because $|\vec{\alpha}(\vec{x})|\rightarrow \infty$ at
$|\vec{x}|\rightarrow0$.} Then, the following theorems are proven: 

\newtheorem{theo}{Theorem}

\begin{theo}
The numbers of lenses satisfy the following relations: 
 (a) $n_{\rm I}\geq 1$, (b) $n<\infty$, 
 (c) $n=n_{\rm I}=1$ for sufficiently large $\vec{y}$.
\end{theo}
{\it Sketch of proof.} (a) The Fermat potential $\phi$ must have the global
minimum, which corresponds to an image of type I. (b) The boundness
implies the images must be contained in the disc
$|\vec{x}|\leq|\vec{y}|+a$, for a fixed $\vec{y}$. Since images are
isolated by assumption, $n<\infty$. (c) For
$|\vec{x}|\rightarrow\infty$, $A\rightarrow I$, and thus a circle of
radius $R$, outside of which there is only one image of type I, exists. 
Again, the boundness implies $|\vec{x}|\geq |\vec{y}|-a>R$ if we choose
$|\vec{y}|>a+R$. Combining these results yields $n=n_{\rm I}=1$.
$\blacksquare$\\ 

To prove another theorem on the number of lenses, we need the index
theorem. First, we need to define the index.\\

{\it Definition.} Let $S$ be a surface $S$, and let $\vec{V}$ be a
tangent vector field on $S$. For each point $p$ of $S$, the {\it index},
denoted as $I(p)$, is defined by 
\begin{equation}
 I(p)=\frac{1}{2\pi}\oint_{c_\epsilon}d\varphi,
\end{equation}
where $\vec{V}=|\vec{V}|(\cos\varphi,\sin\varphi)$, and $c_\epsilon$ is
an oriented, closed curve enclosing the point $p$\\

\begin{figure}[t]
\begin{center}
\includegraphics[width=0.9\hsize]{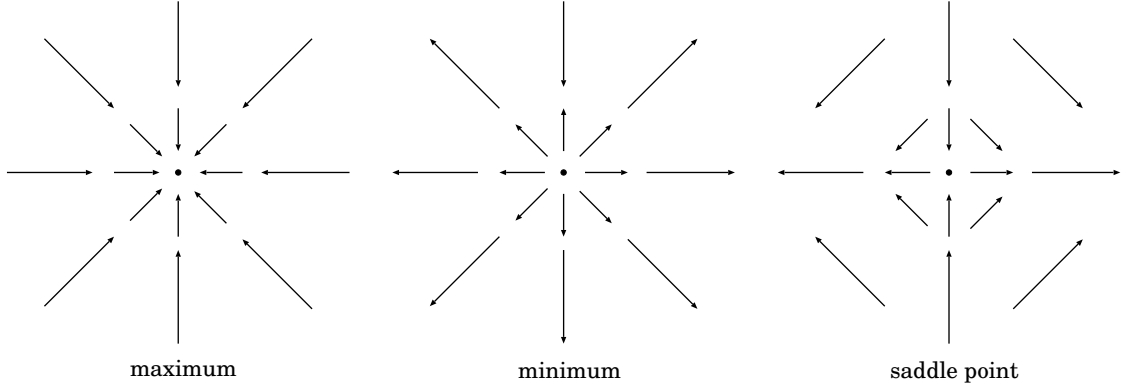}
\caption{Vector fields around three stationary point of $\phi$.
\label{fig:lens_index}}
\end{center}
\end{figure}

Note that the index of a non-critical point $\vec{V}(p)\neq 0$ is $0$,
and only critical points have non-vanishing index. In our specific
example, these points correspond to maxima, minima, and saddle points of
the Fermat potential $\phi$. Figure \ref{fig:lens_index} shows the vector fields
$\vec{\nabla}\phi$ around these critical points; by inspection, we find
that the index of an extremum (i.e., maximum or minimum) is $+1$, and
that of a saddle point is $-1$.

\begin{theo}[Poincar\'{e}-Hopf index theorem]
Let $\vec{V}$ be a continuous tangent vector filed with isolated
 singularities on the compact, connected, orientable 2-dimensional
 surface $S$. If we denote critical points by $p_i$,
\begin{equation}
 \sum I(p_i)=\chi(S),
\end{equation}
where $\chi(S)$ is the Euler characteristic of $S$. 
\end{theo}

The Euler characteristic is a topological invariant, and for polyhedron
it is defined by $V-E+F$, where $V$, $E$, and $F$ are the numbers of
vertices, edges, and faces, respectively. For instance, the Euler
characteristic of $S^2$ is 2. Then we obtain the following theorem:  

\begin{theo}
The numbers of images satisfy the following relation: 
$n_{\rm I}+n_{\rm III}=1+n_{\rm II}$.  
\end{theo}
{\it Sketch of proof.} By considering the standard stereographic
projection, we obtain $\mathbb{R}^2 \approx S^2-\{\rm N\}$, where N is
the sphere's north pole. The north pole corresponds to the infinity, and
the index of the point is $+1$ because the vector filed
$\vec{\nabla}\phi$ is asymptotically radial. Then, the index theorem
gives $n_{\rm I}-n_{\rm II}+n_{\rm III}=1$.  $\blacksquare$\\

The above theorem implies the number of images is odd, the number of
even-parity images exceeds that of odd parity images by
one \citep{burke81}, $n_{\rm II}\geq n_{\rm III}$, and $n>1$ if and only
if $n_{\rm II}\geq 1$. 

\begin{theo}
The image of a source that arrives first at the observer is of type I
 and appears brighter than the source would appear in the absence of the lens. 
\end{theo}
{\it Sketch of proof.} The global minimum of the Fermat potential $\phi$ 
corresponds to an image of type I, and by definition, it arrives first.
The condition of type I is ${\rm tr}A>0$, i.e., $\kappa<1$. This implies
\begin{equation}
 \mu^{-1}=(1-\kappa)^2-\gamma^2<1-\gamma^2 \leq 1,
\end{equation}
therefore we obtain $\mu>1$ \citep{schneider84}.  $\blacksquare$\\

Finally, we comment on the odd-number theorem. In most lensed quasar
systems, the number of images are even \citep[mostly 2 or
4; see][]{kochanek04b}. This implies either the central mass distribution is
so cuspy that the assumption of  $|\vec{\alpha}(\vec{x})|\rightarrow 0$
for $|\vec{x}|\rightarrow0$ becomes wrong (i.e., steeper than
$\rho(r)\propto r^{-2}$) or the central image is demagnified enough not
to be observed.  Since the amount of demagnification depends on the
central mass distribution of the lens so that the steeper density
profile tends to demagnify the central image more, the absence/detection
of the central image is used to constrain the mass distribution in the
innermost region of the lens \citep*[e.g.,][]{rusin01a,keeton03a,winn04}.

\subsection{Fold and Cusp Caustics}

Hereafter we denote $\phi_i\equiv(\vec{\nabla}\phi)_i$, etc. In this notation,
the lens equation is
\begin{equation}
 \phi_i=0.
\end{equation}
As discussed in \S \ref{sec:lens_cri}, critical curves and caustics are
defined by 
\begin{equation}
 D\equiv{\rm det}\,\phi_{ij}=0.
\end{equation}

We consider images around a critical point $\vec{x}^{(0)}$. Suppose that
$A^{(0)}$ has rank 1 and $\vec{\nabla}D\neq 0$. Then we can introduce
the coordinate system around $\vec{x}^{(0)}$ so that $A^{(0)}$ is
diagonal with $A_{11}^{(0)}\neq 0$ and $A_{22}^{(0)}$. Since 
\begin{equation}
 D=\phi_{11}\phi_{22}-\phi_{12}^2,
\end{equation}
the normal vector and tangential vector at $\vec{x}^{(0)}$ become
\begin{gather}
 \vec{N}^{(0)}=\vec{\nabla}D=\phi_{11}^{(0)}(\phi^{(0)}_{221},\phi^{(0)}_{222}),\\
\vec{T}^{(0)}=R(\pi/2)\vec{\nabla}D=\phi_{11}^{(0)}(-\phi^{(0)}_{222},\phi^{(0)}_{221}).
\end{gather}
Thus we consider the following two cases:
\begin{itemize}
 \item Fold: $A^{(0)}\cdot \vec{T}^{(0)}\neq 0$\\
In this case, the Taylor expansion of the Fermat potential $\phi$
becomes\footnote{The rule for truncating the Taylor expansion is not
trivial; see \citet{schneider92}.}
\begin{equation}
 \phi=\phi^{(0)}+\frac{1}{2}y^2-\vec{x}\cdot\vec{y}+\frac{1}{2}\phi_{11}^{(0)}x_1^2+\frac{1}{2}\phi_{112}^{(0)}x_1^2x_2+\frac{1}{2}\phi_{122}^{(0)}x_1x_2^2+\frac{1}{6}\phi_{222}^{(0)}x_2^3.
\end{equation}
Then the lens equation is
\begin{eqnarray}
y_1&=&\phi_{11}^{(0)}x_1+\phi_{112}^{(0)}x_1x_2+\frac{1}{2}\phi_{122}^{(0)}x_2^2,\\
y_2&=&\frac{1}{2}\phi_{112}^{(0)}x_1^2+\phi_{122}^{(0)}x_1x_2+\frac{1}{2}\phi_{222}^{(0)}x_2^2. 
\end{eqnarray}
Then it is straightforward to compute the critical curve and caustic:
\begin{eqnarray}
 \mbox{Critical curve:}&&\displaystyle{\phi_{122}^{(0)}x_1+\phi_{222}^{(0)}x_2=0,}\\
 \mbox{Caustic:}&&\displaystyle{Q\equiv 2\left(\phi_{11}^{(0)}\right)^2\phi_{222}^{(0)}y_2-\left[\phi_{112}^{(0)}\phi_{222}^{(0)}-\left(\phi_{122}^{(0)}\right)^2\right]y_1^2=0.}
\end{eqnarray}
From this, images near the critical curve are derived as
\begin{eqnarray}
 x_1&=&\frac{\phi_{222}^{(0)}y_1-\phi_{122}^{(0)}y_2}{\phi_{11}^{(0)}\phi_{222}^{(0)}},\\
 x_2&=&\frac{-\phi_{122}^{(0)}y_1\pm\sqrt{Q}}{\phi_{11}^{(0)}\phi_{222}^{(0)}}.
\end{eqnarray}
Thus we have two images around a fold caustic.
 \item Cusp: $A^{(0)}\cdot \vec{T}^{(0)}= 0$\\
As in the case of fold caustic, the Taylor expansion of the Fermat potential $\phi$
becomes 
\begin{equation}
 \phi=\phi^{(0)}+\frac{1}{2}y^2-\vec{x}\cdot\vec{y}+\frac{1}{2}\phi_{11}^{(0)}x_1^2+\frac{1}{2}\phi_{112}^{(0)}x_1^2x_2+\frac{1}{2}\phi_{122}^{(0)}x_1x_2^2+\frac{1}{24}\phi_{2222}^{(0)}x_2^4.
\end{equation}
Then the lens equation is
\begin{eqnarray}
y_1&=&\phi_{11}^{(0)}x_1+\phi_{112}^{(0)}x_1x_2+\frac{1}{2}\phi_{122}^{(0)}x_2^2,\\
y_2&=&\frac{1}{2}\phi_{112}^{(0)}x_1^2+\phi_{122}^{(0)}x_1x_2+\frac{1}{6}\phi_{2222}^{(0)}x_2^3. 
\end{eqnarray}
Again it is straightforward to compute the critical curve and caustic:
\begin{eqnarray}
 \mbox{Critical curve:}&&\displaystyle{2\phi_{122}^{(0)}\phi_{11}^{(0)}x_1+\left[\phi_{2222}^{(0)}\phi_{11}^{(0)}-2\left(\phi_{122}^{(0)}\right)^2\right]x_2^2=0,}\\
 \mbox{Caustic:}&&\displaystyle{8\left(\phi_{122}^{(0)}\right)^3y_1^3+9\left(\phi_{11}^{(0)}\right)^2\left[\phi_{2222}^{(0)}\phi_{11}^{(0)}-3\left(\phi_{122}^{(0)}\right)^2\right]y_2^2=0.}
\end{eqnarray}
That is, the caustic is a semicubic parabora. 
From this, it is found that the images near the critical curve are given by
the solution of the following equations:
\begin{gather}
 x_1-\frac{y_1}{\phi_{11}^{(0)}}+\frac{\phi_{122}^{(0)}}{2\phi_{11}^{(0)}}x_2^2=0,\\
 x_2^3+\frac{6\phi_{122}^{(0)}y_1}{\phi_{2222}^{(0)}\phi_{11}^{(0)}-3\left(\phi_{122}^{(0)}\right)^2}x_2-\frac{6\phi_{11}^{(0)}y_2}{\phi_{2222}^{(0)}\phi_{11}^{(0)}-3\left(\phi_{122}^{(0)}\right)^2}=0.
\end{gather}
Thus we have three images around a cusp caustic.
\end{itemize} 

\begin{figure}[t]
\begin{center}
\includegraphics[width=0.68\hsize]{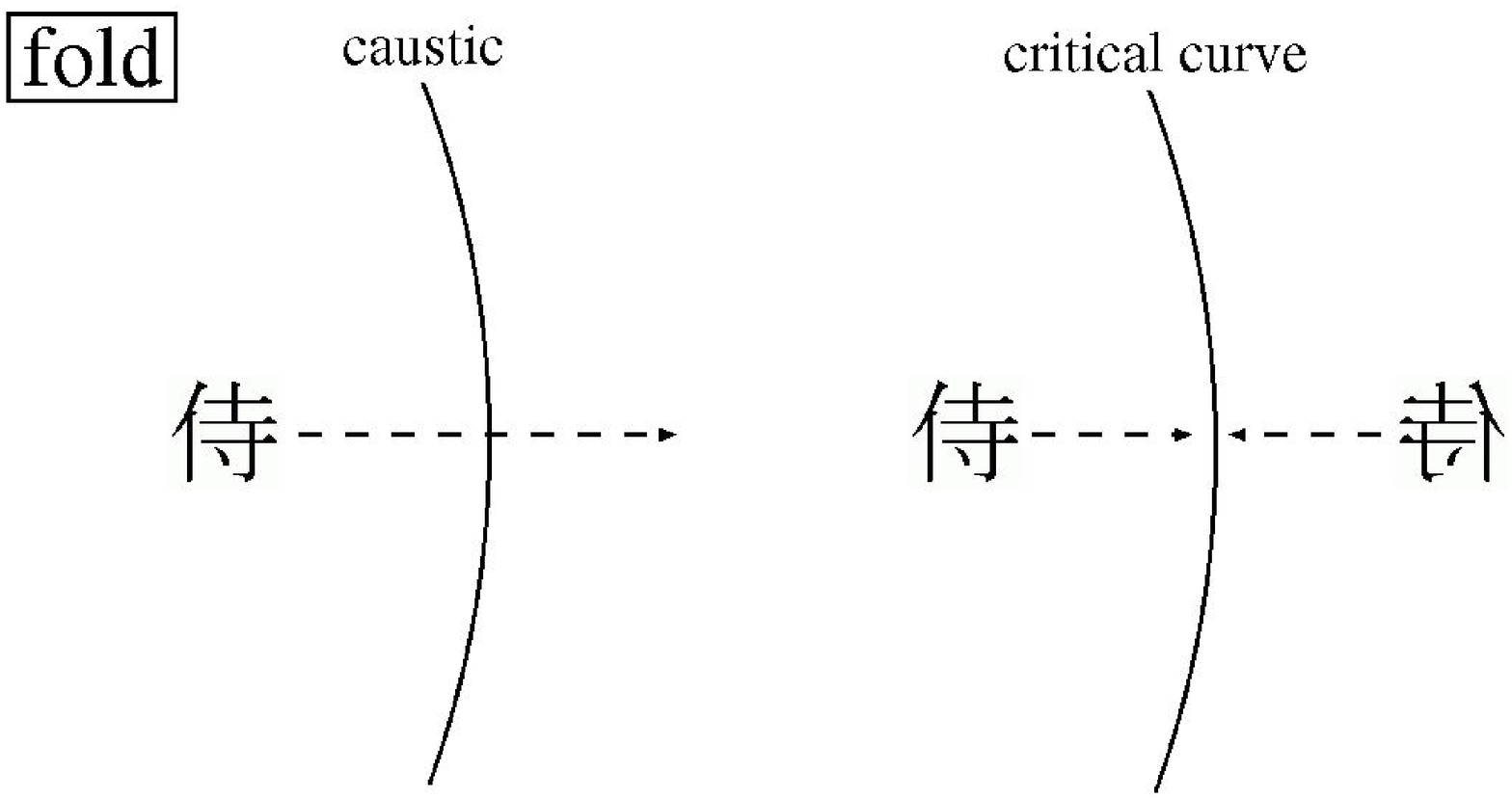}\medskip\\
\includegraphics[width=0.68\hsize]{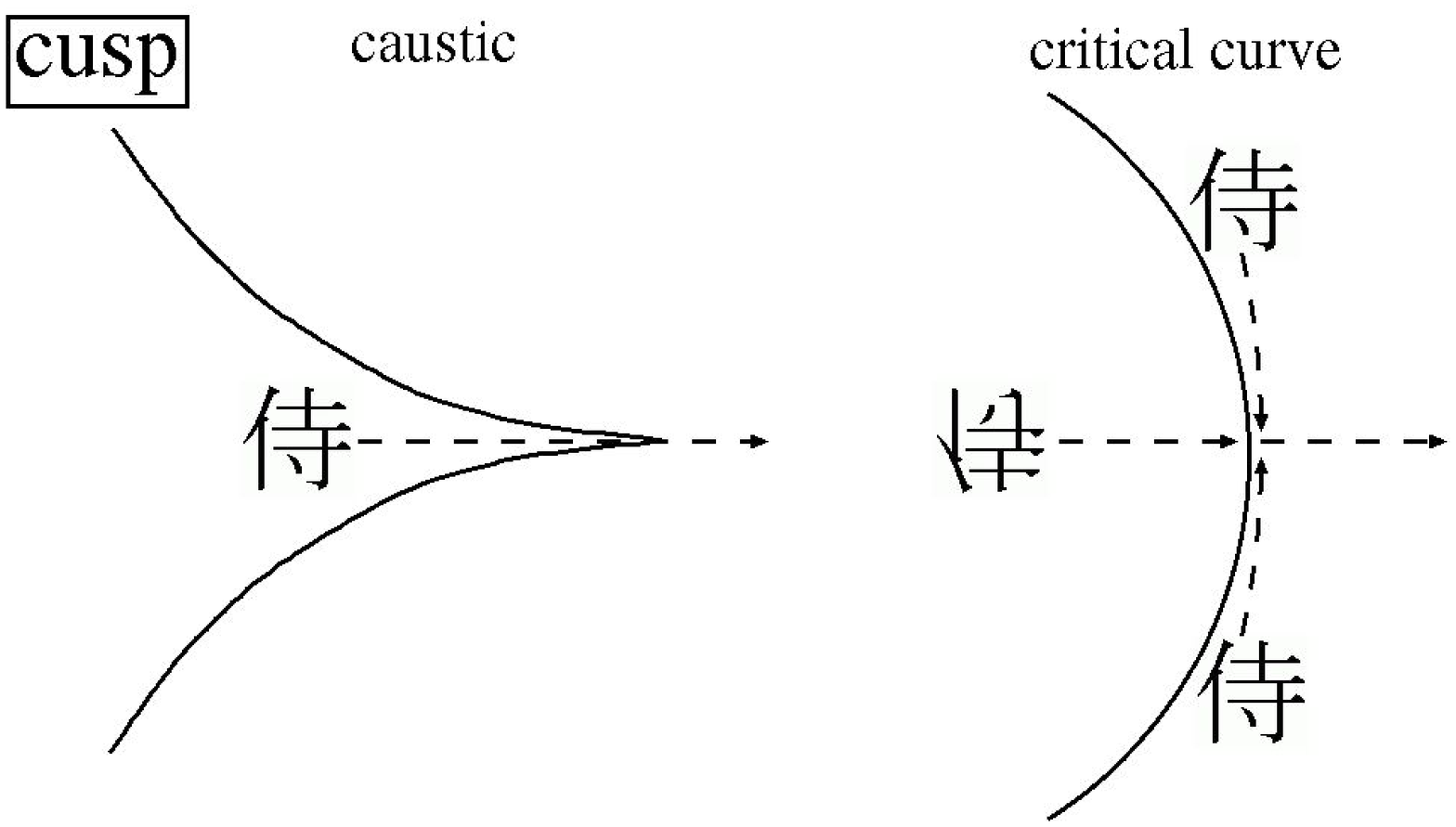}
\caption{Images around fold ({\it upper}) and cusp ({\it lower})
 caustics. As a source moves from inside to outside the fold caustics,
 two images merge and vanish. On the other hand, at the cusp caustic
 three images merge and one image remains. In this Figure, the left hand
 side corresponds to the center of mass distribution. We choose a
 complicated character ({\it samurai} in Japanese) as a source, in order
 to explicitly show parities of images.
\label{fig:lens_foldcusp}}
\end{center}
\end{figure}

Figure \ref{fig:lens_foldcusp} shows images around fold and cusp
caustics. As crossing fold caustics, two images with different parities
are newly created/destroyed. On the other hand, as crossing from outside
to inside cusp caustics, one image is split into three images. 

\chapter{Simulations for Arcs}
\label{chap:arcsim}
\def\mychapheadname{Simulations for Arcs}
\markboth{CHAPTER \thechapter.
{\MakeUppercase\mychapheadname}}{}

Most of this method presented in this Appendix follows the work by
\citet{miralda93b}, \citet{bartelmann94}, and \citet{oguri02b}. 

\section{Lens Mapping}
\markboth{CHAPTER \thechapter.
{\MakeUppercase\mychapheadname}}{\thesection.
\MakeUppercase{Lens Mapping}}

To begin with, we choose a sufficiently large region in the lens plane
in which all the arcs are exist. In this region we prepare
regular grids. Each grid point is denoted by $(x_{1i}, x_{2j})$, where
integers $i$ and $j$ are restricted in $1\leq i, j\leq N_{\rm grid}$. In
the practical calculations, we adopt $N_{\rm grid}=2048$ throughout the
thesis. Given the deflection angle $\vec{\alpha}$, we can calculate the
source point $(y_1(i,j), y_2(i,j))$ which corresponds to $(x_{1i},
x_{2j})$ by using the lens equation. We calculate this for all grid
points and this yields ``mapping table''.

\begin{figure}[t]
\begin{center}
\includegraphics[width=0.5\hsize]{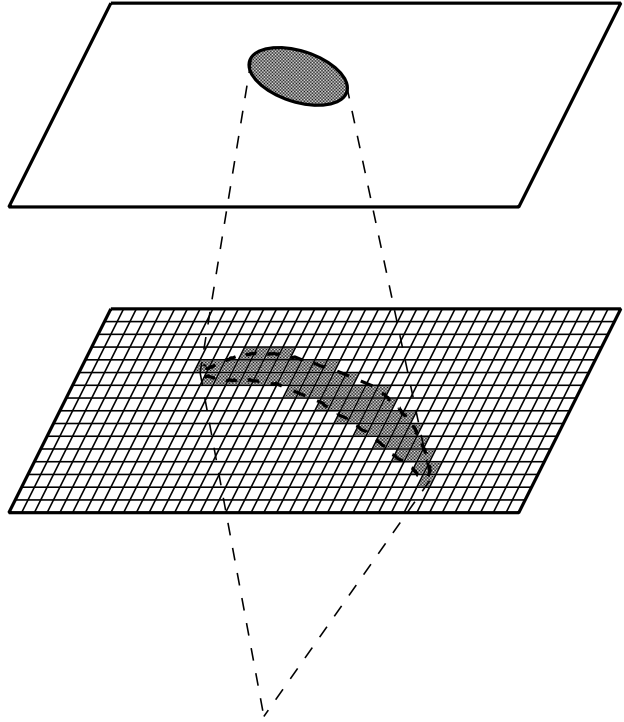}
\caption{Schematic diagram of the lens mapping algorithm. We prepare the
 regular grid in the lens (image) plane, and each grid is checked whether
 it belongs to the source when mapped into the source plane.
\label{fig:arcsim_map}}
\end{center}
\end{figure}

Next we consider the source with center $(y_{\rm 1c}, y_{\rm 2c})$ and
(dimensionless) radius $r_{\rm S}$ and ellipticity $e_{\rm S}$. We
regard the grid point $(x_{1i}, x_{2j})$ is a part of lensed images if
the following condition is satisfied:  
\begin{equation}
 \frac{\left[y_1(i, j)-y_{\rm 1c}\right]^2}{r_{\rm S}^2/(1-e_{\rm S})}+\frac{\left[y_2(i, j)-y_{\rm 2c}\right]^2}{r_{\rm S}^2(1-e_{\rm S})}\leq 1.
\label{arcsim_withinsource}
\end{equation}
For all the grids we check this condition and obtain the pattern of
lensed images. In general, multiple images can be generated by
gravitational lensing. Thus we search each ``image'' grid and recognize
neighboring ``image'' grids as the same image. The magnification of the
image is then proportional to the number of grid points it contains.  
We show a schematic diagram of the method in Figure \ref{fig:arcsim_map}.

\section{Recognition of Arcs}
\markboth{CHAPTER \thechapter.
{\MakeUppercase\mychapheadname}}{\thesection.
\MakeUppercase{Recognition of Arcs}}

To analyze the lens properties, we calculate the magnification $\mu$,
length $l$, and width $w$ of the image. For the multiply lensed system,
we calculate above quantities for each image.  

\begin{itemize}
 \item Magnification $\mu$\\
       Magnification $\mu$ is easily calculated from the number of grid
       points $N_{\rm image}$ which are recognized as the image:
\begin{equation}
 \mu=\frac{N_{\rm image}(\Delta x)^2}{\pi r_{\rm S}^2},
\end{equation}
       where $\Delta x$ is the size of the grid. This gives fairly
       accurate values because images contains many grid points in our
       calculations; typically $N_{\rm image}\gtrsim 50$. 
\item Length $l$\\
       First we search the center of the image (point C) as the point at
       which the value of the left hand side in equation
       (\ref{arcsim_withinsource}) becomes the smallest. Then we find
       the point A in the image which is the farthest away from the
       center. Then we find the point B, also along the image, which is
       the farthest away from the point A. We calculate the length of the image by 
       $l={\rm \overline{AC}+\overline{BC}}$.
\item Width $w$\\
       The width is taken such that $\pi lw=\mu\pi r_{\rm S}^2$.
\end{itemize}

Then, the axis ratio of the arc is defined by
\begin{equation}
 \epsilon=\frac{l}{w},
\end{equation}
and this quantity plays a central role in arc statistics (see Chapter
\ref{chap:arc}). 

\chapter{The Sloan Digital Sky Survey}
\label{chap:appsdss}
\def\mychapheadname{The Sloan Digital Sky Survey}
\markboth{CHAPTER \thechapter.
{\MakeUppercase\mychapheadname}}{}

\section{Introduction}
\markboth{CHAPTER \thechapter.
{\MakeUppercase\mychapheadname}}{\thesection.
\MakeUppercase{Introduction}}

The Sloan Digital Sky Survey (SDSS)\footnote{Funding for the creation
and distribution of the SDSS Archive has been provided by the Alfred P.
Sloan Foundation, the Participating Institutions, the National
Aeronautics and Space Administration, the National Science Foundation,
the U.S. Department of Energy, the Japanese Monbukagakusho, and the Max
Planck Society. The SDSS Web site is http://www.sdss.org/. 

The SDSS is managed by the Astrophysical Research Consortium (ARC) for
the Participating Institutions. The Participating Institutions are The
University of Chicago, Fermilab, the Institute for Advanced Study, the
Japan Participation Group, The Johns Hopkins University, Los Alamos
National Laboratory, the Max-Planck-Institute for Astronomy (MPIA), the
Max-Planck-Institute for Astrophysics (MPA), New Mexico State
University, University of Pittsburgh, Princeton University, the United
States Naval Observatory, and the University of Washington.} 
is the most ambitious astronomical survey project ever undertaken. 
The survey will conduct in both photometric and spectroscopic surveys of
$\sim10,000{\rm deg^2}$ ($1/4$) of the entire sky, using a dedicated
wide-field 2.5-meter telescope at Apache Point Observatory (APO), New
Mexico, USA. The SDSS will determine the positions, redshifts, and absolute
magnitudes of $\sim 10^6$ galaxies and $\sim 10^5$ quasars. The main
purpose of the SDSS is to map the 3-dimensional large-scale structure in 
the universe. 

The striking advantage of the SDSS is the homogeneity of the survey,
despite it covers very large area of the sky. The SDSS will have a
significant impact on astronomical studies as diverse as the large-scale
structure of the universe, the origin and evolution of galaxies, the
relation between dark and luminous matter, the structure of our own
Milky Way, and the properties and distribution of the dust from which
stars like our sun were created.  

\begin{figure}[p]
\begin{center}
\includegraphics[width=0.8\hsize]{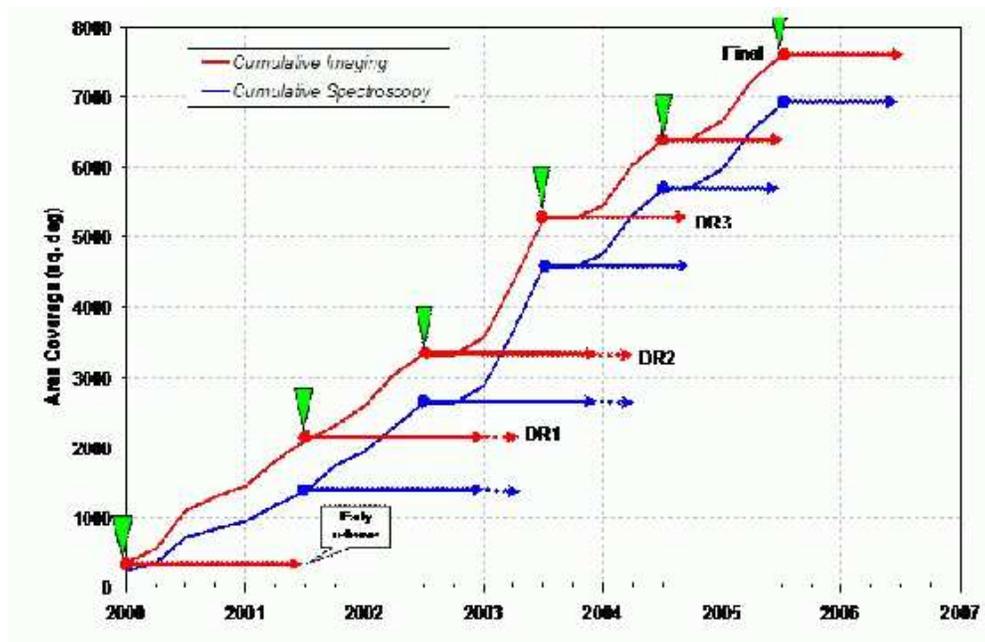}
\caption{The SDSS data release schedule for the 5-year baseline survey. The
 green markers correspond to the date of acquiring the last imaging data
 that will be included in a particular release. Spectroscopic data is
 based on imaging data in the data set and may have been obtained at a
 later date.  Observations that are part of the five-year baseline
 survey are scheduled to end in July 2005.  This
  Figure is taken from a webpage at http://www.sdss.org/science/DRschedule.html.  
\label{fig:appsdss_status}}
\end{center}
\end{figure}

The SDSS started the survey in 2000, and is now ongoing; the SDSS will
be completed in 2005. So far the SDSS data release have been done three
times \citep{stoughton02,abazajian03,abazajian04}, and all the data will
be publicly available till the middle of 2006 (see Table
\ref{table:appsdss_status} and Figure \ref{fig:appsdss_status}). Once
completed, the SDSS will become one of the most fundamental archives for
optical astronomy for next decades. 

\begin{table}[p]
 \begin{center}
  \begin{tabular}{ccccc}\hline\hline
 Data Release & Baseline Schedule$^{\rm a}$ & Actual/Forecast$^{\rm b}$ &
Photometry[deg$^2$]$^{\rm c}$  & Spectroscopy[deg$^2$]$^{\rm c}$ \\
   \hline
 EDR & July 2001 & 5-Jun-2001  & 462  & 386 \\
 DR1 & Jan 2003  & 4-Apr-2003  & 2,099 & 1,360 \\
 DR2 & Jan 2004  & 15-Mar-2004 & 3,324 & 2,627 \\
 DR3 & Oct 2004  & Oct 2004    & 5,300 & 4,600 \\
 DR4 & July 2005 & July 2005   & 6,500 & 5,800 \\
 Final$^{\rm d}$ & July 2006 & July 2006 & 7,700 & 7,000\\
\hline
\end{tabular}
\caption{The SDSS data release schedule for the 5-year baseline survey. This
  Table is taken from a webpage at http://www.sdss.org/science/DRschedule.html.\protect\\
\footnotesize{\hspace*{5mm}${}^{\rm a}$  Baseline schedule as presented in: Data Distribution to the Astronomy Community, Revision 1, September 8, 2000. \hspace{10mm}\protect\\
\hspace*{5mm}${}^{\rm b}$ Dates for the EDR, DR1, and DR2 are actual release dates; dates for releases after April 2004 are current forecasts. \hspace{70mm}\protect\\
\hspace*{5mm}${}^{\rm c}$ The data volumes shown for EDR, DR1, and DR2 are the actual volumes released; data volumes for DR3, DR4 and the final release are estimates based on current forecasts. \hspace{70mm}\protect\\
\hspace*{5mm}${}^{\rm d}$ The final data release corresponds to the release of the last increment of processed data. }}
\label{table:appsdss_status}
 \end{center}
\end{table}

\section{Telescope, Camera, and Spectrograph}
\markboth{CHAPTER \thechapter.
{\MakeUppercase\mychapheadname}}{\thesection.
\MakeUppercase{Telescope, Camera, and Spectrograph}}

The SDSS observing site is at APO in New Mexico, USA. The site is about
2,840-meter above sea level. The SDSS has two telescope; the SDSS main
telescope and the photometric telescope. 

\begin{figure}[t]
\begin{center}
\includegraphics[width=0.29\hsize]{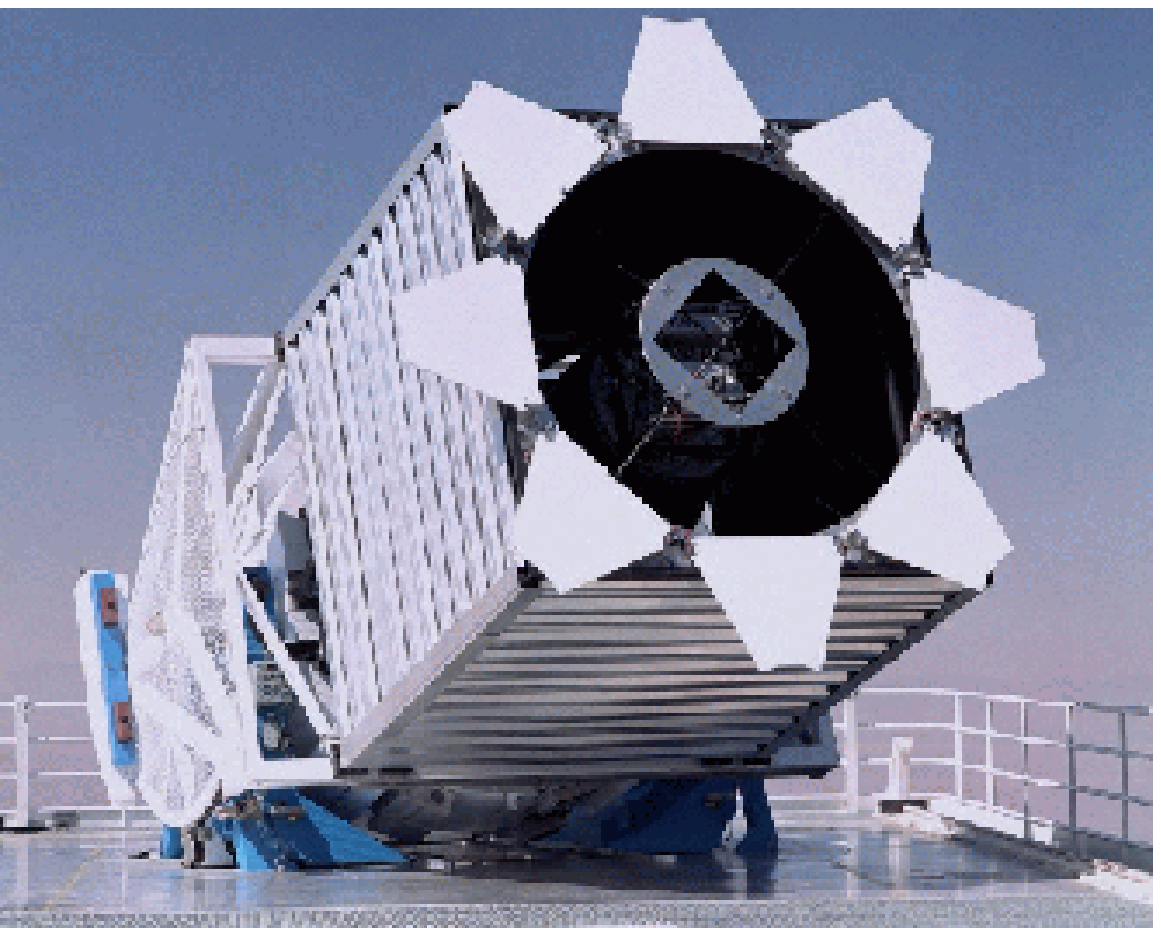}
\includegraphics[width=0.34\hsize]{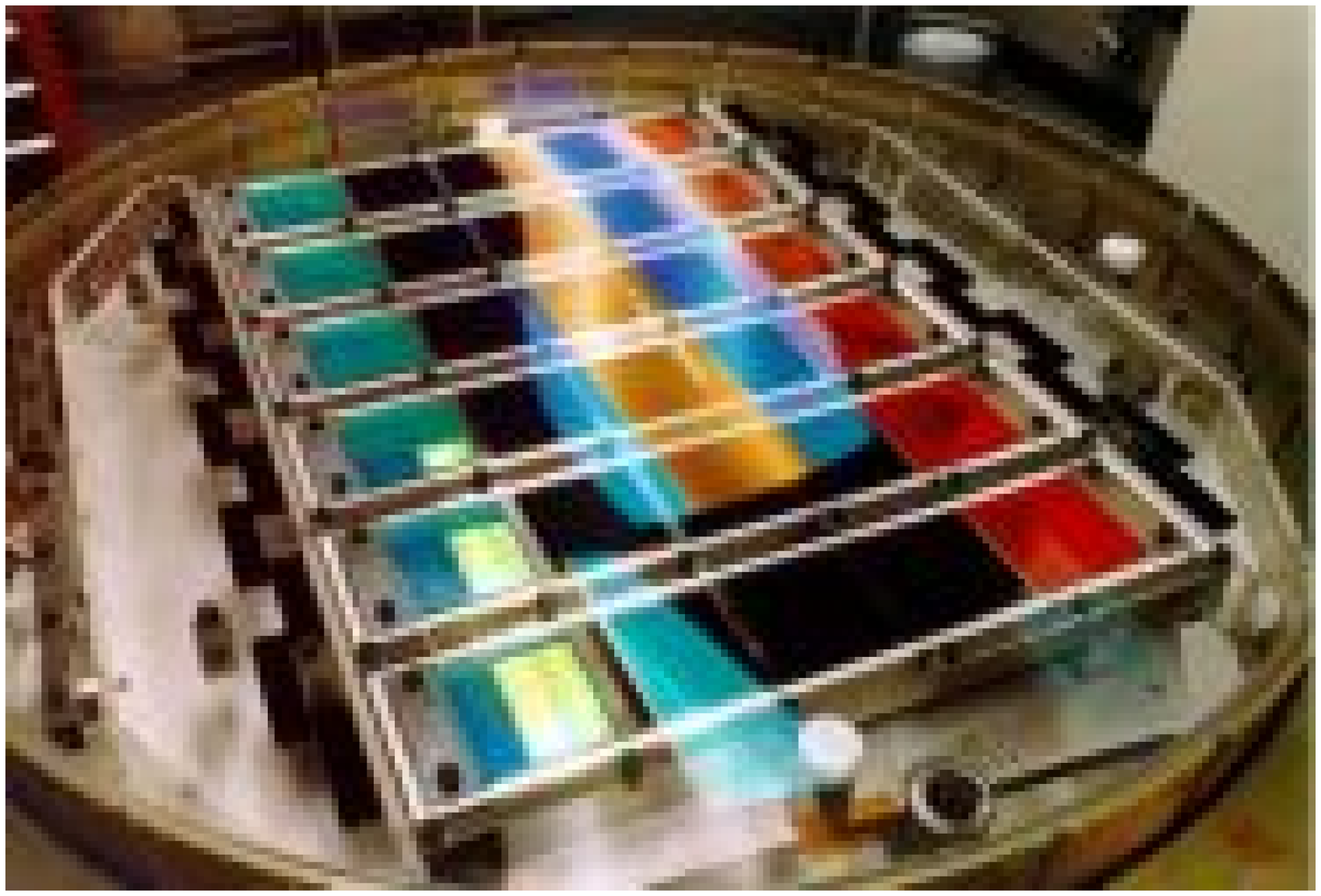}
\includegraphics[width=0.34\hsize]{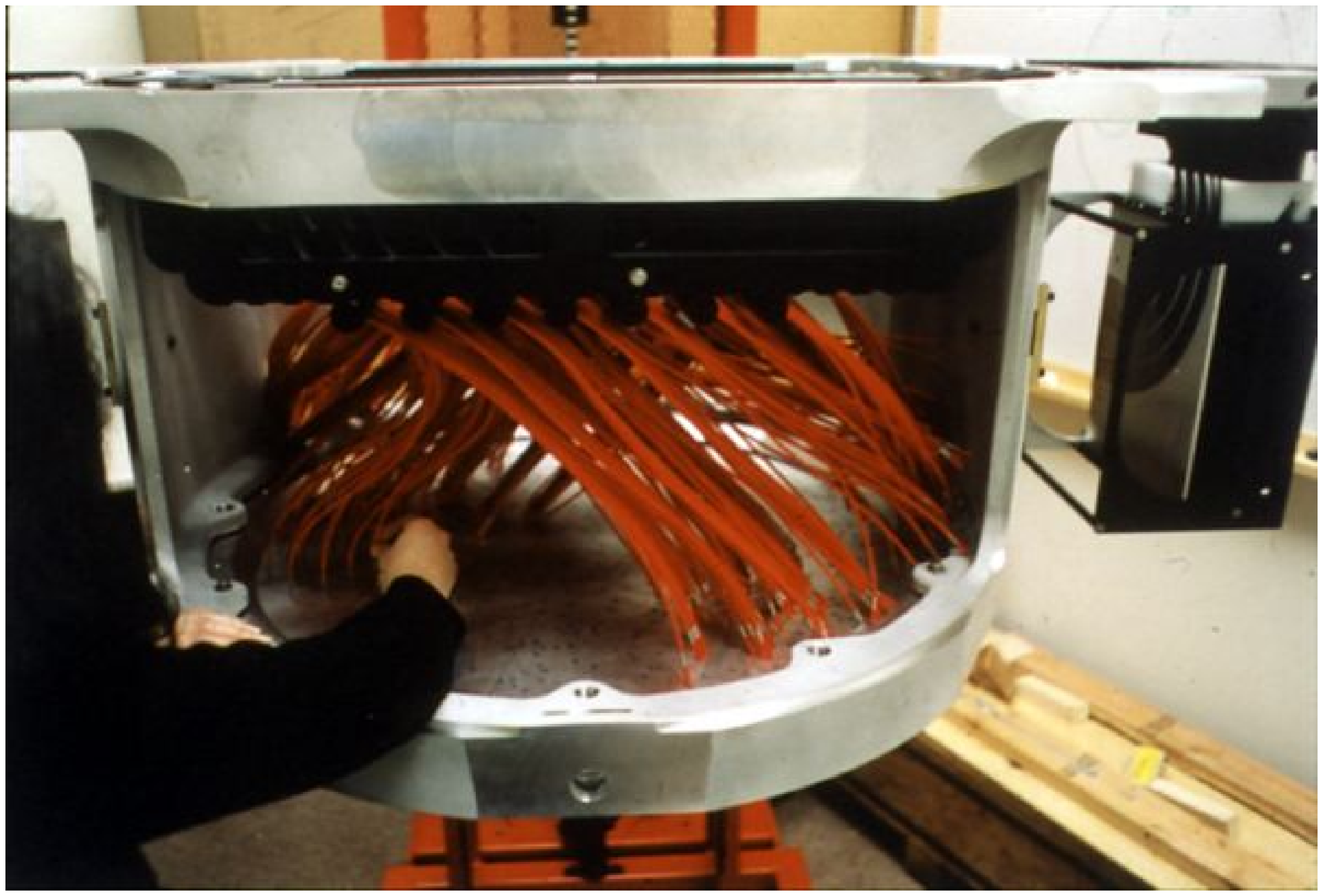}
\caption{Left: The SDSS 2.5-meter main telescope at Apache Point
 Observatory. Center: The SDSS camera assembly. There are 20 photometric
 and 24 astrometric/focus CCDs. Right: A plug plate that has been
 drilled, fitted with optical fibers, and mounted on the spectrograph
 assembly. These Figure are taken from a webpage at http://www.sdss.org/.
\label{fig:appsdss_tel}} 
\end{center}
\end{figure}

\begin{figure}[p]
\begin{center}
\includegraphics[width=0.5\hsize]{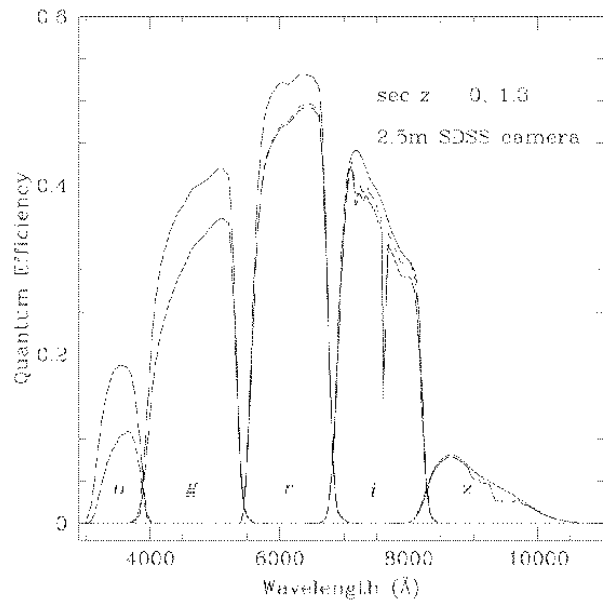}
\caption{The SDSS filter response curves for $u'$, $g'$, $r'$, $i'$, and
 $z'$, without ({\it upper curves}) and with ({\it lower curves}) the
 atmospheric extinction. This Figure is taken from \citet{stoughton02}.
\label{fig:appsdss_res}}
\end{center}
\end{figure}

The SDSS main telescope has a primary mirror with a diameter of
2.5-meter, and has a $3^\circ$ diameter field of view (Figure
\ref{fig:appsdss_tel} Left). The remarkably wide-angle view is to create
a map of the sky, and is realized by the improved Ritchey-Chr\'{e}tien
design. The imaging camera (Figure \ref{fig:appsdss_tel} Center) is set
on the focal plane; it consists of a $5\times 6$ (i.e., 6 chips for
each broad band) large CCD chips ($2048\times2048$ pixels) for the
photometric observations and 24 CCDs ($2048\times400$ pixels) for the
astrometry/focusing \citep{gunn98}.  The scale of the pixel is
$0\farcs4{\rm pisel^{-1}}$, about $1/4$ of the typical seeing size in
observations at APO. The SDSS photometric system uses five broad band,
$ugriz$ \citep{fukugita96}; the SDSS filter response curves are shown in
Figure \ref{fig:appsdss_res}. 
 
The photometric telescope has a primary mirror with a diameter of
0.5-meter, is outfitted with a CCD ($2048\times2048$ pixels), and has
a $1,600{\rm arcmin}^2$ field of view. The photometric telescope is aimed
to observe the secondary standard stars for the photometric calibration.
The secondary standard star network used here \citep{smith02} covers the
whole area of the region the SDSS observes.

The spectroscopic observations are conducted by the double spectrograph
with a dichroic splitter and $2048\times 2048$ CCD chips, by setting a plate fitted
with 320 fibers on the focal plane (Figure \ref{fig:appsdss_tel} Right).
In reality, two identical instruments fed by a single plug-plate are
used to obtain 640 spectra per exposure. The photons come from a fiber
are divided into blue (3800{\AA}--6100{\AA}) and red
(5900{\AA}--9200{\AA}) by dichronic splitter. They are sent to the
grisms; each grating is  640 line mm$^{-1}$ (blue) and 440 line
mm$^{-1}$ (red). The spectral resolution is $\lambda/\Delta
\lambda=1800$. Since each fiber has a diameter of 3-mm on the plate, it
is impossible to observed a pair of objects whose separation is $\leq
55''$ in a single plate. Thus the SDSS adopts the adaptive tiling method to
allocate fibers to spectroscopic targets, and achieves a sampling rate
of $\geq 92\%$ for all targets \citep{blanton03}. 
 
\begin{figure}[p]
\begin{center}
\includegraphics[width=0.8\hsize]{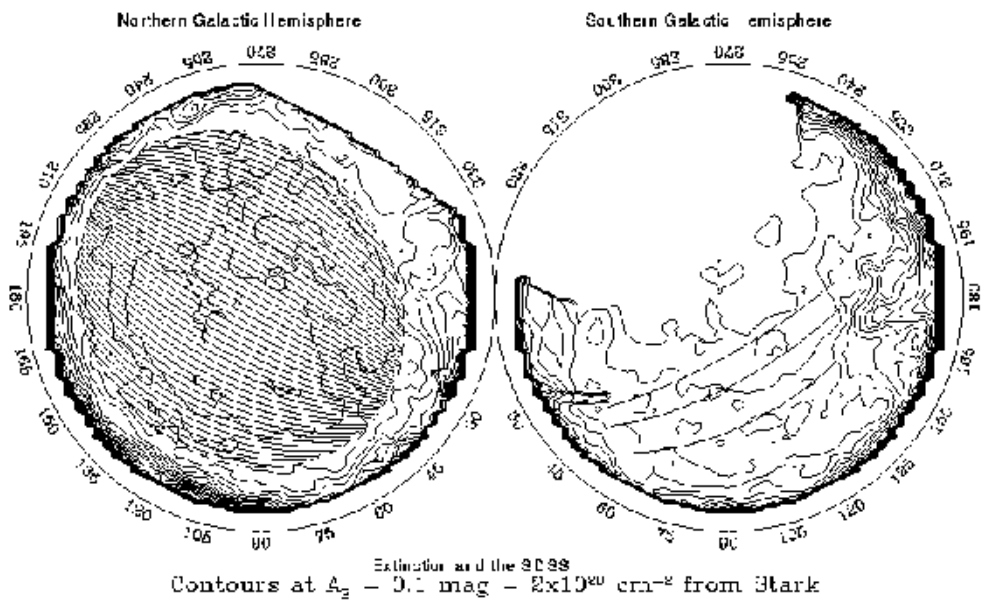}
\caption{The footprints of the Northern and Southern SDSS surveys. The
 tracks for the photometric survey are shown by heavy lines. The
 contours show the extinction measured from the HI column density. 
This Figure is take from a webpage at
 http://archive.stsci.edu/sdss/index.html. 
\label{fig:appsdss_ext}}
\end{center}
\end{figure}

\section{Observations and Data Reductions}
\markboth{CHAPTER \thechapter.
{\MakeUppercase\mychapheadname}}{\thesection.
\MakeUppercase{Observations and Data Reduction}}

Since the SDSS needs to observe the enormous sky area, it adopts the time
delay and integrated mode (or scanning mode); the telescope is fixed
during the observations, and the observing fields are changing as the
earth rotate. In this case, the size of a CCD chip ($13\farcm52$)
determines the exposure time, $55$ sec. The advantages of this mode
include (i) it results in almost 100\% observing efficiency because we
can save the readout time. (ii) it reduces the flat-fielding problem
from 2-dimension to 1-dimension. At the same time, the astrometric CCDs
and photometric telescope observe the standard stars for calibrations.
The astrometric CCDs allow us to determine positions of sources brighter
than $r=20.5$ with accuracy better than $0\farcs1$ \citep{pier03}.

Because of gaps between CCD cameras, one needs a slightly shifted drift
scan observations to fill the gaps. Thus a ``stripe'' consists of two runs.
Although the SDSS mainly observes the north galactic hemisphere, in
addition it observes a small region on the south galactic hemisphere.
The survey region is determined so that the region has minimum galactic
extinction and minimum atmospheric extinction. Figure
\ref{fig:appsdss_ext} shows the survey regions and the extinction.

Targets of spectroscopic observations are chosen according to target
selection algorithms (see below). As seen in Figure \ref{fig:appsdss_tel}
Right, 640 fibers are plugged for positions of targeted objects on each
plate . Spectra are usually obtained with $3\times 15$ minutes
exposures, but it may be changed according to the observing conditions.

The SDSS automated data processing, called pipeline, is used to reduce
the data, to select spectroscopic targets, and to make a object catalog.
There are three main pipelines; photometric pipeline, target selection
pipeline, and spectroscopic pipeline.

The photometric pipeline reduces the photometric data and make a object
catalog. It carries out bias subtraction, flat-fielding, and fixing bad
pixels/columns. The data of the photometric telescope are also reduced
automatically; the pipeline measures the extinction coefficients and
photometric parameters in real time \citep{hogg01}. Then it searches
objects in the data with templates of stars and galaxies, and calculates
several values such as magnitudes\footnote{The SDSS is based on the AB
system, but in reality it uses ``asinh'' magnitudes \citep*{lupton99}
which agrees with the standard definition for large fluxes.}, colors,
object-types, etc. The data are put into a catalog named ``tsObj''. 

After making the object catalog, the target selection pipelines select the
object for the spectroscopic observations. The target selection
algorithms have been developed for galaxies \citep{strauss02}, luminous
red galaxies \citep{eisenstein01}, quasars \citep{richards02}, etc.
These target selection algorithms are determined so that the resulting
spectroscopic samples become approximately homogeneous.

Finally, the spectroscopic pipeline reduces the data and obtains
1-dimensional spectra. It also identifies emission/absorption lines,
obtains accurate radial velocities, and classifies the spectra.

\chapter{Mass Modeling of SDSS J1004+4112}
\label{chap:1004}
\def\mychapheadname{Mass Modeling of SDSS J1004+4112}
\markboth{CHAPTER \thechapter.
{\MakeUppercase\mychapheadname}}{}

\section{One-component Models\label{sec:onemodel}}
\markboth{CHAPTER \thechapter.
{\MakeUppercase\mychapheadname}}{\thesection.
\MakeUppercase{One-component Models}}

To search for mass models that can explain the image configuration of
the large-separation lensed quasar SDSS~J1004+4112 (Chapter
\ref{chap:sdss}), we use standard lens modeling techniques as
implemented in the software of \citet{keeton01d}.  The main constraints
come from the image positions.  We also use the flux ratios as constraints,
although we broaden the errorbars to 20\% to account for possible
systematic effects due to source variability and time delays, micro- or
milli-lensing, or differential extinction (See Table~\ref{table:1004_posflux}
for the full set of constraint data). In particular, the different
colors of the images and the different absorption features seen in
Figure~\ref{fig:sdss_spec} suggest that differential extinction may be a
significant effect.  Here we do {\it not\/} use the position of the main
galaxy as a constraint, because we want to understand what constraints
can be placed on the center of the lens potential from the lens data alone. 

We first consider the simplest possible models for a 4-image lens: an
isothermal lens galaxy with a quadrapole produced either by ellipticity
in the galaxy or by an external shear.  A spherical isothermal lens
galaxy has surface mass density 
\begin{equation}
  \kappa(r) = \frac{\Sigma(r)}{\Sigma_{\rm crit}}
  = \frac{r_{\rm ein}}{2r}\ ,
\end{equation}
where $r_{\rm ein}$ is the Einstein radius of the lens, and
$\Sigma_{\rm crit}$ is the critical surface mass density (eq.
[\ref{lens_critsurf}]). The Einstein radius is related to 
the velocity dispersion $\sigma$ of the galaxy by 
\begin{equation}
  r_{\rm ein} = 4\pi\,\left(\frac{\sigma}{c}\right)^2\,
    \frac{D_{\rm LS}}{D_{\rm OS}}\ .
\end{equation}
For an elliptical model we replace $r$ with $r[1+((1-q^2)/(1+q^2))\cos
2(\theta-\theta_e)]^{1/2}$ in the surface density, where $q$ and
$\theta_e$ are the axis ratio and position angle of the ellipse.

Simple models using either pure ellipticity or pure shear fail
miserably, yielding $\chi^2$ values no better than $2\times10^{4}$ for
$N_{\rm dof} = 4$ degrees of freedom.  This failure is not surprising:
most 4-image lenses require {\it both\/} ellipticity and external shear
\citep*[e.g.,][]{keeton97}, and such a situation is likely in
SDSS~J1004+4112 since the main galaxy is observed to be elongated and
the surrounding cluster surely contributes a shear. 

\begin{table}[tb]
 \begin{center}
  \begin{tabular}{crrcc}\hline\hline
Object & $x$[arcsec]$^{\rm a}$ & $y$[arcsec]$^{\rm a}$ &
Flux[arbitrary]$^{\rm b}$ & PA[deg]$^{\rm c}$ \\
\hline
A  & $ 0.000\pm0.012$ & $ 0.000\pm0.012$ & $1.0  \pm0.2  $ & $\cdots$ \\
B  & $-1.301\pm0.011$ & $ 3.500\pm0.011$ & $0.682\pm0.136$ & $\cdots$ \\
C  & $10.961\pm0.012$ & $-4.466\pm0.012$ & $0.416\pm0.083$ & $\cdots$ \\
D  & $ 8.329\pm0.007$ & $ 9.668\pm0.007$ & $0.195\pm0.039$ & $\cdots$ \\
G1 & $ 7.047\pm0.053$ & $ 4.374\pm0.053$ & $\cdots$        & $-19.9$ \\
\hline
\end{tabular}
\caption{Summary of positions, flux ratios, and position angles
 (PA) of SDSS~J1004+4112 used in the mass modeling.\hspace{70mm}\protect\\
\footnotesize{
\hspace*{5mm}${}^{\rm a}$ The positive directions of $x$ and $y$ are defined by
 West and North, respectively.\hspace{70mm}\protect\\
\hspace*{5mm}${}^{\rm b}$ Errors are broadened to 20\% to account for possible
 systematic effects.\hspace{70mm}\protect\\
\hspace*{5mm}${}^{\rm c}$ Degrees measured East of North.
}} 
\label{table:1004_posflux}
 \end{center}
\end{table}

\begin{figure}[tb]
\begin{center}
\includegraphics[width=0.7\hsize]{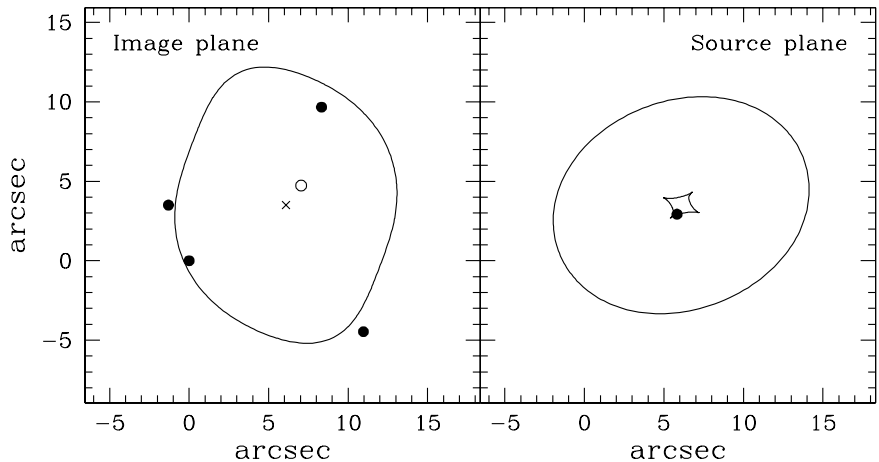}
\caption{Critical curve ({\it left}) and caustic ({\it right}) for the
 best-fit SIE+shear lens model of SDSS~J1004+4112.  In the left panel,
 the filled circles mark the image positions, the open circle indicates
 the observed position of the brightest cluster galaxy G1, and the cross
 marks the best-fit deflector position.  In the right panel the filled
 circle marks the inferred source position.
 \label{fig:1004_sieg-crit}}
\end{center}
\end{figure}

\begin{figure}[tb]
\begin{center}
\includegraphics[width=0.7\hsize]{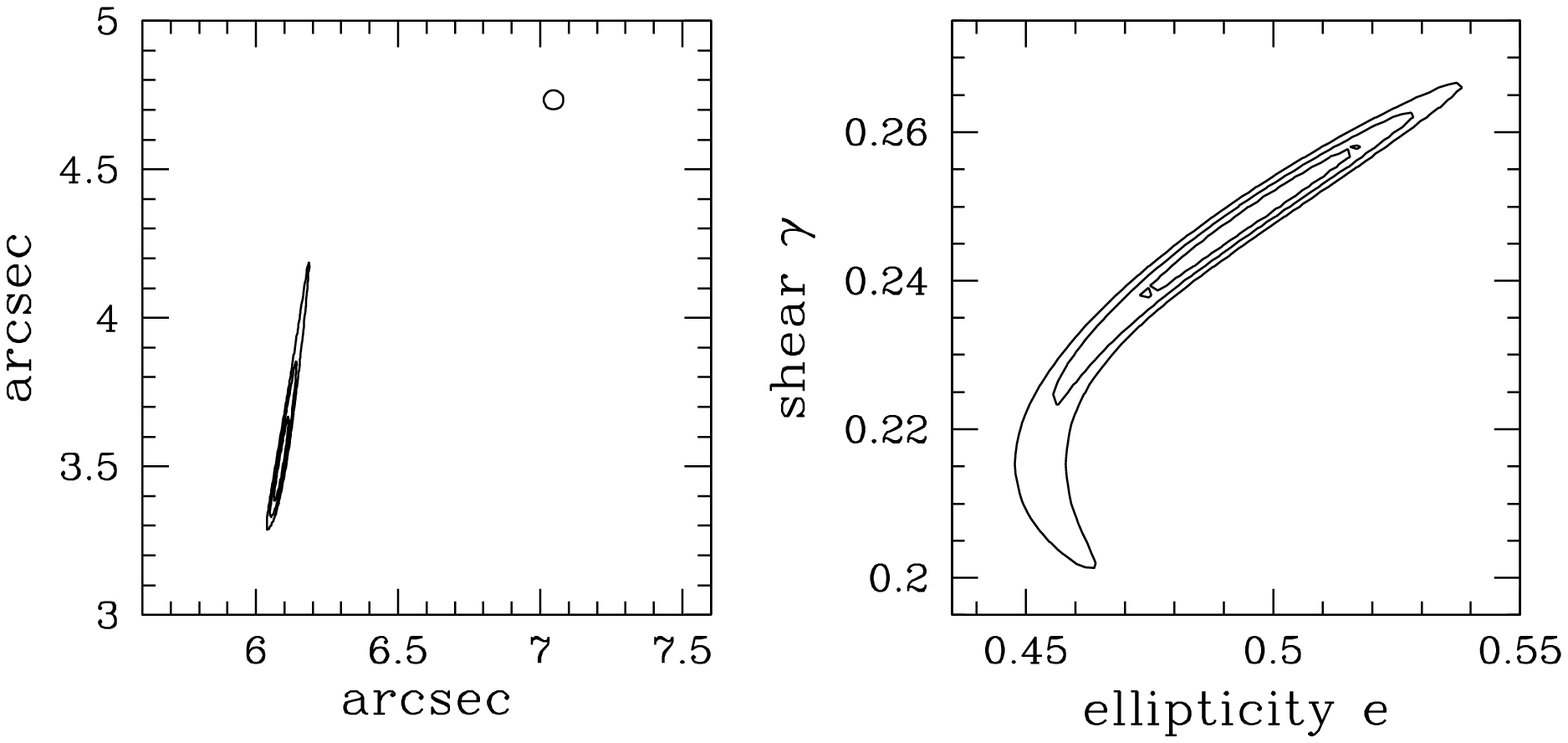}
\caption{Likelihood contours drawn at 1$\sigma$, 2$\sigma$, and
 3$\sigma$ for various parameter combinations in SIE+shear lens models.
 The left panel shows constraints on the position of the deflector;
 the circle marks the observed position of the main galaxy. The right
 panel shows contours in the ellipticity--external shear plane.
 \label{fig:1004_sieg-params}}
\end{center}
\end{figure}

We therefore try models consisting of a singular isothermal ellipsoid
(SIE) plus an external shear $\gamma$.  Even though such models are
still comparatively simple, they can fit the data very well with a
best-fit value of $\chi^2 = 0.33$ for $N_{\rm dof} = 2$.  The best-fit
model has an Einstein radius $r_{\rm ein} = 6\farcs9 = 35\,h^{-1}$~kpc
corresponding to a velocity dispersion of 700~km~s$^{-1}$, an
ellipticity $e=0.50$ at position angle $\theta_e=21\fdg4$, and an
external shear $\gamma=0.25$ at position angle $\theta_\gamma=-60\fdg9$.
Among other known lenses, such a large shear is found only in lenses
lying in cluster environments \citep{burud98,barkana99}.
Figure~\ref{fig:1004_sieg-crit} shows the critical curves and caustics for
the best-fit model.  The inferred source position lies very close to the
caustic and fairly near a cusp, implying that the total magnification is
$\sim$57. Figure~\ref{fig:1004_sieg-params} shows the allowed ranges for the
position of the deflector and the ellipticity and external shear in the
model. The models indicate a small but significant offset of
$1\farcs6 = 7.9\,h^{-1}$~kpc between the center of the lens potential
and the main galaxy, although it remains to be seen whether that
offset is real or an artifact of these still simple lens models.

\section{Two-component Models}
\markboth{CHAPTER \thechapter.
{\MakeUppercase\mychapheadname}}{\thesection.
\MakeUppercase{Two-component Models}}

Even though the simple SIE+shear model provides a good fit to the
data, we believe that it is not physically plausible because the
system clearly has multiple mass components and it seems unlikely
that all of the mass is associated with a single $\sim$700~km~s$^{-1}$
isothermal component.  The next level of complication is to add a
mass component representing the cluster halo.  We still model the
galaxy G1 explicitly, treating it as an isothermal ellipsoid
constrained by its observed position.  At this point we do not
further complicate the model by attempting to treat the other
galaxies within the lens explicitly.

\subsection{Methods}

We model the cluster component with an NFW profile which has been
predicted in cosmological $N$-body simulations:
\begin{equation}
 \rho(r)=\frac{\rho_{\rm crit}(z)\delta_{\rm c}(z)}
{\left(r/r_{\rm s}\right)\left(1+r/r_{\rm s}\right)^2},
\label{nfw}
\end{equation}
where $r_{\rm s}$ is a scale radius, $\delta_{\rm c}$ is a characteristic
overdensity (which depends on redshift), and $\rho_{\rm crit}(z)$ is
the critical density of the universe.  Although the NFW density profile
appears to deviate from the results of more recent $N$-body simulations
in the innermost region, we adopt this form for simplicity.  The lensing
properties of a spherical NFW halo are described by the lens potential
\citep{bartelmann96,golse02,meneghetti03a}
\begin{equation}
\psi(r) = 2\kappa_{\rm s}\,r_{\rm s}^2 \left[
\ln^2\frac{r}{2r_{\rm s}}-\mbox{arctanh}^2\sqrt{1-(r/r_{\rm s})^2} \right],
\end{equation}
where the lensing strength is specified by the parameter $\kappa_{\rm
s}=b_{\rm NFW}/4$ (see eq. [\ref{lat_bnfw}]).
Since asphericity in the cluster potential is important in modeling
this system, we generalize the spherical model by adopting elliptical
symmetry in the potential.  Making the potential (rather than the
density) elliptical makes it possible to compute the lensing properties
of an NFW halo analytically \citep{golse02,meneghetti03a}.  We may
still be over-simplifying the mass model, because the cluster profile
may have been modified from the NFW form by baryonic processes such
as gas cooling \citep{rees77,blumenthal86}, and the cluster may have
a complex angular structure if it is not relaxed
\citep[e.g.,][]{meneghetti03a}. To allow for the latter possibility, we
still include a tidal shear in the lens model that can approximate the
effects of complex structure in the outer parts of the cluster. Overall,
our goal is not to model all of the complexities of the lens potential,
but to make the minimal realistic model and see what we can learn.

Even with our simplifying assumptions, we still have a complex
parameter space with 11 parameters defining the lens potential:
the mass, ellipticity, and position angle for the galaxy G1;
the position, mass, scale radius, ellipticity, and position angle
for the cluster; and the amplitude and position angle of the shear.
There are also three parameters for the source (position and flux).
With just 12 constraints (position and flux for each of four images),
the models are under-constrained.  We therefore expect that there
may be a range of lens models that can fit the data.  To search
the parameter space and identify the range of models, we follow
the technique introduced by \citet{keeton03b} for many-parameter
lens modeling.  Specifically, we pick random starting points in
the parameter space and then run an optimization routine to find
a (local) minimum in the $\chi^2$ surface.  Repeating that process
numerous times should reveal different minima and thereby sample
the full range of models.  Many of the recovered models actually
lie in local minima that do not represent good fits to the data,
so we only keep recovered models with $\chi^2 < 11.8$ \citep[which
represents the 3$\sigma$ limit relative to a perfect fit when
examining two-dimensional slices of the allowed parameter
range; see][]{press92}.

We make one further cut on the models.  From the previous section,
we know that an SIE+shear lens model can give a good fit to the
data.  Thus, there are acceptable two-component models where most
or all of the mass is in the galaxy component and the cluster
contribution is negligible.  To exclude such models as physically
implausible, we impose an upper limit on the velocity dispersion
of the model galaxy.  Specifically, we only keep models with
$\sigma_{\rm gal} < 400$~km~s$^{-1}$, because there are essentially
no galaxies in the observed universe with larger velocity
dispersions, even in rich clusters
\citep[e.g.,][]{kelson02,bernardi03,sheth03}.  Formally, we impose
this cut as an upper limit $r_{\rm ein} < 2\farcs25$ on the
Einstein radius of the galaxy G1.

\subsection{Results\label{sec:modelresult}}

\begin{figure}[tb]
\begin{center}
\includegraphics[width=0.7\hsize]{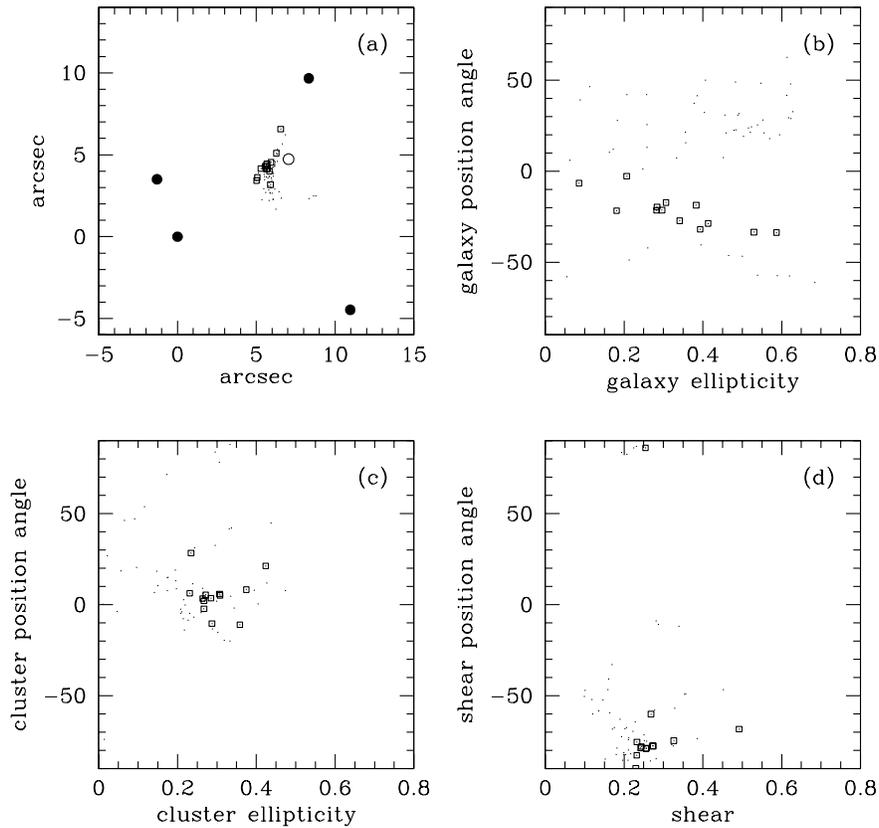}
\caption{Allowed parameter ranges for galaxy+cluster lens models with
a cluster scale radius $r_{\rm s}=40''$. (a) The position of the
cluster component.  The filled circles mark the image positions,
and the open circle marks the observed brightest cluster galaxy G1.
(b) The ellipticity and position angle of the galaxy component.
(c) The ellipticity and position angle of the cluster component.
(d) The amplitude and position angle of the external shear.
Small points show all models, while boxes mark models where the
model galaxy is roughly aligned with the observed galaxy
($\theta_e = -19\fdg9\pm20\fdg0$).
 \label{fig:1004_clus-params}}
\end{center}
\end{figure}

We first consider models where the scale radius of the cluster is
fixed as $r_{\rm s} = 40''$ (we shall justify this choice below).
Figure~\ref{fig:1004_clus-params} shows the allowed parameter ranges for
acceptable models.  First, panel (a) shows that the cluster
component is restricted to a fairly small (but not excessively
narrow) range of positions near the center of the image
configuration.  This is mainly a result of our upper limit on
the mass of the galaxy component; there is a certain enclosed
mass implied by the image separation, and if the galaxy cannot
contain all of that mass then the cluster component must lie
within the image configuration to make up the difference.  It is
interesting to note that even in these more complicated models
there still seems to be a small offset between the center of the
cluster component and the brightest cluster galaxy G1, although the
lower limit implied by our models is just $0\farcs71 = 3.6h^{-1}$kpc.

Figure~\ref{fig:1004_clus-params}b shows that the allowed values for the
ellipticity and position angle of the galaxy G1 basically fill
the parameter space, so these parameters are not constrained by
the lens data.  We might want to impose an external constraint,
however.  Analyses of other lens systems show that the lensing mass is
typically aligned with the projected light distribution
\citep*{keeton98b,kochanek02b}.  We may therefore prefer lens models
where the model galaxy is at least roughly aligned with the observed
galaxy, which has a position angle of $-19\fdg9$. To illustrate this
possible selection, we show all models but highlight those where the
position angle of the model galaxy is in the range $\theta_e =
-19\fdg9\pm20\fdg0$.  The broad $20\arcdeg$ uncertainties prevent this
constraint from being too strong.

Figure~\ref{fig:1004_clus-params}c shows that there are some acceptable
models where the cluster potential is round, but most models have some
ellipticity that is aligned roughly North--South. This is
in good agreement with the distribution of member galaxies which is also
aligned roughly North--South (see Figure~\ref{fig:sdss_colorcut}). The
ellipticity $e \sim 0.2$--0.4 is actually quite large, considering that
this parameter describes the ellipticity of the potential, not that of
the density.  Figure~\ref{fig:1004_clus-params}d shows that all of the
acceptable models require a fairly large tidal shear $\gamma \gtrsim
0.10$, and models where the galaxy is aligned with the observed galaxy
have a strong shear $\gamma \gtrsim 0.23$.  The shear tends to be
aligned East--West.  The fact that the models want both a large
cluster ellipticity and a large tidal shear strongly suggest that
there is complex structure in the cluster potential outside of
the image configuration.  It would be interesting to see whether
there is any evidence for such structure in, for example, X-rays
from the cluster.

\begin{figure}[tb]
\begin{center}
\includegraphics[width=0.75\hsize]{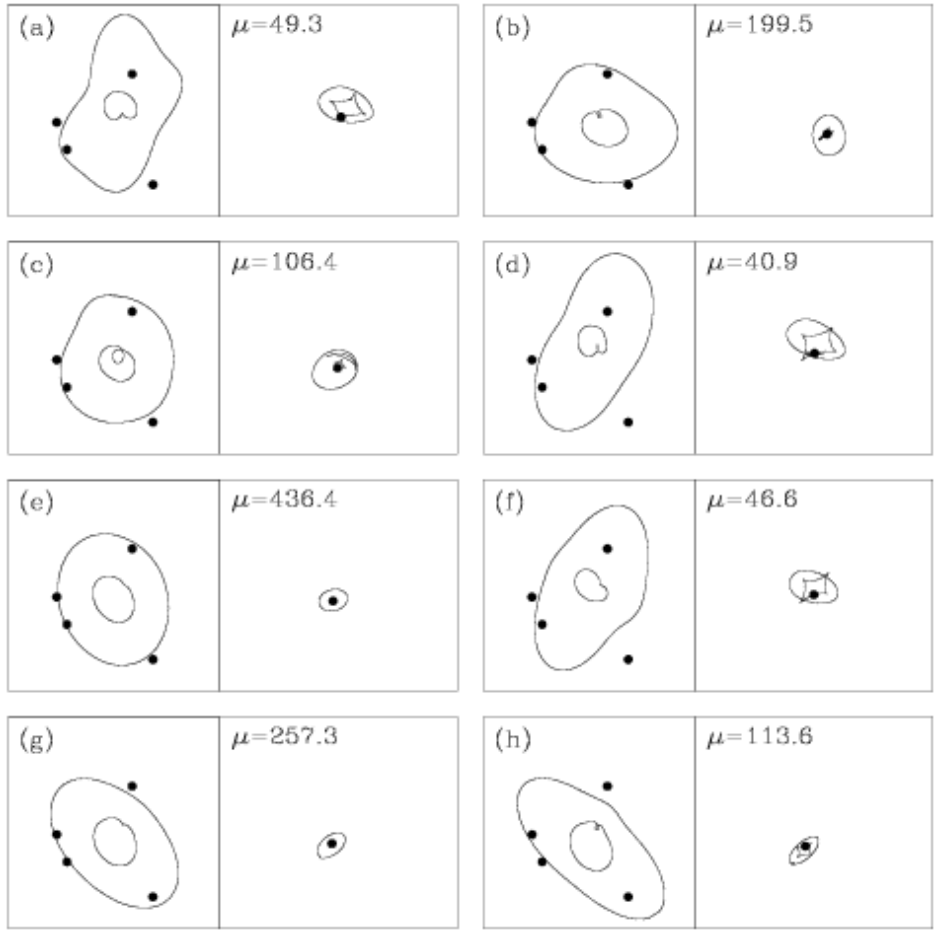}
\caption{Critical curves and caustics for sample galaxy+cluster lens
 models with a cluster scale radius $r_{\rm s}=40\arcsec$.  In each
 panel, the left-hand side shows the critical curves in the image plane,
 and the right-hand side shows the caustics in the source plane on the
 same scale.  The points in the image plane show the observed image
 positions, and the point in the source plane shows the inferred
 source position.  The value of $\mu$ gives the total magnification
 in each model.
 \label{fig:1004_clus-crit}}
\end{center}
\end{figure}

Figure~\ref{fig:1004_clus-crit} shows critical curves and caustics for
sample lens models.  The critical curves are not well determined.  The
outer, tangential critical curve can point either northeast (panel e) or
northwest (panel d), or it can have a 
complex shape (panel a).  Sometimes there is just one inner, radial
critical curve (panel e), but often there are two (panel c).  The
distance of the source from the caustic (and of the images from the
critical curve) varies from model to model, so the total magnification
can range from $\sim$50 to several hundred or even more.  Finally,
perhaps the most interesting qualitative result is that even the
image parities are not uniquely determined.  In most models (e.g.,
panels a--f) images A and D lie inside the critical curve and have
negative parity while B and C lie outside the critical curve and
have positive parity.  However, in some models (e.g., panels g--h)
the situation is reversed.  Having ambiguous image parities is very
rare in lens modeling.

\begin{figure}[tb]
\begin{center}
\includegraphics[width=0.45\hsize]{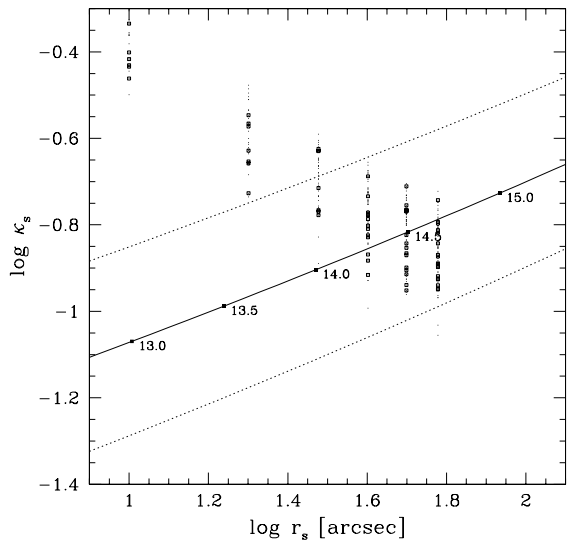}
\caption{Relation between the cluster scale radius $r_{\rm s}$ and
 lensing strength $\kappa_{\rm s}$.  The solid line shows the predicted
 relation for clusters with the canonical median concentration, and the
 dotted lines show the 1$\sigma$ range due to the scatter in
 concentration. The labeled points show the
 value of $\log(M)$ (in units of $h^{-1}M_\odot$) at various points
 along the curve.  The points show fitted values of $\kappa_{\rm s}$ for
 lens models with $r_{\rm s}=(10,20,30,40,50,60)$ arcsec.  As in Figure
 \ref{fig:1004_clus-params}, small points show all models, while boxes mark
 models where the model galaxy is roughly aligned with the observed
 galaxy.
 \label{fig:1004_NFWnorm}}
\end{center}
\end{figure}

So far we have only discussed models where the cluster has a
scale radius $r_{\rm s}=40''$.  We have also computed models with
$r_{\rm s} = (10, 20, 30, 50, 60)$ arcsec and we find that all of the
results are quite similar.  To understand what value of the scale
radius is reasonable, we must consider which (if any) of the models
have physically plausible cluster parameters.  Even though NFW models
are formally specified by two parameters $r_{\rm s}$ and $\kappa_{\rm
s}$, $N$-body simulations reveal that the two parameters are actually
correlated. NFW models therefore appear to form a one-parameter family
of models, although with some scatter which reflects the scatter of the
concentration parameter $c_{\rm vir}=r_{\rm vir}/r_{\rm s}$ ($r_{\rm
vir}$ is a virial radius of the cluster).
Figure~\ref{fig:1004_NFWnorm} shows the predicted relation between
$r_{\rm s}$ and $\kappa_{\rm s}$, including the scatter.  For
comparison, it also shows the fitted values of $\kappa_{\rm s}$ in lens
models with different scale radii.  Models with $r_{\rm s}=10''$ or
$20''$ require $\kappa_{\rm s}$ much larger than expected,
corresponding to a halo that is too concentrated. Models with $r_{\rm s}
\ge 30''$, by contrast, overlap with the predictions and thus are
physically plausible.  We can therefore conclude very roughly that the
cluster component must have $r_{\rm s} \gtrsim 30''$ and a total
virial mass $M \gtrsim 10^{14}\,h^{-1}\,M_\odot$.

Finally, we can use the models to predict the time delays between
the images.  The models always predict that the time delay between
images C and D is the longest and the delay between A and B is the
shortest.  However, there is no robust prediction of the temporal
ordering: most models predict that the sequence should be C--B--A--D,
but a few models predict the reverse ordering D--A--B--C.  This is
a direct result of the ambiguity in the image parities, because the
leading image is always a positive-parity image
\citep[e.g.,][]{schneider92}. We note, however, that all of the models
where the model galaxy is roughly aligned with the observed galaxy have
the C--B--A--D ordering.

\begin{figure}[tb]
\begin{center}
\includegraphics[width=0.45\hsize]{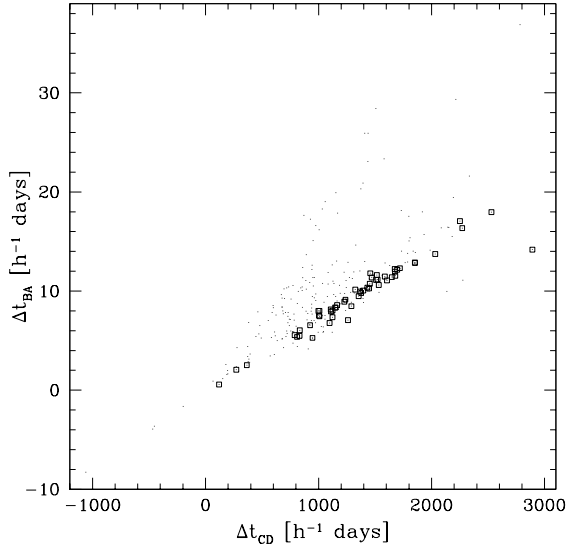}
\caption{Predictions for the longest ($\Delta t_{CD}$) and shortest
($\Delta t_{BA}$) time delays, where $\Delta t_{ij}>0$ means image
$i$ leads image $j$, and vice versa.  Results are shown for models
where the cluster has scale length $r_{\rm s} = (30,40,50,60)$ arcsec.
As in Figure~\ref{fig:1004_clus-params}, small points show all models,
while boxes mark models where the model galaxy is roughly aligned
with the observed galaxy.
 \label{fig:1004_clus-tdel}}
\end{center}
\end{figure}

Figure~\ref{fig:1004_clus-tdel} shows the predictions for the long and short
time delays.  The long delay between C and D can be anything up to
$\sim\!3000\,h^{-1}$~days, while the short delay between A and B can
be up to $\sim\!37\,h^{-1}$~days.  For the models where the galaxy
is roughly aligned with the observed galaxy, the two delays are
approximately proportional to each other with
$\Delta t_{CD}/\Delta t_{BA} = 143\pm16$.  These results have
several important implications.  First, the A--B time delay should
be on the order of weeks or months, so it should be very feasible
to measure it, provided that the source has detectable variations.
Measuring the A--B delay would be very useful because it would
determine the temporal ordering, and thereby robustly determine the
image parities.  In addition, it would allow a good estimate of the
long C--D delay and indicate whether attempting to measure that delay
would be worthwhile.  Second, the enormous range of predicted time
delays means that constraining the Hubble constant with this system
\citep{refsdal64a,refsdal64b} will be difficult because of large systematic
uncertainties in the lens models.  Although \citet{koopmans03} recently
showed that it is possible to obtain useful constraints on the Hubble
constant even in a complex system with two mass components, the analysis
is very complex and requires extensive data including not just the image
positions and all of the time delays, but also an Einstein ring image
and the velocity dispersion of one of the mass components. Even if we
obtain such data for SDSS~J1004+4112 in the near future, it seems likely
that it will be difficult to obtain reliable constraints on the Hubble
constant given the complexity of the lens potential in SDSS~J1004+4112.
The time delays, however, would still be extremely useful, because they
would determine the temporal ordering and hence the image parities, and
they would provide constraints that can rule out many of the models
that are currently acceptable.

\section{Summary}
\markboth{CHAPTER \thechapter.
{\MakeUppercase\mychapheadname}}{\thesection.
\MakeUppercase{Summary}}

We have shown that reasonable mass models can successfully reproduce
the observed properties of the lens.  When we consider models that
include both the cluster potential and the brightest cluster galaxy,
we find a broad range of acceptable models.  Despite the diversity
in the models, we find several general and interesting conclusions.
First, there appears to be a small ($\gtrsim\!4\,h^{-1}$~kpc) offset
between the brightest cluster galaxy and the center of the cluster
potential.  Such an offset is fairly common in clusters
\citep[e.g.,][]{postman95}.  Second, the cluster potential is
inferred to be elongated roughly North--South, which is consistent
with the observed distribution of apparent member galaxies.  Third,
we found that a significant external shear $\gamma\sim 0.2$ is needed
to fit the data, even when we allow the cluster potential to be
elliptical.  This may imply that the structure of the cluster potential
outside of the images is more complicated than simple elliptical
symmetry.  Fourth, given the broad range of acceptable models, we
cannot determine even the parities and temporal ordering of the images,
much less the amplitudes of the time delays between the images.
Measurements of any of the time delays would therefore provide powerful
new constraints on the models.  We note that the complexity of the
lens potential means that the time delays will be more useful for
constraining the mass model than for trying to measure the Hubble
constant.

Our modeling results suggest that further progress will require new
data (rather than refinements of current data).  The interesting
possibilities include catalogs of confirmed cluster members, X-ray
observations, and weak lensing maps, not to mention measurement of
time delays and confirmation of lensed arcs (either the possible
arclets we have identified, or others).  For instance, with an
estimated cluster mass of $M \sim 3\times 10^{14}\,h^{-1}\,M_\odot$,
the estimated X-ray bolometric flux is
$S_X \sim 10^{-13}\,{\rm erg\,s^{-1}cm^{-2}}$, which means that the
cluster should be accessible with the {\it Chandra} and
{\it XMM-Newton} X-ray observatories; the excellent spatial of
{\it Chandra} may be particularly useful for separating the diffuse
cluster component from the bright quasar images (which have a total
X-ray flux $S_X \sim 2\times 10^{-12}{\rm erg\,s^{-1}cm^{-2}}$ in
the {\it ROSAT} All Sky Survey).  The confirmation of lensed arclets
would be very valuable, as they would provide many more pixels'
worth of constraints on the complicated lens potential.  In
principle, mapping radio jets in the quasar images could unambiguously
reveal the image parities \citep[e.g.,][]{gorenstein88,garrett94},
but unfortunately the quasar appears to be radio quiet as it is not
detected in radio sky surveys such as the FIRST survey \citep*{becker95}. 


\end{document}